%% file: CorrBubbles.tex
\documentclass[letterpaper,titlepage,11pt]{article}
\usepackage[T1]{fontenc}
\usepackage[utf8]{inputenc}
\usepackage{lmodern}
\usepackage{enumerate}
\usepackage[english]{babel}
\usepackage{comment}
\usepackage[numbers,sort&compress]{natbib}
\usepackage{bm}
\usepackage[usenames,dvipsnames]{xcolor}
\usepackage[a4paper]{geometry}
\usepackage{float}
\usepackage{titlesec}

\titleclass{\part}{top} 
\titleformat{\part}
[display]
{\raggedright\normalfont\large\bfseries}
{\vspace{3pt}\MakeUppercase{\partname} \thepart}
{0pt}
{\vspace{1pc}\huge\MakeUppercase}
\titlespacing*{\part}{0pt}{0pt}{20pt}

\usepackage[colorlinks=true,urlcolor=blue,citecolor=magenta]{hyperref}
\usepackage{amsmath}
\usepackage{amsfonts}
\usepackage{amssymb}
\usepackage{graphicx}
\usepackage[dvipsnames]{xcolor}
\usepackage{mathtools}
\usepackage{simplewick}
\usepackage{hyperref}
\usepackage{multirow}
\usepackage{enumitem}
\usepackage{physics}
\usepackage[compat=1.1.0]{tikz-feynman}
\usepackage{subfigure}
\usepackage{booktabs}
\usepackage{xcolor}
\usepackage{tikz}
\usetikzlibrary{decorations.pathmorphing}

\allowdisplaybreaks[1]

\definecolor{ccqqqq}{rgb}{1,0.5,0}
\definecolor{uuuuuu}{rgb}{0.26666666666666666,0.26666666666666666,0.26666666666666666}
\definecolor{qqwwzz}{rgb}{0,0.3,0.9}

\usepackage{makecell}

\newcommand{\nl}{\nonumber \\} 
\def\p{\partial}

\newcommand{\beq}{\begin{equation}}
\newcommand{\eeq}{\end{equation}}
\newcommand{\bea}{\begin{eqnarray}}
\newcommand{\eea}{\end{eqnarray}}

\def\le{\left(}
\def\ri{\right)}

\usepackage{adjustbox}
\hypersetup{   colorlinks=true,  citecolor=blue, linkcolor=blue, urlcolor=red }

\usepackage{xcolor,pifont}
\newcommand*\colourcheck[1]{%
  \expandafter\newcommand\csname #1check\endcsname{\textcolor{#1}{\ding{52}}}%
}
\colourcheck{blue}
\colourcheck{green}
\colourcheck{red}

\newcommand*\colourmark[1]{%
  \expandafter\newcommand\csname #1mark\endcsname{\textcolor{#1}{\ding{56}}}%
}
\colourmark{blue}
\colourmark{green}
\colourmark{red}

\usepackage{tabularx}
\usepackage[skip=1ex]{caption}
\usepackage{textcomp}

\usepackage{changepage} 

\numberwithin{equation}{section}

\def\b{\beta}
\def\l{\lambda}
\def\s{\sigma}

\newcommand{\eg}{{e.g.,}\ }
\newcommand{\ie}{{i.e.,}\ }
\newcommand\blfootnote[1]{%
  \begingroup
  \renewcommand\thefootnote{}\footnote{#1}%
  \addtocounter{footnote}{-1}%
  \endgroup
}
\usepackage[normalem]{ulem}

\newcommand{\rep}[1]{ {#1}}

\begin{document}

\begin{titlepage}
\rightline{\vbox{   \phantom{ghost} }}

\begin{flushright}
YITP-25-150
\end{flushright}

\vskip 1 cm
\begin{center}
{\LARGE \bf
Probing the bubble interior
with entanglement \\ 
 entropy and bulk-cone singularities  
}
\end{center}
\vskip 0.5 cm

\title{}
\date{\today}
\author{Roberto Auzzi}
\author{Stefano Baiguera}
\author{Lihan Guo}
\author{Giuseppe Nardelli}
\author{Nicolò Zenoni}

\centerline{\large {{\bf Roberto Auzzi$^{1,2}$, Stefano Baiguera$^{1,2}$, Lihan Guo$^{1,2,3}$, }}}
\centerline{\large {{\bf Giuseppe Nardelli$^{1,4}$, Nicolò Zenoni$^{5}$}}}

\vskip 0.5cm

\begin{center}
\sl ${}^1$ Dipartimento di Matematica e Fisica, Università Cattolica del Sacro Cuore, \\
Via della Garzetta 48, 25133 Brescia, Italy
\\[1mm]
\sl ${}^2$ INFN Sezione di Perugia, Via A. Pascoli, 06123 Perugia, Italy \\[1mm]
\sl ${}^3$ Institute for Theoretical Physics, KU Leuven, \\
Celestijnenlaan 200D, B-3001 Leuven, Belgium \\[1mm]
\sl ${}^4$ TIFPA - INFN, c/o Dipartimento di Fisica, Università di Trento, \\ 38123 Povo (TN), Italy  \\[1mm]
\sl ${}^5$ Yukawa Institute for Theoretical Physics, Kyoto University, \\ Kyoto 606-8502, Japan
\end{center}

\vskip 1.3cm \centerline{\bf Abstract} \vskip 0.2cm 
\noindent

In the thin wall approximation,
we study a class of asymptotically AdS black holes 
which contain a spherically symmetric vacuum  bubble with a different (positive or negative) cosmological constant.  
Collapsing, expanding, and static bubble solutions are considered.
Among these, expanding bubbles with positive cosmological constant
can provide a way to apply the AdS/CFT correspondence to describe the physics of an expanding universe. We systematically study the causal structure of the solutions 
as a function of 
the cosmological constant, the mass of the black hole, and the tension of the bubble.
We then compute the holographic entanglement entropy for a class of boundary subregions using extremal codimension-two surfaces as a probe. For collapsing bubbles, we find examples in which the entanglement entropy also explores the geometry inside the black hole bifurcation surface.
As a complementary way to probe the interior of the bubble,
we investigate 
almost-null radial geodesics related to the bulk-cone singularities of boundary two-point correlators.
While the bulk-cone singularities for collapsing and expanding bubbles are consistent with thermalization at late time, static bubbles violate thermalization and exhibit properties similar to those of scar states.

\blfootnote{ \scriptsize{ \texttt{roberto.auzzi@unicatt.it,stefano.baiguera@pg.infn.it,lihan.guo@unicatt.it, \newline
giuseppe.nardelli@unicatt.it,nicolo@yukawa.kyoto-u.ac.jp}} }

\end{titlepage}

\tableofcontents

\noindent
\hrulefill

\input{Sections/Introduction}

\input{Sections/Setup}

\input{Sections/Entropy}

\input{Sections/Singularities_corr}

\input{Sections/Conclusions}

\section*{Acknowledgements}
NZ acknowledges Tarek Anous and Tadashi Takayanagi for valuable comments during the presentation of some preliminary results in the workshop "Quantum Gravity and Information in Expanding Universe" (YITP-W-24-19) at Yukawa Institute for Theoretical Physics, Kyoto University.
NZ also thanks APCTP, Pohang, for its hospitality during the External Program [APCTP-2025-E09] held at Duy Tan University, Da Nang, where the results were presented.
RA,  LG and GN are supported in part by
INFN through the \textit{Gauge and String Theory} (GAST) research project.
The work of SB is supported by the INFN grant \textit{Gauge Theories and Strings} (GAST) via a research grant on \textit{Holographic dualities, quantum information and gravity}.
The work of NZ is supported by MEXT KAKENHI Grant-in-Aid for Transformative Research Areas A ”Extreme Universe” No. 21H05184.

\appendix

\input{Sections/Additional_mat}

\addcontentsline{toc}{section}{References}

\bibliography{CorrBubbles}
\bibliographystyle{CorrBubbles}

\end{document}

%% file: Sections/Introduction.tex
\section{Introduction}

One of the most remarkable insights in theoretical physics over the past decades is the holographic principle, which relates gravitational quantities in a bulk geometry to the observables of a quantum system living on its boundary~\cite{tHooft:1993dmi,Susskind:1994vu}.
The idea that the gravitational degrees of freedom scale with the area rather than the volume of the spacetime was first hinted by the Bekenstein-Hawking formula~\cite{PhysRevD.7.2333,Hawking:1975vcx}. 
Later, a better quantitative understanding of the holographic principle has been achieved with the advent of the Anti-de Sitter/Conformal Field Theory (AdS/CFT) correspondence~\cite{Maldacena:1997re,Gubser:1998bc,Witten:1998qj}.
In this work, we will investigate two entries of the AdS/CFT dictionary: the entanglement entropy of a state in a bipartite system, and the singularities of the two-point functions of CFT operators.

Motivated by the precious insights obtained within the AdS/CFT duality, many efforts have been made to extend the holographic principle to other settings.
An interesting case is undoubtedly de Sitter (dS) spacetime (\eg see the reviews ~\cite{Spradlin:2001pw,Anninos:2012qw,Galante:2023uyf}), since it describes with good approximation the early and late inflationary stages of evolution of our universe.
While the area of cosmological horizons admits a thermodynamic interpretation similar to the Bekenstein-Hawking entropy of black holes~\cite{PhysRevD.15.2738}, identifying a precise holographic dual for dS spacetime has proven far more challenging.
Nevertheless, the physical importance of this problem has stimulated several lines of research, and recent years have seen a revival of attempts to extend holography to dS spacetime, \eg see ~\cite{Geng:2019bnn,Susskind:2021omt,Susskind:2021esx,Coleman:2021nor,Shaghoulian:2021cef,Banihashemi:2022htw,Silverstein:2022dfj,Chandrasekaran:2022cip,Narovlansky:2023lfz,Batra:2024kjl,Verlinde:2024znh,Verlinde:2024zrh,Tietto:2025oxn,Irakleous:2025trr}.

\vskip 2mm

In this context, an interesting setup is provided by geometries in which an inflating interior spacetime is glued to an exterior 
asymptotically AdS region, equipped with a timelike boundary where the AdS/CFT dictionary can be applied.
Such configurations can be realized by coupling gravity to a scalar field theory whose potential admits two minima, characterized by different values of an order parameter and the cosmological constant, separated by a domain wall. 
In the analytically-tractable thin wall approximation, the thickness of the domain wall is neglected~\cite{Blau:1986cw,Farhi:1986ty,Farhi:1989yr,Freivogel:2005qh,Fu:2019oyc,Simidzija:2020ukv}, so that
the order parameter is discontinuous and the domain wall dynamics can be studied using the Israel junction conditions~\cite{israel1966singular}.
We will refer to this class of gravitational backgrounds as \textit{bubble geometries}.
An alternative framework, free of discontinuities in the order parameter, is provided by the so-called \textit{centaur geometries}~\cite{Anninos:2017hhn,Anninos:2018svg,Anninos:2020cwo,Anninos:2022qgy}.
These were introduced in two bulk dimensions, but are forbidden in higher dimensions by the null energy condition~\cite{Lowe:2010np}
\footnote{Nonetheless, certain generalizations to higher-dimensions that evade these obstructions have been proposed in Ref.~\cite{Anninos:2022ujl}.}. 
In the present work, we focus on bubble geometries, 
which can be consistently defined in spacetime dimensions greater than two.
Related theoretical frameworks have also been explored in~\cite{Mirbabayi:2020grb,Antonini:2022blk,Biasi:2022ktq,Antonini:2022fna,Sahu:2023fbx,Balasubramanian:2023xyd,Gupta:2025jlq}.

We will analyze $(d+1)$--dimensional bubble geometries invariant under time reversal of the bulk coordinate time $t \rightarrow -t$, constructed by gluing a spherically symmetric bubble of vacuum with an external AdS-Schwarzschild black hole along a domain wall with negligible thickness.\footnote{Here and in the following, we refer to dS and AdS spacetimes in $d+1$ dimensions, such that the dual boundary CFT is $d$--dimensional.}
The geometry is characterized by three parameters $(\lambda, \kappa, m)$ proportional to the cosmological constant of the interior spacetime, the tension of the domain wall, and the mass of the black hole, respectively.\footnote{The precise relations between the parameters $(\lambda, \kappa, m)$ and the above physical quantities are given in eqs.~\eqref{eq:cosmological_constant_lambda}, \eqref{general-V-bubble-0}, and \eqref{eq:BH_mass}, respectively.}
The null energy condition forces $\kappa \geq 0$. 
Denoting the trajectory of the domain wall with $R(\tau)$, where $R$ is a radial coordinate and $\tau$ the proper time (as measured by an observer living
on top of the domain wall), we refer to the possible bubble solutions with $t \geq 0$ as follows:
\begin{itemize}
    \item A \textit{collapsing} bubble is a solution with $\dot{R} \leq 0$, whose radius collapses to zero in a finite proper time. 
\item An \textit{expanding} bubble is a solution with $\dot{R} \geq 0$, whose radius increases for an infinite proper time. 
    \item A \textit{static} bubble is a solution such that the shell radius $R(\tau)$ is constant.    
    This situation, when it exists, corresponds to a fine-tuned value of the mass parameter $m$ that separates the expanding from the collapsing solutions.
\end{itemize}
While for a collapsing bubble the interior is a finite spacetime region, for expanding and static ones the interior contains an infinite spacetime volume.
In particular, for expanding bubbles with $\l>0$, the geometry provides a way to embed an infinite universe in accelerated expansion inside an asymptotically AdS spacetime.

\vskip 2mm

In gravitational physics, it is particularly interesting to study geometric probes that enter the region behind a black hole horizon~\cite{Louko:2000tp,Kraus:2002iv}, to clarify how the interior degrees of freedom are encoded by the quantum theory located at the AdS boundary.
\rep{Since both expanding and collapsing bubbles may lie entirely inside the black hole bifurcation surface, it is not obvious how a boundary observer can discriminate between such different bulk geometries.} A useful probe in this context is
the entanglement entropy of a boundary subregion, which is holographically dual to the area of the Hubeny-Rangamani-Takayanagi (HRT) surface~\cite{Ryu:2006bv,Hubeny:2007xt}, namely the extremal codimension-two bulk surface with minimal area which is homologus to the given boundary subregion.
This prescription has been extensively applied to explore critical points, strongly-coupled systems, thermalization, wormholes, quantum information and much more, \eg see~\cite{Nishioka:2009un,Balasubramanian:2011ur,Rangamani:2016dms,Harlow:2016vwg}.
A common prejudice is that the HRT surfaces does not
probe the geometry beyond the bifurcation surface.
We will provide explicit counterexamples showing that, for collapsing bubbles, the interior can indeed be probed by an HRT surface, even when the domain wall lies inside the bifurcation surface. 
By contrast, we will present evidence that this does not occur for expanding bubbles.

Other probes are therefore desirable for a boundary observer 
aiming to explore the interior of an expanding bubble.
A powerful tool is provided by the singularity structure of the two point functions of boundary operators
$\langle \mathcal{O}(x) \mathcal{O} (y) \rangle$.
For a free theory, these singularities arise 
when $(x-y)^2=0$, \ie when the insertion points are null-separated.
In holographic theories, a new class of singularities arises due to null geodesics extending into the bulk and connecting the two boundary points~\cite{Fidkowski:2003nf,Festuccia:2005pi,Hubeny:2006yu,Maldacena:2015iua,Dodelson:2020lal,Gary:2009ae,Horowitz:2023ury,Kolanowski:2023hvh,Dodelson:2023nnr}.
Such singularities also occur when $x$ and $y$ are connected 
by the infinite-energy limit of a spacelike geodesic \cite{Hubeny:2006yu}.
We will refer to such trajectory as an \textit{almost-null geodesic}.
This behaves similarly to a null geodesic, with the addition of a possible bounce off the black hole singularity (for $d \geq 3$) and off the dS infinity. 
We will collectively denote these classes of singularities
as \textit{bulk-cone singularities}. 

\vskip 2mm

The main novelties of this work are summarized as follows:
\begin{itemize}
\item We systematically explore the causal structure of Lorentzian bubble solutions across the parameter space $(\lambda, \kappa, m)$, considering both expanding and collapsing bubbles with both signs of $\l$. The $\l>0$ case was already studied in Ref.~\cite{Freivogel:2005qh}, while the Euclidean version was examined in~\cite{Fu:2019oyc}.
\item
 We compute the holographic entanglement entropy at vanishing boundary time in $(2+1)$--dimensional bubble geometries.
 In certain regimes, there are several competing extremal surfaces among which we have to pick the minimal one.
 For the collapsing bubble, we find regimes where the HRT surface enters the region behind the black hole bifurcation surface and probes the interior of the bubble.
 \item
We study bulk-cone singularities in $(3+1)$-dimensional bubble geometries, the lowest-dimensional case where a spacelike almost-null geodesic is pushed away from a black hole singularity.
A qualitative sketch of the bulk-cone singularities
in bubble geometries was briefly discussed in some special cases in Ref.~\cite{Freivogel:2005qh}. 
In this paper, we perform a more detailed and quantitative  study. Specifically, we consider a radial (almost-)null geodesic that leaves the AdS boundary at time $t_{\rm in}$, propagates in the bulk, and then comes back to the AdS boundary at time $t_{\rm fin}$.
We classify the functional dependence $t_{\rm fin} (t_{\rm in})$ for all the possible bubble geometries, identifying the 
characteristic features that distinguish their causal structure.
\end{itemize}

\vskip 2mm

The paper is organized as follows.
We introduce general features of the bubble solutions in section~\ref{sec:preliminaries}, providing the equations of motion of the domain wall. We then discuss the phase diagram  of the causal structure of the geometries.
This analysis allows us to determine the Penrose diagram of the bubble geometries in each region of the parameter space.
In section~\ref{sec:EE_bubbles} we compute the holographic entanglement entropy of a boundary arc 
at vanishing boundary time
in a three-dimensional bubble geometry.
Section~\ref{sec:sing_corr} contains the analysis of the bulk-cone singularities in four-dimensional bubble geometries.
The main results are summarized
in section~\ref{sec:conclusions},
where future directions are also outlined. Technical details are deferred to appendices.

%% file: Sections/Setup.tex
\section{Theoretical setting and parameter space}
\label{sec:preliminaries}

In this section, we introduce the geometric setup that describes a 
spherically symmetric and time-reversal invariant bubble of (A)dS spacetime contained within an exterior black hole (BH) solution in asymptotically AdS spacetime.
The two regions of spacetime are separated by a shell, carrying non-trivial energy, whose trajectory is determined by imposing Israel junction conditions~\cite{israel1966singular}.
Throughout this work, we will only consider the case of a shell represented by a \textit{thin domain wall}
with negligible thickness.
In other words, the width of the domain wall is much smaller than the AdS curvature scale; other than the discontinuity at the domain wall, the geometry is smooth.
 
The dynamics of the false vacuum bubble was initially studied in Refs.~\cite{Blau:1986cw,Farhi:1986ty,Farhi:1989yr} for an exterior geometry given by a BH in asymptotically flat spacetime, as possible models to describe the universe in a laboratory. 
The case of a geometry composed by a dS bubble inside an asymptotically AdS BH was studied in Ref.~\cite{Freivogel:2005qh}. 
In this paper we will study a larger region of parameter space, including a bubble with arbitrary cosmological constant in the interior. 
Euclidean solutions were previously studied in 
Ref.~\cite{Fu:2019oyc}, while Ref.~\cite{Simidzija:2020ukv} investigated their relation with conformal defects.

For simplicity, we will only restrict to spherically symmetric solutions (see Ref.~\cite{Aguirre:2005xs} for an analysis of the stability of the spherically symmetric solution in the case of a flat exterior). One can study more general setups that allow for transitions between AdS and dS vacua, for instance by considering a scalar field theory with non-trivial potential. Nonetheless, the assumptions that we listed above are convenient because they allow for an analytic treatment of various problems, as investigated in, \eg Refs.~\cite{Sahu:2023fbx,Balasubramanian:2023xyd,Gupta:2025jlq,Auzzi:2023qbm}.

We introduce in section~\ref{ssec:bubble_metric} the metric and the causal properties of the geometry.
We determine the shape of the domain wall by imposing Israel junction conditions. This problem can be recast into the classical motion of a particle in an effective potential, see section~\ref{ssec:effective potential}. Different choices of the parameters of the geometry lead to different solutions for the bubble, as described in section~\ref{ssec:parameter-space}.
The causal structure of the bubble geometry is presented in section~\ref{ssec:causal_structure_expanding_bubble} for an expanding solution, and in section~\ref{ssec:causal_structure_collapsing_bubble} for a collapsing one.
One configuration of particular interest -- mainly due to the possibility to perform analytic investigations -- is the static bubble, in which case the radius of the domain wall is time-independent. We focus on this latter case in section~\ref{ssec:mass_static_bubble}.

\subsection{Interior and exterior geometries}
\label{ssec:bubble_metric}

Let us consider a $(d+1)$--dimensional spacetime composed by an interior and an exterior parts, glued together on the surface of a thin domain wall. Denoting with $i,o$ the region inside (outside) the shell and assuming spherical symmetry, the infinitesimal line element takes the form
\beq
ds^2_{i,o}=(g_{i,o})_{\mu \nu} dx_{i,o}^{\mu} dx_{i,o}^{\nu} =
-f_{i,o}(r) \, dt_{i,o}^2+ \frac{dr^2}{f_{i,o} (r)} + r^2 d \Omega_{d-1}^2 \, ,
\label{metric-zero}
\eeq
where $g_{i,o}$ denotes the metric, $f_{i,o}(r)$ the blackening factor, and $d\Omega_{d-1}^2$ the line element of a unit $(d-1)$--dimensional sphere, which we choose to characterize the spatial sections of the geometry. 
From now on, we will refer to eq.~\eqref{metric-zero} as the metric of the \textit{bubble geometry}, such that it is implicitly understood that the interior and exterior parts are joined together through a domain wall with world-volume line element
\beq
ds^2_{\rm bubble} = - d\tau^2 + R^2(\tau) d\Omega_{d-1}^2 \, ,
\eeq
where $\tau$ is a proper time parameter, and $R(\tau)$ the size of the shell.
We stress that the radial coordinate $r$ is continuous across all the geometry, while the time coordinate is generally discontinuous when crossing the domain wall.

\rep{The metric \eqref{metric-zero} is a general spherically symmetric ansatz for the solutions of the vacuum Einstein's equations}; from now on, we will specialize to the case where the interior geometry is an empty (A)dS 
background, characterized by the blackening factor
\beq
f_i(r)=1-\l \, r^2 \, ,
\label{f-dS}
\eeq
where $\l>0$ corresponds to dS, and $\l<0$ to AdS spacetime (in global coordinates).
The quantity $\lambda$ is related to the cosmological constant $\Lambda$ as follows
\beq
\Lambda=\frac{d(d-1)}{2} \l \, ,
\label{eq:cosmological_constant_lambda}
\eeq
and to the (A)dS radius $L$ via the identity
\beq
L = \frac{1}{\sqrt{|\lambda|}} \, .
\eeq
Next, we assume that the exterior geometry is a BH in asymptotically AdS$_{d+1}$ spacetime. This is parametrized by the following blackening factor
\beq
f_o(r) =r^2 +1 - \frac{m}{r^{d-2}}  
\, ,
\label{f-BH}
\eeq
where $m$ is called mass parameter, and we set the AdS radius in the exterior region to $1$. 
The asymptotic mass $M$ of the BH is proportional to the mass parameter $m$ via 
\beq
M = \frac{(d-1) \Omega_{d-1}}{16 \pi G_N} \, m \, ,
\label{eq:BH_mass}
\eeq
where $\Omega_{d-1}$ 
denotes the dimensionless volume of the spherical geometry along the $(d-1)$ orthogonal directions.
The mass parameter $m$ can be recast in terms of the horizon radius $r_h$, defined by the condition $f_o(r_h)=0$, by means of the relation
\beq
m =r_h^{d-2}  \,  \le r_h^2 + 1 \ri  \, .
\label{mass expression}
\eeq
In the remainder of the paper, we will mostly focus on the three- and four-dimensional cases (in these conventions, $d=2$ and $d=3$).
To compare the above general formulae with the results previously obtained in Ref.~\cite{Auzzi:2023qbm}, we stress that in $d=2$ the blackening factor becomes
\beq
f_o(r)|_{d=2} = r^2 + 1-m = r^2 - \mu \, , 
\label{f-BTZ}
\eeq
where we introduced another parameter $\mu$ such that
\beq
\mu \equiv m-1 \, .
\label{eq:mass_mum}
\eeq
This is sometimes referred to as the mass parameter of the three-dimensional Ba\~{n}ados-Teitelboim-Zanelli (BTZ) black hole~\cite{Banados:1992wn}. In this case, the horizon radius is related to $\mu$ as follows:
\beq
r_h = \sqrt{\mu} \, .
\eeq
In the remainder of this work, we will restrict to $m \geq 1$ in $d=2$, such that the geometry admits a black hole horizon.

A convenient tool to describe the causal structure of bubble geometries consists in using Eddington-Finkelstein (EF) (or \textit{null}) coordinates
\beq
v_{i,o}= t_{i,o} +r^*_{i,o}(r) \, , \qquad
u_{i,o}= t_{i,o} -r^*_{i,o}(r) \, ,
\label{u-v-def}
\eeq
where $r_{i,o}^*(r)$ is the \textit{tortoise coordinate}.
We refer to appendix~\ref{ssec:tortoise} for more details on the tortoise coordinate.

\subsection{Junction conditions and effective potential}
\label{ssec:effective potential}

The domain wall carries a non-vanishing energy density. 
The extrinsic curvature $K_{a b}$ is discontinuous when crossing the domain wall, and the difference between the two sides is quantified by the energy momentum tensor localized on the shell.
For spherically symmetric geometries, it is customary to introduce the quantities (\eg see sections 3.7--3.9 of Ref.~\cite{Poisson:2009pwt})
\beq
\b_{i,o}=\le K^{\theta}_{\theta} \ri_{i,o} \, R(\tau) \, ,
\eeq
where $R(\tau)$ is the shell's radius (from now on, we will omit the explicit dependence on the proper time $\tau$), and $K^{\theta}_{\theta}$ is a particular component of the extrinsic curvature along an angular direction $\theta$ of the transverse $(d-1)$--dimensional sphere. 
The jump between the values of $\b_i$ and $\b_o$ is governed by the Israel junction conditions \cite{israel1966singular}
\beq
\b_i - \b_o= \kappa \, R \, , 
\label{general-V-bubble-0}
\eeq
which in our case are characterized by the quantities
\beq
\b_i=\pm \sqrt{\dot{R}^2+f_i(R)}  \, , \qquad \b_o=\pm \sqrt{\dot{R}^2+f_o(R)} \, , \qquad
\kappa=\frac{8 \pi G_N \, \s}{d-1} \, .
\label{general-V-bubble-1}
\eeq
In the latter set of equations, $\dot{R} \equiv dR/d\tau$ denotes the derivative with respect to the proper time, $\s$ is the domain wall's tension, and $G_N$ the Newton's constant.\footnote{With an abuse of notation, we will often refer to $\kappa$ itself as the domain wall's tension. The differential equations~\eqref{general-V-bubble-0} with parameters~\eqref{general-V-bubble-1} assume that the energy-momentum tensor is normalized as $T^{ij} = - \sigma h^{ij} \delta(\eta) +  \text{regular terms}$, where $\eta=0$ identifies the location of the domain wall.}
The sign of $\b_{i,o}$ is positive if either the radial coordinate $r$ increases as the domain wall is approached from the interior, or if $r$ increases as one moves away from the domain wall in the exterior.
If both $\b_{i,o}$ have the same sign, $r$ is monotonic near the wall. 
If  $\b_{i,o}$ have different signs, $r$ is locally extremized at the location of the wall.

For later convenience, we re-express the curvature parameters $\beta_i, \beta_o$ as follows:
\beq
\b_i=\frac{f_i(R)-f_o(R) + \kappa^2 R^2}{2 \kappa R} \, , \qquad
\b_o=\frac{f_i(R)-f_o(R) - \kappa^2 R^2}{2 \kappa R} \, .
\label{beta-i-beta-o-gen}
\eeq
By plugging eq.~\eqref{beta-i-beta-o-gen} inside~\eqref{general-V-bubble-0}, we find that the dynamics of the domain wall is governed by the following equations of motion
\begin{subequations}
    \beq
\dot{R}^2+V_{\rm eff} (R)=0 \, , 
\label{general-V-bubble}
\eeq
\beq
V_{\rm eff}(R) \equiv f_o(R) - \frac{(f_i(R) -f_o(R) -\kappa^2 \, R^2)^2}{4 \, \kappa^2 \, R^2}  \, .
\label{general-V-bubble-potenziale}
\eeq
\end{subequations}
The differential equations~\eqref{general-V-bubble} describe the dynamics of a point particle in the effective potential~\eqref{general-V-bubble-potenziale}. 
In the geometry~\eqref{metric-zero} with blackening factors \eqref{f-dS} and \eqref{f-BH}, the latter potential can be recast into the form \cite{Fu:2019oyc}
\beq
V_{\rm eff}(R) =
- \mathcal{A}  \, R^2 
 +1
-\frac{\mathcal{B}}{ R^{d-2}}
-\frac{\mathcal{C}}{ R^{2d-2}} \, ,
\label{eq:effective_potential_ABC}
\eeq
where we introduced the convenient quantities
\beq
\mathcal{A} \equiv \frac{(\l+\kappa^2-1)^2 + 4 \l }{4 \kappa^2} \, , \qquad
\mathcal{B} \equiv  m \, \frac{ \left(\kappa ^2-\lambda -1\right)}{2 \kappa ^2  } \, , \qquad
\mathcal{C} \equiv \frac{{m}^2}{4 \kappa ^2 } \, .
\label{eq:functions_ABC}
\eeq
Similarly, we can plug the blackening factors \eqref{f-dS} and \eqref{f-BH} inside the extrinsic curvature parameters~\eqref{beta-i-beta-o-gen} to get
\beq
\b_i=\frac{\kappa^2-1 -\lambda }{2 \kappa} \, R
+\frac{m}{2 \kappa} \, \frac{1}{R^{d-1}}
 \, , \qquad
\b_o=-\frac{\kappa^2+1 +\lambda }{2 \kappa} \, R
+\frac{m}{2 \kappa} \, \frac{1}{R^{d-1}} \, .
\label{beta-i-beta}
\eeq
The values of $\beta_{i,o}$ in the above identity determine which overall signs to pick in the defining relations~\eqref{general-V-bubble-1}.
The signs of $\beta_{i,o}$ play an important role to distinguish the various configurations of bubble solutions that we analyze in subsection~\ref{ssec:parameter-space}.\footnote{Additional details on the relation between the various solutions and the signs of $\beta_{i,o}$ are discussed in appendices~\ref{app:small_large_mass_limits} and~\ref{ssec:sign_curvature_parameters}.}
Following Ref.~\cite{Blau:1986cw}, to identify the above signs, it is useful to solve eq.~\eqref{general-V-bubble} in terms of the mass parameter $m$, \ie 
\beq
m=(1+\lambda-\kappa^2) R^d 
\pm 2 \kappa R^{d-1} \sqrt{1+\dot{R}^2 -\lambda R^2} \, .
\label{eq-massa-bubble}
\eeq
In the following, let us denote by $R_0$ the initial size of the bubble at $\tau=0$, when $\dot{R}=0$.\footnote{Notice that the horizontal middle line in the Penrose diagram of any bubble geometry lies at constant $t=0$, where $t$ is the bulk time. In particular, on this line we also have $\tau=0$ ($\tau$ proper time on the domain wall) and $t_{\infty}=0$, where $t_{\infty}$ is the time on the right AdS boundary.}
We denote with $m=m_{\pm}(R_0)$ the functional relation between $R_0$ and the mass of the bubble, determined by 
\beq
m_\pm(R_0)=(1+\lambda-\kappa^2) R_0^d 
\pm 2 \kappa R_0^{d-1} \sqrt{1 -\lambda R_0^2} \, .
\label{eq-massa-bubble-2}
\eeq
Plugging eq.~\eqref{eq-massa-bubble-2} inside~\eqref{beta-i-beta}, we then find
\beq
\b_i(R_0)=\pm \sqrt{1-\l R_0^2 } \, ,
\qquad
\b_o(R_0)=-\kappa R_0
\pm \sqrt{1-\l R_0^2 }
\label{eq:beta:discrimine}
\eeq
Therefore, the two allowed values of the mass $m_\pm(R_0)$
in eq.~(\ref{eq-massa-bubble-2}) correspond to the two different signs of the curvature parameter $\b_i(R_0)$, respectively.

An important property of the collapsing bubbles follows. 
\rep{By definition, a collapsing bubble is characterized by $\dot{R}<0$ for bulk time $t >0$, and the radius of the bubble collapses to zero in a finite proper time 
$\tau = \tau_0$.}
Equation~\eqref{beta-i-beta} tells us that when $R \rightarrow 0$, then $\beta_i>0$. As we increase $R$ (and decrease $\tau$), the only way for $\beta_i$ to change sign would be that the domain wall crosses an event horizon. 
This cannot happen in the case of a global AdS interior, since it does not admit any horizon. 
On the contrary, an empty dS interior admits a cosmological horizon, which the bubble may cross. 
For the sake of the argument, let us assume that the bubble crosses the cosmological horizon. Since the shell's trajectory is timelike, the bubble would eventually meet past timelike infinity $\mathcal{I}^-$ (located at $R \to \infty$) in the dS region. 
On the other hand, in the case of time-reversal symmetric solutions, the bubble must satisfy $R \to 0$ also when $\tau \to -\tau_0$.
In other words, the bubble is confined inside the dS static patch. 
This fact is in contradiction with the bubble approaching $\mathcal{I}^-$. 
We thus conclude that, for time-reversal symmetric collapsing bubbles, the identity $\beta_i(R(\tau))>0$ holds for any $\tau$, and in particular $\beta_i(R_0)>0$.
As a direct consequence, for time-reversal symmetric collapsing bubbles we must choose the $+$ sign in eq.~\eqref{eq:beta:discrimine}, which corresponds to the solution $m_+(R_0)$. 
Instead, time-reversal symmetric expanding bubbles can in principle admit both signs for $\b_i(R_0)$. 

\subsection{Parameter space}
\label{ssec:parameter-space}

We discuss the qualitative behavior of the solutions to the differential equation \eqref{general-V-bubble}, as determined by the effective potential \eqref{eq:effective_potential_ABC}, for various choices of the parameters defining the geometry. 
While this investigation was carried out in Euclidean signature in Ref.~\cite{Fu:2019oyc}, here we perform a classification of the parameter space in Lorentzian signature.
We mostly present the results for general dimension $d$, focusing on $d=2,3$ only when explicitly stated.
Moreover, we restrict in our analysis to bubble geometries invariant under the time reversal symmetry $t \to -t$. 

In order to classify the main features of the solutions,
it is useful to look at the sign of some of the quantities
in eq.~\eqref{eq:functions_ABC}.
Since $\mathcal{C} >0$, the effective potential~\eqref{general-V-bubble-potenziale} is an increasing function of the radial coordinate for small enough $r$.
We distinguish two cases according to the sign of $\mathcal{A}$ as follows:
 \begin{itemize}
 \item When $\mathcal{A}>0$, the effective potential $V_{\rm eff}$ is a decreasing function of the radial coordinate for large enough $r$.  Therefore, $V_{\rm eff}$ must have at least a local maximum.
 As proven in appendix A of Ref.~\cite{Fu:2019oyc}, in this case there are no local minima of the potential, because $V''_{\rm eff}(R)$ computed at the extremal point $R=R_e$ (for which $V'_{\rm eff}(R_e)=0$)
 is always strictly negative.\footnote{Here, we use the notation $V' \equiv dV/dR$.}
This implies that the maximum is unique, and there are now three sub-cases.
\textbf{(1)} If $V(R)>0$ at the maximum, then the equations of motion \eqref{eq:effective_potential_ABC} admit two solutions, one expanding and one contracting at positive time, each of them separately invariant under time reversal.
\textbf{(2)} If $V(R)<0$ at the maximum, then the equations of motion do not admit any solution invariant under time reversal. 
\textbf{(3)} In the fine-tuned case where $V(R)=0$ at the location of the maximum, the only solution invariant under time reversal is a static bubble with constant $R=R_e=R_0$ (where $R_0$ was defined below eq.~\eqref{eq-massa-bubble}). 
 \item
When $\mathcal{A}<0$, the effective potential $V_{\rm eff}$ is an increasing function of the radial coordinate for large enough $r$.
In this case, the equation $V_{\rm eff}(R)=0$ admits a single solution  $R=R_0$.\footnote{The existence of a single solution $R_0$ can be proven by plugging the value of $\mathcal{B}$ obtained from the identity $V_{\rm eff}(R_0)=0$ inside the expression for $V'_{\rm eff}(R)$.}
\rep{Since $V'_{\rm eff}(R_0) >0$ (see also appendix B of Ref.~\cite{Fu:2019oyc}), then in this regime the expanding solution does not exist, while we will show that for any value of the mass parameter $m$, there exists one collapsing bubble}.
\end{itemize}

\begin{figure}[ht]
\center
\includegraphics[scale=0.6]{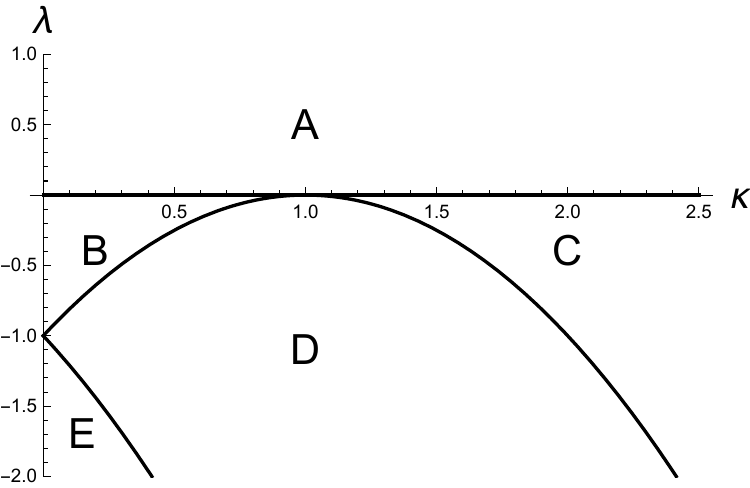}
\caption{Phase diagram of the bubble solutions as a function of $(\kappa, \l)$.
}
\label{Phase-Diagram}
\end{figure}

The above discussion showed that the sign of $\mathcal{A}$ plays a relevant role to classify the solutions to the equations of motion.
The quantity $\mathcal{A}$
vanishes in correspondence of the cosmological constants
\beq
 \l_1 \equiv - (\kappa+1)^2 \, ,
 \qquad
\l_2 \equiv - (\kappa-1)^2 \, .
\label{eq:critical_lambda}
\eeq
The curves in eq.~\eqref{eq:critical_lambda},
in addition with the line $\l=0$, are depicted in fig.~\ref{Phase-Diagram}, where they determine the boundaries of the various regions of the phase diagram.
We find that $\mathcal{A} < 0$ for $\l_1<\l<\l_2$, which corresponds to region $D$.
Instead, we find that $\mathcal{A}>0$ in two cases: either $ \l > \l_2$ (regions $A, B, C$), or $\l<\l_1$ (region $E$). 
In particular, the interior geometry is dS spacetime in region $A$ ($\l >0$), while it is empty AdS in regions $B$ ($C$), characterized by $\l_2 < \l < 0$ and $\kappa < 1$ ($\kappa >1$).

The expanding bubbles in regions $A, B, C$ satisfy the useful property  $\beta_o(R_{0})<0$, which can be proven as follows.
The radial coordinate of an expanding bubble reaches $R \to \infty$.  In this limit, the first term in eq.~\eqref{beta-i-beta} dominates.
Since in regions $A,B, C$, the cosmological constant satifies $\l > \l_2 \geq -\kappa^2 -1$, we get $\beta_o(R \to \infty) \approx \frac{-\kappa^2 -1-\l}{2\kappa} R <0$. 
This result implies that the domain wall must hit the left boundary of the AdS BH spacetime (see configurations $A, B, C$ in fig.~\ref{Pen-Dia-Expanding-Bubble}, that we will describe in subsection~\ref{ssec:causal_structure_expanding_bubble}). 
For the sake of the argument, let us assume that the bubble originated inside the white hole horizon at a certain time $\tilde{t}$. If this was true, then the bubble would have fallen inside the BH horizon at time $-\tilde{t}$ due to the time reversal symmetry, but this is not the case. Therefore, we conclude that the bubble did \textit{not} originate inside the white hole horizon. Since the shell's trajectory is timelike, the bubble always remains 
behind the BH bifurcation surface without entering the region connected to the right AdS black hole boundary, and we have  $\beta_o(R_{0})<0$. 
This argument was originally used to characterize region $A$ in Ref.~\cite{Freivogel:2005qh}.

For a fixed value of the mass parameter $m$, one can have several possible solutions for the bubble geometry: collapsing, expanding, static (for a fine-tuned value of the mass parameter $m=m_{\rm static}$), or no time-reversal solutions at all. Let us describe the main features of the various regions in fig.~\ref{Phase-Diagram}:
\begin{itemize}
\item
In regions $A$, $C$ and $E$ there exist two distinct time reversal-symmetric solutions (the expanding and the collapsing one) only when $m\leq m_{\rm static}$.
While in region $E$ the static and the expanding bubbles both lie outside the BH horizon, in regions $A$ and $C$ they are located inside the BH horizon.
In region $A$, the interior geometry is a portion of dS spacetime (which is finite for the collapsing bubble, and infinite for the expanding one).
In region $C$ and $E$, the interior of both bubble solutions is a portion of the global AdS spacetime which contains the center of AdS.
\item In region $B$, both the expanding and collapsing bubbles exist for arbitrary values of $m$.
The collapsing solution includes the center of AdS.
The expanding bubble resembles a thermofield double state (TFD), because it includes two disconnected AdS boundaries.
There is no static bubble in this region of parameters.
\item In region $D$, there is a single collapsing bubble -- which includes the center of AdS spacetime in its interior -- for any value of $m$. There are neither expanding, nor static bubbles in this regime. 
\rep{As argued in Ref.~\cite{Simidzija:2020ukv}, 
the physics of the bubble geometry in this region is described by the insertion of  a quench operator in a conformal field theory.
We will further discuss this relation in section \ref{ssec:conformal_defects}.}
\end{itemize}
For convenience, a summary of the various possibilities is collected in table~\ref{tab:parameter_space}, see the summary of results in section~\ref{sec:conclusions}.
The limits of small and large mass in the parameter space are discussed in appendix~\ref{app:small_large_mass_limits}. 
Finally, let us mention that at fixed $m$ and $\mathcal{A} \to 0^+ $, the expanding bubble becomes infinitely large. 
When $\mathcal{A}=0 $, at large $R$ we find that $V_{\rm eff}$ approaches a constant value. In other words, in this limit the expanding solution is not allowed anymore.

\subsubsection{Relation between parameter space and conformal interfaces}
\label{ssec:conformal_defects}

Interestingly, the curves in eq.~\eqref{eq:critical_lambda} have an interpretation in the context of conformal interfaces.
Let us consider the -- apparently unrelated -- setting composed by two regions of empty AdS$_3$ spacetime, joined together through the worldsheet of a two-dimensional domain wall with negligible thickness~\cite{Randall:1999vf}.
This bottom-up model is conjectured to be holographically dual to a two-dimensional conformal interface separating two CFTs.
In this framework, one can show that the dual geometry describes a well-defined ground state of an interface CFT if the string tension lies within the range~\cite{Karch:2000ct,Bachas:2001hpy}
\beq
\kappa_- \leq \kappa \leq \kappa_+ \, , 
\qquad 
\kappa_- \equiv \left| 1 - \sqrt{-\lambda}  \right| \, , \qquad
\kappa_+ = 1 + \sqrt{-\lambda} \, ,
\label{eq:range_tensions_AdS}
\eeq
which are precisely the curves $\lambda = \lambda_{1,2}$ defined in eq.~\eqref{eq:critical_lambda}.
The lower bound on the tension forbids the nucleation of Coleman-De Luccia bubbles~\cite{Coleman:1980aw,Barbon:2010gn}.
The upper bound is the critical value above which the world-volume metric of the domain wall has a positive curvature~\cite{Karch:2000ct,Banerjee:2018qey,Banerjee:2019fzz}.
The same inequalities also arise in other contexts, \ie from BPS bounds in supergravity~\cite{Cvetic:1992bf}, in the computation of the boundary entropy~\cite{Affleck91,Simidzija:2020ukv}, in the study of the transmission of energy across a conformal interface~\cite{Quella:2006de,Meineri:2019ycm,Bachas:2020yxv,Bachas:2021tnp,Baig:2022cnb,Bachas:2022etu,Baig:2023ahz,Baig:2024hfc,Gutperle:2024yiz,Liu:2025khw,Afxonidis:2025jph}, \rep{or in the study of the vacuum energy of $d$--dimensional interface CFTs~\cite{May:2021xhz}.}
The bounds \eqref{eq:range_tensions_AdS} can also be interpreted in terms of the boundary entropy $\log g$, since this quantity is related to the transmission coefficient in the case of a geometry with a single brane.
In particular, the window of tensions corresponds to an excursion in $\log g$ from $-\infty$ and $+\infty$, see eq.~(3.15) of Ref.~\cite{Simidzija:2020ukv}.

For the purpose of studying bubble geometries, it is interesting to observe that the conditions~\eqref{eq:range_tensions_AdS} are satisfied in the region denoted with $D$ in fig.~\ref{Phase-Diagram}, where the only allowed solution (for any value of the mass parameter $m$) is a collapsing bubble.
In this region, the physics of the bubble geometry can be understood in terms of a boundary theory where two conformal field theories CFT$_1$ 
(which corresponds to the bubble interior)
and CFT$_2$ (which corresponds to the exterior)
are glued across a conformal interface with negligible thickness~\cite{Simidzija:2020ukv}. 
The dual boundary state of CFT$_2$ is created via the action of a quench operator (associated with the specific choice of the interface) on the vacuum $| \psi_1 \rangle $ of CFT$_1$, together with an evolution in Euclidean time which regulates the UV divergences in the energy of the state. 
Schematically, the state of CFT$_2$  at $t=0$
can be built as follows
\beq
| \psi_2 \rangle=\lim_{\epsilon_1 \to 0}
e^{- \epsilon \, H_2} \, \mathcal{Q}
\,  e^{-\epsilon_1 H_1} | \psi_1 \rangle \, ,
\label{eq:Holo-ween}
\eeq
where $ \mathcal{Q}$ is the operator which performs the quench, $H_{1,2}$ are the Hamiltonians of CFT$_{1,2}$, and $\epsilon$ is a regulator that makes the energy of $| \psi_2 \rangle$ -- as a state of CFT$_2$ -- finite.\footnote{In the limit $\epsilon \to 0$ with fixed $\l$ and $\kappa$, the black hole's mass scales as $m \propto 1/\epsilon^2$, see Ref.~\cite{Simidzija:2020ukv} for details).}
Following the notation introduced by the authors of~\cite{Simidzija:2020ukv}, we will refer to the $t=0$ state in eq.~(\ref{eq:Holo-ween}) as a \textit{Holo-ween quench}.

\subsection{Causal structure of expanding bubbles}
\label{ssec:causal_structure_expanding_bubble}

\begin{figure}[h!]
\center
\subfigure[region $A$]{\label{subfig:regionA_Pen_Exp} \includegraphics[scale=0.3]{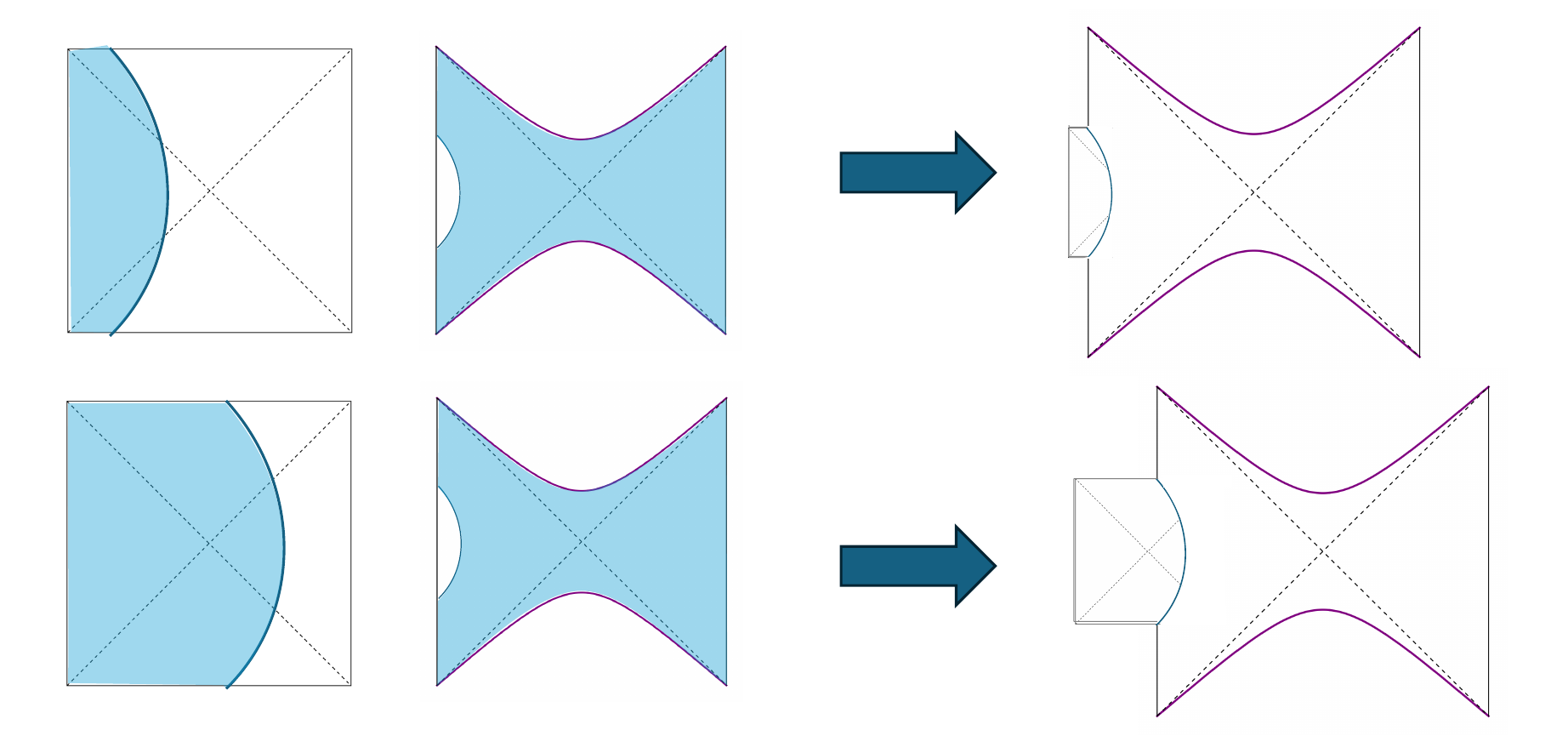}} \quad
\subfigure[region $B$]{ \includegraphics[scale=0.3]{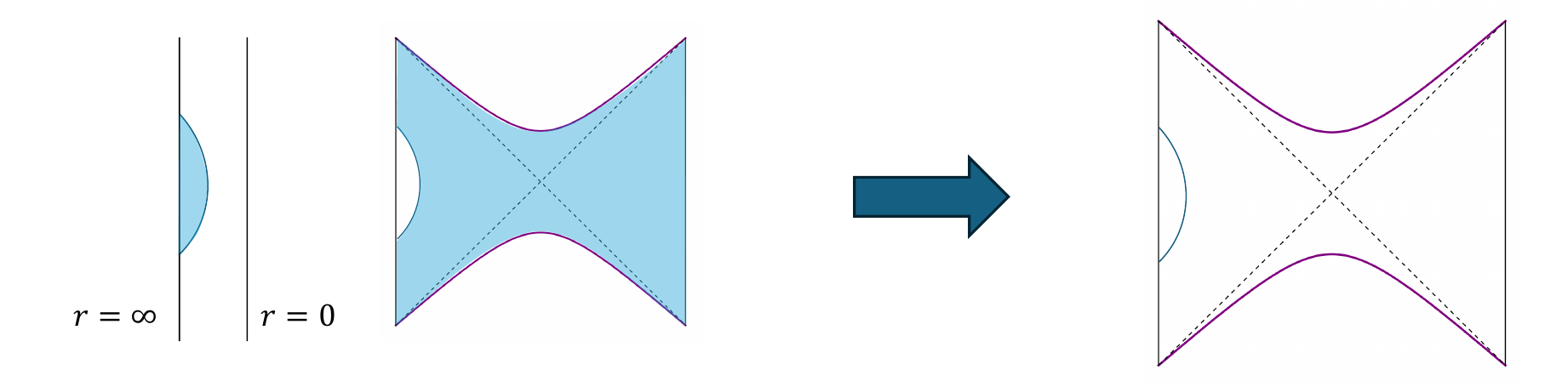}}
\subfigure[region $C$]{\label{subfig:regionC_Pen_Exp} \includegraphics[scale=0.3]{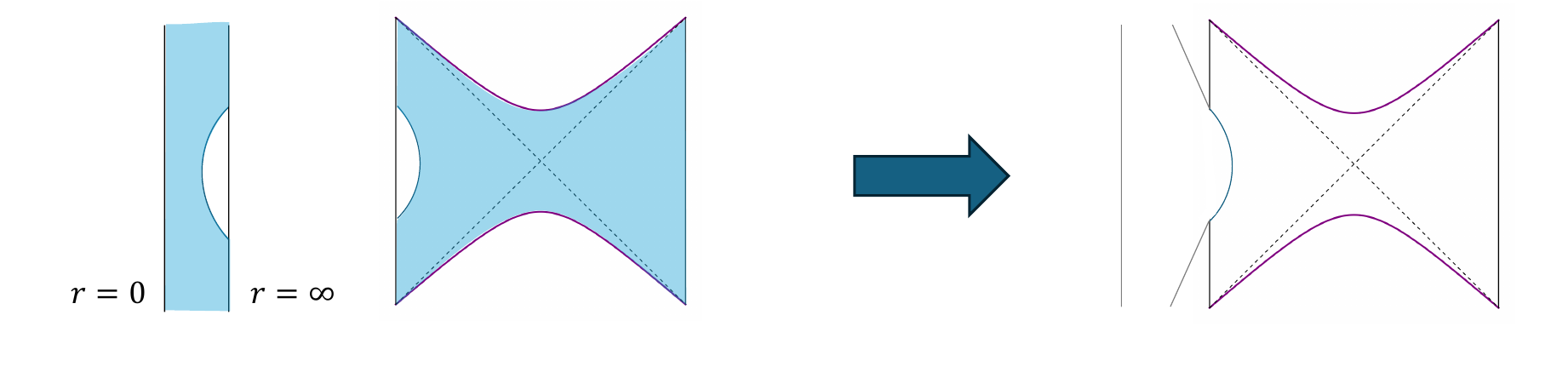}}
\subfigure[region $E$]{ \includegraphics[scale=0.3]{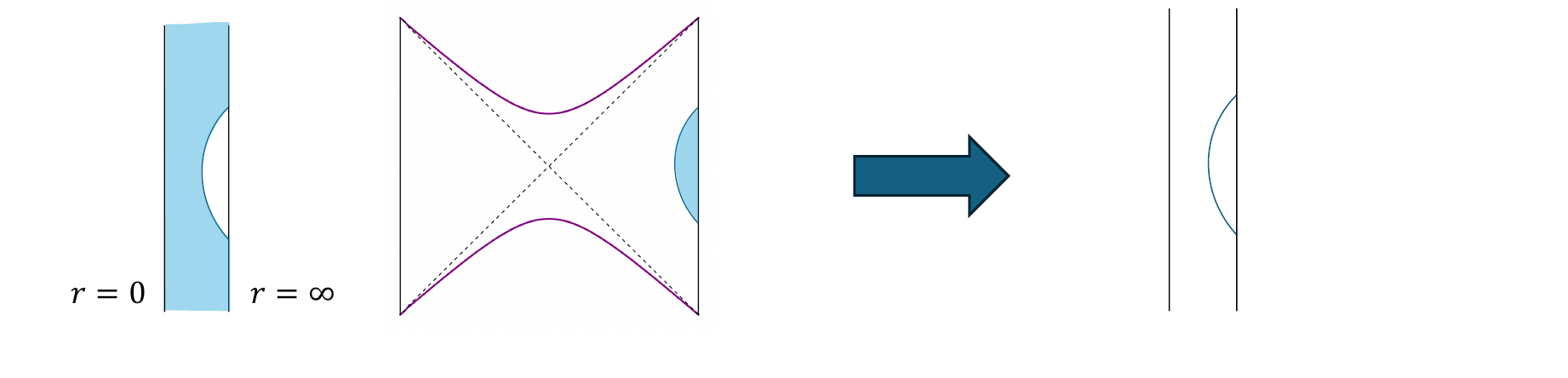}}
\caption{Penrose diagrams for expanding bubbles in the regions of parameter space where they exist.
}
\label{Pen-Dia-Expanding-Bubble}
\end{figure}

We draw in fig.~\ref{Pen-Dia-Expanding-Bubble} the Penrose diagram of the bubble geometries, starting from the case of expanding solutions.
The causal structure depends on the region of the parameter space in fig.~\ref{Phase-Diagram}, as determined by the choice of parameters $(\lambda, \kappa)$ and by the sign of the curvature parameters $\b_i(R_0)$ and $\b_o(R_0)$ (see appendix~\ref{ssec:sign_curvature_parameters} for more details).

We summarize the main features of the Penrose diagram case by case:
\begin{itemize}
\item In region $A$, the interior geometry is an infinite portion of the dS spacetime.
In the thin wall approximation, a curious feature of the expanding bubble is that a portion of an AdS boundary (on the left side of the Penrose diagram) is attached to the dS timelike infinities $\mathcal{I}^{\pm}$, that appear as the shell attains an infinite size.
\rep{It is not clear whether  
this pathological feature might be an artifact of the thin wall approximation or
if it is also present 
in a full solution of the scalar-gravity system~\cite{Bousso:2004tv,Freivogel:2005qh}.}
\item In region $B$, the geometry admits two distinct AdS boundaries: one on the left, and the other on the right side of the Penrose diagram.
This case is similar to an eternal AdS black hole, but here with the inclusion of a false bubble of vacuum (characterized by a cosmological constant with $-1<\l<0$) that reaches the left boundary.
The bubble geometry describes a vacuum instability mediated by a Coleman-de Luccia (CdL) instanton~\cite{Coleman:1980aw}. 
\item In region $C$, two  distinct AdS boundaries bifurcate from the intersection of the shell with the asymptotic AdS boundary of the BH background.
\rep{Similarly to  region $A$, it is not clear if this
counterintuitive feature is also present in the
full solution of the scalar-gravity system.}
\item In region $E$, there is a single AdS boundary on the right side of the Penrose diagram.
This case is similar to an AdS spacetime with fixed cosmological constant that includes a false 
CdL bubble of vacuum with $\l=-1$.
In terms of the boundary theory, the CdL bubble corresponds to a Fubini instanton of the boundary CFT~\cite{Barbon:2010gn}.
\end{itemize}

\begin{figure}[h!]
\center
\subfigure[between region A and B]{ \includegraphics[scale=0.3]{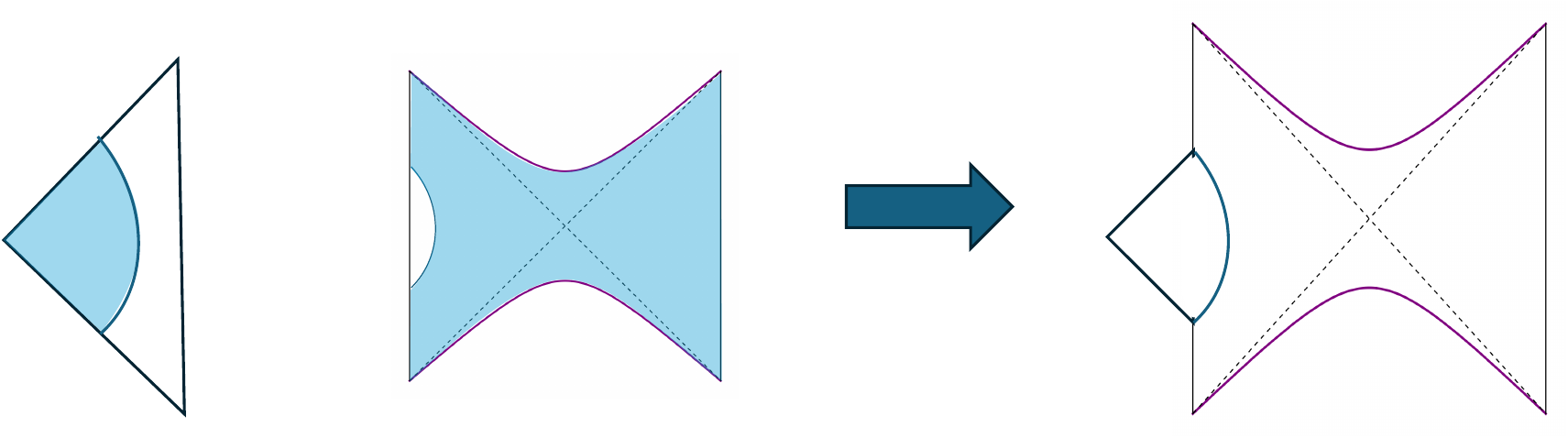}} \quad
\subfigure[between region A and C]{ \includegraphics[scale=0.3]{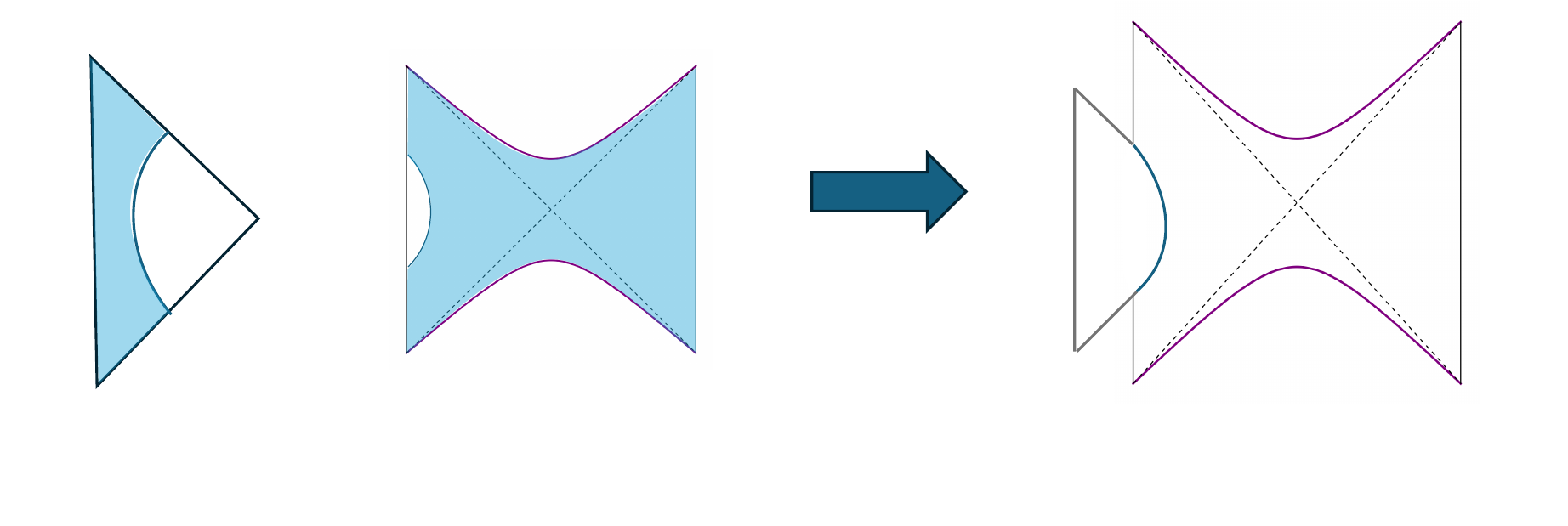}}
\caption{Penrose diagrams of an expanding bubble for an interior geometry with $\l=0$. }
\label{fig:Penrose-diagram-lambda-0}
\end{figure}

The structure of the Penrose diagram of the expanding bubble is special for $\l=0$: when $0<\kappa<1$, the expanding bubble is characterized by $\b_i(R_0)<0$; when $\kappa>1$, we have $\b_i(R_0)>0$
(see appendix~\ref{ssec:sign_curvature_parameters}).
In both cases, one of the AdS timelike boundaries is attached to the null infinity of the flat space, as depicted in the Penrose diagrams sketched in fig.~\ref{fig:Penrose-diagram-lambda-0}.
We believe that this is an artifact of the thin wall approximation, too.
  
\subsection{Causal structure of collapsing bubbles}
\label{ssec:causal_structure_collapsing_bubble}

\begin{figure}[h!]
\center
\subfigure[case I]{ \includegraphics[scale=0.23]{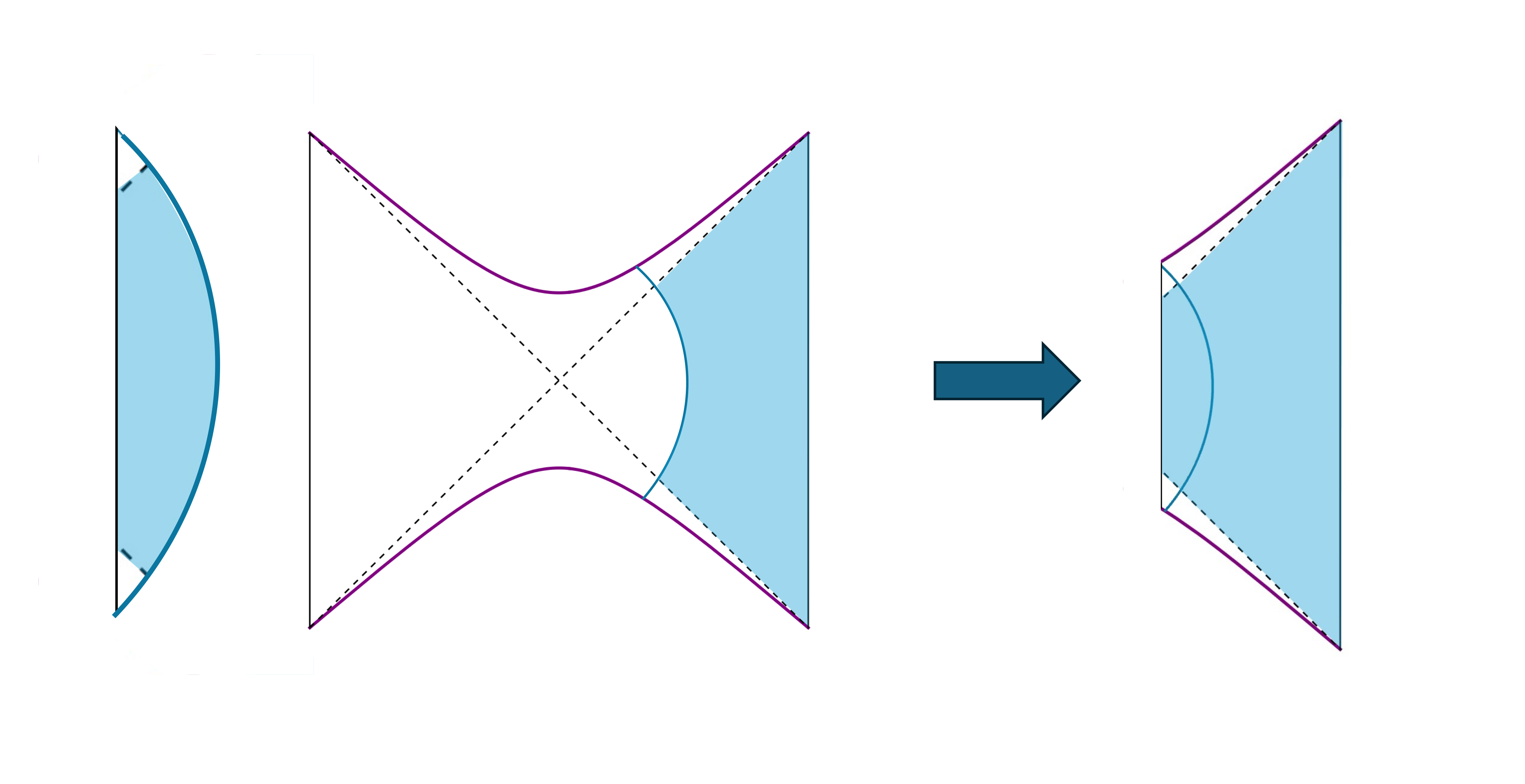}} 
\subfigure[case II]{ \includegraphics[scale=0.23]{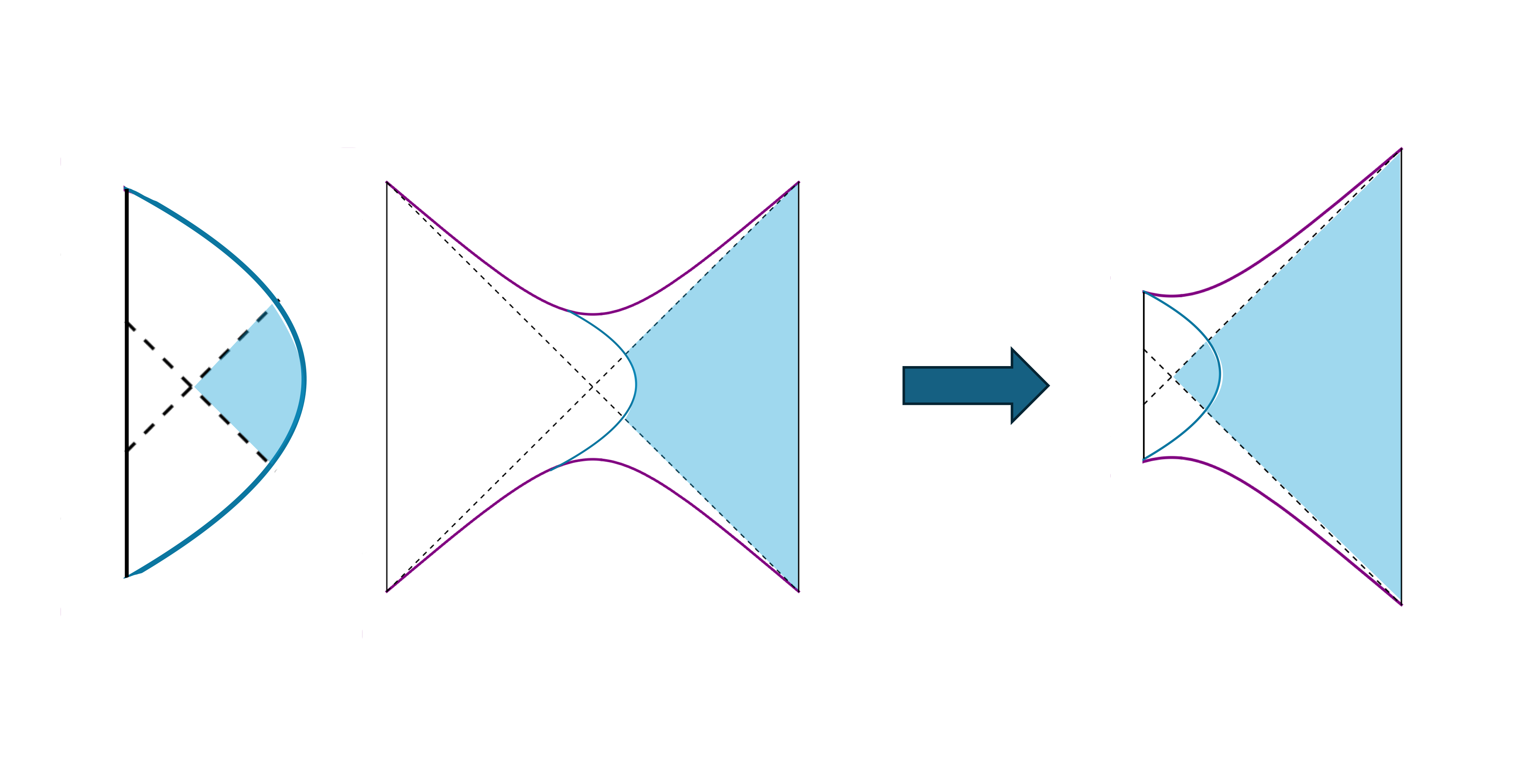}}
\subfigure[case III]{ \includegraphics[scale=0.23]{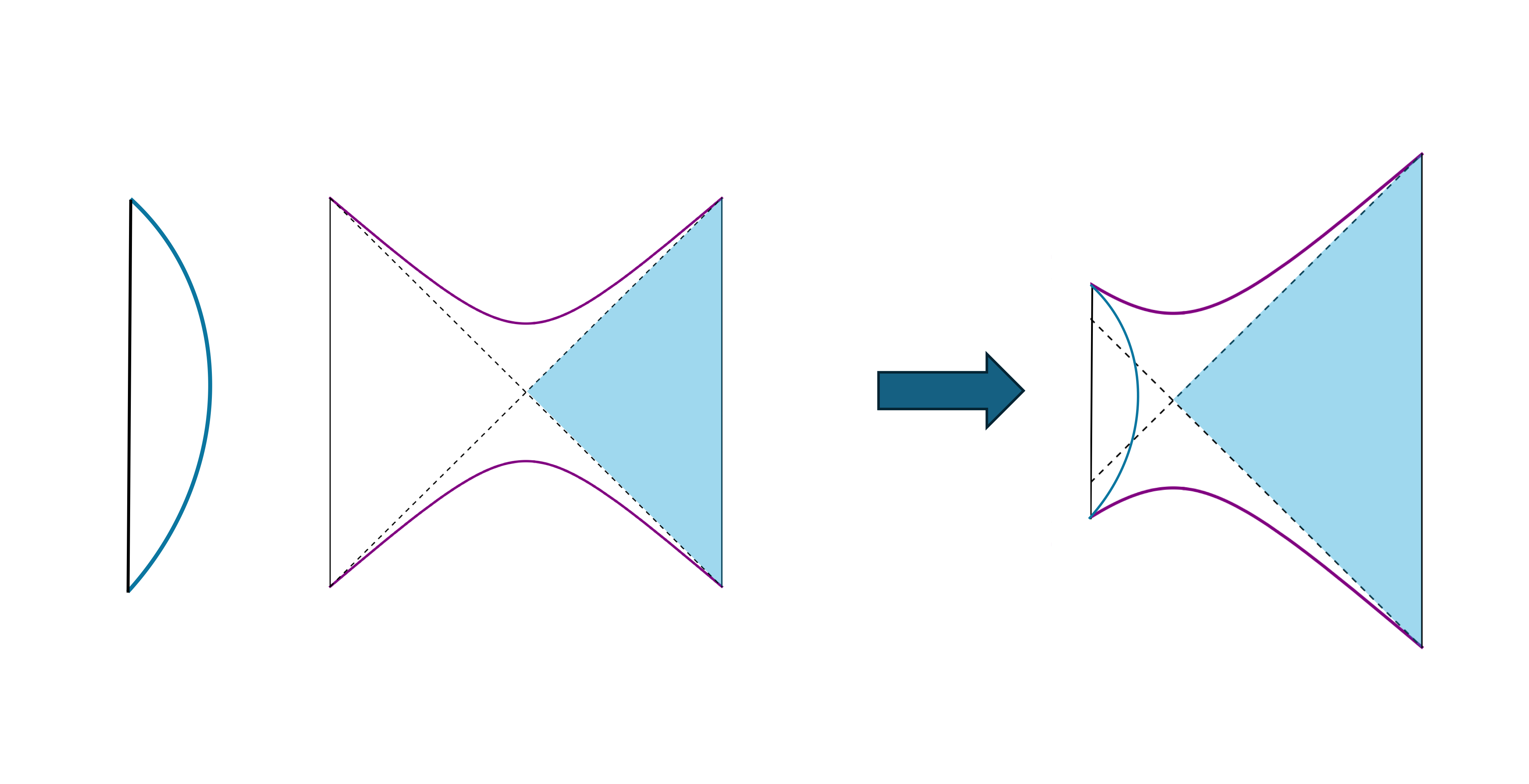}}
\caption{Penrose diagrams for collapsing bubbles.
 We show in blue the trajectory of the domain wall, both in the 
 interior geometry (left panel) and in the exterior AdS black hole (right panel). The interior background can be either a portion of empty AdS, dS or flat spacetime. 
}
\label{Pen-Dia-Collapsing-Bubble}
\end{figure}

In contrast to the expanding bubble investigated in subsection~\ref{ssec:causal_structure_expanding_bubble}, the Penrose diagram of a collapsing bubble is qualitatively independent of the sign of $\lambda$. 
There exist three different structures, which depend on the shape of the causal wedge of the AdS boundary, see fig.~\ref{Pen-Dia-Collapsing-Bubble}.
In case I, the center of the bubble is causally connected
with the right AdS boundary. In case II, the center of the bubble is casually disconnected from the AdS boundary. However, the bubble is located outside the BH horizon, therefore part of the interior geometry is still causally connected with the AdS boundary. 
In case III, all the interior of the bubble is causally disconnected from the right AdS boundary.

\begin{figure}[ht]
\center
\includegraphics[scale=0.4]{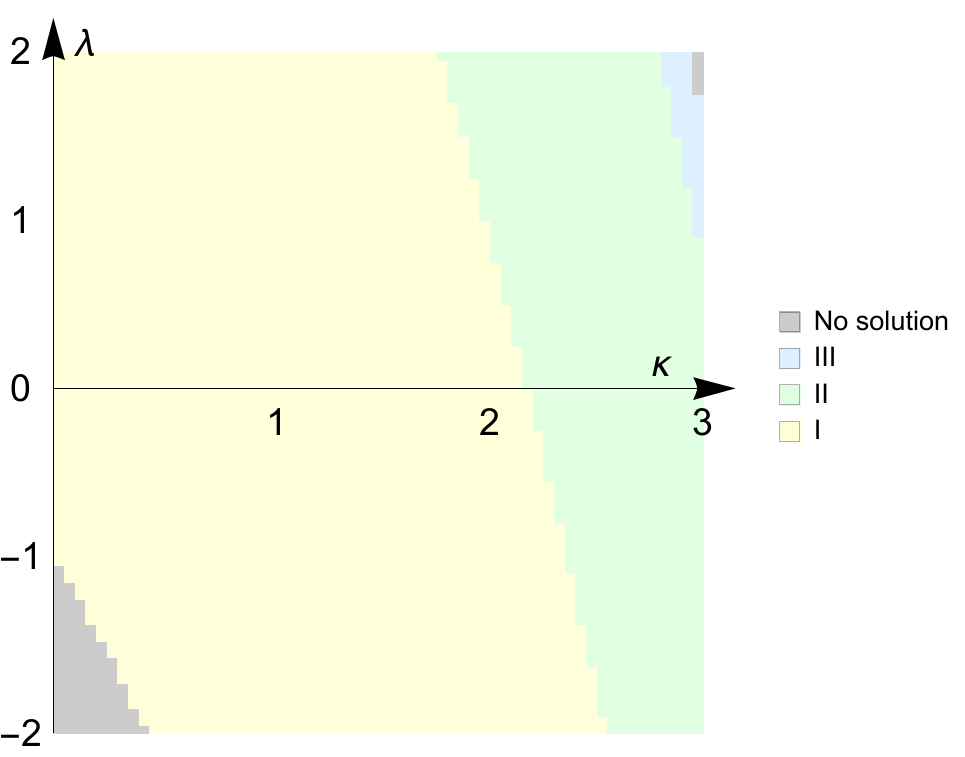}
\qquad
\includegraphics[scale=0.4]{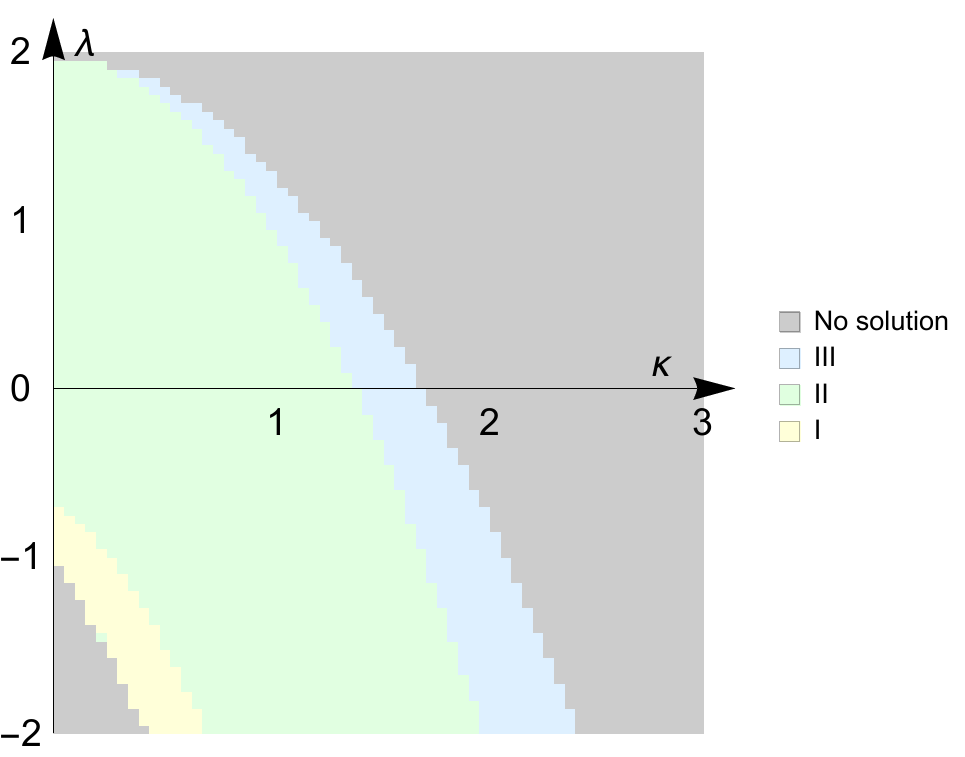}
\qquad
\includegraphics[scale=0.4]{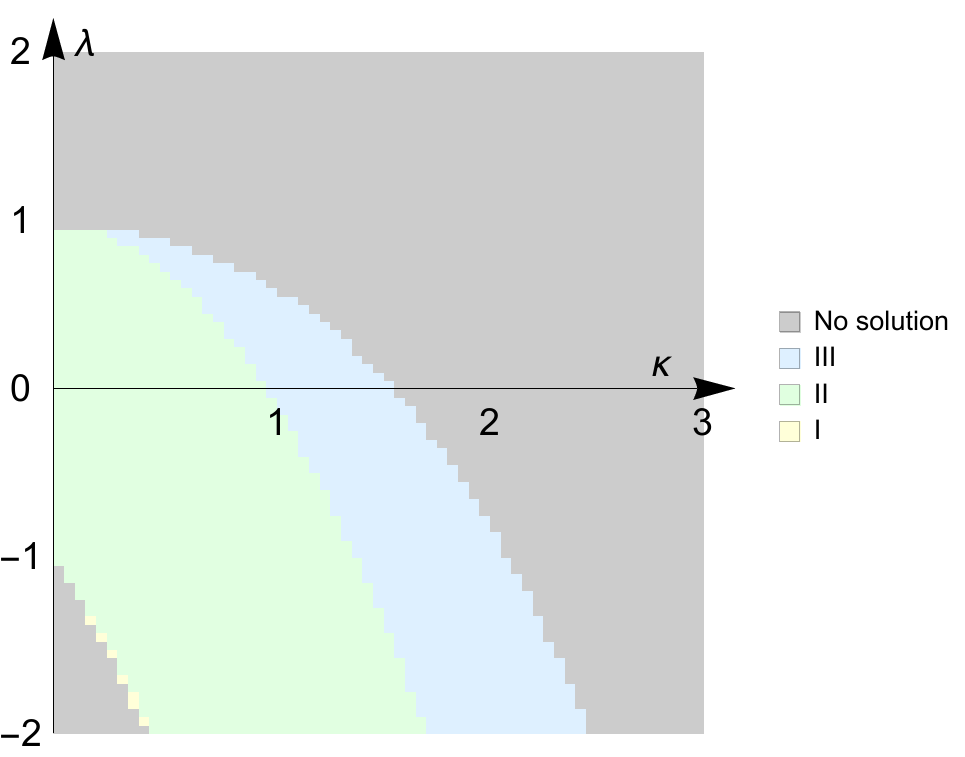}
\qquad
\includegraphics[scale=0.4]{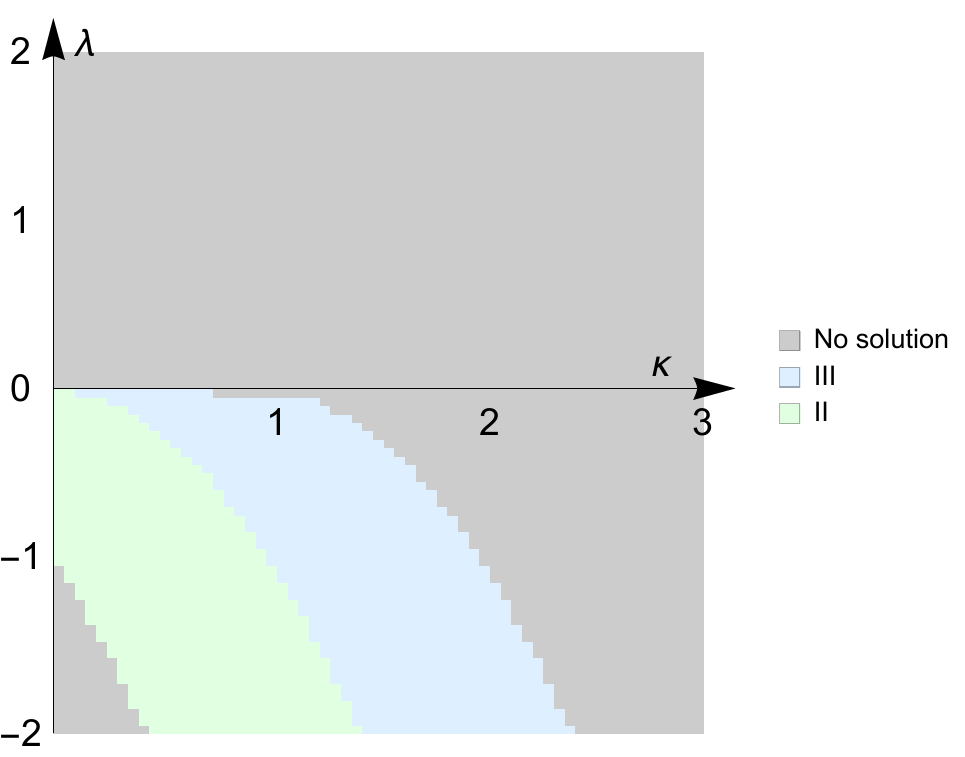}
\caption{ 
Numerical scan in parameter space of the collapsing bubble
for $d=2$, in the $(\kappa,\l)$
plane and for different values of $m=1.1$ (top-left),
$m=1.5$ (top-right),
$m=2$ (bottom-left) and $m=100$ (bottom-right).
In the gray region, there is no time-reversal
symmetric bubble. The classification I, II and III refers to the shape of the causal wedge, see fig.~\ref{Pen-Dia-Collapsing-Bubble}.
}
\label{figure-beta-outside-d2}
\end{figure}

\begin{figure}[ht]
\center
\includegraphics[scale=0.4]{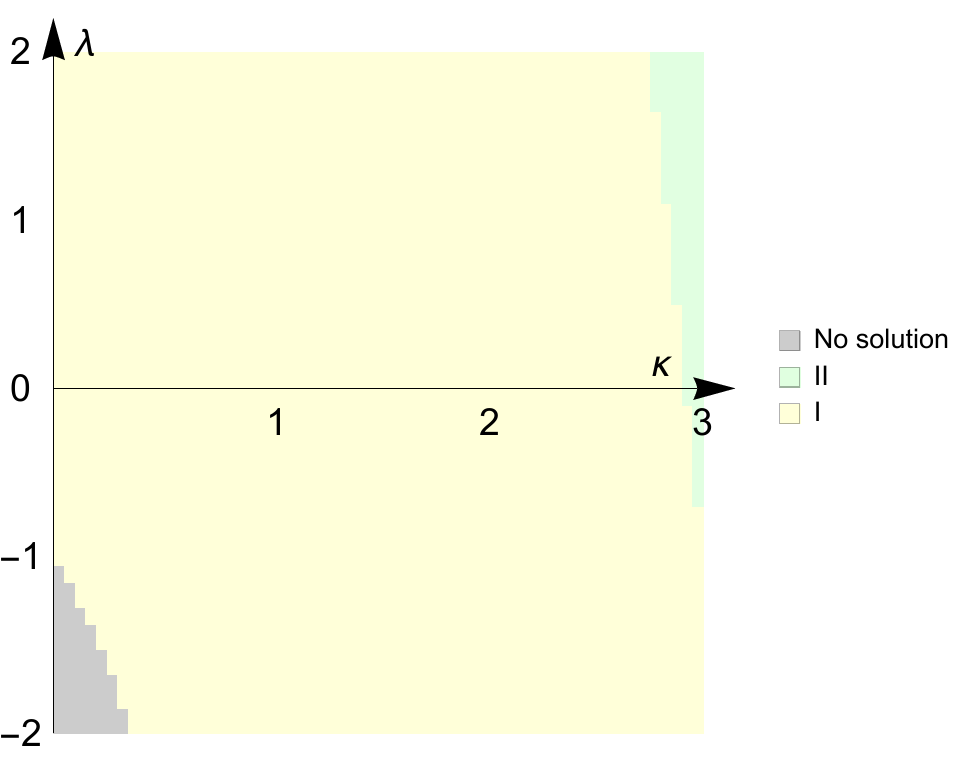}
\qquad
\includegraphics[scale=0.4]{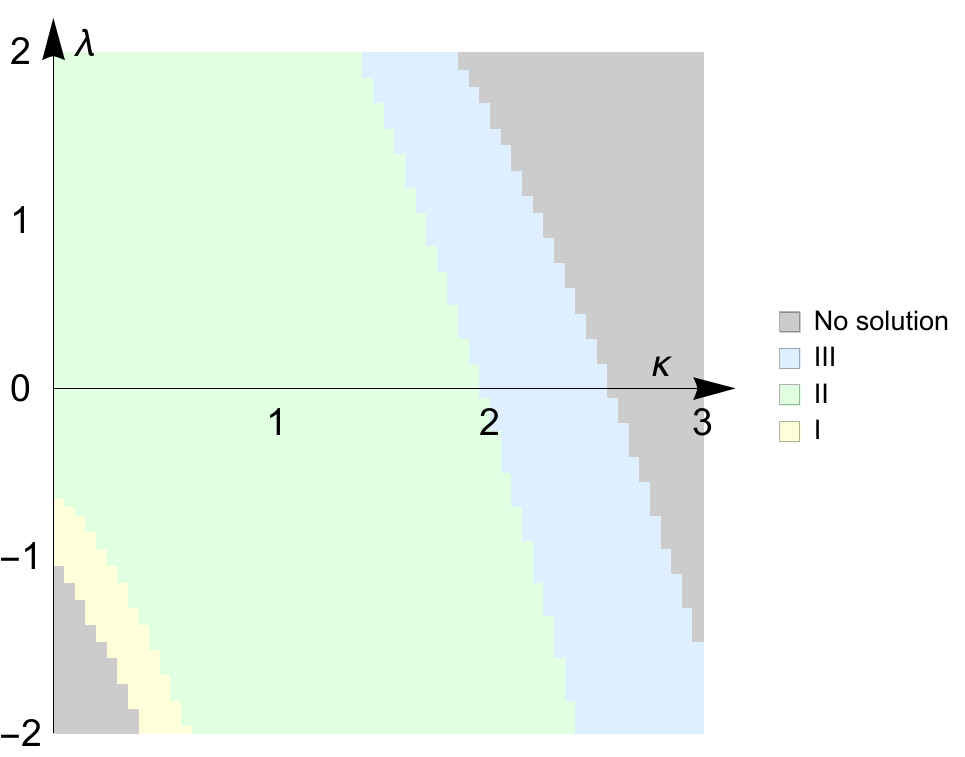}
\qquad
\includegraphics[scale=0.4]{Figures/ScanCollapsingM2d0.pdf}
\qquad
\includegraphics[scale=0.4]{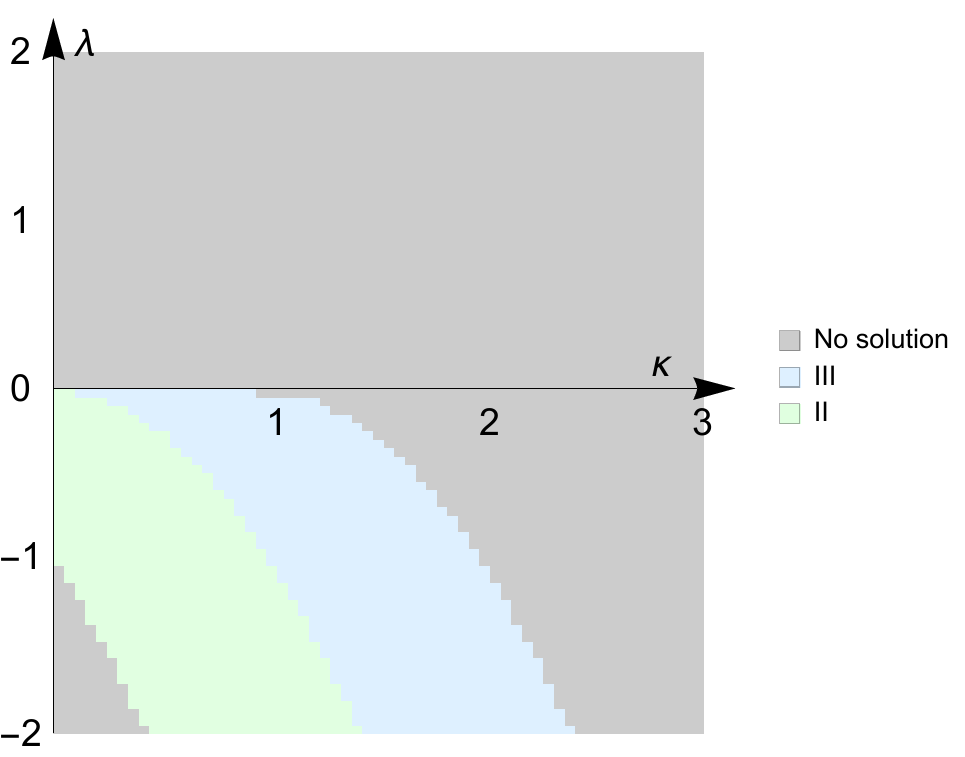}
\caption{ 
Numerical scan in parameter space of the collapsing bubble
for $d=3$, in the $(\kappa,\l)$
plane and for different values of $m=0.1$ (top-left),
$m=0.625$ (top-right),
$m=2$ (bottom-left) and $m=500$ (bottom-right).
In the gray region, there is no time-reversal
symmetric bubble. The classification I, II and III refers to the shape of the causal wedge, see fig.~\ref{Pen-Dia-Collapsing-Bubble}.
}
\label{figure-beta-outside-d3}
\end{figure}

Let us now scan the parameter space of the collapsing bubble, that we depict in figs.~\ref{figure-beta-outside-d2} and \ref{figure-beta-outside-d3} for $d=2$ and $d=3$, respectively. 
Case III is characterized by the constraint $\b_o(R_0)<0$, while cases I and II by the condition $\b_o(R_0)>0$.
To discriminate between the latter two cases, we numerically compute the radial coordinate $r=r_I$ of the intersection between the bubble trajectory and an outgoing light ray which starts from the center of the bubble at $t=0$. The inequality $r_I>r_h$ identifies case I, while $r_I<r_h$ corresponds to case II.
The parameter space of a collapsing bubble presents a curious behavior in correspondence of the Hawking-Page transition (\eg see Ref.~\cite{Hawking:1982dh}). 
In our conventions (see subsection~\ref{ssec:bubble_metric}), the transition occurs when the horizon radius satisfies $r_h=1$ (corresponding to $m=2$) in any dimension.
When $m=2$, the region corresponding to case I is almost excluded (except for a tiny region), see the bottom-left panel of figs.~\ref{figure-beta-outside-d2} and \ref{figure-beta-outside-d3}.

At the threshold of the black hole mass, which is at $m=1$ in $d=2$ and at $m=0$ in $d \geq 3$, the condition $\b_o(R_0)>0$ always holds  for the collapsing bubble. This fact can be proven as follows:
\begin{itemize}
\item
For $d \geq 3$, the limit $m \to 0$ implies $R_0 \to 0$, see eq.~\eqref{eq-massa-bubble-2}. Then, it is enough to notice that collapsing bubbles necessarily have $\beta_i(R_0)>0$. This selects the solution $\beta_o(R_0) = -\kappa R_0 + \sqrt{1- \lambda R_0^2}$ in eq.~\eqref{eq:beta:discrimine}. In the limit $R_0 \to 0$, we get $\beta_o(R_0) \to 1$.
\item For $d=2$ and $m=1$ (equivalently, $\mu=0$), we find, by solving eq.~(\ref{eq-massa-bubble-2}):
\beq
R_0=\frac{1}{\sqrt{\l+(\kappa \pm 1)^2}} \, .
\label{eq-mu-equal-1}
\eeq
The value of $R_0$ with the $+$ sign in the denominator of eq.~(\ref{eq-mu-equal-1}) always corresponds to the collapsing solution, because it is smaller than the root with the $-$ sign. 
We can then use eq.~(\ref{eq:beta:discrimine}) to check that the relation $\b_o(R_0)>0$ holds for the collapsing solution.
\end{itemize}

Finally, we show that when $\l \leq -\kappa^2$, the collapsing bubble is located outside the BH bifurcation surface. 
Combining eqs.~\eqref{mass expression} and \eqref{beta-i-beta}, we find 
\beq
\b_o(r_h)=\frac{1-r_h^2 (\l+\kappa^2)}{2 \kappa r_h} \, .
\eeq
The above expression vanishes for $r_h = (\lambda + \kappa^2)^{-1/2}$.
When $\l \leq -\kappa^2$, this solution is imaginary and therefore must be discarded, implying that $\b_o(r_h)$ does not vanish for any real value of $r_h$. 
When $\l<-\kappa^2-1$, we have $\b_o (r_h)>0$.\footnote{The identity $\lambda = -\kappa^2 - 1$ corresponds the blue curve in fig.~\ref{region-parameters-betai-betao}.}
Combining this fact with the above observation that $\b_o(R_0)>0$, we conclude by continuity that collapsing bubbles are initially located outside the BH horizon for $\l \leq -\kappa^2$.
This is consistent with the numerical study reported in figs.~\ref{figure-beta-outside-d2} and \ref{figure-beta-outside-d3}.


\subsection{Static bubbles}
\label{ssec:mass_static_bubble}

\begin{figure}[ht]
    \centering
\subfigure[]{\includegraphics[scale=0.3]{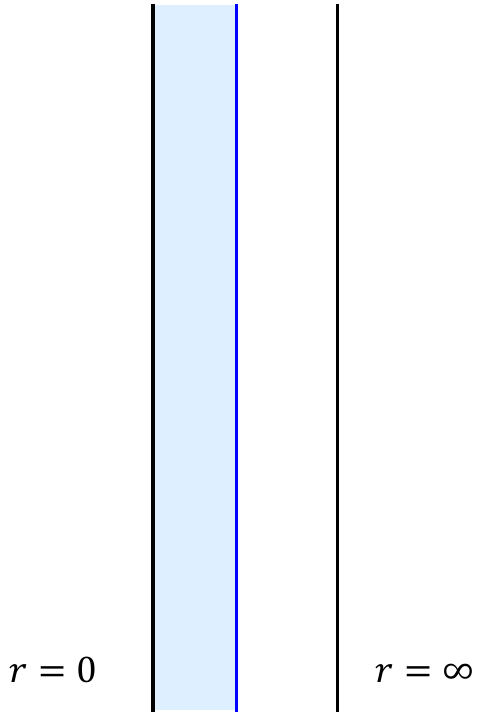}}\qquad \qquad
\subfigure[]{\includegraphics[scale=0.2]{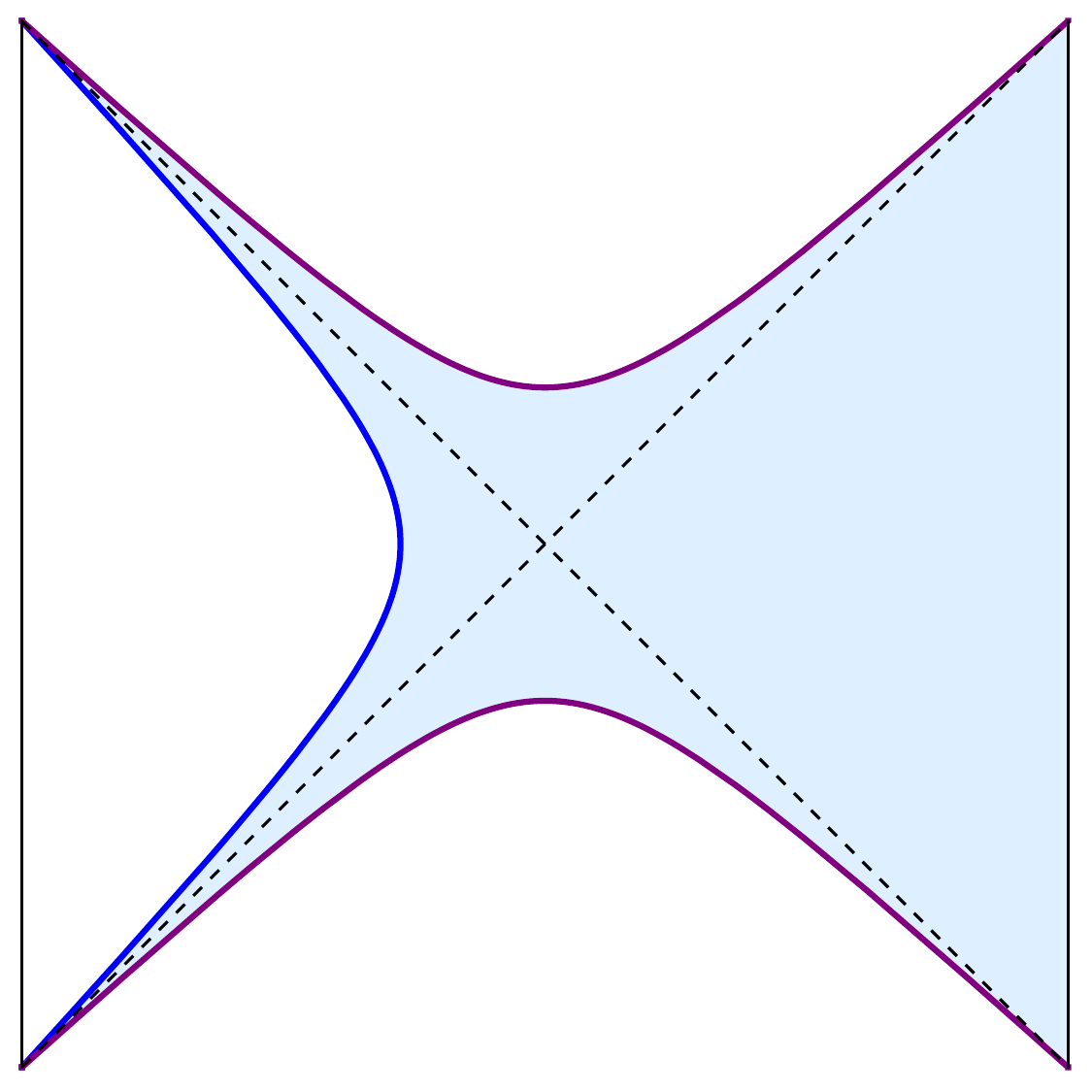}}
    \caption{Penrose diagram of the bubble geometry in the static bubble setting with (a) AdS interior and (b) external black hole background.}
    \label{fig:Penrose_static_AdS}
\end{figure}

\begin{figure}[ht]
    \centering 
    \subfigure[]{\includegraphics[scale=0.2]{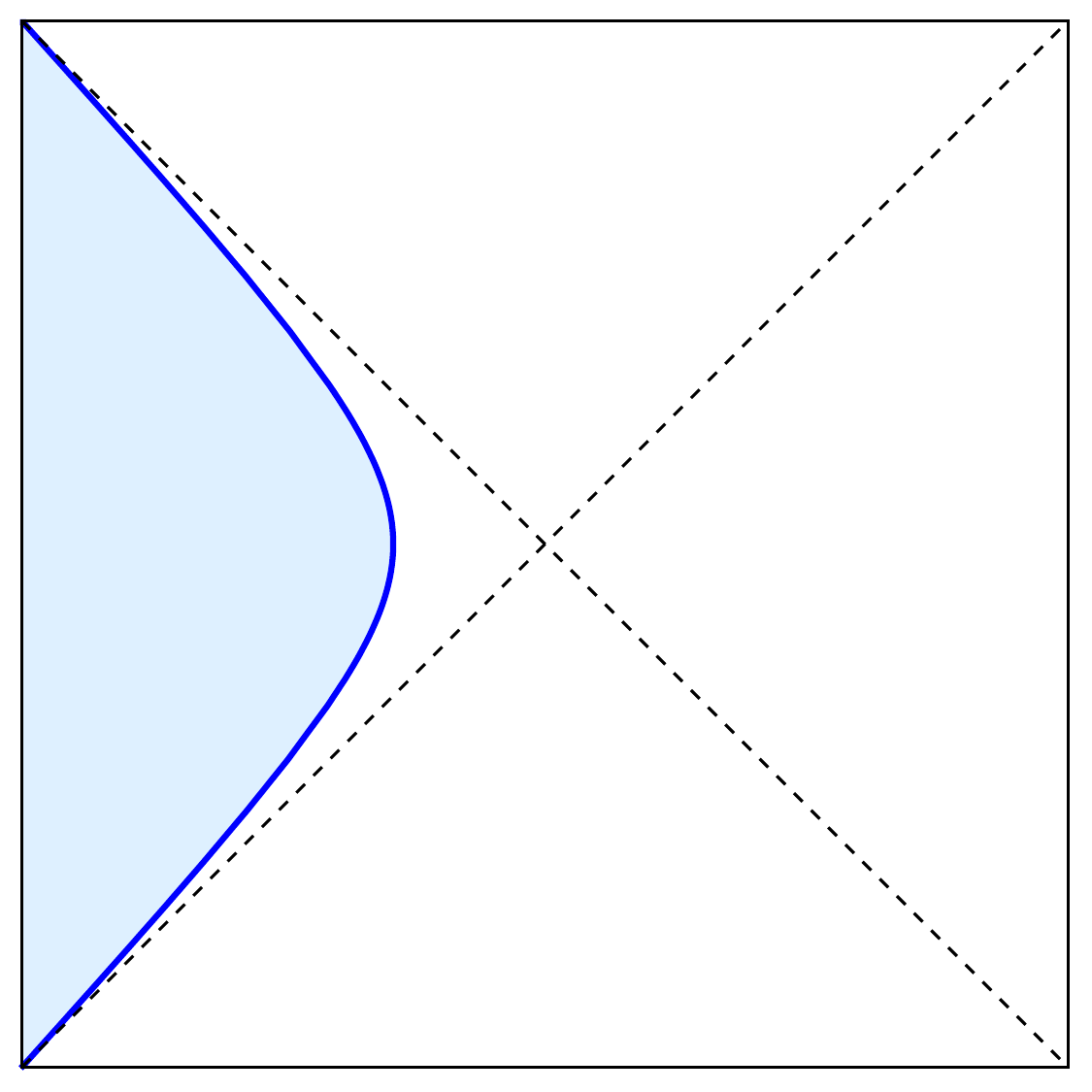}}\qquad \qquad
    \subfigure[]{\includegraphics[scale=0.2]{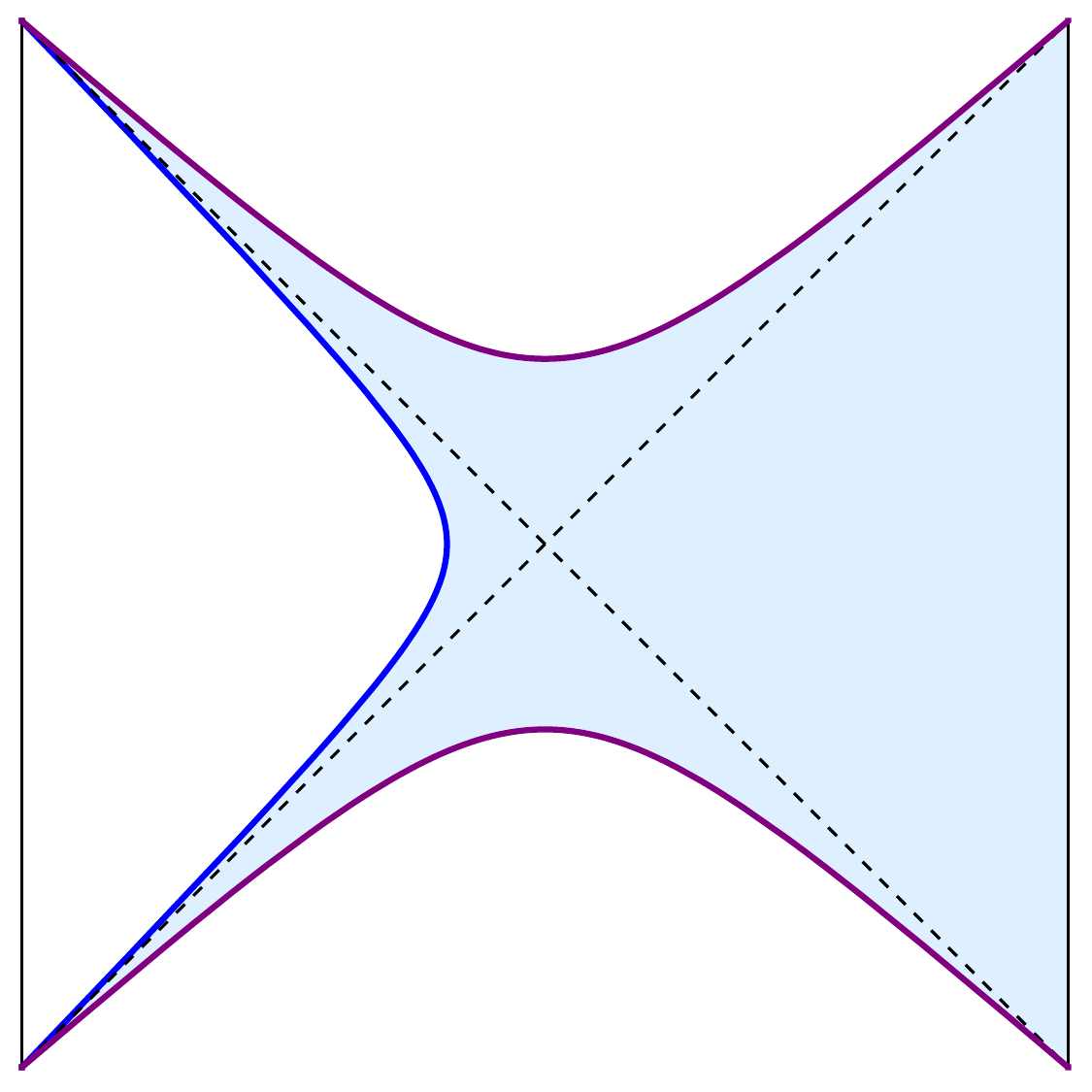}}
     \caption{Penrose diagram of the bubble geometry in the static bubble setting with (a) dS interior and (b) external black hole background.}
\label{fig:Penrose_static_dS}
\end{figure}

As anticipated in subsection~\ref{ssec:parameter-space}, the static bubble configuration only exists in regions $A, C$ and $E$ of the parameter space illustrated in fig.~\ref{Phase-Diagram}.
In regions $A, C$ the domain wall is located inside the BH horizon, while in region $E$ it sits outside the BH horizon.
As representative examples, we depict the Penrose diagram for a static bubble with an interior AdS geometry (referring to region $C$) in fig.~\ref{fig:Penrose_static_AdS}, and with a dS interior (case $A$) in fig.~\ref{fig:Penrose_static_dS}.

\begin{figure}[ht]
    \centering
\includegraphics[scale=0.5]{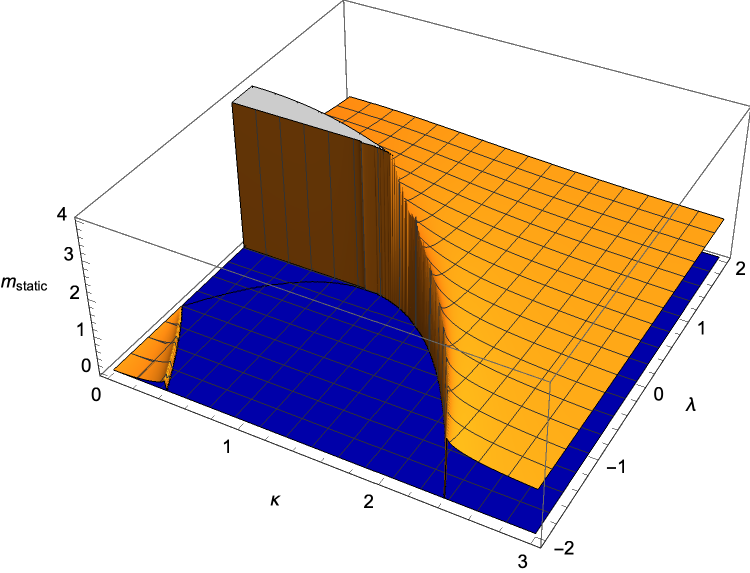}
\qquad
\includegraphics[scale=0.5]{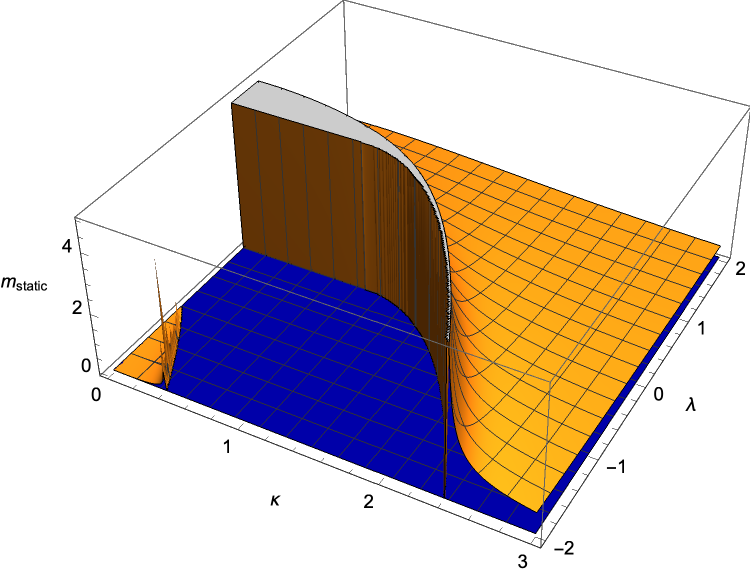}
    \caption{Mass of the static bubble in $d=2$ (left panel)
    and in $d=3$ (right panel)
    as a function of $(\kappa, \lambda)$ in regions $A, C, E$ 
    of the phase diagram in fig.~\ref{Phase-Diagram}. 
    \rep{The static bubble does not exist in regions $B$    and $D$.
    For $d=2$, the mass diverges
    just at the boundary between regions   $(A,B)$.
    For $d=3$, the mass diverges
    at the boundaries between 
    regions $(A,B)$, $(C,D)$ and $(D,E)$.}
    }
    \label{fig:mass_static_2d3d}
\end{figure}

Next, let us determine the mass parameter $m_{\rm static}$ of the BH solution and the fixed radius $R_{\rm static}$ of the shell as functions of the cosmological constant $\lambda$ and of the domain wall's tension $\kappa$.
Imposing the definition of static bubble $\dot{R}=0$ inside the equations of motion~\eqref{general-V-bubble} leads to the equivalent condition $V_{\rm eff}(R_{\rm static})=0$.
Moreover, in order to be a static solution, we require that the bubble is an extremum of the potential $(V'_{\rm eff}(R_{\rm static})=0)$.
Combining the above constraints, we get
\begin{subequations}
    \beq
V_{\rm eff}(R)= 0 \, \Rightarrow \,
\le (1+\l+\kappa^2)^2-4\kappa^2 \ri R^{2d}
    - 4 \kappa^2 R^{2(d-1)}
    -2m(1+\l-\kappa^2)R^{d}+m^2=0 \, ,
     \label{eq:Rstatic_1}
    \eeq
    \beq
V_{\rm eff}'(R)=0 \,  \Rightarrow \,
\le (1+\l+\kappa^2)^2-4\kappa^2 \ri R^{2d}+(d-2)m(1+\l-\kappa^2)R^{d}+(1-d)m^2=0 \, .
\label{eq:Rstatic_2}
    \eeq
\end{subequations}
By subtracting eq.~\eqref{eq:Rstatic_1} from \eqref{eq:Rstatic_2}, we get the simpler identity
\beq
d m \le 1 + \lambda - \kappa^2 \ri R^d + 4 \kappa^2 R^{2(d-1)} 
-d m^2 = 0 \, .
\label{eq:diff_Rstatic_bubble}
\eeq
There is no closed form solution to the previous set of equations for unspecified $d$.
We will present explicit results for $d=2,3$
\begin{itemize}
\item
{\bf $d=2$.}
The mass of the three-dimensional static bubble reads
\cite{Auzzi:2023qbm} 
\beq
m_{\rm static}^{d=2}
=\frac{\sqrt{(\kappa^2+\l-1)^2 +4 \l}-(\kappa^2+\l-1)}{2 \l} +1\, .
\label{eq:mstatic_d2}
\eeq
By plugging the expression~\eqref{eq:mstatic_d2} inside
eq.~\eqref{eq:diff_Rstatic_bubble}, we determine the constant radius of the shell
\beq
 R_{\rm static}^{d=2} 
= \frac{m_{\rm static}^{d=2}}{ 
\sqrt{m_{\rm static}^{d=2}  \le 1 + \lambda - \kappa^2 \ri 
+ 2 \kappa^2}}  \, .
\label{eq:Rstatic_sol2_2d}
\eeq
\item
{\bf $d=3$.}
In four spacetime dimensions, eqs.~\eqref{eq:Rstatic_2} and \eqref{eq:diff_Rstatic_bubble} become
\begin{subequations}
\beq
\le (1+\l+\kappa^2)^2-4\kappa^2 \ri R^{6}+ m(1+\l-\kappa^2)R^{3} -2 m^2=0 \, ,
\label{eq:Rstatic2_3d}
\eeq
\beq
 4 \kappa^2 R^{4} + 3 m \le 1 + \lambda - \kappa^2 \ri R^3 
-3 m^2 = 0 \, .
\label{eq:diff_Rstatic_3d}
\eeq
\end{subequations}
The second identity admits two possible solutions for $R_{\rm static}$, but one of them needs to be discarded because it is always negative.
The only allowed root is given by
\beq
(R_{\rm static}^{d=3})^{3} = \frac{4m}{1+\lambda-\kappa^2 + \sqrt{9 \kappa^4 + 9 (\lambda+1)^2 +2 \kappa^2 (7\lambda -9) } } \, .
\label{eq:Rstatic_sol2_3d}
\eeq
Plugging eq.~\eqref{eq:Rstatic_sol2_3d} inside~\eqref{eq:diff_Rstatic_3d}, we obtain the mass of the AdS black hole for a static shell
\beq
m_{\rm static}^{d=3} = \frac{128 \kappa^3}{3 \sqrt{3}} \frac{1}{\sqrt{\le 1+\lambda-\kappa^2  + \mathcal{M}(\kappa, \lambda) \ri \le -3-3 \lambda +3 \kappa^2 + \mathcal{M}(\kappa, \lambda)  \ri^{3}  }} \, ,
\label{eq:mstatic2_3d}
\eeq
where we introduced the short-hand notation
\beq
\mathcal{M} (\kappa, \lambda) \equiv  \sqrt{9 \kappa^4 + 9 (\lambda+1)^2 +2 \kappa^2 (7\lambda -9) }   \, .
\label{eq:quantityMcal}
\eeq
\end{itemize}
We plot the mass of the static bubble as a function of the domain wall's tension $\kappa$ and the cosmological constant $\lambda$ in fig.~\ref{fig:mass_static_2d3d}.

It is interesting to consider a few limiting cases, in particular:
\begin{itemize}
\item The mass and the radius of the static bubble
diverge for $\l \to 0$ and $0<\kappa<1$, corresponding to the segment separating regions $A$ and $B$ of the phase diagram in fig.~\ref{Phase-Diagram}, approached from above. In this limit, using eqs.~\eqref{eq:Rstatic_sol2_2d} and \eqref{eq:Rstatic_sol2_3d} and the approximation $r_h\approx m^{1/d}$ valid for large $m$, we get
    \beq
\frac{R_{\rm static}}{r_h} \approx \frac{1}{(1-\kappa^2)^{1/d}} \, .
\label{eq:property-static-AB}
\eeq
In the special case where $\kappa \to 1$, we further find that  $R_{\rm static} \gg r_h$. This latter case was investigated in section 2.3 of Ref.~\cite{Freivogel:2005qh}.
\item
At the boundary between regions $C, D$ described by the curve $\lambda = \lambda_2$ defined in eq.~\eqref{eq:critical_lambda} with $\kappa>1$
we find that
\beq
m_{\rm static}^{d=2}=\frac{\kappa}{\kappa-1} \, ,
\label{eq:mstatic_d2_case1}
\eeq
which is finite and positive for $\kappa>1$.
On the contrary, $m_{\rm static}^{d=3}$ diverges. For both $d=2,3$ we find 
    \beq  
\lim_{\lambda \to \lambda_2}
\frac{R_{\rm static}}{r_h} \to \infty \, .
\label{eq:property-static-CD}
    \eeq
    \item
    On the curve $\lambda=\lambda_1$ 
defined in eq.~\eqref{eq:critical_lambda}, delimiting regions $D, E$ we obtain
\beq
m_{\rm static}^{d=2}=\frac{\kappa}{\kappa+1} \, ,
\label{eq:mstatic_d2_case2}
\eeq
which corresponds to a negative value of $\mu_{\rm static}$.\footnote{A negative value of $\mu$ is allowed in region $E$ because the horizon is cut out from the gluing to the interior geometry.}
On the contrary, $m_{\rm static}^{d=3}$ diverges in this limit.
The radius $R_{\rm static}$ is divergent both in $d=2$
and in $d=3$.
\end{itemize}

It is worth observing that the mass of the static bubble is never divergent when the interior geometry is empty dS space.
The reason is that when $\lambda >0$, the black hole horizon $r_h$ is bounded by the dS radius $L_{\rm dS}$ as follows~\cite{Freivogel:2005qh},
\beq
r_h \leq L_{\rm dS} = \frac{1}{\sqrt{\lambda}} \, ,
\eeq
thus implying that the mass is bounded from above, too.

%% file: Sections/Entropy.tex
\section{Entanglement entropy}
\label{sec:EE_bubbles}

Having described the properties of a generic bubble solution in section~\ref{sec:preliminaries}, our next goal is to employ the AdS/CFT correspondence to compute physical properties that correspond to geometric objects probing the (A)dS interior of the shell.
In this section,
we compute the holographic entanglement entropy (EE)  of a boundary subregion using the HRT formula \cite{Hubeny:2007xt}
\begin{equation}
    S_{\mathcal{R}} = \frac{\mathcal{A}_{\mathcal{R}}}{4G} \, ,
    \label{eq:RT_formula}
\end{equation}
where $\mathcal{A}_{\mathcal{R}}$ is the area of the extremal codimension-two hypersurface with minimal area anchored at the boundary subregion ${\mathcal{R}}$ and homologous to it. 

Non-minimal extremal surfaces also play an important role in the holographic duality, for instance in the 
\textit{Python's lunch conjecture}~\cite{Brown:2019rox}, see also \cite{Engelhardt:2021mue,Engelhardt:2021qjs,Arora:2024edk,Bak:2021qbo,Akers:2022qdl,Akers:2024wre,May:2024epy}.
This conjecture states that if a geometry admits a local minimal surface that is not globally minimal, then it is exponentially hard 
to reconstruct information on the spacetime region between the local minimal and the global minimal solution.
This hardness is quantitatively measured by the \textit{restricted complexity} $\mathcal{C}_R$, a concept originally introduced to compute the difficulty of decoding the information on the BH's interior contained within the Hawking radiation~\cite{Harlow:2013tf,Balasubramanian:2022fiy,Gyongyosi:2022vaf}.
Reference~\cite{Brown:2019rox} conjectured that the complexity to reconstruct a spacetime region is geometrically encoded by the \textit{Python's lunch}, \ie a bulge inside a wormhole connecting different asymptotic regions.
The bulge is identified through a \textit{maximinimax} procedure as follows (see, \eg Ref.~\cite{Arora:2024edk}).
Let us consider a smooth path $x_{\Sigma}(s)$ that interpolates between the global and the local minima surfaces on a given spatial slice $\Sigma$, where $s$ is a continuous parameter.\footnote{Alternative definitions of the bulge in terms of level set functions are given, \eg in Refs.~\cite{Brown:2019rox,May:2024epy}.}
We first define
\beq
\mathcal{A}_{\rm{bulge}, \Sigma}=
\min_{x_{\Sigma}(s)} \, \max_{s} \, \mathcal{A}(x_{\Sigma}(s))
\eeq
where $\mathcal{A}(p)$ is the area of the minimal surface $p=x_{\Sigma}(s)$, which is homologous to the boundary region.
The bulge surface, whose area we denote by $\mathcal{A}_{\rm bulge}$, is then defined by a maximization of $\mathcal{A}_{\rm{ bulge}, \Sigma}$ over the choice of the spatial slice $\Sigma$.
In turn, the holographic restricted complexity is given by
\beq 
\mathcal{C}_R \propto \exp \le \frac12 \frac{\mathcal{A}_{\rm bulge}-\mathcal{A}_{\rm ext}}{4 G} \ri \, ,
\label{eq:restricted-complexity}
\eeq
where $\mathcal{A}_{\rm ext}$ is the area of the external (false minimum) extremal surface.

Let us now consider possible extremal, but non-minimal, surfaces associated with the full AdS boundary $S^{d-1}$. 
The vanishing surface at $r=0$ may not be the only extremal solution satisfying the homology condition (see also the discussion in \cite{Sahu:2023fbx} for a similar setting). 
Examples in which the bulge surface (and local minimal surfaces) break the time reversal symmetry are studied, \eg in Ref.~\cite{Engelhardt:2023bpv}. 
On top of that, it has been explicitly shown that bulge surfaces on time-reflection symmetric slices can spontaneously break the spherical symmetry~\cite{Arora:2024edk}, as we discuss below.
Let us consider extremal surfaces at $t=0$ with spherical symmetry. If the bubble is initially on the left side of the Penrose diagram, the bulk slice at $t=0$ includes the BH bifurcation surface, which is a locally minimal surface. In this situation, a Python's lunch is realized. 
If the bubble does not contain a full dS static patch, a bulge surface is given by the intersection between the bulk slice $t=0$ and the domain wall, since the radius of the transverse sphere grows from the BH horizon to the domain wall and decreases from the domain wall to the center of the (A)dS bubble.
If the bubble contains a full dS static patch, the surface of the bubble cannot be a bulge surface, since the radius of the transverse sphere is monotonic nearby the domain wall.
The cosmological bifurcation surface also cannot be a bulge surface, because it is an extremal surface which locally maximizes the area functional. \rep{In this case, either the spherical symmetry or the assumption that the surface lies at $t=0$ must fail.}

\rep{At fixed $G_N$, when $\mathcal{A}_{\rm bulge} \gg \mathcal{A}_{\rm ext}$ (in our case, $\mathcal{A}_{\rm ext}$ is the area of the horizon), the restricted complexity $\mathcal{C}_R$ is exponentially large in the difference of the areas.}
In particular, due to eq.~\eqref{eq:property-static-CD}, we find that this scenario is realized for the static bubble at the boundary between region $C$ and $D$.

In the remainder of this section we specialize to $d=2$, where the exterior geometry is a BTZ black hole and the HRT surface reduces to a geodesic.
We will consider extremal surfaces associated with a segment.
We stress that, while collapsing bubbles initially located outside the BH do not present a Python's lunch associated with the whole AdS boundary at $t=0$, this is not necessarily true in the case of a generic subregion.
Indeed, we will observe below that a Python's lunch may take place for such bubble solutions depending on the size of the boundary subregion. 

This section is organized as follows.
We explore the construction of geodesics at constant time in
the interior and exterior parts of the geometry in
section~\ref{ssec:const_t_geod}. We then discuss the conditions for
gluing geodesics across the domain wall of a bubble geometry in section~\ref{ssec:joining_geo}.
In section~\ref{subsec:geod-in-bubble} we determine the
full solution, obtained by gluing the two parts of the geodesic.
We present the solution with minimal length (the holographic EE) in section~\ref{subsec:geod-minimal}.


\subsection{Spacelike geodesics at $t=0$}
\label{ssec:const_t_geod}

We consider a one-dimensional boundary subregion ${\mathcal{R}}$ given by an arc with opening angle $\Delta\theta \in [0, 2\pi]$ at constant boundary time $t_{b} =0$. 
In this subsection, we investigate the (possibly multiple) extremal surfaces that are homologous to this boundary region, among which the HRT surface is the spacelike geodesic with minimal length.
The total geodesic solution in the full bubble geometry can either lie in the BTZ part of spacetime only, or cross the domain wall, thus exploring the interior of the bubble as well. 

First of all, we can prove that the globally minimal HRT surface attached at $t_{b}=0$ on the boundary lies entirely on the $t=0$ slice in the bulk, which we denote by $\Sigma_0$.
In general, not all the extremal surfaces which are attached at the boundary slice $t_{b}=0$ lie on $\Sigma_0$, see the discussion in Ref.~\cite{Engelhardt:2023bpv}.
The hypersurface $\Sigma_0$ has vanishing extrinsic curvature tensor
$K_{ab}=0$, therefore it is a \textit{totally geodesic} surface \cite{Hubeny:2007xt}. This means that minimal surfaces on $\Sigma_0$ are also extremal surfaces in the whole spacetime. For the sake of the argument, let us assume that a globally minimal codimension-two surface $\mathcal{A}$ that lies outside of the hypersurface $\Sigma_0$ exists. Using the methods described in Ref.~\cite{Wall:2012uf}, one can shoot out from $\mathcal{A}$
a codimension-one null hypersurface $\mathcal{N}$ along one of the null directions, which will intersect $\Sigma_0$ in a codimension-two surface $\mathcal{B}$, defined as $\mathcal{B}=\mathcal{N} \cap \Sigma_0$.
Following~\cite{Wall:2012uf}, if the null energy condition holds, then the area of $\mathcal{B}$ is smaller than the area of $\mathcal{A}$. 
Therefore, we found that $\mathcal{A}$ must have a bigger area than the minimal surface on $\Sigma_0$ (which is also an extremal surface in the full spacetime, as we argued above).

In light of this argument, we can restrict our analysis to extremal surfaces that lie on $\Sigma_0$.
Due to the compactness of the boundary subregion, there exist pairs of spacelike geodesics $\gamma_1$ and $\gamma_2$ anchored at the edges of a given subregion $\Delta\theta$, whose associated holographic EE is denoted with $S(\Delta\theta)$ and $S(2\pi - \Delta\theta)$, respectively.
In the full Kruskal extension of the BTZ black hole (without any bubble), because of the BH horizon, only the geodesic $\gamma_1$ is homologous to $\Delta\theta$. On the other hand, the disconnected curve given by the union of $\gamma_2$ and the geodesic wrapping the BH horizon also satisfies the same homology condition \cite{Ryu:2006bv,Ryu:2006ef,Headrick:2007km,Azeyanagi:2007bj}. The disconnected curve is shorter for a critical opening angle $\Delta\theta_{\rm crit}>\pi$ of the boundary subregion \cite{Blanco:2013joa,Hubeny:2013gta}.

With the exception of expanding bubbles in region $B$ of the parameter space (see fig.~\ref{Phase-Diagram}), in the bubble geometry~\eqref{metric-zero} the situation differs in that both $\gamma_1$ and $\gamma_2$ are homologous to the boundary subregion $\Delta\theta$. 
Consequently, we have that
\beq
S(\Delta \theta)=S(2 \pi- \Delta \theta) \, .
\label{eq:S_mirrored}
\eeq
Thus, it is not restrictive to focus on boundary subregions with $\Delta\theta \in \left[ 0, \pi \right]$.

The action functional  for a geodesic at constant bulk time reads
\begin{equation}
    l_{i,o}= \int \left( \frac{r'^2_{i,o}}{f_{i,o}(r)} +r^2 \theta'^2_{i,o} \right) dp 
    \equiv \int \mathcal{L} (r, r', \theta') \, dp \, ,
    \label{eq:length_general_geod_spacelike}
\end{equation}
where $p$ is a spacelike affine parameter running along the geodesic trajectory $( r(p), \theta(p) )$, and the \textit{prime} denotes a derivative with respect to such a parameter. 
\rep{The action $ l_{i,o}$
in eq. (\ref{eq:length_general_geod_spacelike})
coincides with the length 
of the geodesic
 in the case of a proper length parametrization, as we will choose in eq.~\eqref{eq:rprime_geod} below.}
Since the above Lagrangian $\mathcal{L}$ does not depend on $\theta_{i,o}(p)$ due to spherical symmetry, there is a conserved momentum
\begin{equation}
\label{eq:j-def}
    j_{i,o} \equiv \frac{1}{2} \frac{\partial \mathcal{L}}{\partial \theta'_{i,o}}
    = r^2 \, \theta'_{i,o} \, .
\end{equation}
Since we are considering a spacelike geodesic, choosing $\mathcal{L} =1$
we can identify the affine parameter with the length of the curve. This normalization requirement allows to solve for $r'_{i,o}$ as a function of the conserved momentum: 
\begin{equation}
    \mathcal{L} = \frac{r'^2_{i,o}}{f_{i,o}(r)} + \frac{j_{i,o}^2}{r^2} = 1 
    \quad  \Rightarrow \quad
    r'_{i,o} = \pm \frac{\sqrt{\left( r^2 - j_{i,o}^2 \right)f_{i,o}(r)}}{r} \, .
    \label{eq:rprime_geod}
\end{equation}
In the above solution, the $\pm$ signs describe different branches of the geodesic. Note that along the slice $t_{i,o}=0$ the (cosmological or BH) horizon, when it exists, reduces to the bifurcation point. Consequently, $f_{i,o}(r) \geq 0$ at any point along the geodesic. For consistency of the above solution, we then need to take $r \geq j_{i,o}$. 
We denote by $r_* = j_{i,o}$ the value of the $r$-coordinate at the turning point, where $r'=0$.

It is useful to paramereterize the geodesic  by the function $\theta_{i,o}(r)$. A geodesic solution, which extremizes the functional~\eqref{eq:length_general_geod_spacelike}, 
satisfies the following equations of motion
\begin{equation}
\label{eq:geod-eq}
    \frac{d\theta_{i,o}(r)}{dr} = 
    \frac{\theta'_{i,o}}{r'_{i,o}} =
    \pm \frac{j_{i,o}}{r \sqrt{\left( r^2 - j_{i,o}^2 \right)f_{i,o}(r)}} \, ,
\end{equation}
where in the last step we plugged in the identities~\eqref{eq:j-def} and \eqref{eq:rprime_geod}.

In particular, in the exterior geometry, the differential equation~\eqref{eq:geod-eq} with blackening factor~\eqref{f-BTZ} is solved by
\begin{align}
    \theta_o^{\pm}(r) &= \pm \frac{1}{r_h} \log\left( \frac{j_o \, \sqrt{r^2 - r_h^2} + r_h \sqrt{r^2-j_o^2}}{r \sqrt{|j_o^2 - r_h^2|}} \right) + \varphi_o \, ,
    \label{eq:geod_sol_exterior_RT}
\end{align}
where $\varphi_o$ is an integration constant.
For different values of $j_o$ compared to the horizon radius $r_h$ of a BTZ black hole, various configurations of the geodesics -- attached to the right boundary of the fully extended BTZ background -- can exist.
If $j_o>r_h$, the solution in eq.~(\ref{eq:geod_sol_exterior_RT})
stays outside the BH bifurcation surface. If $j_o<r_h$, the HRT geodesic enters the BH horizon without coming back to the right boundary of the BTZ black hole. 
However, in the presence of a bubble, geodesics with $j_o<r_h$ might come back to the right boundary of the BTZ black hole.

In the interior geometry, by plugging the blackening factor~\eqref{f-dS} inside eq.~\eqref{eq:geod-eq}, we find the solution  
\begin{equation}
    \theta_i^{\pm}(r) = \pm \mathrm{arctan} \, \left( \sqrt{ \frac{r^2 -j_i^2}{j_i^2 (1- \lambda r^2)} } \right)
    + \varphi_i^\pm \, ,
    \label{eq:theta-i-geod}
\end{equation}
where $\varphi_i^\pm$ are integration constants.

The length of the geodesic can be found by evaluating
the action functional~\eqref{eq:length_general_geod_spacelike} on-shell as follows:
\begin{equation}
\label{eq:geod-length}
    l_{i,o} = \int \frac{dr}{r'_{i,o}}
    = \int \frac{r}{\sqrt{\left( r^2 - j_{i,o}^2 \right)f_{i,o}(r)}} \, dr \, .
\end{equation}
We will use this general formula to compute the length of the HRT geodesic in bubble geometries. To this aim, we will need to first compute the length of an extremal trajectory in each side of the geometry separately, and then minimize the sum of lengths across the full geometry.

\subsection{Joining geodesics}
\label{ssec:joining_geo}

To find a global geodesic in the full bubble geometry~\eqref{metric-zero}, we need to glue the geodesics determined in each region of the spacetime across the domain wall.
We denote by $x^\mu_{i,o}(p)$ the coordinates of the geodesic trajectory in the internal and external regions of spacetime, respectively.
The tangent vector to the geodesic is 
\beq
\frac{d x^\mu_{i,o}}{d p} = (t'_{i,o}(p) , r'_{i,o}(p), \theta'_{i,o}(p)) \, .
\eeq
We denote by $X^\mu_{i,o}(\tau)$ the trajectory of the domain wall
at constant $\theta$, with $\tau$ a proper time. The tangent directions to the domain wall are given by
\beq
n^\mu_{i,o}=\frac{d X^\mu_{i,o}}{ d \tau} = \le \dot{T}_{i,o}(\tau), \dot{R}(\tau),0 \ri \, ,
\qquad
m^\mu_{i,o}=\le 0,0,\frac{1}{R(\tau)} \ri \, .
\eeq
As discussed in Ref.~\cite{IglesiasZemmour2019RefractionAR},
spacelike geodesics need to satisfy a \textit{refraction law} at the location of the domain wall
\beq
\label{snell-law-covariante}
(g_i)_{\mu \nu} \frac{d x^\mu_i}{d p} n^\nu_i= 
(g_o)_{\mu \nu} \frac{d x^\mu_o}{d p} n^\nu_o \, ,
\qquad
(g_i)_{\mu \nu} \frac{d x^\mu_i}{d p} m^\nu_i= 
(g_o)_{\mu \nu} \frac{d x^\mu_o}{d p} m^\nu_o \, ,
\eeq
which in the background~\eqref{metric-zero} give
\beq
- f_i \dot{T}_i t'_i + \frac{1}{f_i} \dot{R} r' =
- f_o \dot{T}_o t'_o + \frac{1}{f_o} \dot{R} r'\, , \qquad
\theta'_i=\theta'_o \, . 
\label{eq:congiungere-geodesics}
\eeq
Note that the first identity in eq.~(\ref{eq:congiungere-geodesics})
is solved by a fixed-time geodesic at $t=0$, because the condition
$\dot{R}=0$ holds. From the second identity
in eq.~(\ref{eq:congiungere-geodesics}), we find the constraint  $j_i=j_o$.
Equivalent conditions have been found in Refs.~\cite{Balasubramanian:2011ur,Antonelli:2018qwz} for different setups, and a similar refraction law was determined in the case of codimension-one extremal spacelike surfaces in Ref.~\cite{Auzzi:2023qbm}.

\subsection{Geodesics in the bubble geometry}
\label{subsec:geod-in-bubble}

Let us denote by $R_0$ the radius of the bubble at the initial time $t=0$ and by $j \equiv j_i = j_o$ the conserved momentum along the full geodesic. 
The quantity $j$ is not, in general, a monotonic function of the opening angle $\Delta \theta$ of the boundary subregion to which the geodesic is attached. For a given $\Delta \theta$, we should find the value of $j$ which corresponds to the geodesic with minimum length. Nonetheless, we can parameterize the full family of geodesics sitting at $t=0$ by the real positive quantity $j$.
For $j \geq R_0$, the geodesic is completely outside the bubble, while for $0 \leq j < R_0 $, the geodesic is composed by an interior and an exterior part, joined at $r=R_0$.
There are four cases of interest for a complete classification of the geodesic solutions:
\begin{itemize}
\item {\bf{The geodesic is completely outside the bubble.}} 
In this case, ${j \geq R_0 \geq r_h}$ and we can choose the boundary condition $\theta_o^{\pm}(j)=0$ in eq.~(\ref{eq:geod_sol_exterior_RT}), 
 which fixes $\varphi_o^{\pm} = 0$. The geodesic length is
     \beq
        l_{\rm ext}(j)
        =  \log \left( \frac{4 R_c^2}{j^2 -r_h^2} \right) + \mathcal{O}(R_c^{-1}) \, ,
        \label{eq:l-ext}
    \eeq
where $R_c$ is a UV cutoff, and here we performed a series expansion for $R_c \to \infty$.
By imposing in eq.~\eqref{eq:geod_sol_exterior_RT} that the subregion opening angle at the boundary is $\Delta \theta$, we find a relation between this latter quantity and the conserved momentum:
    \begin{align}
        \Delta\theta(j) &= 
        \frac{1}{r_h} \log \left( \frac{j +r_h}{j - r_h} \right)
        = \frac{2}{r_h} \, \mathrm{arctanh} \, \left( \frac{r_h}{j} \right) \, .
        \label{eq:Delta-theta-j-out} 
    \end{align}
The total geodesic length thus reads 
    \begin{equation}
        l_{\rm ext}(\Delta\theta) = 2 \log \left( \frac{2 R_c}{r_h} \sinh \left( \frac{r_h \, \Delta\theta}{2} \right) \right) \, ,
        \label{eq:geodesic_length_case1ext}
    \end{equation}
where we plugged the relation~\eqref{eq:Delta-theta-j-out} inside eq.~\eqref{eq:l-ext}.

By means of eq.~\eqref{eq:Delta-theta-j-out}, we can derive a condition under which a fully external geodesic does not exist for $\Delta\theta \approx \pi$. Namely, by defining $\Delta\theta(j=\tilde{R}) = \pi$ and inverting eq.~\eqref{eq:Delta-theta-j-out}, we get
    \beq 
   \tilde{R}=r_h \coth \le \frac{\pi}{2} r_h \ri \, .
   \label{Rtilde}
    \eeq
When $R_0 >\tilde{R}$, a fully external HRT surface does not exist nearby $\Delta \theta \approx \pi$.
When $R_0 <\tilde{R}$, a candidate external HRT surface always exists, but it is not guaranteed to have minimal length. The identification of the geodesic solutions with minimal length will be performed in subsection~\ref{subsec:geod-minimal}.
 \item { \bf{The geodesic enters a  bubble that is outside the BH bifurcation surface.}} In this case we have ${0 \leq j < R_0}$, and part of the geodesic is inside the bubble.
Since by construction $R_0>r_h$, we can have either $j>r_h$, or $j<r_h$. The solution for the geodesic in the internal region is given by eq.~(\ref{eq:theta-i-geod}) with $\varphi_i^\pm =0$, so that $\theta_i^{\pm}(r=j)=0$. The geodesic length \eqref{eq:geod-length} in such region reads
\beq
  l_i(j) = \frac{2}{\sqrt{\lambda}} \arcsin \left( \sqrt{\frac{\lambda \left(R_0^2 -j^2 \right)}{1- j^2 \lambda}} \right) \, .
  \label{eq:li-ds-ads}
\eeq
The exterior part of the geodesic is parametrized by eq.~\eqref{eq:geod_sol_exterior_RT}. In this case, the additional ingredient is that we need to glue the trajectory outside the bubble with the part inside the shell.
In order for the full geodesic to be continuous across the domain wall, we impose $\theta_o^+(R_0) = \theta_i^+(R_0)$, which fixes the integration constants in eq.~\eqref{eq:geod_sol_exterior_RT} to
\beq
\varphi_o =  \arctan \left( \sqrt{ \frac{R_0^2 -j^2}{j^2 (1- \lambda R_0^2)} } \right) - \frac{1}{r_h} \log\left( \frac{j \, \sqrt{R_0^2 - r_h^2} + r_h \sqrt{R_0^2-j^2}}{R_0 \sqrt{|j^2 - r_h^2|}} \right) \, .
\label{eq:theta-o-const}
\eeq
The length of the external part thus reads
    \begin{equation}
    l_o(j) = 2 \int_{R_0}^{R_c} \frac{r}{\sqrt{\left( r^2 - j^2 \right)\left( r^2 - r_h^2 \right)}} \, dr =
         2 \log \left( \frac{2 R_c}{\sqrt{R_0^2 -j^2} + \sqrt{R_0^2 -r_h^2}} \right) \, ,
          \label{eq:lo}
    \end{equation}
    where $R_c$ is the UV cutoff.
Combining eqs.~\eqref{eq:geod_sol_exterior_RT} and \eqref{eq:theta-o-const}, we obtain the opening angle of the boundary subregion as a function of the conserved momentum $j$:
\beq
\Delta\theta(j)=2
\,
 \arctan \left( \sqrt{ \frac{R_0^2 -j^2}{j^2 (1- \lambda R_0^2)} } \right)+
 \frac{2}{r_h} \log \le
\frac{R_0 (j+r_h)}{j \sqrt{R_0^2-r_h^2} +r_h \sqrt{R_0^2-j^2}}
 \ri \, .
  \label{eq:Delta-theta-j-full}
\eeq
Summing eqs.~\eqref{eq:li-ds-ads} and \eqref{eq:lo}, we get the total length of the geodesic
\begin{align}
    \frac{l_{\rm bub}(j)}{2} =  
    \frac{1}{\sqrt{\lambda}} \arcsin \left( \sqrt{\frac{\lambda \left(R_0^2 -j^2 \right)}{1- j^2 \lambda}} \right) + \log \left( \frac{2 R_c}{\sqrt{R_0^2 -j^2} + \sqrt{R_0^2 -r_h^2}} \right)  \, .  
\end{align}


\item  {\bf{The geodesic enters a bubble that is inside
the BH bifurcation surface, with $\b_i(R_0)>0$.}} 
In this case we have ${0 \leq j < r_h < R_0}$, where the constraint $j<r_h$ guarantees that the external part of the geodesic penetrates beyond the BH bifurcation surface.
Notice that these inequalities exclude the following cases:  \textbf{(1) }expanding bubbles in region $A$ of fig.~\ref{Phase-Diagram} that contain the cosmological bifurcation surface, and \textbf{(2)} expanding bubbles in region $B$ that contain a left AdS boundary. The former setting will be considered in the next bullet point.

For clarity, we sketch the geodesic with $j=0$ in the left panel of fig.~\ref{fig:geod_inside_bubble}.
Let us scan the diagram from left to right.
The part of the geodesic solution inside the bubble is still given by eq.~\eqref{eq:theta-i-geod} with $\varphi_i^\pm=0$, as in the previous case. 
Correspondingly, the length of this part is given by eq.~\eqref{eq:li-ds-ads}. 
However, contrary to the previous case, the geodesic crosses the bubble before reaching the BH bifurcation surface.
The portion of the geodesic trajectory located between the domain wall and the BH bifurcation surface (identified by the red curve in fig.~\ref{fig:geod_inside_bubble1}) is described by the branch $\theta_o^-(r)$ of eq.~\eqref{eq:geod_sol_exterior_RT}, with boundary condition given by the continuity requirement $\theta_o^-(R_0)=\theta_i^+(R_0)$. This constraint gives
\beq
\Delta\theta(j)=2
\,
 \arctan \left( \sqrt{ \frac{R_0^2 -j^2}{j^2 (1- \lambda R_0^2)} } \right)+
 \frac{2}{r_h} \log \le
\frac{  \, j \sqrt{R_0^2-r_h^2} +r_h \sqrt{R_0^2-j^2} }{R_0 \, (r_h-j)} 
 \ri \, .
  \label{eq:Delta-theta-j-other-side}
\eeq
The region of the Penrose diagram external to the BH bifurcation surface is described by the same solution $\theta_o^+(r)$ used in the previous bullet points, joined continuously with $\theta_o^-(r)$ at the BH bifurcation surface.
The total length of the trajectory in the exterior region is given by
 \beq
 l_o(j) = 2 \left( \int_{r_h}^{R_c} + \int_{r_h}^{R_0} \right) \frac{r}{\sqrt{\left( r^2 - j^2 \right)\left( r^2 - r_h^2 \right)}} \, dr \, .
 \label{eq:bullet3_ext_length}
 \eeq
Therefore, summing the contributions~\eqref{eq:li-ds-ads} and \eqref{eq:bullet3_ext_length}, we get the length of the full geodesic:
\beq
\frac{l_{\rm bub}(j)}{2}
= \frac{1}{\sqrt{\lambda}} \arcsin \left( \sqrt{\frac{\lambda \left(R_0^2 -j^2 \right)}{1- j^2 \lambda}} \right) + \log \left( \frac{2 R_c}{\sqrt{R_0^2 -j^2} - \sqrt{R_0^2 -r_h^2}} \right) \, .
  \label{eq:Length-j-other-side}
\eeq

\begin{figure}[t]
\center
\subfigure[]{\label{fig:geod_inside_bubble1} \includegraphics[scale=0.3]{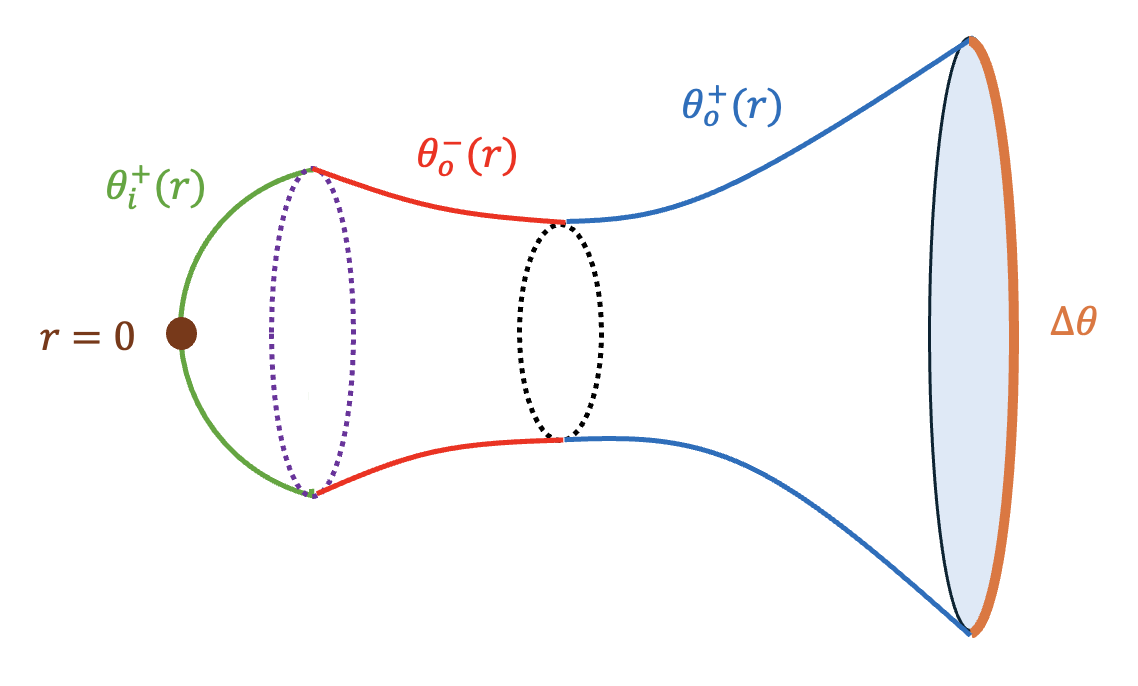}}
\qquad
\subfigure[]{\label{fig:geod_inside_bubble2} \includegraphics[scale=0.3]{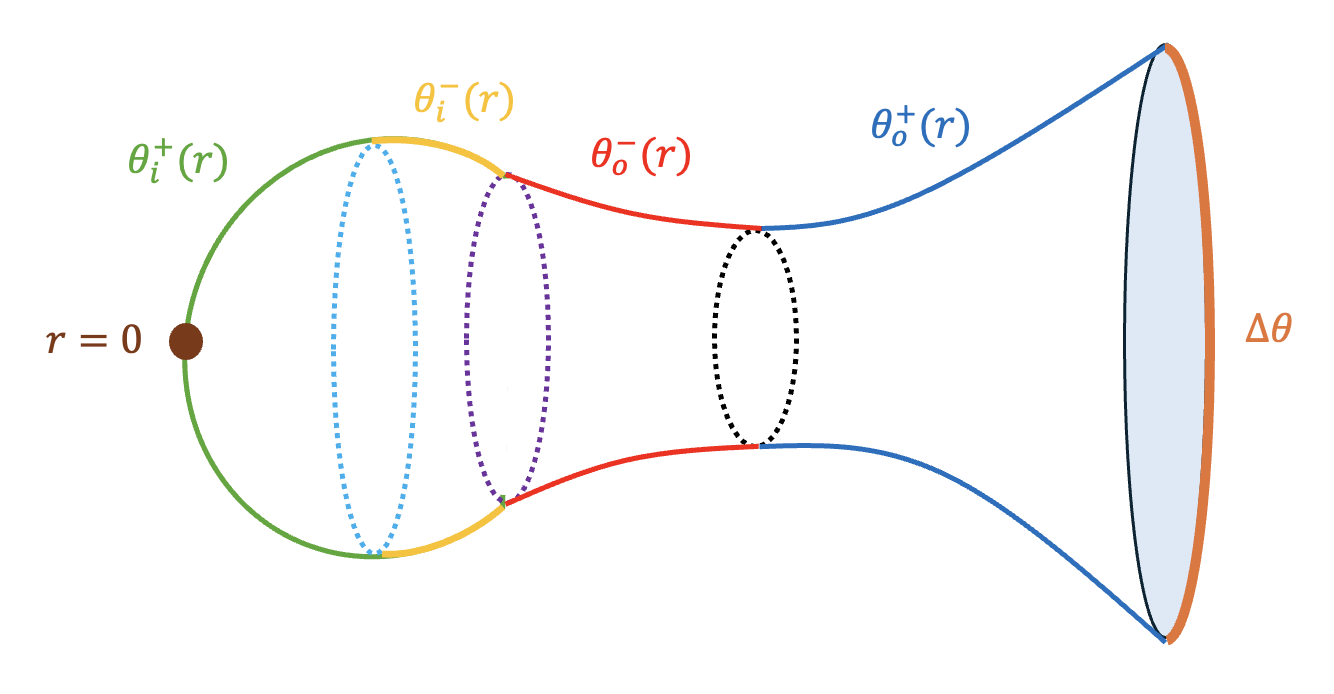}}
\caption{Sketch of the geodesic with $j=0$ and $\Delta\theta=\pi$ for a bubble inside the BH bifurcation surface with $\beta_i(R_0)>0$ (left) and $\beta_i(R_0)<0$ (right). 
 The black circumference represents the bifurcation surface $r=r_h$, the purple circumference the domain wall $r=R_0$, and the light blue one the cosmological bifurcation surface $r=1/\sqrt{\lambda}$.
 The branches $+$ ($-$) of the external solution $\theta_o(r)$ are shown in blue (red), while the branches $+$ ($-$) of the internal solution $\theta_i(r)$ are displayed in green (yellow).
 The solutions in the lower half of the plot are obtained from the ones in the upper half by $\theta_{i,o}^\pm(r) \to -\theta_{i,o}^\pm(r)$.
 }
\label{fig:geod_inside_bubble}
\end{figure}

\item  {\bf{The geodesic enters a bubble that is inside
 the BH bifurcation surface, with $\b_i(R_0)<0$.}} 
This case is realized by expanding bubbles in region $A$ of fig.~\ref{Phase-Diagram} that contain a full dS static patch. We do not consider in this work expanding bubbles in region $B$ that include a left AdS boundary.

Let us scan the Penrose diagram from left to right following the representative depiction in fig.~\ref{fig:geod_inside_bubble2} for the case $j=0$.
Starting from the turning point at $r=0$, the interior part of the geodesic trajectory differs from the previous case in that it crosses a cosmological bifurcation surface before reaching the domain wall.
Instead, the external portion of the geodesic is the same as in the previous bullet point.

More precisely, the most internal part of the trajectory (green curve in fig.~\ref{fig:geod_inside_bubble2}) is given by the branch $\theta_i^+(r)$ with $\varphi_i^+=0$, whereas the branch $\theta_i^-(r)$ with $\varphi_i^- =\pi$ extends in the right static patch and reaches the domain wall (yellow curve in fig.~\ref{fig:geod_inside_bubble2}).
The latter trajectory then glues with the external part of the geodesic (red curve in fig.~\ref{fig:geod_inside_bubble2}) under the continuity condition $\theta_o^-(R_0) = \theta_i^-(R_0)$. Finally, this is glued with the blue trajectory $\theta^+(r)$, that extends from the BH bifurcation surface until the boundary subregion.
This fixes
\beq
\Delta\theta(j)=2
\,
 \left(\pi -\arctan \left( \sqrt{ \frac{R_0^2 -j^2}{j^2 (1- \lambda R_0^2)} } \right) \right)+
 \frac{2}{r_h} \log \le
\frac{  \, j \sqrt{R_0^2-r_h^2} +r_h \sqrt{R_0^2-j^2} }{R_0 \, (r_h-j)} 
 \ri \, .
  \label{eq:Delta-theta-j-other-side-full-static-patch}
\eeq
The length outside the bubble is the same as in the previous case, whereas the length inside the bubble now reads
\beq
 l_i(j) = 2 \left( \int_{R_0}^{1/\sqrt{\lambda}} + \int_0^{1/\sqrt{\lambda}} \right) \frac{r}{\sqrt{\left( r^2 - j^2 \right)\left( 1 - \lambda r^2 \right)}} \, dr \, .
\eeq
Summing all the contributions, we get the total geodesic length
\beq
 \frac{l_{\rm bub}(j)}{2}
= \frac{1}{\sqrt{\lambda}} \left( \pi - \arcsin \left( \sqrt{\frac{\lambda \left(R_0^2 -j^2 \right)}{1- j^2 \lambda}} \right) \right) + \log \left( \frac{2 R_c}{\sqrt{R_0^2 -j^2} - \sqrt{R_0^2 -r_h^2}} \right)  \, .
\eeq
\end{itemize}
Surprisingly, there are choices of parameters for which geodesics exploring the interior of a bubble beyond the BH bifurcation surface are shorter than geodesics anchored at the same subregion, but fully located outside the BH. However, we have numerically checked that this phenomenon never happens for static or expanding bubbles, but only for contracting ones.


\subsection{Geodesics with minimal length}
\label{subsec:geod-minimal}

According to the HRT prescription, the entanglement entropy $S(\Delta\theta)$ of a spatial subregion in a two-dimensional field theory is dual to the length of the shortest geodesic homologous to the subregion itself. 
The existence of several candidate HRT surfaces for a given $\Delta\theta$ makes the computation non-trivial.\footnote{Reference~\cite{Sahu:2023fbx} discussed a similar problem, with the following differences with our setting. There, the interior geometry was given by a FRW universe with the same (negative) cosmological constant as the AdS exterior, and the energy-momentum tensor contained extra contributions due to dust and radiation. Moreover, surface of the bubble was not a proper domain wall, but rather a shell of non-relativistic matter with negligible thickness and vanishing pressure.}
In appendix~\ref{app:entropy}, we describe in detail the strategy to find the shortest geodesic, and provide numerical results for $S(\Delta\theta)$. We also discuss the occurrence of Python's lunches for certain boundary subregions.
In this subsection, we focus on the case of opening angle $\Delta\theta = \pi$, which despite being simpler, is fundamental to discriminate among different behaviors of $ S(\Delta\theta)$.

\begin{figure}[t]
\center
\includegraphics[scale=0.6]{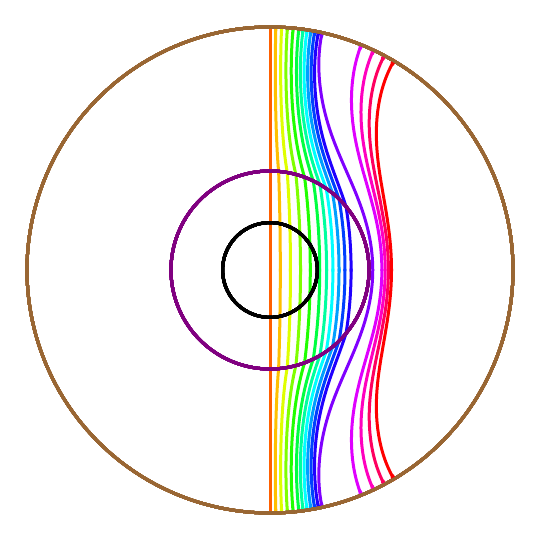}
\qquad
\includegraphics[scale=0.9]{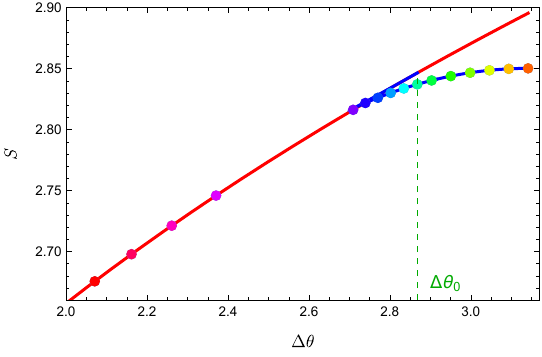}
\caption{Left: Geodesics with minimal length for various choices of the boundary opening angles $\Delta \theta$. The purple (black) circumference represents the domain wall (BH horizon) at $t=0$.
Right: Comparison between the entanglement entropy obtained by minimization (colored dots corresponding to the left panel) and the one obtained without minimization. The red curve is the length of the geodesic outside the bubble, while the blue curve corresponds to geodesics exploring the bubble. For a certain range of $\Delta\theta$, three extremal geodesics exist.
 Here we set $\l=0.1, \, \mu=0.1, \, \kappa=0.4, \, R_c = 10^3$.
 }
\label{fig:geod_entropy_pos_lambda}
\end{figure}

As a general property, we have that $\Delta \theta (j=0)=\pi$.
The corresponding geodesic is given by the constant trajectory $\theta(r) = \pm \pi/2$, which enters the bubble, turning around $r=0$.
Depending on the choice of parameters $(\kappa, \lambda, m)$ and for a given $j$, there can exist two more candidates for the minimal HRT surface subject to the condition $\Delta\theta(j) =\pi$. 
The first candidate is an internal geodesic characterized by $0< j=j_{\pi,i} < R_0$. The second candidate is a geodesic fully located outside the BH horizon, characterized by $j = j_{\pi,o} > R_0$. Both of them, only one, or none of these geodesics can exist, depending on the value of $(\kappa, \lambda, m)$, see Appendix~\ref{app:entropy}.

From here on, we denote by $l$ both the length of the geodesic exploring the interior of the bubble $l_{\rm bub}$ and the length of the fully external geodesic $l_{\rm ext}$. 
When the bubble is outside the BH bifurcation surface, we recognize three qualitatively different regimes:
\begin{itemize}
\item \textbf{Case $a$.}
The fully external geodesic with $j = j_{\pi,o}$ exists,
and it is the minimal HRT surface.
Consequently, the globally minimal HRT surface never enters the bubble for any value of $\Delta\theta$, even if the bubble is outside the BH horizon.
\item \textbf{Case $b$.}
The fully external geodesic with $j = j_{\pi,o}$ exists,
but it is not the minimal HRT surface.
In this case, the minimal geodesic for
$\Delta \theta=\pi$ explores the interior of the bubble,
and it may have $j=0$ or $j=j_{\pi,i}$.
Therefore, a phase transition between these two solutions occurs at a critical angular opening $\Delta\theta_{\rm crit, b} < \pi$. In this case, there exists a Python's lunch associated with the boundary subregion with $\Delta \theta=\pi$.

 \item \textbf{Case $c$.}
There is no fully external geodesic with $j = j_{\pi,o}$.
A necessary and sufficient condition for this to happen is $R_0 > \tilde{R}$, where these quantities were defined around eqs.~\eqref{eq-massa-bubble} and~\eqref{Rtilde}. A phase transition between the internal geodesics occurs at a certain $\Delta\theta_{\rm crit, c} < \pi$, and a Python's lunch takes place for an opening angle $\Delta\theta < \pi$.
\end{itemize}

When the bubble is located inside the BH bifurcation surface, we recognize the following two regimes:
\begin{itemize}
\item \textbf{Case $d$.} The minimal HRT surface for $\Delta \theta=\pi$ is outside the BH horizon.
\item \textbf{Case $e$.} The minimal HRT surface for $\Delta \theta=\pi$ enters the BH horizon and probes the bubble interior.
\end{itemize}

Minimal geodesics with $\Delta\theta \leq \pi$ and the corresponding holographic EE are shown in fig.~\ref{fig:geod_entropy_pos_lambda}
for a solution belonging to case $c$, namely a contracting dS bubble initially outside the BH horizon.
In the picture, $\Delta\theta_0 \equiv \Delta\theta (j=R_0)$.
We may naively expect that the phase transition takes place at this value of the subregion opening angle, $\Delta\theta_{\rm crit, c} = \Delta\theta_0$.
However, as it is clear from the figure, the critical angle satisfies $\Delta\theta_{\rm crit, c} < \Delta\theta_0$.
We have observed the same property for collapsing AdS bubbles initially outside the BH horizon.
A similar result has been found in Ref.~\cite{Antonelli:2018qwz} for solutions corresponding to a special limit of expanding bubbles in region $E$ and $d=2$, identified by $\lambda \to -1$ and $\kappa, m \to 0$. 
This is a case where both the internal and external regions are pure AdS spacetimes, and the trajectory of the expanding bubble is given by a lightcone centered at $r=0$ in empty AdS. 

\begin{figure}[ht]
\center
\includegraphics[scale=0.4]{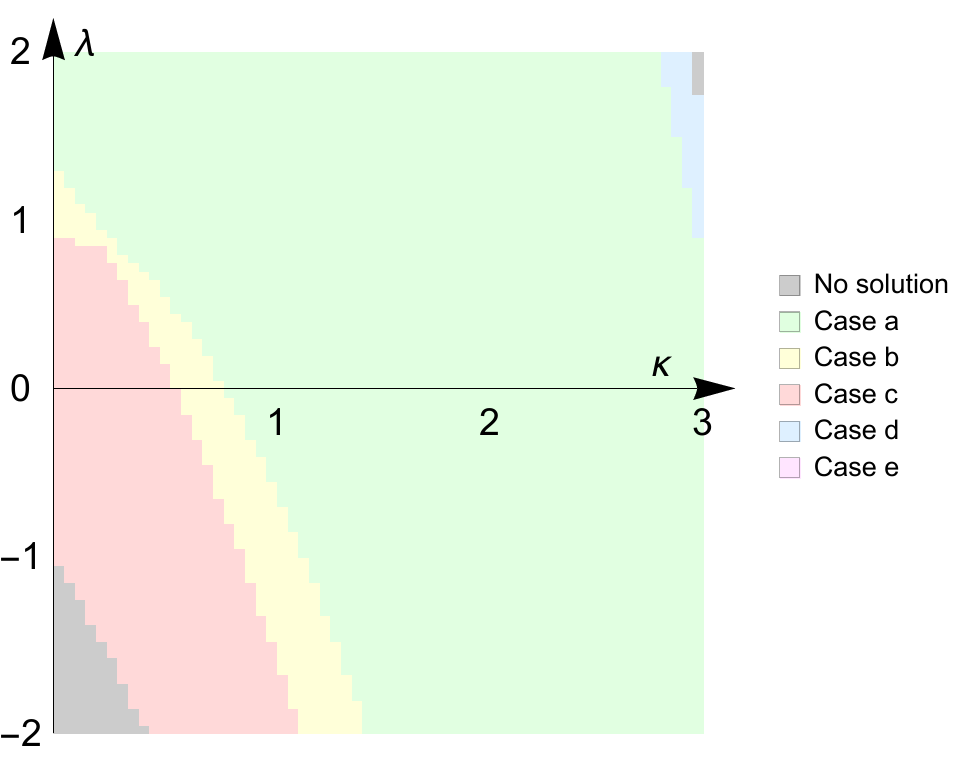}
\qquad
\includegraphics[scale=0.4]{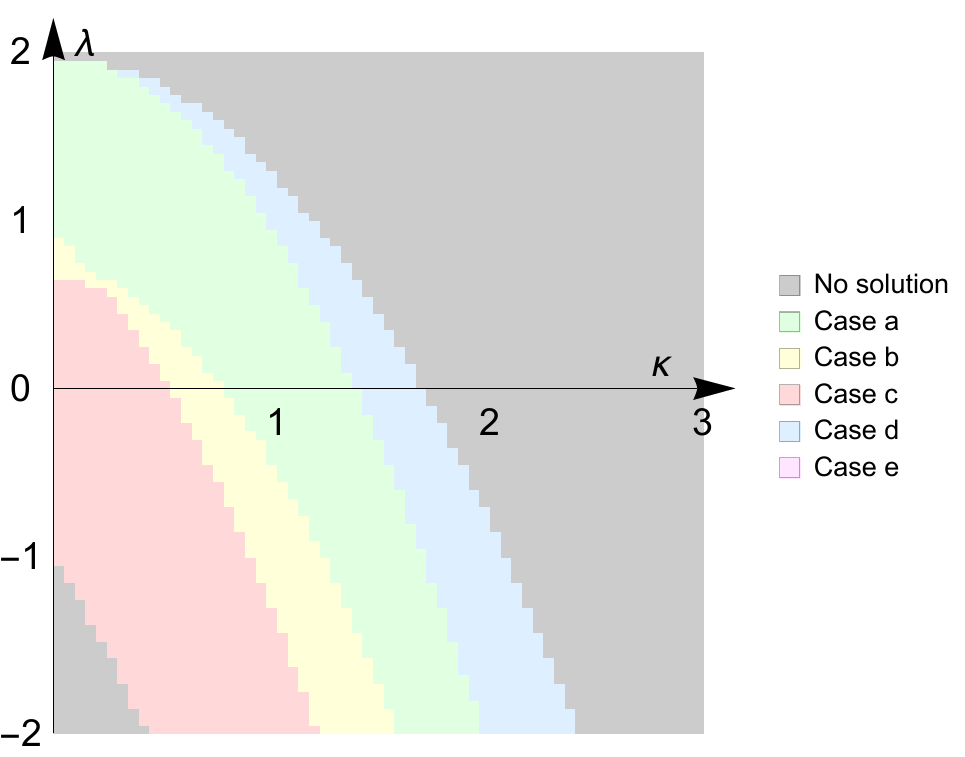}
\qquad
\includegraphics[scale=0.4]{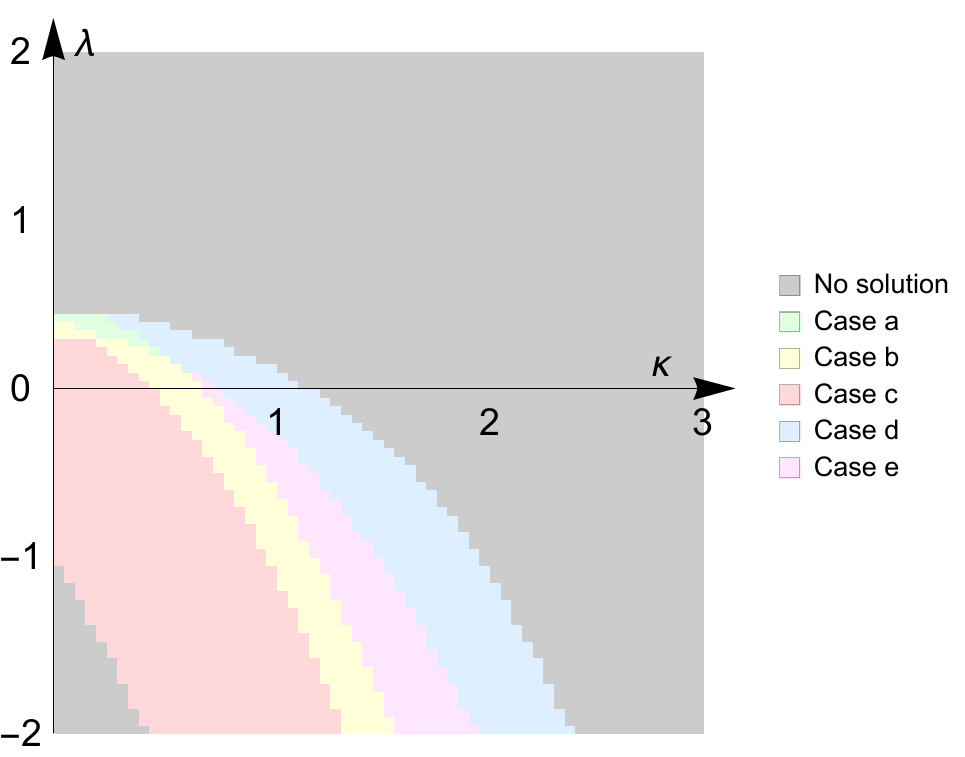}
\qquad
\includegraphics[scale=0.4]{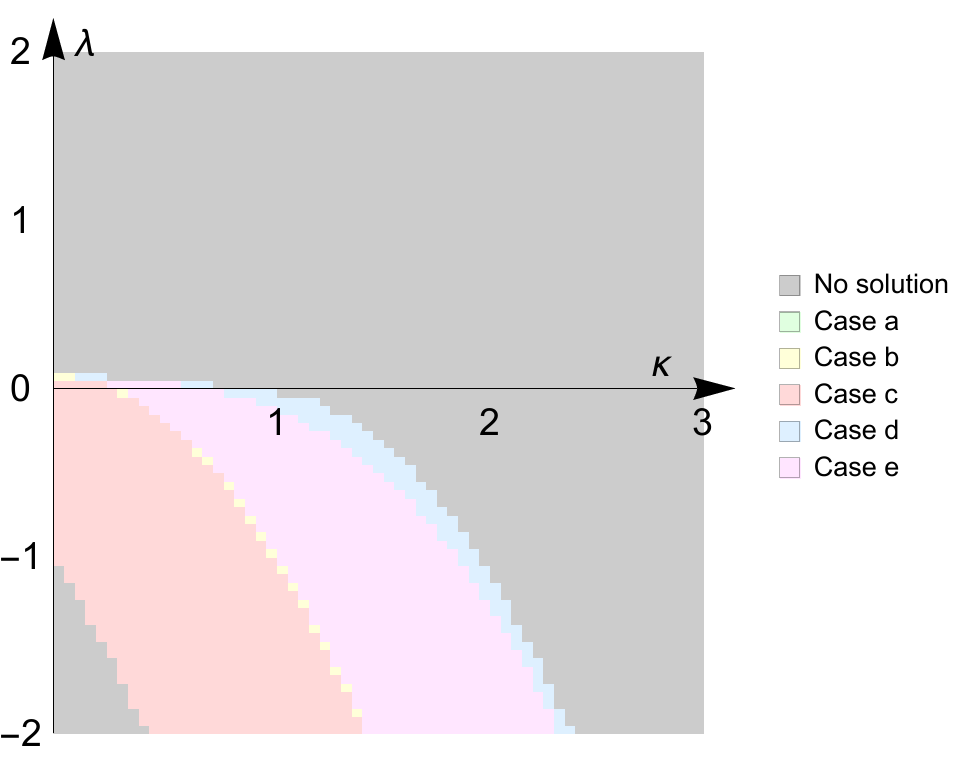}
\caption{ 
Numerical scan of the behavior of holographic EE in parameter space of the collapsing bubble in $d=2$, as a function of $(\kappa,\l)$ and for various mass parameters $m=1.1$ (top-left), $m=1.5$ (top-right),
$m=3$ (bottom-left) and $m=10$ (bottom-right).
}
\label{figure-scan-EE}
\end{figure}

We show in fig.~\ref{figure-scan-EE} a numerical scan of the parameter space $(\kappa,\lambda)$ for collapsing bubbles with various values of the BH mass parameter $m$.
We find that there is no direct relation between the behavior of EE (cases $a$, $b$ and $c$) and the fact that we are in configurations I or II of the collapsing bubble, see fig.~\ref{Pen-Dia-Collapsing-Bubble}.
As we will see in section~\ref{sec:sing_corr}, the structure of the bulk-cone singularities will instead differ for the configurations I and II of the Penrose diagram.

Another interesting property of the minimal geodesic with $\Delta\theta = \pi$ is that it can access the interior region of collapsing bubbles even when they are inside the BH bifurcation surface, \ie for the configuration III of fig.~\ref{Pen-Dia-Collapsing-Bubble}. 
As it can be seen in fig.~\ref{figure-scan-EE},
when increasing of the mass $m$, the region of parameter space for which the HRT surface passes the BH bifurcation surface enters the bubble gets larger. This can be heuristically understood from the fact that for large horizons, the external geodesic is repelled by the BH and is pushed close to the right AdS boundary, where the length is bigger. 
It is therefore more convenient in terms of length for the geodesic to go the other way around, wrapping the bubble around $r=0$. Interestingly, even for intermediate values of the mass parameter $m$, this behavior happens more likely for AdS bubbles ($\lambda<0$) rather than dS ones ($\lambda>0$).
On the other hand, we found that for static and expanding bubbles inside the BH bifurcation surface, the HRT with $\Delta\theta = \pi$ never enters the bubble and is always confined in the exterior of the BH. This is consistent with the intuition that such bubbles are larger compared to the contracting ones.

Our analysis of the holographic EE focused on a boundary subregion at $t_{b}=0$. Nevertheless, we can use the continuity of the time evolution to infer some information about the EE at early times.
When the HRT surface enters the BH bifurcation surface, we expect a non-trivial time dependence, where the geodesic explores the interior of the bubble inside the BH.
In particular, the holographic EE interpolates between the $t_{b}=0$ result and the late time regime, where we expect to approach the equilibrium thermal value, see, \eg \cite{Abajo-Arrastia:2010ajo,Balasubramanian:2011ur,Hartman:2013qma,Cooper:2018cmb}.
On the contrary, when the HRT surface is located outside the BH bifurcation surface, we expect a trivial time dependence of the geodesic generated by the boundary timelike Killing vector.
In this case, the holographic EE is time-independent, and it does not probe the geometry behind the BH horizon.

%% file: Sections/Singularities_corr.tex
\section{Bulk-cone singularities}
\label{sec:sing_corr}

In this section, we focus on another geometric tool that probes bubble geometries: almost-null spacelike geodesics that leave and come back to the AdS boundary.
According to the AdS/CFT dictionary, these objects are related to the singularities of two-point functions of CFT scalar operators $\mathcal{O}(x)$ with large scaling dimension $\Delta \gg 1.$
In the bulk, such operators are dual to scalar fields with large mass $m \sim \Delta$.
By using the semiclassical \textit{geodesic approximation}, the two-point functions of such boundary operators read~\cite{Louko:2000tp,Fidkowski:2003nf,Kraus:2002iv,Festuccia:2005pi,Balasubramanian:2011ur}
\beq
\langle \mathcal{O}(x)  \mathcal{O}(y) \rangle \sim e^{-m \ell(x,y)} \, ,
\eeq
where $\ell(x,y)$ is the proper length -- appropriately regularized -- of a geodesic trajectory connecting two boundary points.

In the limit when the spacelike geodesic becomes almost-null, the two-point function connecting two points on the same AdS boundary exhibits a singular behavior~\cite{Freivogel:2005qh,Hubeny:2006yu,Dodelson:2020lal,Dodelson:2023nnr,Ceplak:2024bja}.
The divergences of the above two-point functions are referred to as \textit{bulk-cone singularities}.
If the two points connected by the almost-null geodesic lie on disconnected AdS boundaries, there is a subtlety, because the singularity could belong to a secondary sheet of the analytically-continued correlator \cite{Fidkowski:2003nf}.\footnote{Given a multi-valued complex function, a secondary (or higher) sheet is a copy of the complex plane that differs from the principal branch, corresponding instead to other values of the function.}
In this section, we will always consider correlators
involving the same AdS boundary.

Let us parametrize the boundary as $(t, \Omega)$ where $t$ is a time coordinate and $\Omega$ denotes the angular directions.
We will consider the case of spacelike almost-null geodesics
connecting opposite points on the boundary sphere.
The two-points function singularities exhibit the following
behavior
\beq
\langle \mathcal{O}(t_{\rm in},\Omega)  \mathcal{O}(s, -\Omega) \rangle \sim 
\frac{1}{\le s  - t_{\rm fin}(t_{\rm in})  \ri^{2 \Delta}} \, ,
\label{eq:divergence_2pt_function}
\eeq
where $t_{\rm fin}(t_{\rm in})$ is the \textit{final} time at which a geodesic, starting from the AdS boundary at an \textit{initial} time $t_{\rm in}$, reaches the boundary after traveling into the bulk, and $\Delta \sim m$. 
When employing the formula~\eqref{eq:divergence_2pt_function}, we assume that the boundary points where the divergence occurs are connected by a radial spacelike and almost-null geodesics, in which case the two CFT operators are located at antipodal points.
This happens because the  geodesics necessarily pass through the center of AdS (pole of dS) interior spacetime, where they emerge to the opposite side of the transverse sphere.\footnote{When the boundary points are connected by non-radial geodesics, the angular coordinates of the two CFT operators can be related in a non-trivial way. For instance, see appendix~A.2 of Ref.~\cite{Hubeny:2006yu}.}

The goal of this section is to determine the functional dependence $t_{\rm fin} (t_{\rm in})$ by varying the parameters of the background, and understand how the information of CFT correlators is encoded by the various regions of spacetime.
We then perform a numerical analysis of the dynamical bubble for several configurations. After general remarks on the solutions in section~\ref{ssec:sing_dynamical_bubble}, we consider the cases of collapsing (section~\ref{ssec:collapsing_bubbles}) or expanding (section~\ref{ssec:expanding_AdS}) bubbles.
We conclude with the analysis of the static bubble in section~\ref{ssec:sing_static_bubble}, where analytic results can be achieved. 

In this section, we will work with dimensionality $d=3$, which provides the lowest-dimensional non-trivial setting to compute bulk-cone singularities. In $d=2$, almost-null spacelike geodesics do not bounce off the AdS BH singularity, therefore this kind of analysis would be less interesting~\cite{Fidkowski:2003nf}.
The latter property can be checked as follows.
From the radial geodesic equations in the external BH spacetime
(\ref{metric-zero}), we get
\beq
\dot{t}_o =\frac{E}{f_o(r)} \, , \qquad \dot{r}^2 = f_o(r) +E^2  \, ,
\label{geo-spacelike-radial}
\eeq
where $E$ denotes the conserved quantity associated with the Killing vector $\p_{t_o}$ and the \textit{dot} denotes a derivative with respect to the affine parameter along the geodesic trajectory.
When $E \to \infty$, we recover the null limit.
Using the identities in eq.~(\ref{geo-spacelike-radial}) and the blackening factor~\eqref{f-BH}, we find
\beq
\dot{r}^2 =E^2 +1+r^2 -\frac{m}{r^{d-2}} \, .
\eeq
As a consequence, the geodesic is repelled by the singularity at $r=0$ only when $d \geq 3$, in correspondence of which the latter term in the above equation diverges.
In the remainder of this section, we will mostly investigate \textit{almost-null} geodesics, which differ from \textit{exactly-null} geodesics in that the former are repelled by BH singularities, while the latter are not. We will consider exactly-null geodesics only where explicitly mentioned.


\subsection{General remarks}
\label{ssec:sing_dynamical_bubble}

In the following subsections, we study the divergences in the two-point function~\eqref{eq:divergence_2pt_function} of CFT operators for a bubble geometry.
In this framework, the profile of the domain wall is determined by the differential equations \eqref{eq:dynamical bubble differential equation 1} and \eqref{eq:dynamical bubble differential equation 2} (see appendix~\ref{ssec:domain_wall_traj} for more details).

In view of the upcoming analysis, we point out that almost-null geodesics in a bubble geometry enjoy various discrete symmetries.
We consider a radial trajectory starting from the right AdS boundary at generic $t_{\rm in}$ and coming back to the same boundary (possibly after multiple reflections) at the time $t_{\rm fin}$. We denote with $(t_{\rm in}, t_{\rm fin})$ a time interval that corresponds to a bulk-cone singularity connecying these two points.
First of all, we notice that it is always possible to follow a trajectory by evolving in the opposite direction. As a consequence, if a singularity of the two-point function occurs in correspondence of a bulk geodesic that intersects the AdS boundary at the initial and final boundary times $(t_{\rm in}, t_{\rm fin})$, then there is also another radial almost-null geodesic that intersects the boundary at times 
\beq
(\tilde{t}_{\rm in},\tilde{t}_{\rm fin})= (t_{\rm fin},t_{\rm in}) \, .
\label{eq:symm_time_exchange}
\eeq
Secondly, the invariance of the bulk geometry under time reflection $t \to -t$ implies that there is also a radial almost-null geodesic associated with the initial and final times
\beq
(\hat{t}_{\rm in},\hat{t}_{\rm fin})=
(-t_{\rm in},-t_{\rm fin}) \, .
\label{eq:symm_time_rev}
\eeq
We expect that the structure of bulk-cone singularities probes the internal geometry of the bubble.
Since the dynamical bubble is not invariant under time shifts (except for the static case), it is relevant to study the dependence of $t_{\rm fin}$ on the initial time $t_{\rm in}$ and discover how this function encodes information about the boundary correlators.

The general strategy is the following:
\begin{enumerate}
    \item Consider an almost-null
future-oriented geodesic 
    sent at time $t_{\rm in}$ from the AdS boundary. Follow its evolution (which possibly includes reflections at the BH singularity) until it reaches the domain wall.
    \item 
    \label{step2_sing_dyn}
    Use eq.~\eqref{eq:dynamical_bubbles_diffeq} to determine the relation between the time coordinate inside and outside the shell. Since the radial coordinate is continuous, this step relates the null coordinates $U_o, V_o$ outside the bubble with the ones $U_i, V_i$ inside.\footnote{It is convenient to work with the null coordinates $u, v$ because there always exists one of them that is continuous when crossing an event horizon.}
    \item Follow the trajectory of the geodesic inside the bubble, until it hits the domain wall a second time. 
    \item Repeat step~\ref{step2_sing_dyn}. The geodesic propagates in the exterior geometry, possibly reflecting at the BH singularity, and finally reaches the AdS boundary again at a time $t_{\rm fin}$.
\end{enumerate}
The above steps define the function $t_{\rm fin}(t_{\rm in})$, which corresponds to the case of an initially future-oriented geodesic leaving the right AdS boundary. 
In the forthcoming plots, this function will be drawn in color. Using the symmetry in eq.~(\ref{eq:symm_time_rev}), we find the corresponding result for an initially past-oriented
geodesic, that will be denoted in gray in the plots. It turns out that the symmetry in eq.~\eqref{eq:symm_time_exchange} does not give rise to new bulk-cone singularities.

For a dynamical solution, the above steps will be performed numerically, and we will report the results in the form of the functional dependence of of $t_{\rm fin}$ on $t_{\rm in}$.
The only exception is the static bubble, where the geodesic trajectory can be analytically computed (see appendix~\ref{app:bulk_cone_sing} for an explicit calculation).
In the following, we will separately study various examples of collapsing and expanding bubbles, by fixing come illustrative values of the parameters of the bulk spacetime.
As a guide for the reader to navigate among the plethora of cases, we summarize the list of the figures in table~\ref{tab:results}.

\begin{table}[h!]   
\begin{center}   
\begin{tabular}  {|p{35mm}|c|c|} \hline  & \textbf{Penrose diagram} 
 & \textbf{$t_{\rm fin} (t_{\rm in})$}  \\ \hline
\rule{0pt}{4.9ex}
Case I & Fig.~\ref{fig:Penrose diagram very small AdS bubble 0}  & Fig.~\ref{fig:very_small_AdS_a_cased}  \\
\rule{0pt}{4.9ex} Case II & Fig.~\ref{fig:Penrose diagram very small AdS bubble 1} & Figs.~\ref{Collapsing-Bubble-AdS-A}, \ref{fig:tfti_dS_CaseII}  \\ 
\rule{0pt}{4.9ex} Case III & Fig.~\ref{fig:Penrose diagram very small AdS bubble 1}  & Fig.~\ref{bulk-singo-caseIII}  \\
\rule{0pt}{4.9ex}  
Transition I-II & Fig.~\ref{fig:transitionIandII1} & Fig.~\ref{fig:transitionIandII2} \\
\rule{0pt}{4.9ex}  
 Expanding bubble & Figs.~\ref{fig:Penrose_expanding_AdS}, \ref{fig:Expanding_geodesic_trajectory_dS_interior} & Figs.~\ref{fig:expanding_tfti_AdS}, \ref{fig:tf_notsolarge} 
 \\  \rule{0pt}{4.9ex}
Static bubble  & Figs.~\ref{fig:nullgeo_static_AdS}, \ref{fig:nullgeo_static_dS} & Fig.~\ref{fig:plots_deltat_AdS3d_static}   
\\[0.2cm]
\hline
\end{tabular}   
\caption{Location in this paper of the Penrose diagrams and bulk-cone singularities for various bubble solutions.
} 
\label{tab:results}
\end{center}
\end{table}

\subsection{Collapsing bubbles}
\label{ssec:collapsing_bubbles}

Let us study the bulk-cone singularities for a collapsing bubble.
It turns out that by varying the parameters of the model $(\kappa, \l, m)$, the causal wedge of the right boundary of the AdS BH can present three different structures, reported in fig.~\ref{Pen-Dia-Collapsing-Bubble} and further discussed in subsection~\ref{ssec:causal_structure_collapsing_bubble}.
A numerical scan of the parameter space in $d=3$ was reported in fig.~\ref{figure-beta-outside-d3}.
We will show below that the qualitative behavior of the bulk-cone singularities is different in configuration I compared to configurations II and III.

\subsubsection{Configuration I}
\label{ssec:configurationI_collapsing}

The causal wedge corresponding to configuration I of fig.~\ref{Pen-Dia-Collapsing-Bubble} is attained in the light yellow region of the parameter space in fig.~\ref{figure-beta-outside-d3}.
We notice that this configuration can be realized for a large portion of the $(\l, \kappa)$ parameter at fixed and small $m$ (including both signs of $\lambda$), but cannot be attained in the large-$m$ regime.

\begin{figure}[h!]
\centering
\subfigure[case 1]{\label{subfig:case1_confI} \includegraphics[scale=0.18]{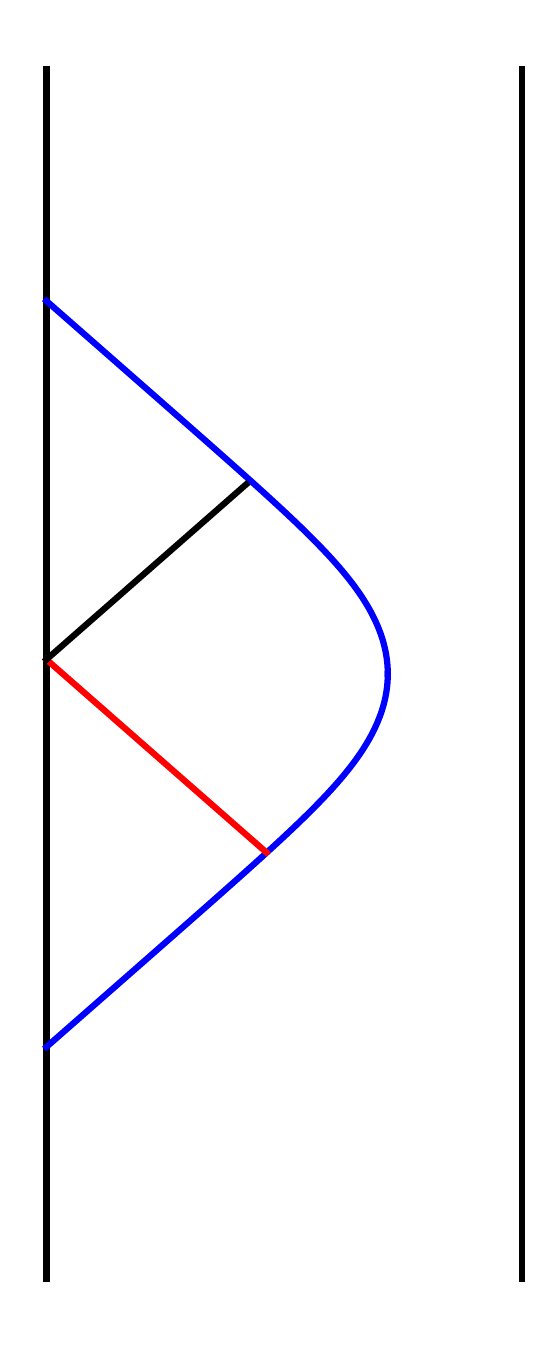} \quad
\includegraphics[scale=0.2]{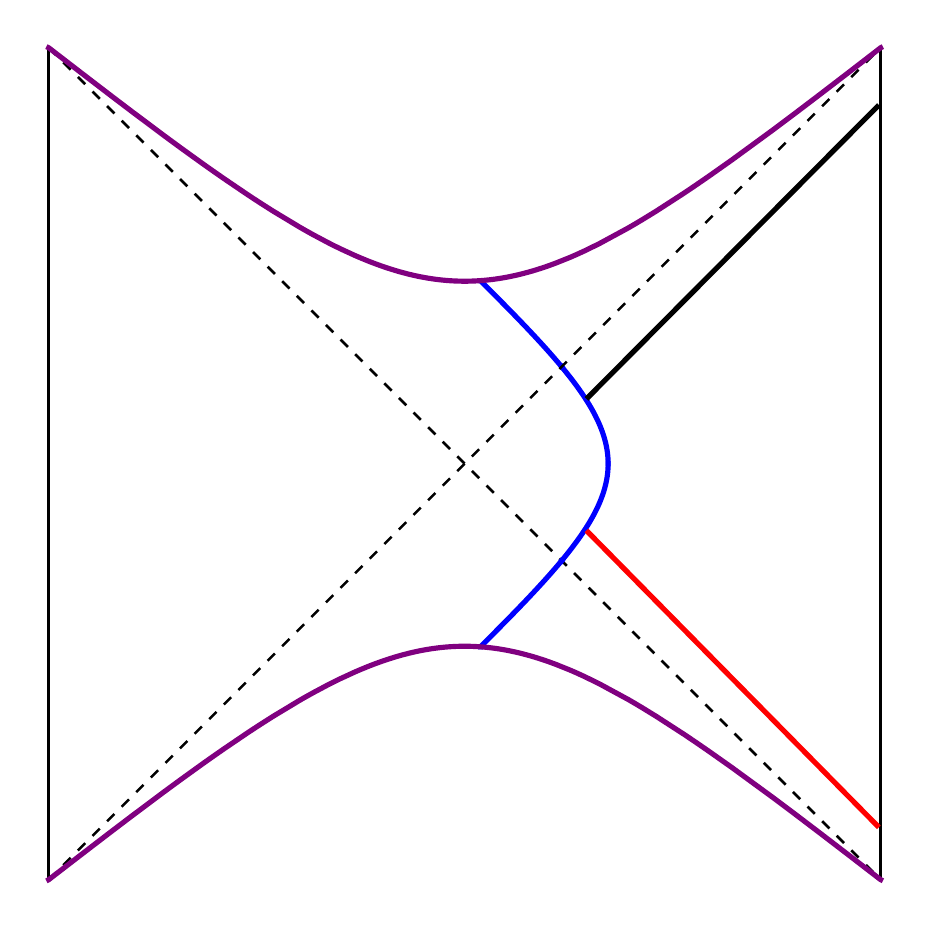}} \qquad \qquad
\subfigure[case 2]{\label{subfig:case2_confI}  \includegraphics[scale=0.18]{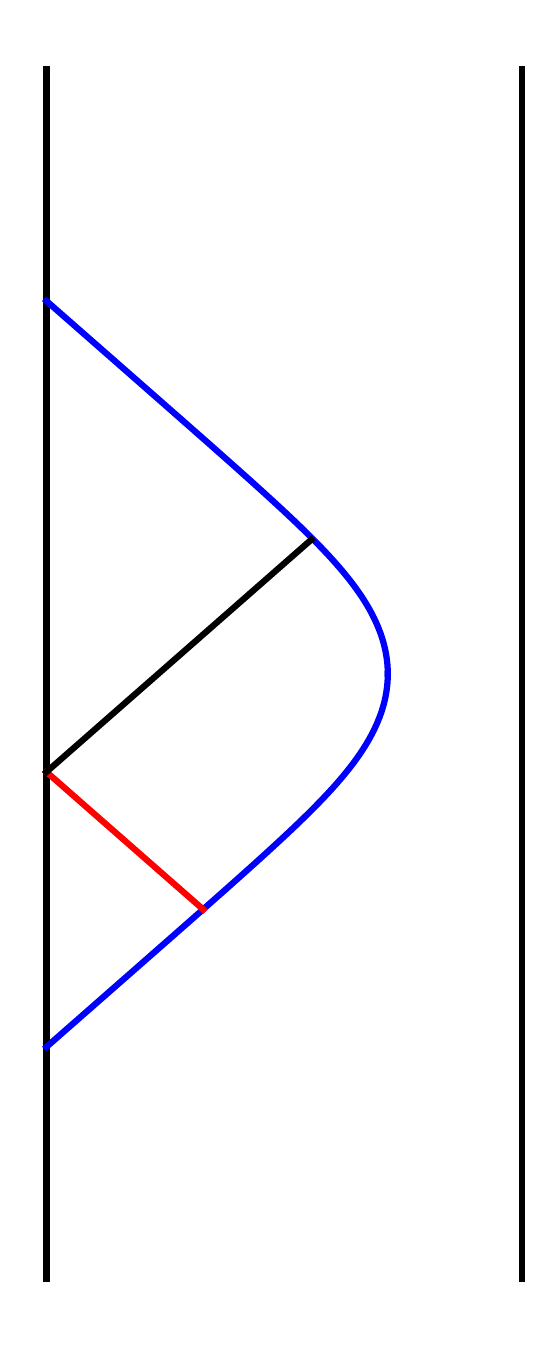} \quad
\includegraphics[scale=0.2]{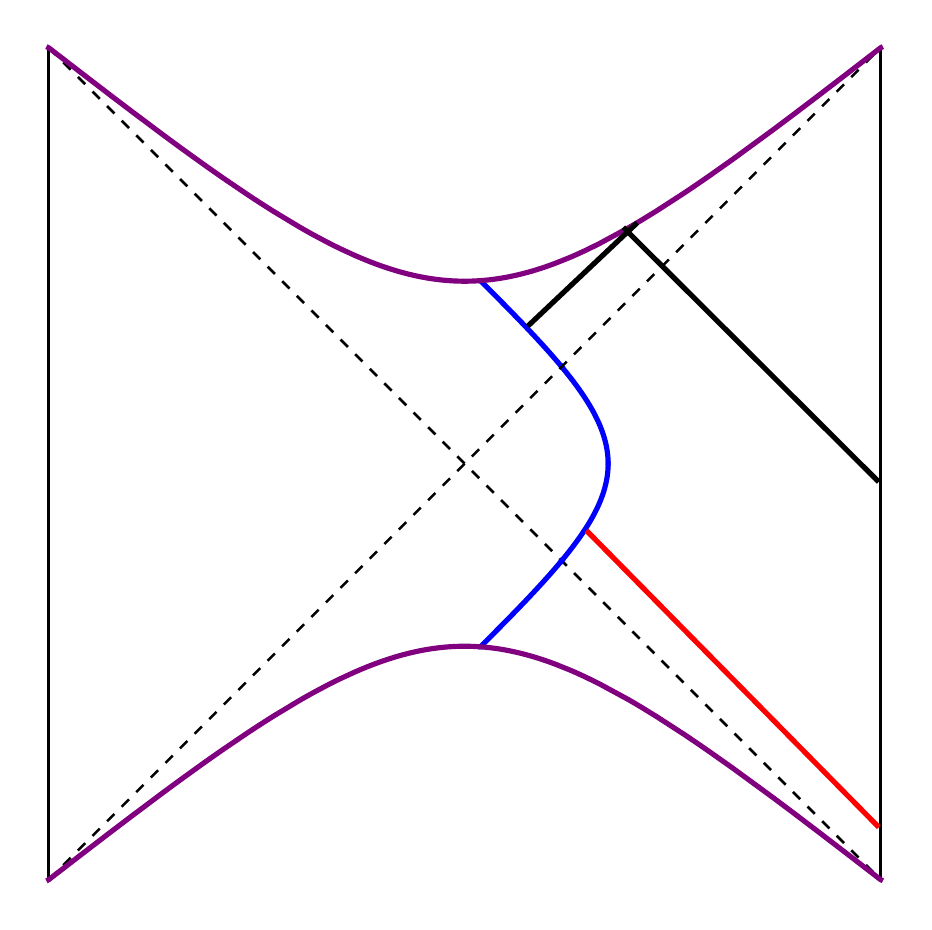}}\\
\subfigure[case 3]{ \includegraphics[scale=0.18]{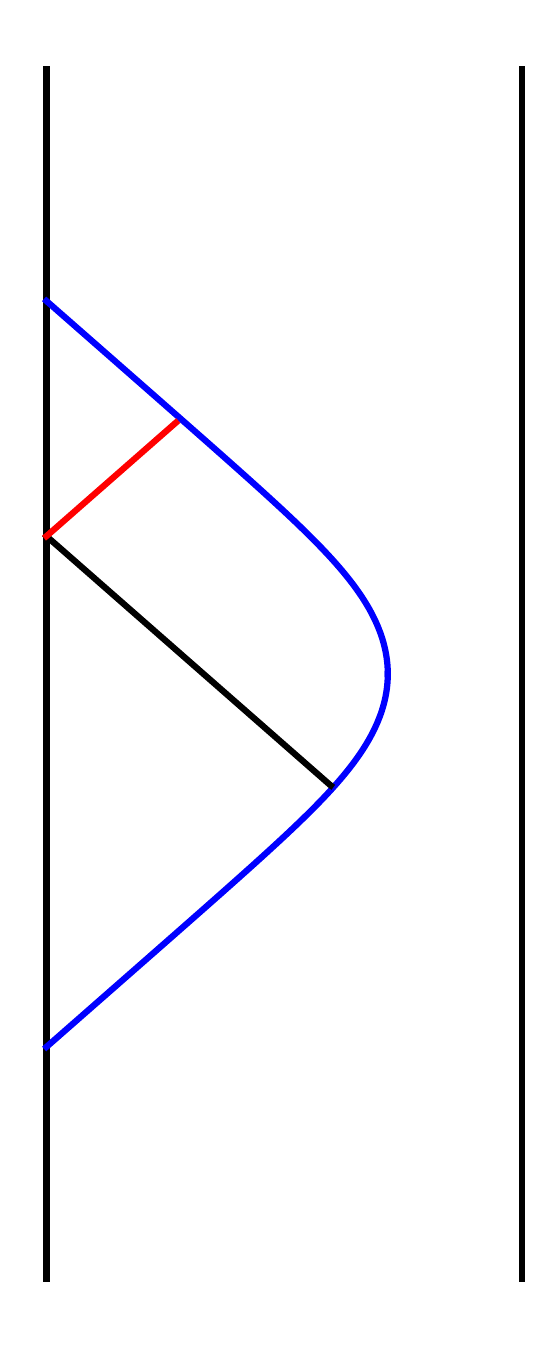} \quad
\includegraphics[scale=0.2]{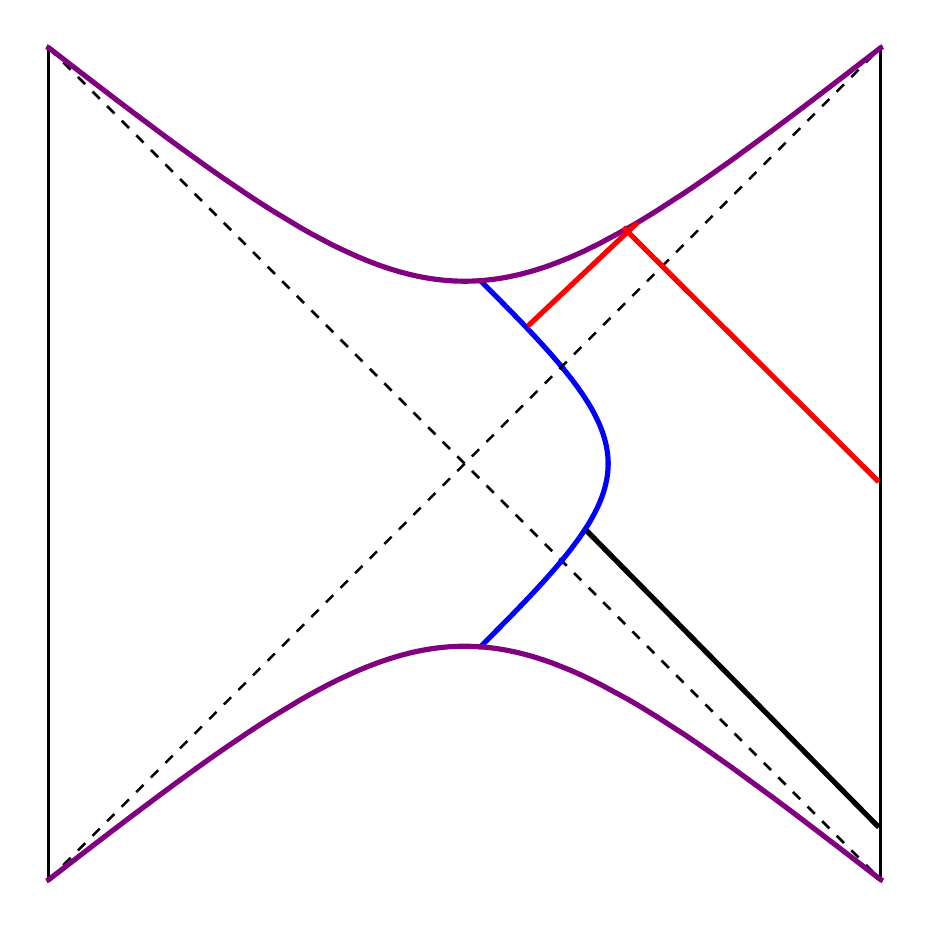}}
    \caption{
    Penrose diagram and trajectories of almost-null geodesics for a collapsing bubble with interior AdS geometry (configuration I in fig.~\ref{Pen-Dia-Collapsing-Bubble}).
     The color code is the same as in fig.~\ref{fig:nullgeo_static_AdS}.
    }
    \label{fig:Penrose diagram very small AdS bubble 0}
\end{figure}

\begin{figure}[h!]
    \centering
  \subfigure[$\lambda <0$]{\label{confI_AdS_tfin1} \includegraphics[scale=0.35]{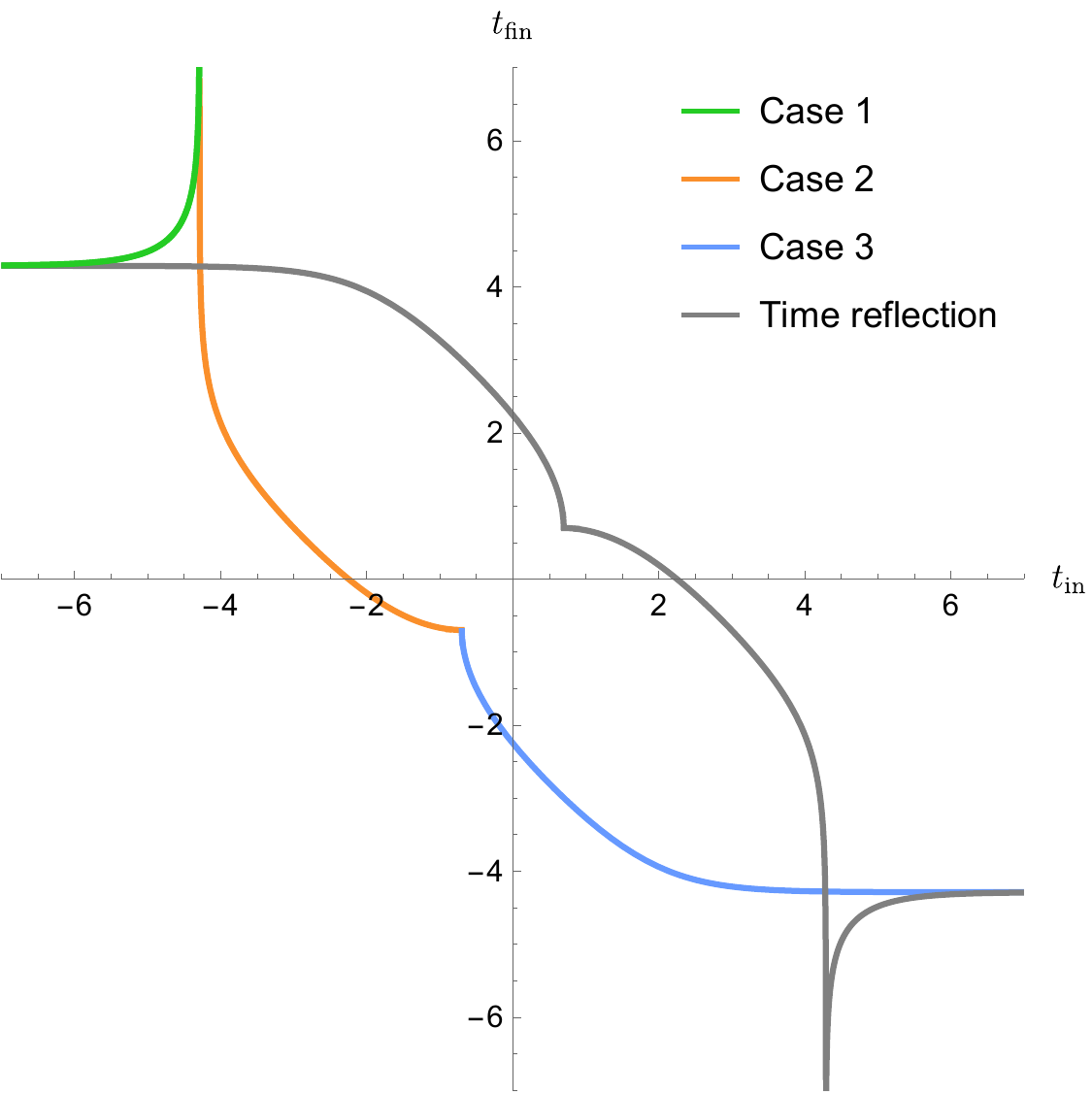}}
\qquad
\subfigure[$\lambda >0$]{\label{confI_AdS_tfin2} \includegraphics[scale=0.35]{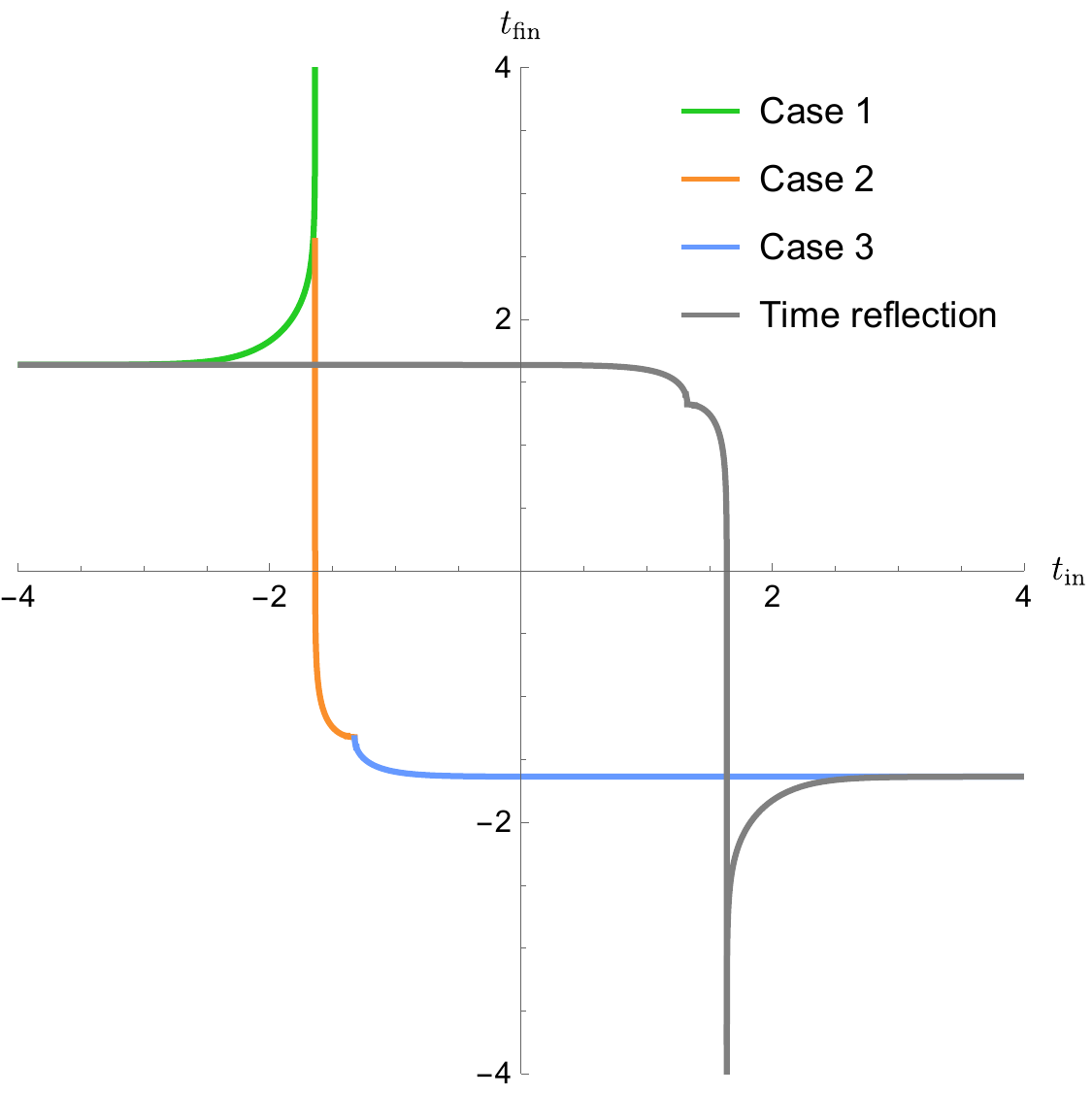} }
\caption{Plot of $t_{\rm fin}$ as a function of $t_{\rm in}$ for an illustrative numerical example  of  configuration I of fig.~\ref{Pen-Dia-Collapsing-Bubble}. 
On the left panel we fix $\kappa=0.25, \l=-1, r_h=0.5$, while in the right
panel we set $\kappa=0.5, \l=0.5, r_h=0.12$.
The green, orange, and blue curves refers to cases 1,  2, and  3 in table \ref{tab:verysmall_AdS_caseI}, respectively. The gray curve is obtained by applying the transformation~\eqref{eq:symm_time_rev} to the above mentioned curves. The divergence of the function $t_{\rm fin}(t_{\rm in})$ corresponds to an event horizon formation.
}
\label{fig:very_small_AdS_a_cased}
\end{figure}

Let us consider an almost-null spacelike geodesic starting from the right AdS boundary and directed towards the positive time direction. The qualitative details of the reflections are the same for both signs of $\l$. Depending on the initial boundary time, we can identify the following three cases, reported for increasing $t_{\rm in}$ in fig.~\ref{fig:Penrose diagram very small AdS bubble 0}:
\begin{enumerate}
    \item The ingoing geodesic at constant $v$ in the exterior geometry (depicted in red) intersects the domain wall. 
    Then the geodesic propagates in the interior AdS geometry, where it passes through the center of empty AdS spacetime, emerges at the antipodal point on the spherical section, and
    hits the domain wall again. Finally, the geodesic (depicted in black) propagates in the exterior geometry as a curve at constant $u$, going back to the AdS boundary without obstacles.  We refer to this configuration as case 1. 
    Here, the boundary times satisfy $t_{\rm fin} > t_{\rm in}$.
    \item In case 2, the ingoing red geodesic at constant $v$ encounters the domain wall, passes through the center of the interior AdS spacetime, and then emerges as an outgoing geodesic at constant $u$ (depicted in black) in the exterior geometry, where it is reflected at the future BH singularity before reaching the AdS boundary.
    The boundary times satisfy $t_{\rm fin} \geq t_{\rm in}$.\footnote{There is a critical time when the ingoing and outgoing almost-null geodesics in the AdS BH geometry coincide, and $t_{\rm fin} = t_{\rm in}$. As we will see in fig.~\ref{fig:very_small_AdS_a_cased}, this transition between cases 2 and 3 is characterized by a kink in the functional dependence $t_{\rm fin}(t_{\rm in})$.}
    \item The ingoing red geodesic at constant $v$ is reflected at the BH singularity, and then hits the domain wall. In the interior geometry, the trajectory goes through the center, and then encounters the shell again. In case 3, the outgoing black geodesic at constant $u$ propagates without obstacles until it reaches again the right AdS boundary.
    The boundary times satisfy $t_{\rm fin} \leq t_{\rm in}$.
\end{enumerate}

For convenience, we summarize in table~\ref{tab:verysmall_AdS_caseI} the behavior of the radial almost-null geodesics across the geometry in the case of configuration I.
In these and in other similar figures in the paper, the red line denotes an ingoing almost-null geodesic at constant $v$, while the black line is an outgoing geodesic. 
From now on, we will avoid repeating the full description of the trajectories of the almost-null geodesics in the case of other bubble geometries, and simply refer to similar tables.

\begin{table}[h!]   
\begin{center}   
\begin{tabular}  {|p{35mm}|c|c|c|} \hline  
 & 
 \makecell{\textbf{Reflection before}  \\ \textbf{meeting the bubble}} 
& \makecell{\textbf{Reflection after}  \\ \textbf{meeting the bubble}}  \\ \hline \rule{0pt}{4.9ex}
Case 1 & \redmark   & \redmark  \\
\rule{0pt}{4.9ex} Case 2 &  \redmark & \greencheck  \\
\rule{0pt}{4.9ex} Case 3 &  \greencheck & \redmark  \\[0.2cm]
\hline
\end{tabular}   
\caption{
Configurations of the radial almost-null geodesics in the bubble geometry with either AdS or dS interior (case I in fig.~\ref{Pen-Dia-Collapsing-Bubble}). From top to bottom, we increase the initial boundary time $t_{\rm in}$. 
} 
\label{tab:verysmall_AdS_caseI}
\end{center}
\end{table}

Numerical plots of the final time $t_{\rm fin}$ as a function of the initial $t_{\rm in}$ are shown in  fig.~\ref{fig:very_small_AdS_a_cased}.
In fig.~\ref{confI_AdS_tfin1} we consider a point in parameter space
with negative $\l$, while in fig.~\ref{confI_AdS_tfin2} we take a positive
$\l$. The qualitative structure of the function $t_{\rm fin}(t_{\rm in})$ is the same for both signs of $\l$.
The green curve covers case 1 of table~\ref{tab:verysmall_AdS_caseI}, with boundary times satisfying $t_{\rm in} < 0 < t_{\rm fin}$.
The function $ t_{\rm fin}( t_{\rm in})$ monotonically increases until there is a positive divergence, which corresponds to the propagation of the outgoing almost-null geodesic along the future BH horizon. 
This phenomenon can be interpreted as a consequence of the event horizon formation in a gravitational collapse~\cite{Hawking:1975vcx,Hubeny:2006yu}.
We will comment the physical meaning of this divergence in more detail below. 
After the divergence of $t_{\rm fin}$, the orange curve describes case 2, where the inequality $t_{\rm fin} > t_{\rm in}$ is still valid, but the final time monotonically decreases.
The transition to case 3 is marked by the appearance of a kink at a negative value of $t_{\rm in} = t_{\rm fin}$, where the function is continuous.
The kink corresponds to the limit in which the infalling geodesic
hits the intersection between the BH singularity and the bubble.
This might be an artifact of the thin wall approximation.
After the kink, the time $t_{\rm fin}<0$ monotonically decreases until it approaches a negative constant value for late times $t_{\rm in}$ (blue curve in fig.~\ref{fig:very_small_AdS_a_cased}).

For configuration I, we can shoot future-oriented radial null geodesics which come back to the right AdS boundary, corresponding to case 1 in fig.~\ref{fig:Penrose diagram very small AdS bubble 0}.\footnote{We remind that cases 2 and 3 in fig.~\ref{fig:Penrose diagram very small AdS bubble 0} only exist for spacelike geodesics in the infinite energy limit. Exact-null geodesics are not repelled by the BH singularity, therefore they only exist in case 1.} 
We denote by $t_h$ the critical time above which the case 1 geodesics no longer exist
(which corresponds to the limit in which the black line in fig.~\ref{subfig:case1_confI} lies on top of the horizon). If we take the limit $t_{\rm in} \to t_h$ from the left, the function $t_{\rm fin}(t_{\rm in})$ diverges to $+\infty$. Looking at fig.~\ref{subfig:case2_confI}, we infer that the same behavior is achieved also taking the 
limit from the right.
As anticipated above, we can interpret the quantity $t_h$ as the time of formation of the event horizon~\cite{Hubeny:2006yu}.
Using a combination of the symmetries in eqs.~(\ref{eq:symm_time_exchange}) and (\ref{eq:symm_time_rev}), we find
\beq
 \lim_{t_{\rm in} \to -\infty} t_{\rm fin}(t_{\rm in}) =-t_h \, ,
\qquad
 \lim_{t_{\rm in} \to +\infty} t_{\rm fin}(t_{\rm in}) =t_h \, .
\label{eq:definizione-th-generale}
\eeq
Let us analyze in more detail the limit $t_{\rm in} \to - \infty$, which belongs to the regime of case 1 in fig.~\ref{fig:Penrose diagram very small AdS bubble 0}.
The trajectory of a null infalling geodesic shot from the AdS boundary at time $t_{\rm in}$ is described by the curve at constant null coordinate $v=v_1 = t_{\rm in}$.
Denoting with $t_{o1}$ and $r_1$ the time and radial coordinates of the null geodesic at the intersection with the domain wall, we can express
\beq
t_{\rm in} = v_1=t_{o1}+r^*_o (r_1) \, .
\label{eq:vv1}
\eeq
Differentiating eq.~(\ref{eq:vv1}) with respect to $r_1$,
we get
\beq
\frac{d t_{\rm in}}{d r_1}  = 
\frac{d t_{o1}}{dr_1}+\frac{1}{f_o (r_1)} \, ,
\label{eq-diff-approx}
\eeq
where we used the definition~\eqref{tortoise} of tortoise coordinate.
In the limit $t_{\rm in} \to -\infty$, it is clear from the geometric setup that $r_1 \to r_h^+$ (from above).
Since $t_{o1}$ and $r_1$ are also coordinates on the domain wall, they satisfy eq.~\eqref{eq:dTdR}. Nearby $r_1\to r_h$, we approximate at leading order this identity to $\frac{d t_{o1}}{dr_1} \approx  \frac{1}{f_o(r_1)}$, and the blackening factor to $f_o(r_1) \approx (r_1-r_h) \, f'(r_h)$.
Plugging these approximations inside~(\ref{eq-diff-approx}), we can solve the differential equation as follows:
\beq
r_1-r_h \approx  W_1 \, \exp \le 2 \pi T \, t_{\rm in} \ri \, ,
\label{eq:asintoto-rh}
\eeq
where $W_1>0$ is an integration constant, and the Hawking temperature is
given by $T=f'(r_h)/(4 \pi)$.
Next, we evaluate the final time at which the geodesic goes back to the right AdS boundary. A numerical analysis reveals that $t_{\rm fin}$ is an increasing function of $r_1$, with $\frac{d t_{\rm fin}}{d r_1} >0$.
By approximating at linear order the series expansion of $t_{\rm fin}$ around $t_{\rm in } = -\infty$ and using the above results, we obtain
\beq
t_{\rm fin} \approx
- t_{h}+ W_2 \, \exp \le 2 \pi T \, t_{\rm in} \ri \, ,
\label{eq:asintoto-temperatura}
\eeq
where $W_2>0$ is an integration constant.
Equation~(\ref{eq:asintoto-temperatura}) shows that, in the limit
$t_{\rm in} \to - \infty$, the function $t_{\rm fin}$ approaches its asymptotic value exponentially fast in $t_{\rm in}$, where the argument of the exponential is proportional to the Hawking temperature, thus encoding physical information about the BH.

\subsubsection{Configurations II and III}

Configurations II and III present the same qualitative structure of bulk-cone singularities.
The structure of the causal wedge corresponding to configurations II and III of fig.~\ref{Pen-Dia-Collapsing-Bubble} can be realized in the light green and blue regions of the parameter space in fig.~\ref{figure-beta-outside-d3}, respectively. 
Both cases can be also attained in the large $m$ regime.
For these configurations, it is not possible 
for a radial exactly-null geodesic to go back to the boundary in a finite time. This means that an event horizon exists for any value of the boundary time. 

\begin{figure}[ht!]
    \centering
\subfigure[case 4]{ \includegraphics[scale=0.15]{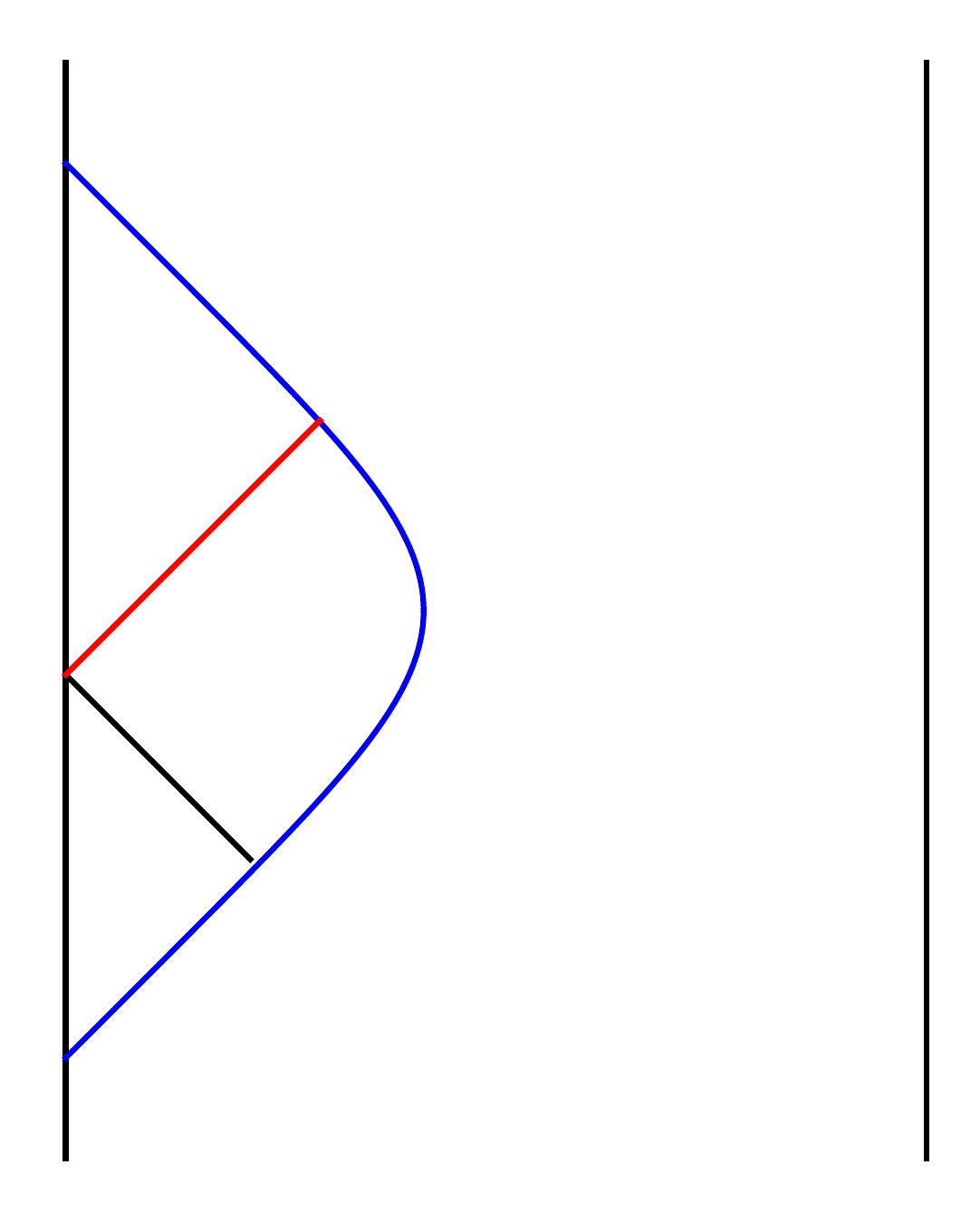} \quad
\includegraphics[scale=0.22]{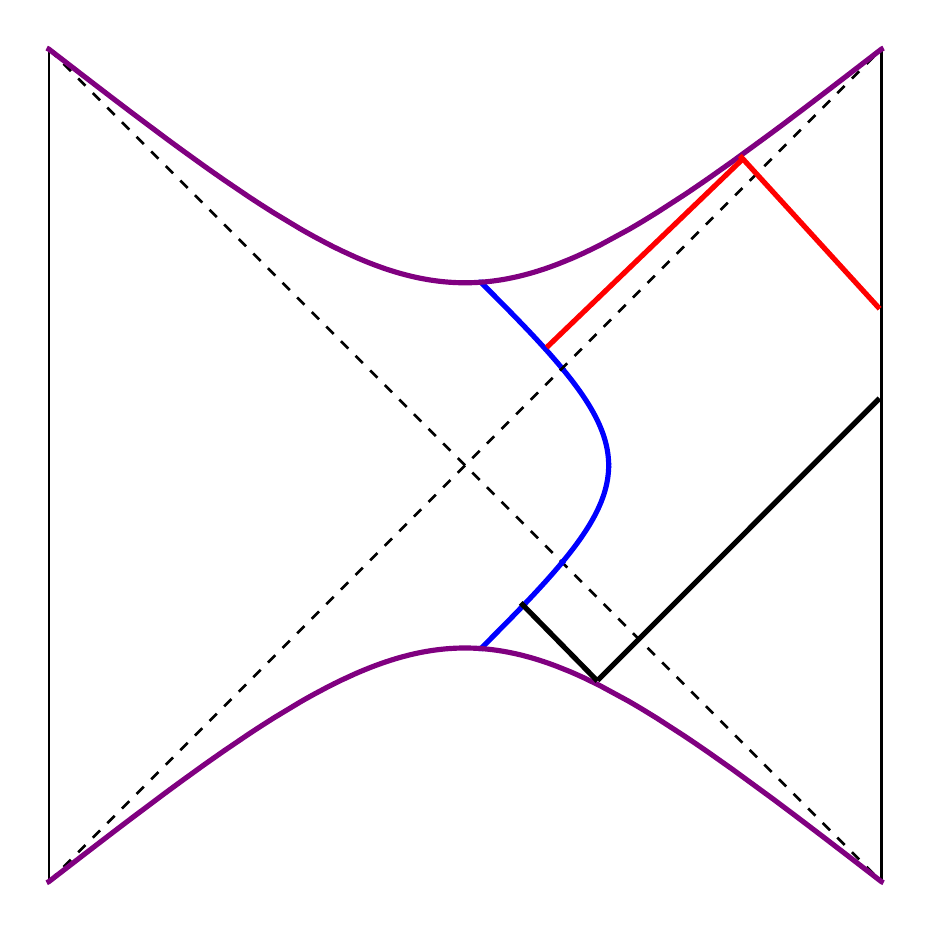}}
    \caption{
    Penrose diagram for the case 4 of a collapsing bubble with interior AdS geometry (configuration II in fig.~\ref{Pen-Dia-Collapsing-Bubble}). }
    \label{fig:Penrose diagram very small AdS bubble 1}
\end{figure}

The possible trajectories of a radial spacelike almost-null geodesic are described in table~\ref{tab:verysmall_AdS_caseI}, where the labeling is chosen such that cases 2 and 3 coincide with the analogous settings in subsection~\ref{ssec:configurationI_collapsing}.
Case 4 corresponds to almost-null geodesics that are reflected by both the future and past BH singularities, as depicted in fig.~\ref{fig:Penrose diagram very small AdS bubble 1} (for definiteness, in the case of an AdS interior geometry).

\begin{table}[ht!]   
\begin{center}   
\begin{tabular}  {|p{35mm}|c|c|c|} \hline  
 & 
 \makecell{
 \textbf{Reflection before}  \\ \textbf{meeting the bubble}}
& \makecell{\textbf{Reflection after}  \\ \textbf{meeting the bubble}}  \\ \hline \rule{0pt}{4.9ex}
Case 2 & \redmark & \greencheck  \\
\rule{0pt}{4.9ex} Case 3 &  \greencheck & \redmark  \\
\rule{0pt}{4.9ex} Case 4 &  \greencheck & \greencheck  \\[0.2cm]
\hline
\end{tabular}   
\caption{
Configurations of the radial almost-null geodesics in the collapsing bubble geometry with either AdS or dS interior for cases II and III in fig.~\ref{Pen-Dia-Collapsing-Bubble}. From top to bottom, we increase the initial boundary time $t_{\rm in}$. } 
\label{tab:verysmall_AdS}
\end{center}
\end{table}

\begin{figure}[ht!]
    \centering
    \subfigure[$\kappa=0.26$]{\label{subfig:tfti_caseIIa}
    \includegraphics[scale=0.35]{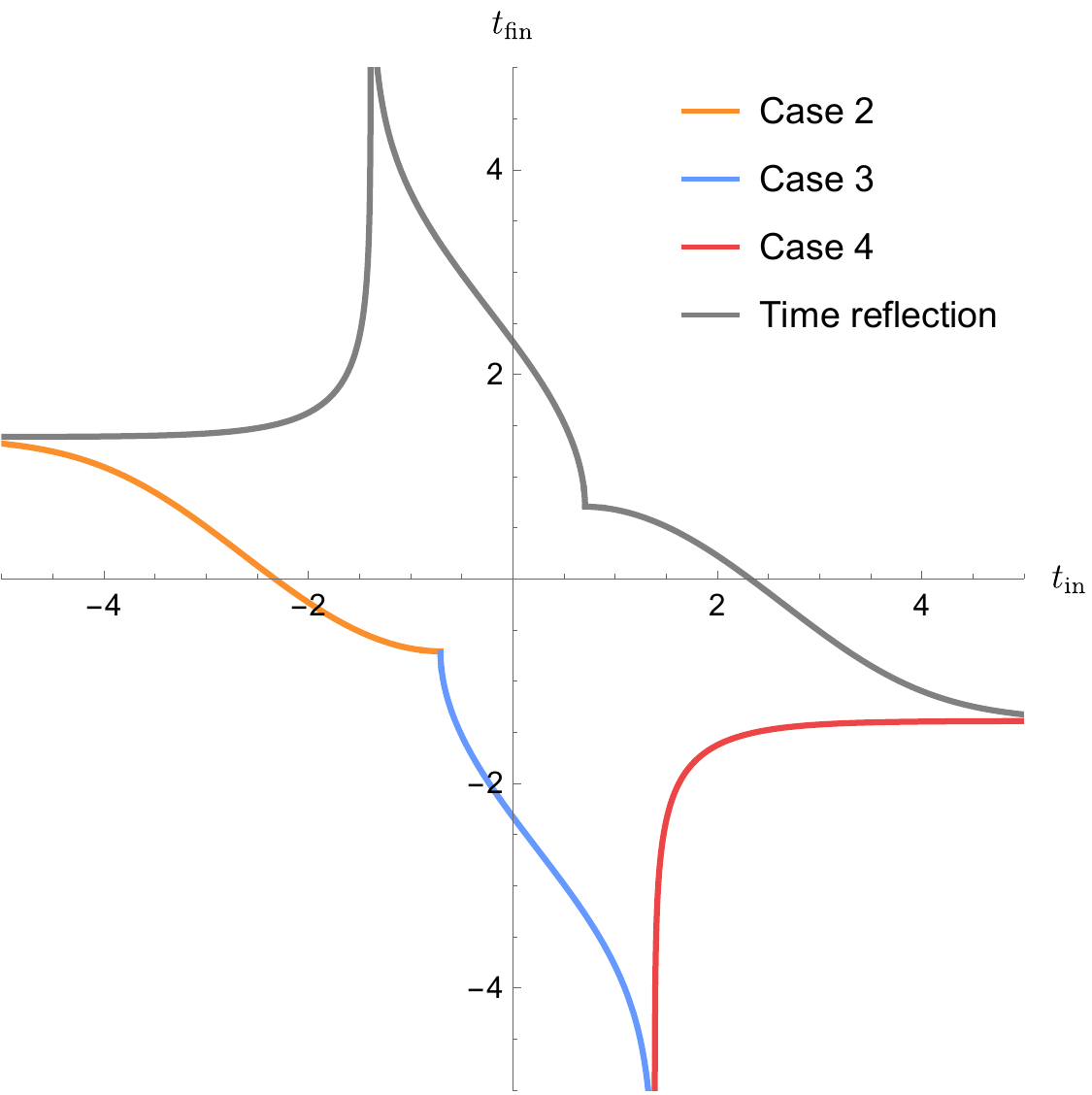}} \qquad
\subfigure[$\kappa=0.5$]{\label{subfig:tfti_caseIIb}
\includegraphics[scale=0.35]{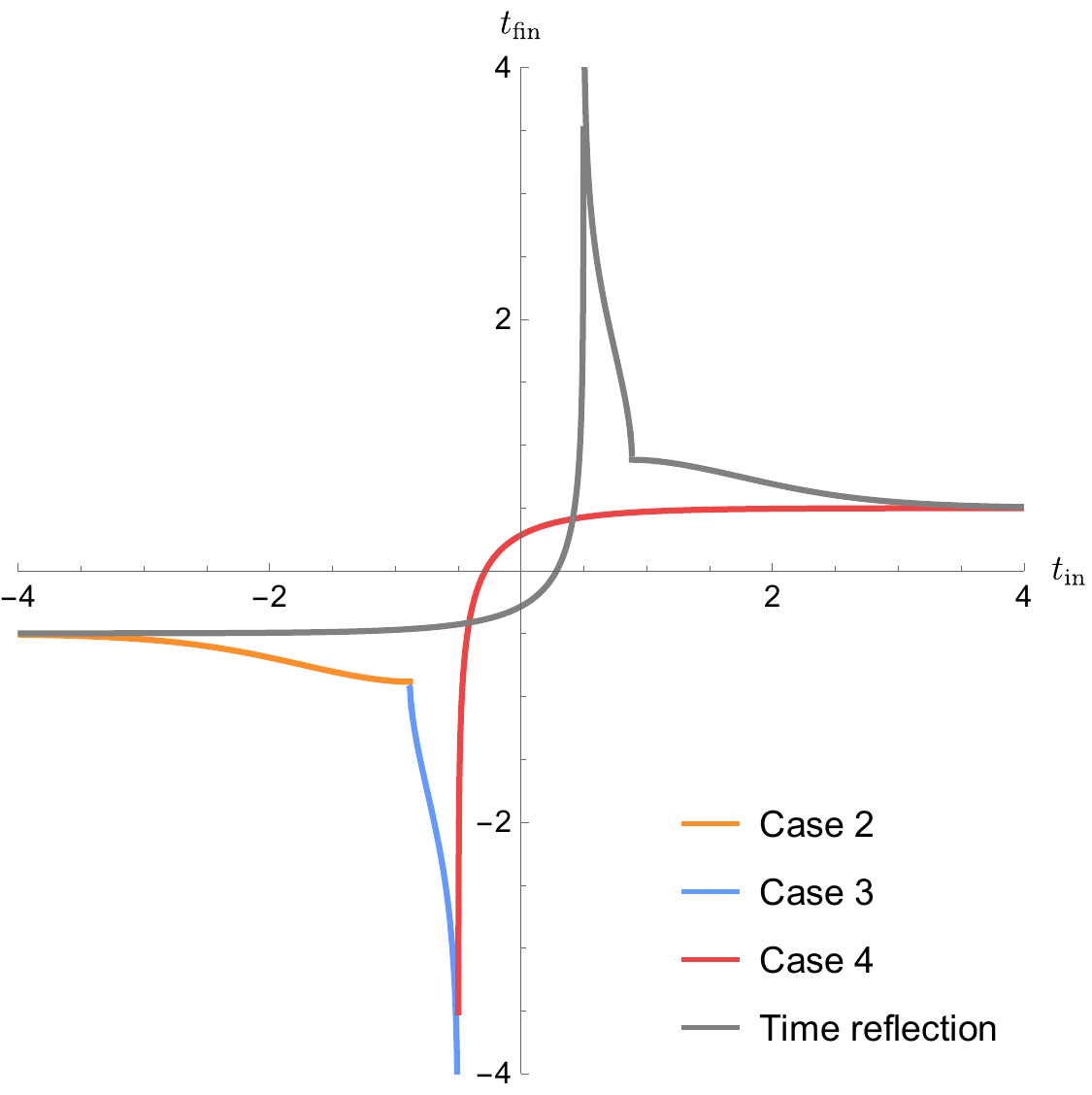}} 
\caption{Plot of $t_{\rm fin}$ as a function of $t_{\rm in}$ in the case II of fig.~\ref{Pen-Dia-Collapsing-Bubble}. 
We fix $\l=-1, r_h=0.5$, and we vary the tension $\kappa$ as reported below each picture.}
    \label{Collapsing-Bubble-AdS-A}
\end{figure}

Let us discuss the functional dependence of $t_{\rm fin} (t_{\rm in})$,
referring to fig.~\ref{Collapsing-Bubble-AdS-A} for the depiction of some representative examples with interior AdS geometry. 
Scanning the plot from left to right, we note that the orange curve initially describes the configuration in case 2 of table~\ref{tab:verysmall_AdS}, where $t_{\rm fin} > t_{\rm in}$, until the kink marks the transition to case 3.
Here, we observe that the function $t_{\rm fin} (t_{\rm in})$
has a negative divergence, occurring when the outgoing light ray propagates along the past BH horizon.
In contrast with the plot reported for case I, this divergence is not related the formation of an event horizon.
Depending on the point in the parameter space, this divergence can happen at a positive or at a negative value of $t_i$, as can be seen by comparing the two plots reported in fig.~\ref{Collapsing-Bubble-AdS-A}.

\begin{figure}[ht!]
    \centering
   \label{fig:very small bubble case result} \includegraphics[scale=0.3]{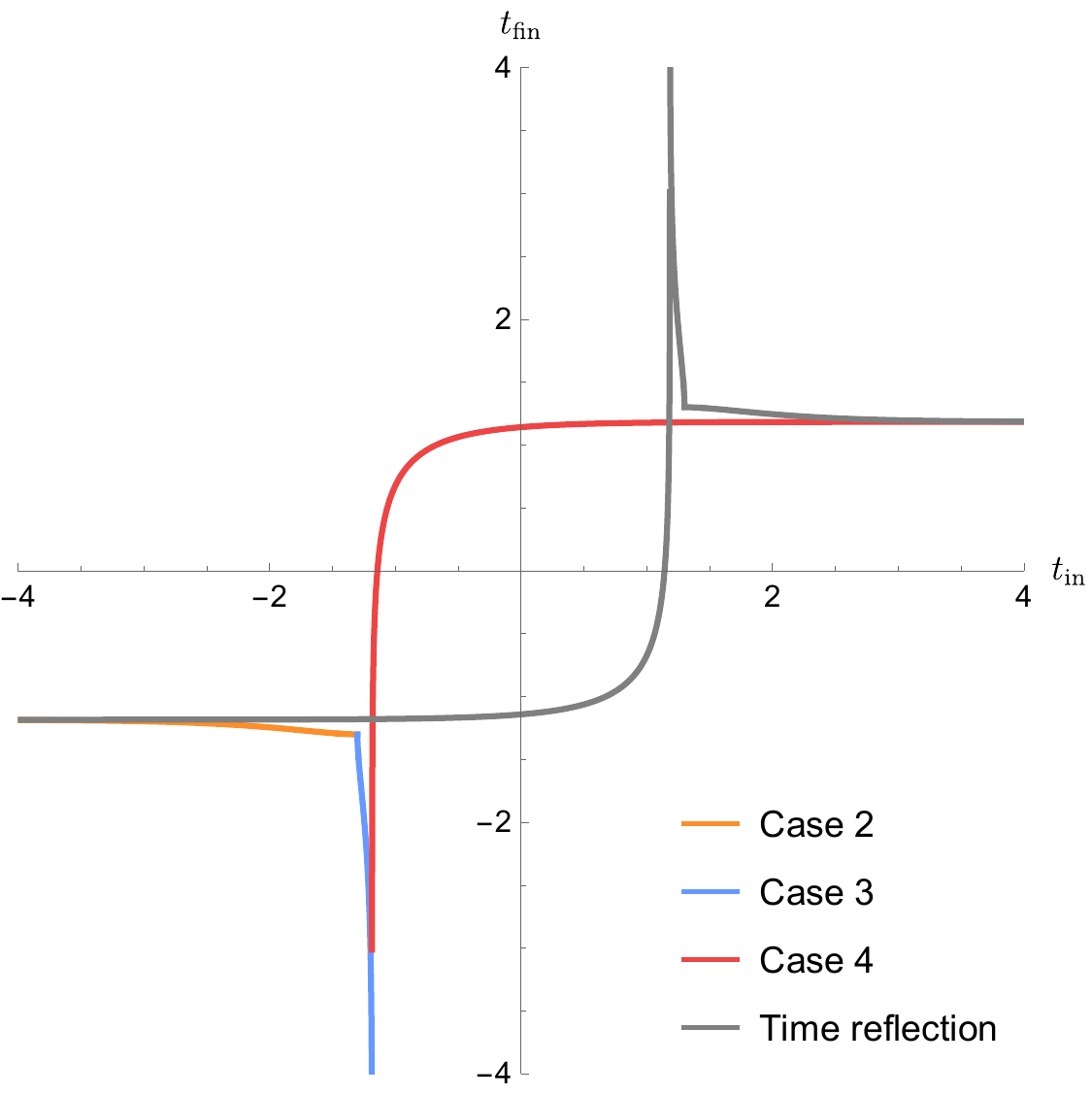}
    \caption{Plot of the final boundary time $t_{\rm fin}$ as a function of the initial time $t_{\rm in}$ for configuration II. We fix $\kappa=0.5, \l=1, r_h=0.5$.}
  \label{fig:tfti_dS_CaseII}
\end{figure}

The case of a bubble in the configuration II with a dS interior is similar: the trajectory of an almost-null geodesics with increasing boundary time follows the three sub-cases reported in table~\ref{tab:verysmall_AdS}, and the behavior of bulk-cone singularities is depicted in fig.~\ref{fig:tfti_dS_CaseII}.

\begin{figure}[ht!]
    \centering
\subfigure[]{ \label{fig:not-so-small AdS bubble result} 
\includegraphics[scale=0.3]{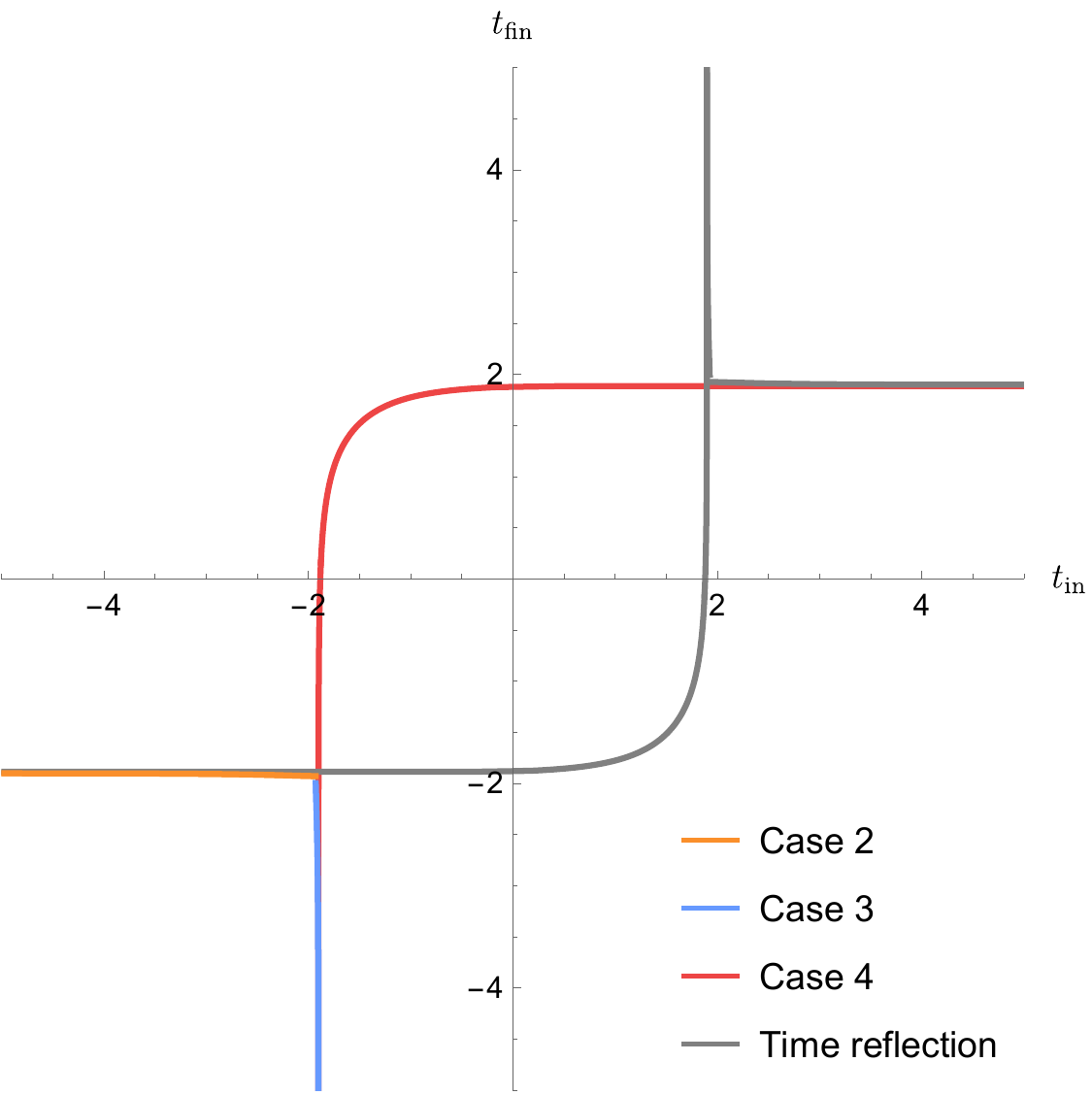}}
 \quad
\subfigure[]{\label{fig:not so small bubble case result} \includegraphics[scale=0.3]{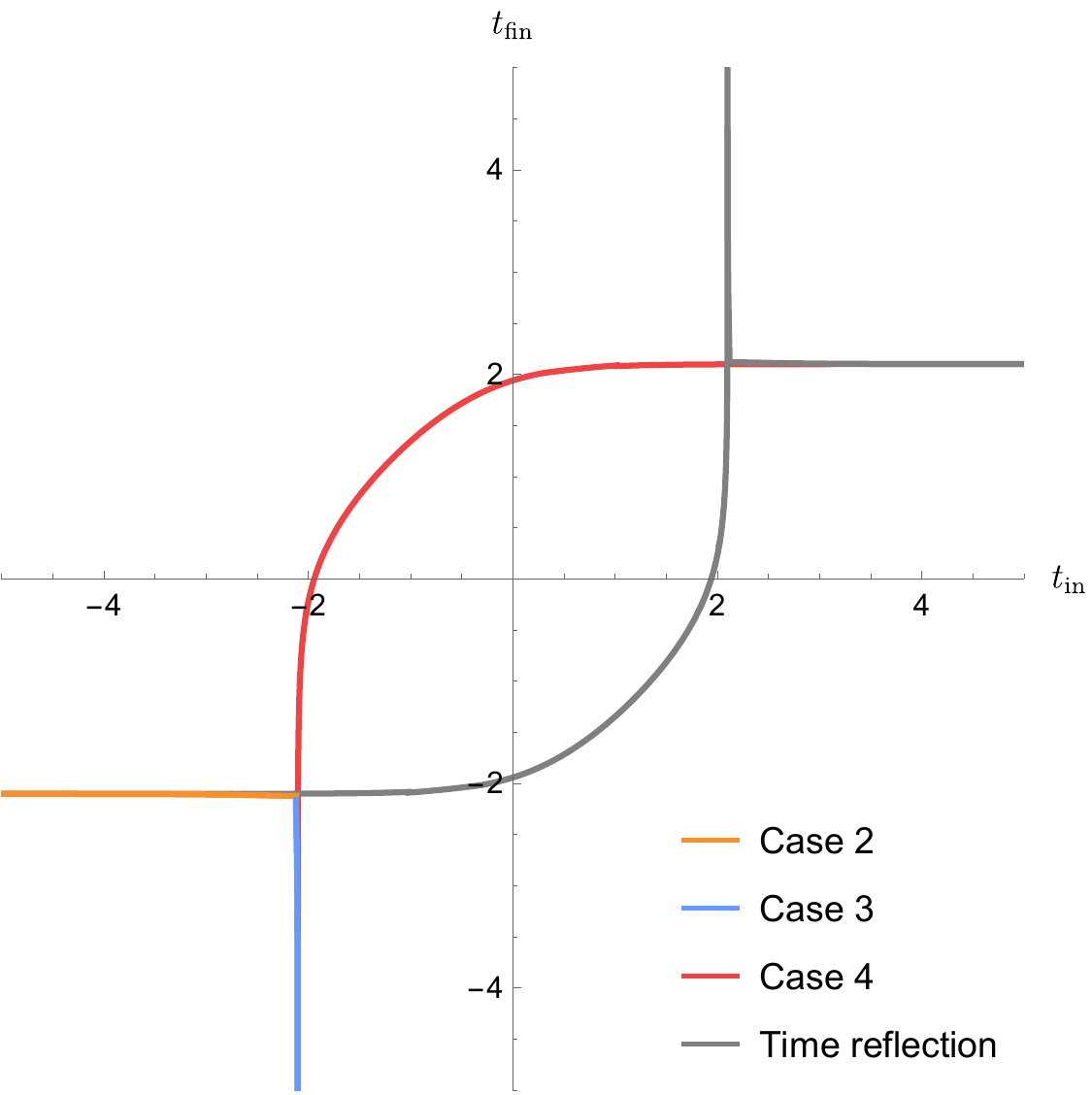}}
    \caption{
     Plot of $t_{\rm fin}$ as a function of $t_{\rm in}$ for some examples of configuration III. 
    (a) We fix $\kappa=2.5, \l=-1, r_h=0.5$.
    (b) We set $\kappa=0.5, \l=1, r_h=0.95$.
   }
   \label{bulk-singo-caseIII}
\end{figure}

There is no neat difference in the structure of the bulk-cone singularities between configurations II and III, see fig.~\ref{bulk-singo-caseIII}.
We stress that in cases II and III the value of $t_{\rm in}$ at which the function $t_{\rm fin}(t_{\rm in})$ has a kink (transition from case 2 to 3 in table~\ref{tab:verysmall_AdS}) is necessarily strictly less than the value of $t_{\rm in}$ at which the function is divergent (transition from case 3 to 4). In particular, for the two values to coincide, the ingoing geodesic should hit the future BH singularity at the intersection with the domain wall, the outgoing one should propagate along the past BH horizon, and the identity $t_{\rm in}=t_{\rm fin}$ (defining the kink) should hold. Putting this facts together, both the ingoing and outgoing geodesics should propagate along the past BH horizon, which means $t_{\rm in} = t_{\rm fin} \to -\infty$.
This can never happen for a collapsing bubble, whose radius eventually shrinks to the singularity $R \to 0$.

For configurations II and III, no future-oriented radial exactly-null geodesic which comes back to the right AdS boundary exists.
Nonetheless, eq.~\eqref{eq:definizione-th-generale} can be used as a definition of the time $t_h$.
In particular, the limit $t_{\rm in} \to -\infty$ is still described by eq.~\eqref{eq:asintoto-temperatura}, but now with $W_2<0$ because $t_{\rm fin}$ is a decreasing function of $r_1 - r_h$ (see case 2 in fig.~\ref{fig:Penrose diagram very small AdS bubble 0}).
While this formula is not associated in this case to the formation of an event horizon, it still reveals physical information about the BH background in that $t_h$ is approached with an exponentially-fast function in the Hawking temperature.

\subsubsection{Criticality at the border between configurations I and II}

\begin{figure}[h!]
\centering
\includegraphics[scale=0.35]{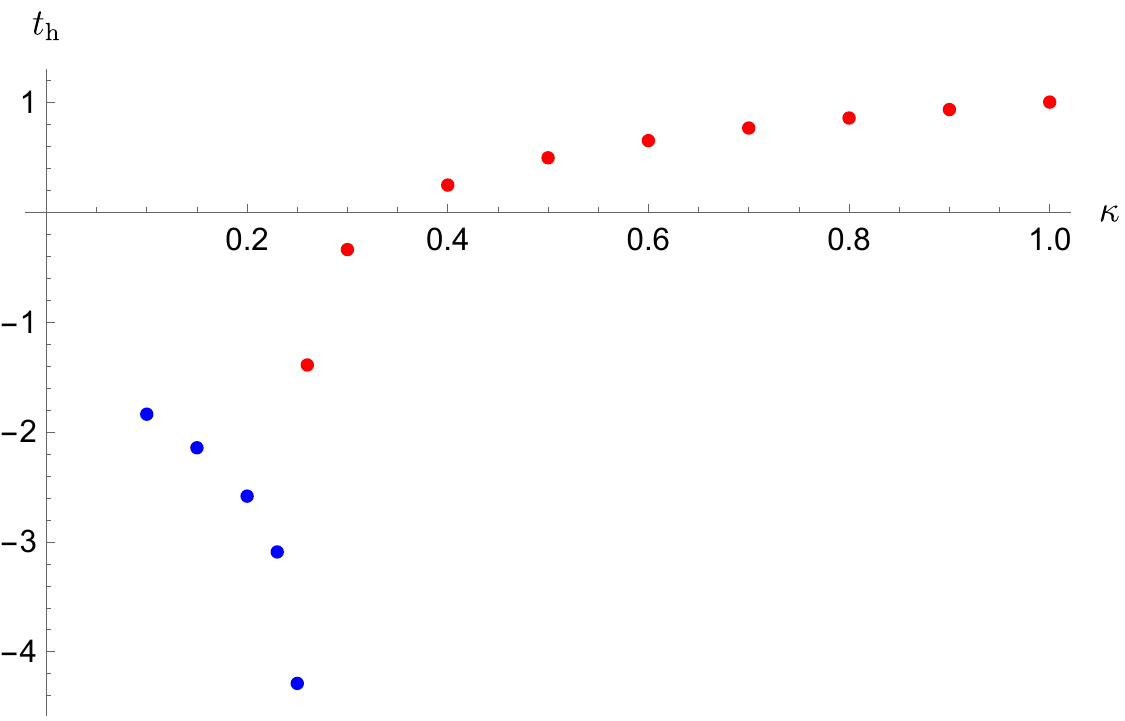}
\caption{ Plot of  $t_h$ as a function of $\kappa$, for fixed $\l=-1$ and $r_h=0.5$.
The critical time $t_h$ diverges around $\kappa \simeq 0.2529$, which is at the boundary between configurations I and II in parameter space.}
\label{fig:tempo_s}
\end{figure}

\begin{figure}[h!]
\centering
\subfigure[]{\label{fig:transitionIandII1}  \includegraphics[scale=0.5]{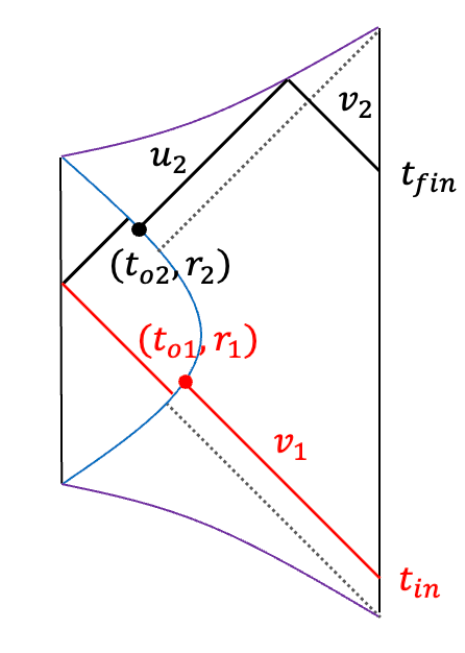}}
\qquad \qquad
\subfigure[]{\label{fig:transitionIandII2} \includegraphics[scale=0.3]{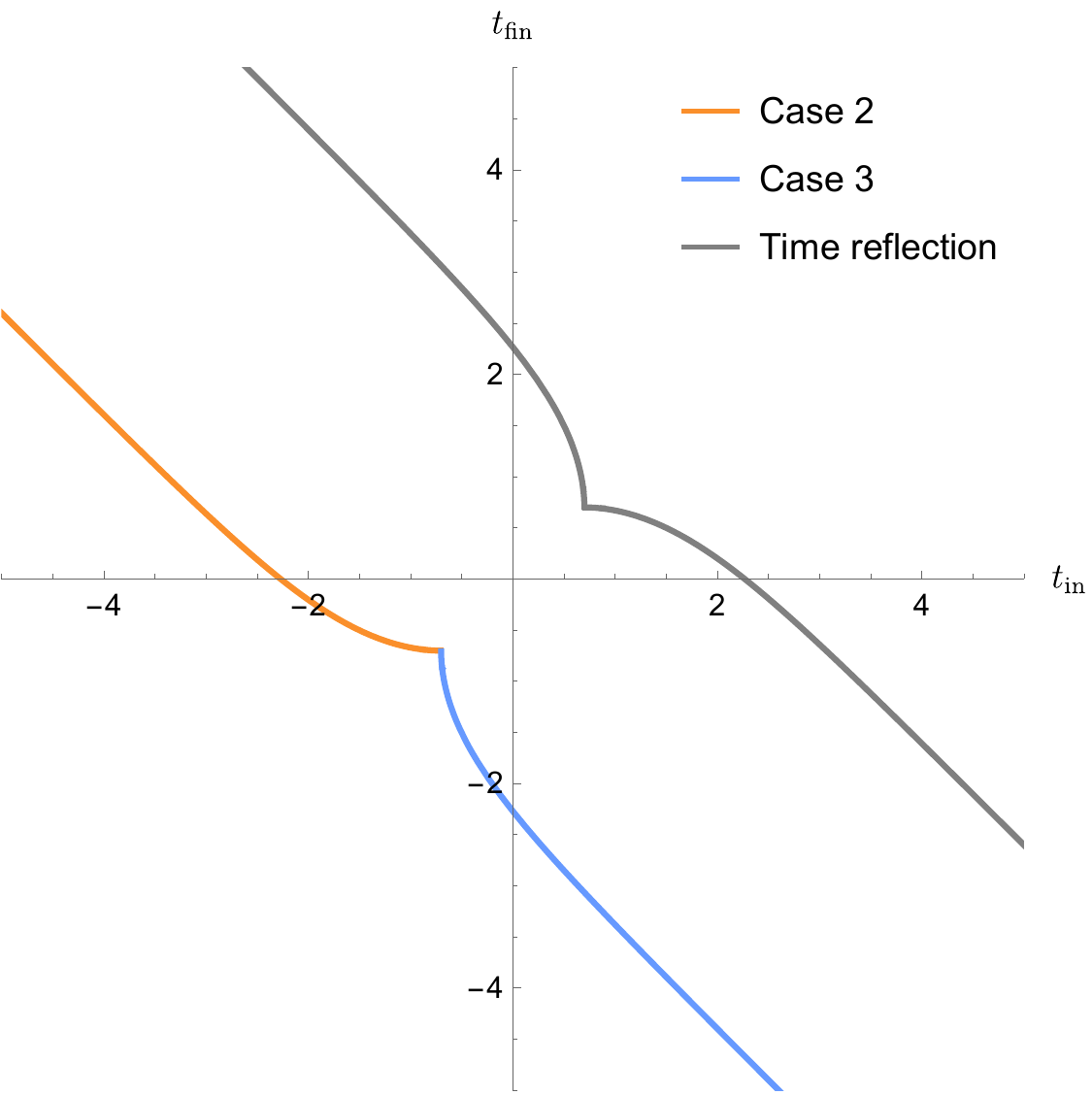}}
\caption{Left: Sketch of the Penrose diagram at the transition between cases I and II.
Right: Plot of $t_{\rm fin}$ as a function of $t_{\rm in}$ 
for an illustrative numerical value at the border between configurations I
and II in parameter space.
Here we fix  $\kappa \simeq 0.2529, \l=-1, r_h=0.5$.
}
\label{fig:very_small_AdS_a_casec}
\end{figure}

Bulk-cone singularities present a peculiar behavior in correspondence of the transition in parameter space between configurations I and II.
In this fine-tuned case, the critical time $t_h$ -- defined in eq.~(\ref{eq:definizione-th-generale}) -- diverges to $-\infty$, as confirmed by the numerical analysis reported in  fig.~\ref{fig:tempo_s}.
The function $t_{\rm fin}(t_{\rm in})$ approaches a linear behavior at large $t_{\rm in}$ (see fig.~\ref{fig:transitionIandII2}), as opposed to the plots in figs.~\ref{fig:very_small_AdS_a_cased} and \ref{Collapsing-Bubble-AdS-A} for the previous settings.
In the limit under consideration, case 1 in table~\ref{tab:verysmall_AdS_caseI} no longer exists, and the event horizon formation (which happens at the border between cases 1 and 2)
is pushed away at $t_{\rm in} \to -\infty$.

In the limit $t_{\rm in } \to - \infty$, the final time can be approximated by
\beq
t_{\rm fin}(t_{\rm in} )\approx -t_{\rm in} + 2 r_o^*(0) \, ,
\label{limite-speciale-I-II}
\eeq 
as we show below.
Consider an infalling radial almost-null geodesic sent from the boundary at time $t_{\rm in}$ lying at constant $v_1=t_{\rm in}$, see fig.~\ref{fig:transitionIandII1}.
This curve intersects the domain wall at the coordinates $(t_{o1}, r_{1})$, satisfying the condition in eq.~\eqref{eq:vv1}. In the limit $t_{\rm in} \to -\infty$, a direct inspection of the Penrose diagram reveals that $r_1 \to r_h^+$ from above. 
After entering the bubble, the geodesic passes through the center (pole) of the AdS (dS) interior, and then intersects the domain wall again at the time $t_{o2}$, with corresponding radial coordinate $r_2$. Due to the time-reversal symmetry, we find 
\beq
r_h-r_2 \approx r_1-r_h \, , 
\qquad
t_{o2}\approx -t_{o1} \, .
\label{eq-approx-2}
\eeq
The first relation implies $r^*_o(r_1) \approx r^*_o(r_2)$.
After the intersection with the domain wall, the almost-null geodesic is
described by an outgoing curve at constant $u_2 =t_{o2}-r^*_o(r_2)$, propagating in the external BH background.  
Combining the parametrization of $u_2$ with eq.~(\ref{eq-approx-2}), we get $u_2 \approx-t_{\rm in}$. 
This almost-null radial geodesic bounces off at the future BH singularity, thus becoming a curve at constant $v_2 = u_2 + 2 r^*_o(0)  = t_{\rm fin}$.
Using $u_2 \approx - t_{\rm in}$, this finally yields the result~\eqref{limite-speciale-I-II}.
Indeed, this analytic result is confirmed by the numerical analysis reported in fig.~\ref{fig:transitionIandII2}.

\subsection{Expanding bubbles}
\label{ssec:expanding_AdS}

Let us now consider the case of an expanding bubble, whose causal structure in various regions of parameter space was depicted in fig.~\ref{Pen-Dia-Expanding-Bubble}.
The setting with a dS interior refers to region $A$ of the parameter space in fig.~\ref{Phase-Diagram}, while the configurations with AdS interior can only occur in regions $B, C, E$.
Let us briefly analyze the main features of the geometries in each case.
In region $A$, the interior dS spacetime always contains timelike infinities $\mathcal{I}^{\pm}$, and a radial almost-null geodesic entering the interior of the bubble may or may not reach this region.
The structure of bulk-cone singularities will be different in these two cases.
In region $B$ of parameter space, the interior AdS spacetime includes another asymptotic boundary.
As a consequence, any radial almost-null geodesic that reaches the left boundary does not come back to the domain wall, and cannot reach the right AdS boundary again. In such case, there are no bulk-cone singularities.
In region $C$ of parameter space, the domain wall is located on the left side of the Penrose diagram, and the interior geometry contains the center of empty AdS. This structure ultimately allows radial almost-null geodesics to come back to the right AdS boundary. 
We expect that the propagation of radial almost-null geodesics encodes interesting information about the BH interior, which is part of the full geometry.
In region $E$ of parameter space, the domain wall is located on the right side of the Penrose diagram, therefore the BH interior is cut away from the full geometry.

For the above reasons, we will focus on studying regions $A$ (for a dS interior) and $C$ (for an AdS interior) of the parameter space, since they are the most interesting cases.
It is important to note that for both geometries
there are regions where the thin wall approximation might fail:
\begin{itemize}
\item For region $A$, in the Penrose diagram
there is a corner where the left AdS boundary of the exterior
touches the dS infinity of the interior, see fig.~\ref{subfig:regionA_Pen_Exp}.
\item
For region $C$, in the Penrose diagram
there is a corner where the left AdS boundary of the exterior
touches the AdS boundary in the interior, see fig.~\ref{subfig:regionC_Pen_Exp}.
\end{itemize}
It would be desirable to perform further studies
without relying on the thin wall approximation.

\subsubsection{AdS interior (region $C$ of parameter space)}

\begin{figure}[ht!]
    \centering
    \subfigure[]{ \label{fig:Penrose_expanding_AdS} \includegraphics[scale=0.28]{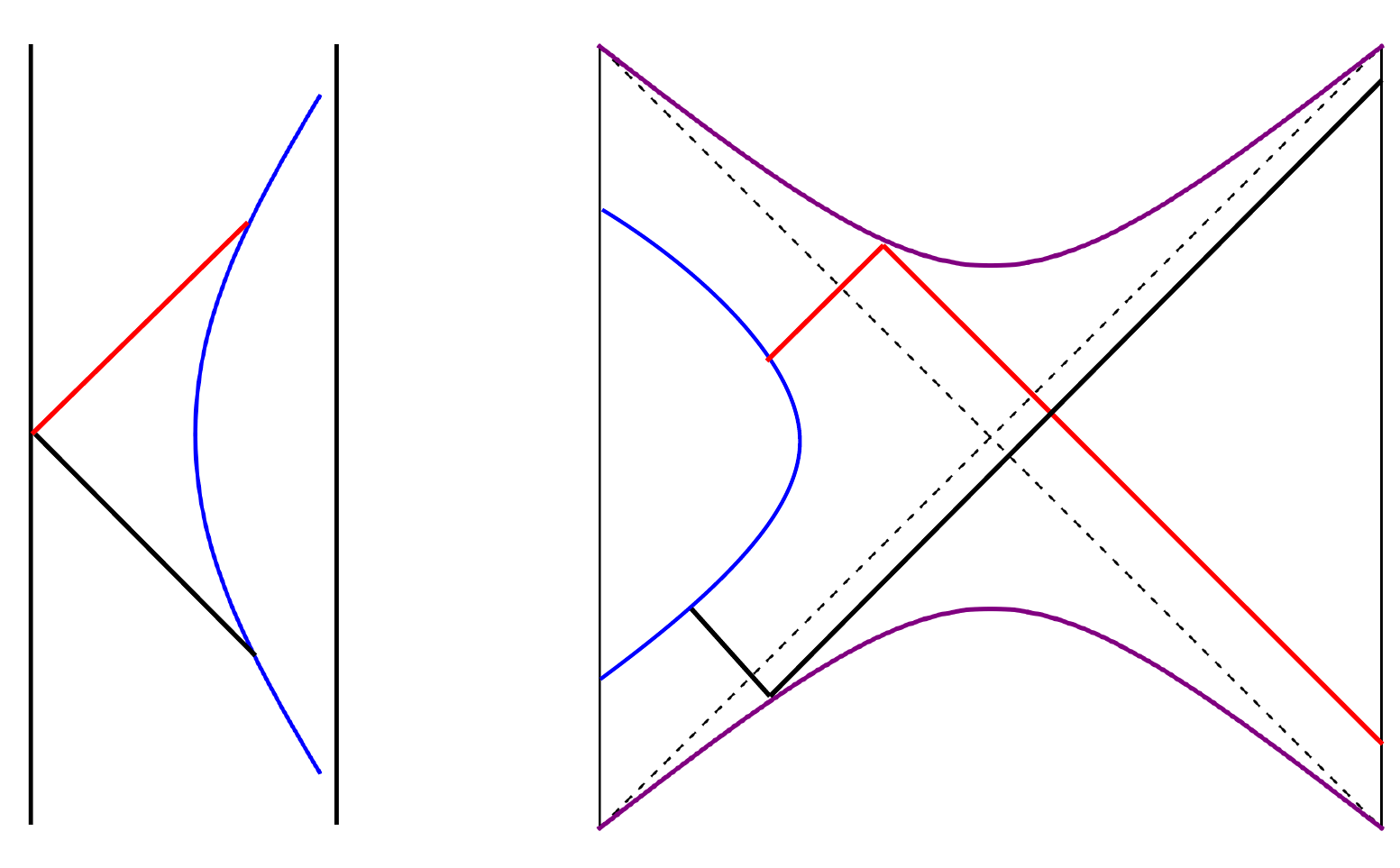}} \quad
    \subfigure[]{\label{fig:expanding_tfti_AdS} \includegraphics[scale=0.28]{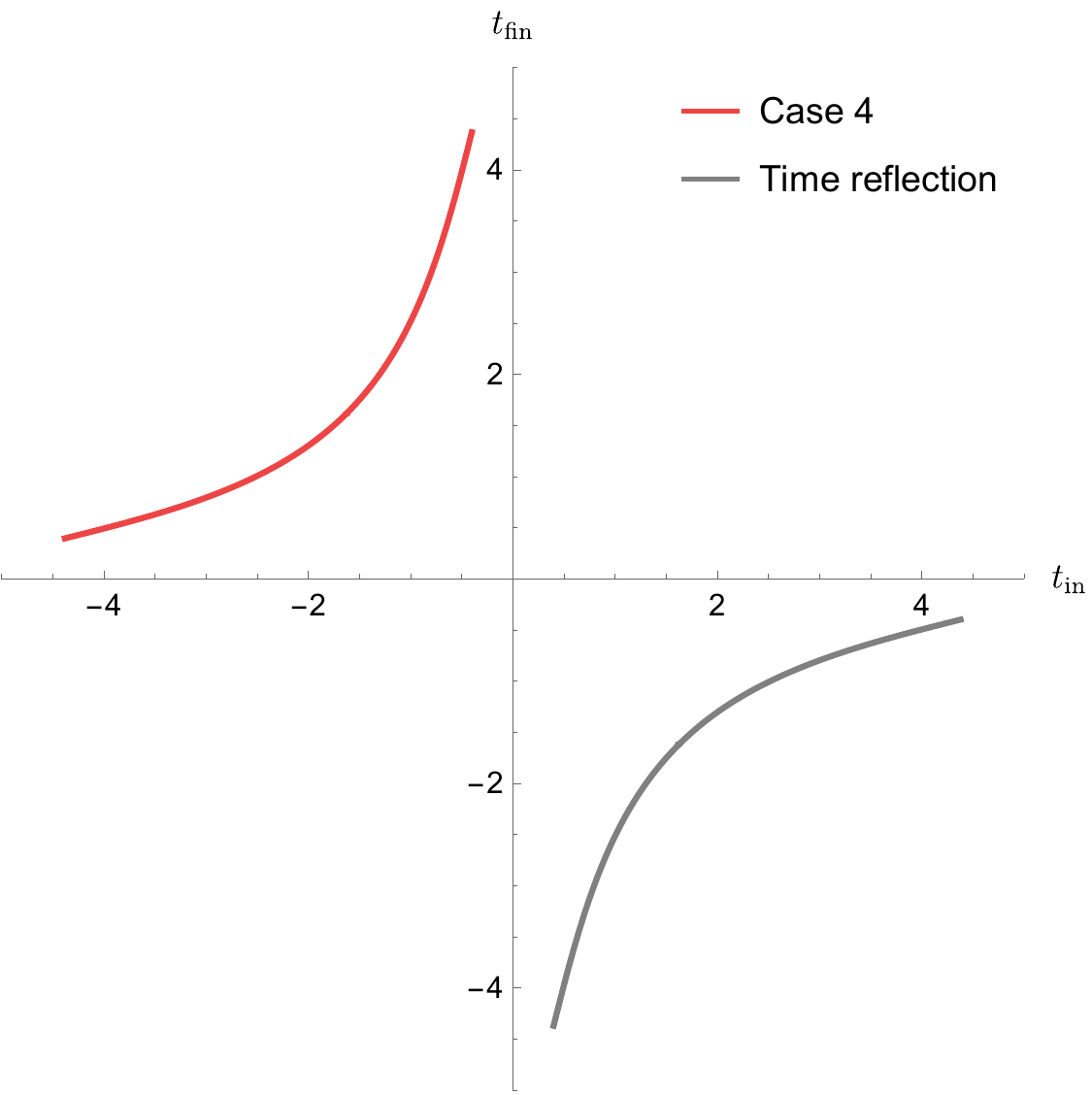}}
    \caption{(a) A cartoon of an almost-null geodesic trajectory for an expanding bubble geometry with AdS interior.
    (b) Plot of $t_{\rm fin}$ as a function of $t_{\rm in}$ in the case of an expanding bubble with AdS interior. 
    We set $\kappa=2.5, \l=-1, r_h=0.5$.
}
\end{figure}

\begin{figure}[ht!]
    \centering
    \subfigure[time close to $t_{\rm in}=0$]
    {\label{fig:t0_collapsing} \includegraphics[scale=0.25]{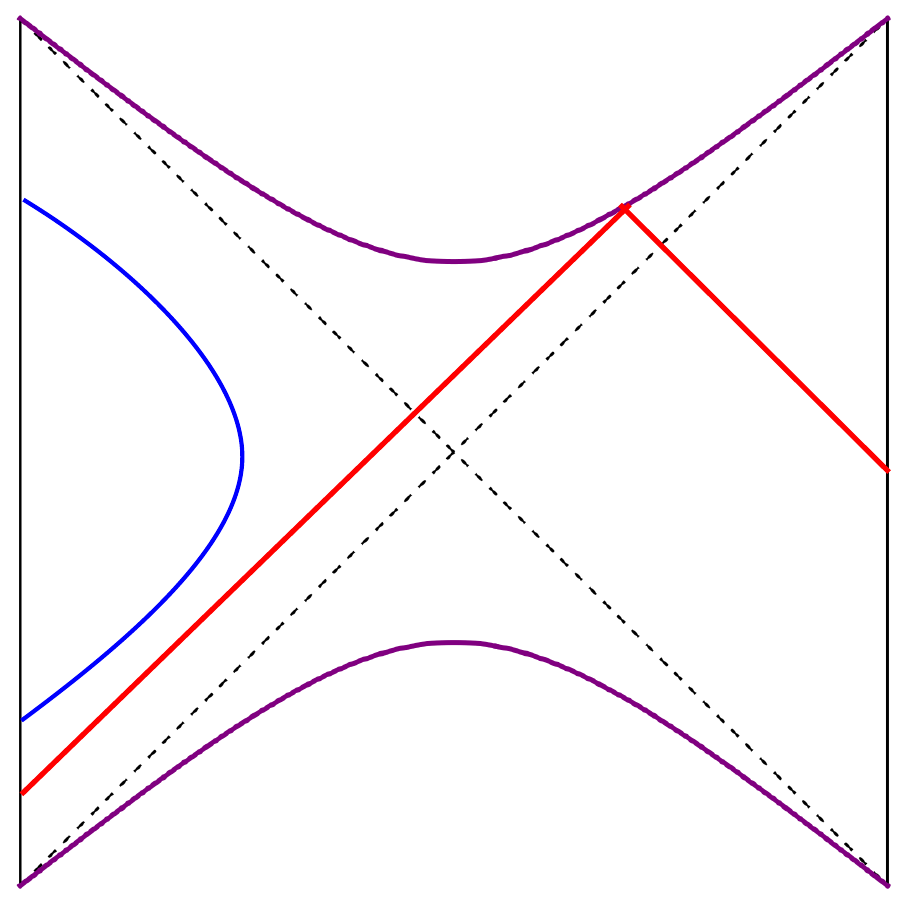}}
    \qquad \qquad
    \subfigure[early times]{ \label{fig:early_times_collapsing} \includegraphics[scale=0.25]{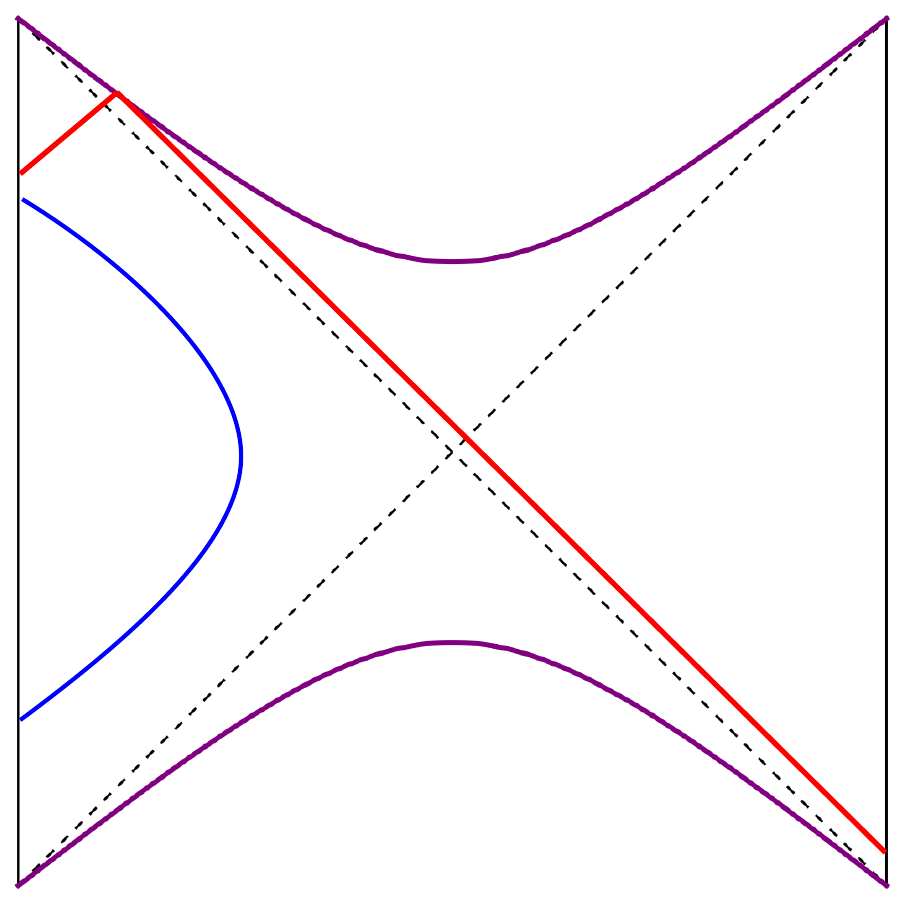}} 
    \caption{A cartoon of an almost-null geodesic trajectory (a) close to $t_{\rm in}=0$ and (b) at early times. In both cases, the geodesic does not enter the interior of the bubble. } 
\label{fig:Expanding_geodesic_trajectory}
\end{figure}

Let us begin with the case of an AdS interior geometry.
The Penrose diagram describing the causal structure of the full background is given in fig.~\ref{fig:Penrose_expanding_AdS}.
In the case when the almost-null spacelike geodesic comes back to the original boundary, it always reflects at the BH singularity twice.
There are cases in which the geodesic
reflects once on the singularity
and it does not come back to the original
AdS boundary.
The reason for this phenomenon can be understood from the diagram displayed in fig.~\ref{fig:Expanding_geodesic_trajectory}.
After the bounce at the BH singularity, the almost-null geodesic reaches the left AdS boundary and gets lost, without coming back to the right AdS boundary.
Due to this reason, the function
$t_{\rm fin}(t_{\rm in})$
is defined just in a finite interval,
i.e. $t_{\rm in1}<t_{\rm in}<t_{\rm in2}$.

We report an example of the dependence of $t_{\rm fin}$ on $t_{\rm in}$ in fig.~\ref{fig:expanding_tfti_AdS}.
Contrary to the collapsing bubbles studied in subsection~\ref{ssec:collapsing_bubbles},
the function $t_{\rm fin}(t_{\rm in})$ does not diverge anywhere.
While $t_{\rm in1}$ is always negative,
$t_{\rm in2}$ can have both signs.
For instance, in the example 
in fig.~\ref{fig:expanding_tfti_AdS} we have $t_{\rm in2}<0$,
while in the one that we will 
discuss later in  fig.~\ref{fig:expanding_bubble_nearstatic}
$t_{\rm in2}>0$.

\subsubsection{dS interior (region $A$ of parameter space)}
\label{ssec:expanding_dS}

In view of the following analysis, we show that a spacelike almost-null geodesic is repelled not only by the BH singularity, but also by the dS future and past timelike infinities $\mathcal{I}^{\pm}$ \cite{Freivogel:2005qh}. Starting from eq.~(\ref{geo-spacelike-radial}) applied to the interior geometry (\ie replacing $o \to i$), we find
\beq
\dot{r}^2 =E^2 +1 - \l r^2 \, .
\eeq
This equation shows that the spacelike almost-null geodesic has a turning point at $r = \sqrt{(E^2+1)/\l} $, in other words it is repelled by $\mathcal{I}^{\pm}$ in the high-energy limit $E \rightarrow \infty$.

\begin{figure}[ht]
    \centering
    \subfigure[case 4]{\includegraphics[scale=0.22]{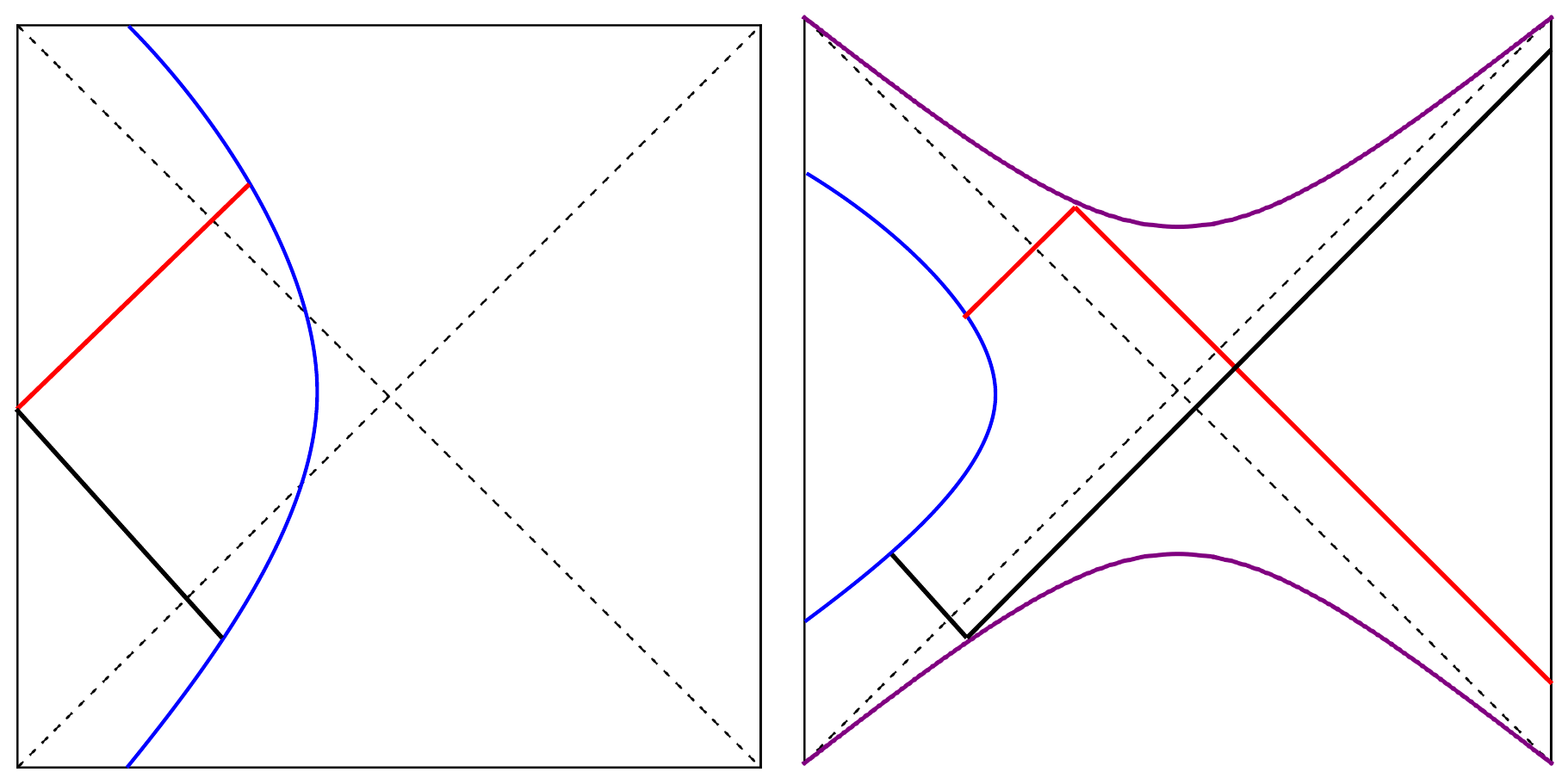}} \qquad \qquad
    \subfigure[case 5]
    {\includegraphics[scale=0.22]{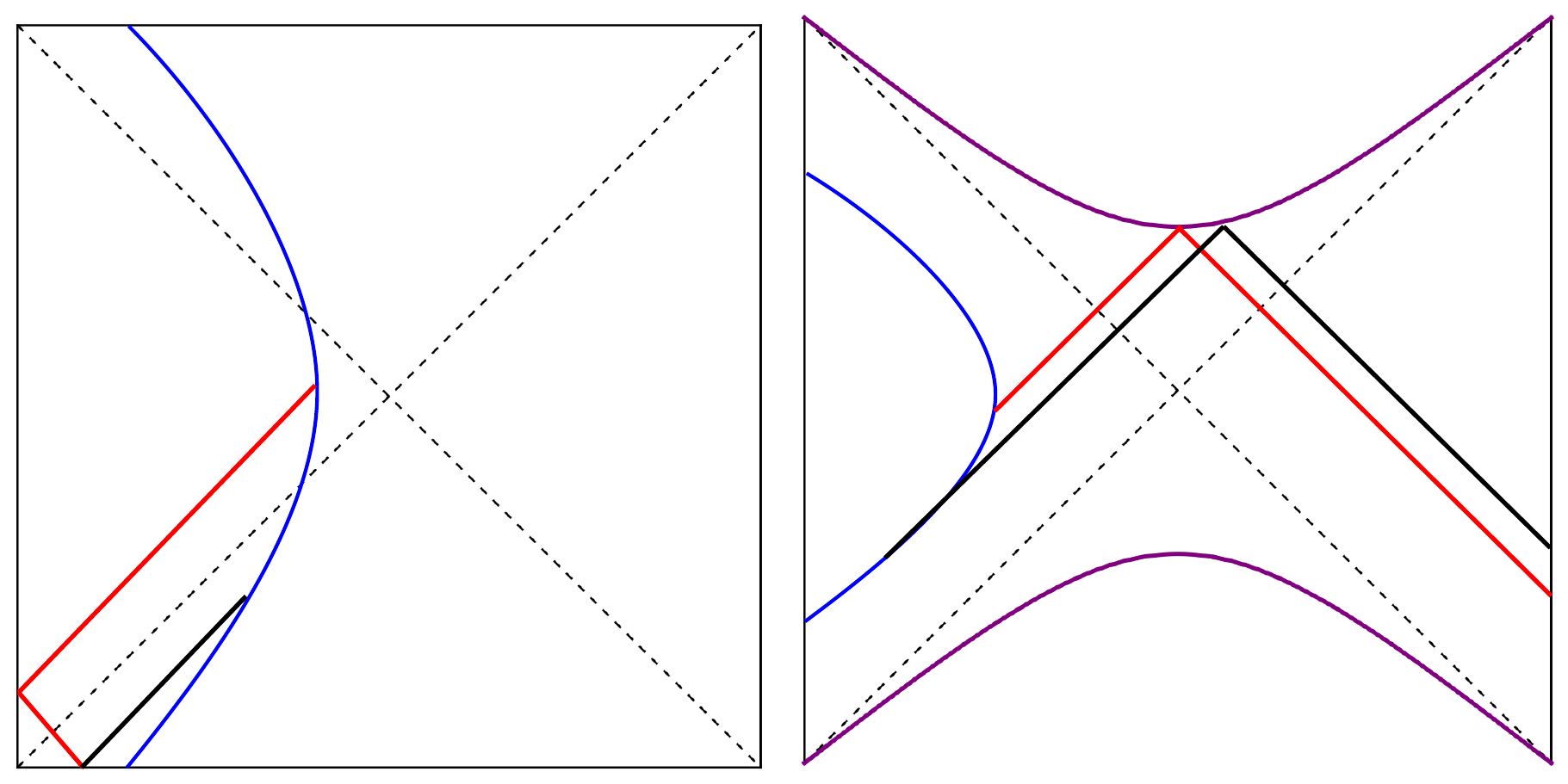}}
    \caption{A cartoon of the geodesic trajectory for (a) case 4 and (b) case 5 in the expanding bubble geometry with dS interior. Red lines denote an ingoing geodesic, while the black refers to the outgoing one.} 
\label{fig:Expanding_geodesic_trajectory_dS_interior}
\end{figure}

The domain wall of an expanding bubble is always located behind the BH horizon.
In region $A$ of parameter space, two additional sub-cases can occur: either the dS bifurcation surface is included inside the bubble geometry, or not.
In both settings, the trajectories of radial almost-null geodesics are characterized by their reflections at timelike infinities $\mathcal{I}^{\pm}$, and they admit a similar time dependence. For increasing time $t_{\rm in}$, the evolution is summarized by table~\ref{tab:expanding_dS}, and the trajectory is plot in fig.~\ref{fig:Expanding_geodesic_trajectory_dS_interior}.

\begin{table}[h!]   
\begin{center}   
\begin{tabular}  {|p{35mm}|c|} \hline   
  & \makecell{\textbf{Reflection at timelike}  \\ \textbf{infinity $\mathcal{I}^{\pm}$ of dS spacetime}} 
 \\ \hline \rule{0pt}{4.9ex}
Case 4 &   \redmark  \\
\rule{0pt}{4.9ex} Case 5  & \greencheck  \\[0.2cm]
\hline
\end{tabular}   
\caption{Configurations of a radial almost-null geodesic for an expanding bubble geometry with dS interior. From top to bottom: we increase the initial boundary time $t_{\rm in}$.} 
\label{tab:expanding_dS}
\end{center}
\end{table}

For similar reason as in the case with the AdS interior, the function $t_{\rm fin}(t_{\rm in})$ is only defined in a finite interval (that we denote $t_{\rm in1}<t_{\rm in}<t_{\rm in2})$ and does not diverge anywhere.
As a new feature, there is a discontinuity that corresponds to the transition between cases 4 and 5.
This is related to the geodesic passing nearby the corner where the left AdS boundary of the exterior BH touches the dS infinity of the interior geometry, see fig.~\ref{subfig:regionA_Pen_Exp}.
This discontinuity might be an artifact of the thin wall approximation.

\begin{figure}[ht!]
\centering
\subfigure[]{\label{subfig:tf_tin_exp_dS1}  \includegraphics[scale=0.3]{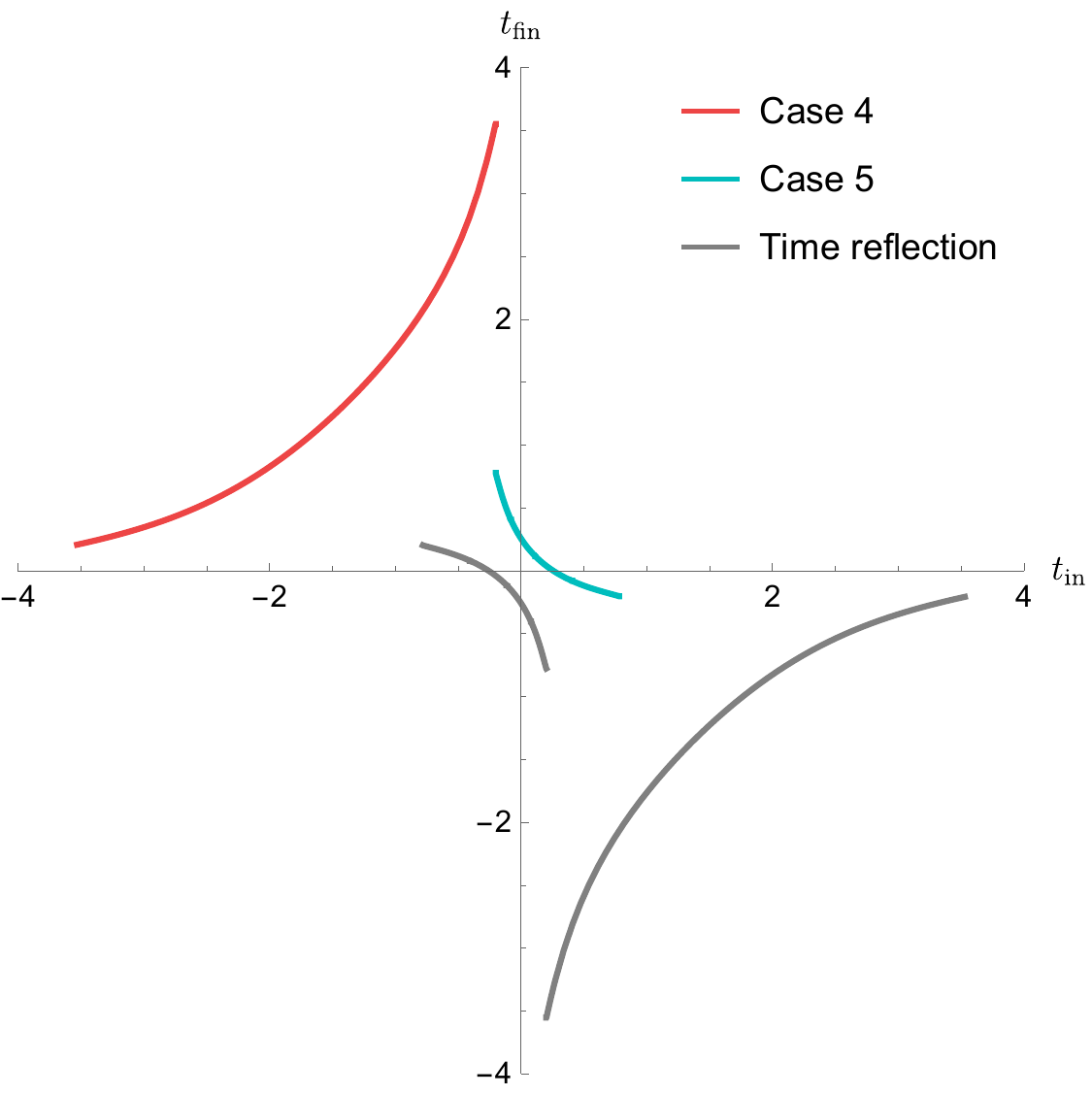}}
\qquad \qquad
\subfigure[]{\label{subfig:tf_tin_exp_dS2}  \includegraphics[scale=0.3]{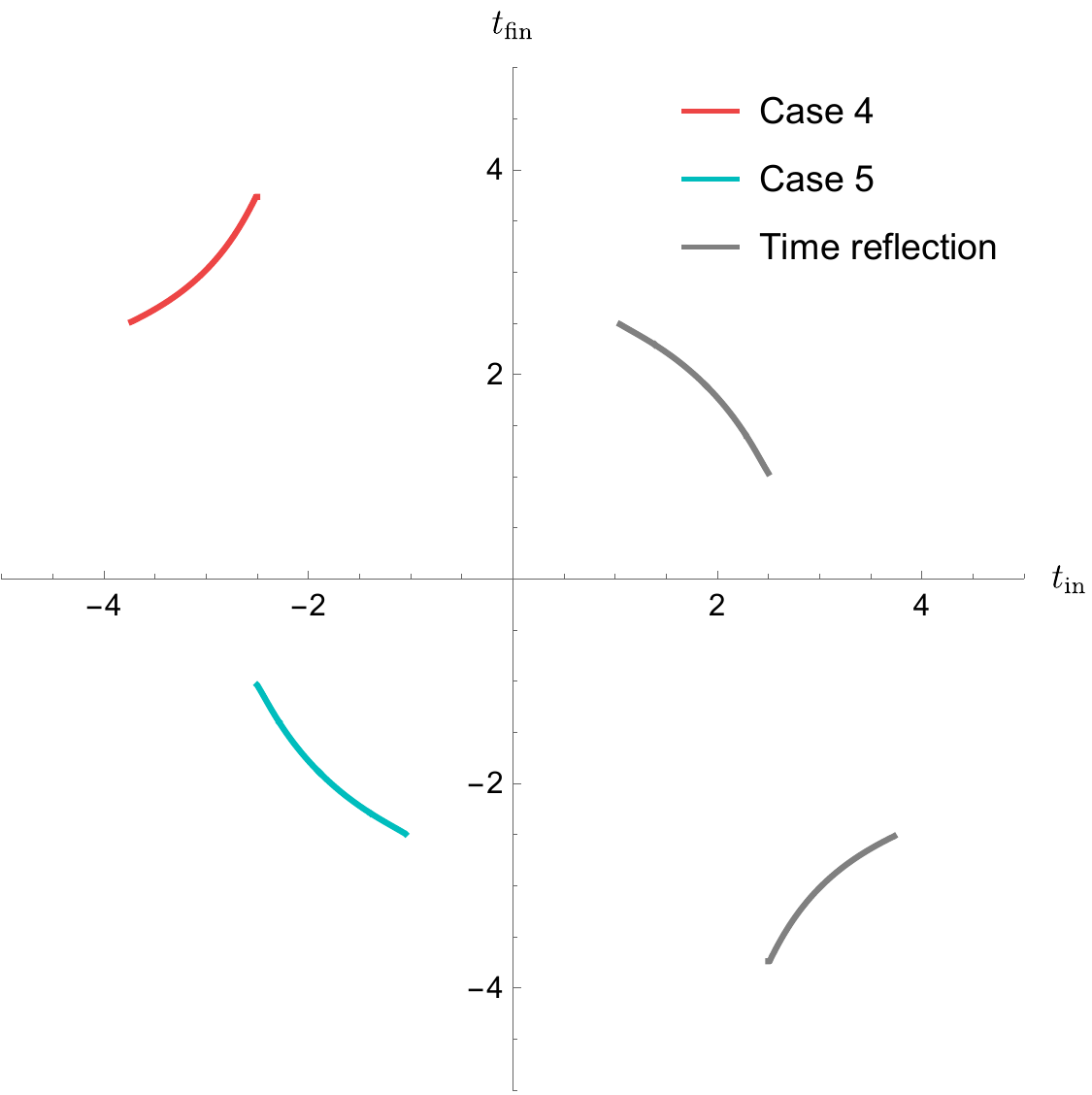}}
\caption{  Plot of $t_{\rm fin}$ as a function of $t_{\rm in}$ for an expanding bubble in region $A$ of parameter space (see fig.~\ref{Phase-Diagram}).
(a)  Example in which the interior does not include the dS bifurcation surface.
Here we fix $\kappa=0.5, \l=1, r_h=0.95$.
(b) Example in which the interior includes the dS bifurcation surface.
We fix $\kappa=0.5, \l=1, r_h=0.5$.
}
\label{fig:tf_notsolarge}
\end{figure}

We show in fig.~\ref{subfig:tf_tin_exp_dS1} the functional dependence of $t_{\rm fin}$ on $t_{\rm in}$, in a example where the interior geometry does not include the dS bifurcation surface, and in fig.~\ref{subfig:tf_tin_exp_dS2} an example where the interior geometry includes the dS bifurcation surface.

\subsection{Static bubbles and scars}
\label{ssec:sing_static_bubble}

The Eigenstate Thermalization Hypotesis (ETH) (see, \eg Refs.~\cite{Srednicki:1999bhx,DAlessio:2015qtq,Lashkari:2016vgj}) states that the time evolution of non-integrable many-body quantum systems leads to a thermalization.
The bulk dual of the thermofield double state is provided by the eternal AdS black hole. In this case, there are no bulk-cone singularities related to radial almost-null spacelike geodesics which return back to the boundary. So, if the assumption of thermalization is correct, we expect that the bulk-cone singularities related to the radial geodesics disappear in the late-time limit of bubble geometries, \ie there are no bulk-cone singularities when $t_{{\rm in}}$ and $t_{{\rm fin}}$ both approach to infinity.  For this phenomenon to happen, one possibility is that  $|t_{{\rm fin}}-t_{{\rm in}}| \to \infty$ when $t_{{\rm in}} \to \infty$.
This condition is satisfied by collapsing bubbles in all configurations I, II, and III, and even at the transition between them. For the expanding bubble case, the bulk-cone singularities do not exist at large $t_{\rm in}$, in a way that is also consistent with thermalization.

In this subsection, we show that the static bubble provides a counterexample to the expected asymptotically thermal behavior, with a finite difference $|t_{\rm fin} - t_{\rm in}|$ at any boundary time.
Static bubbles are an analytically treatable solution to Einstein-scalar gravity where the domain wall sits at constant radial coordinate. 
Let us consider a radial almost-null geodesic starting from the AdS boundary in the outside region, defined in terms of the EF coordinates~\eqref{u-v-def} by the curve at constant $v = v_1$. 
Differently from the dynamical bubble solutions, in the static case the geometry enjoys a time-translation symmetry.
Therefore, without loss of generality, we assume that $v_1 = 0$, \ie the almost-null geodesic intersects the AdS boundary at time $t_{\rm in}=0$.\footnote{We remind that $v=0$ implies $t=0$ at the boundary because we defined the integration constant of the tortoise coordinate in subsection~\ref{ssec:tortoise} such that $r^*_o(\infty)=0$.}
In the following, we will determine the full trajectory of this radial almost-null geodesic across the bubble geometry, until it intersects the AdS boundary again at time $t_{\rm fin}$.

\begin{figure}[ht]
    \centering
\subfigure[]{  \label{fig:nullgeo_static_AdS} \includegraphics[scale=0.2]{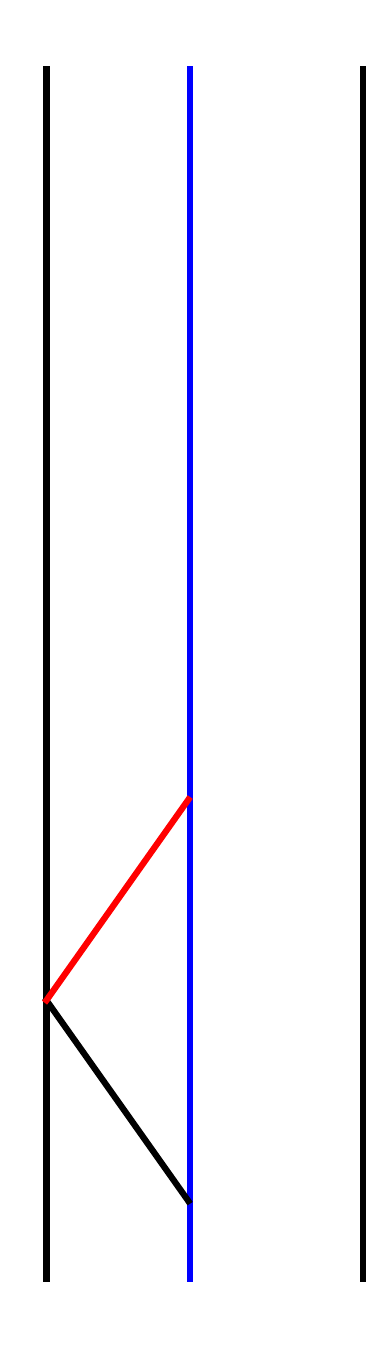}
\includegraphics[scale=0.2]{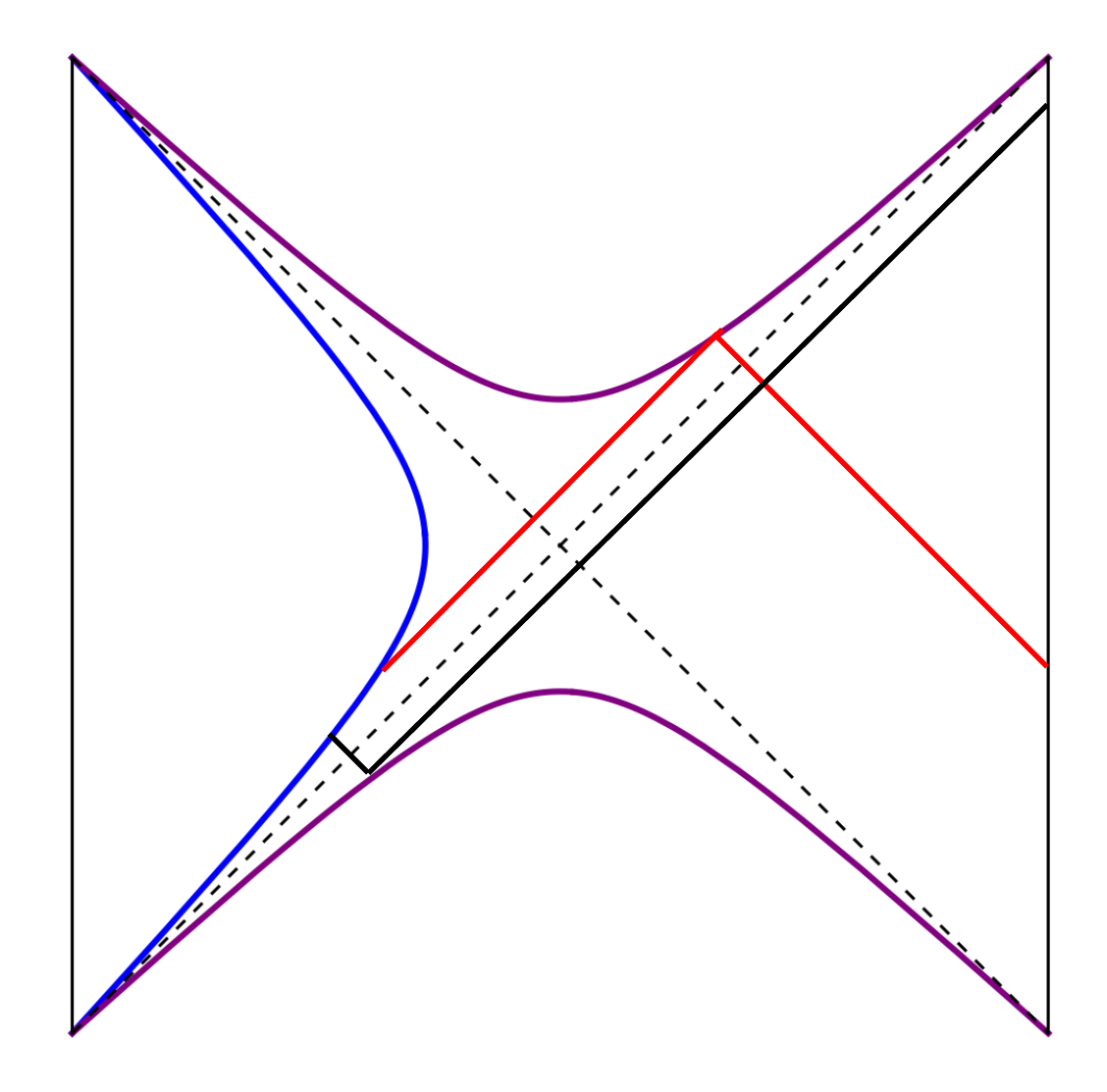}}
\qquad \qquad
\subfigure[]{ \label{fig:nullgeo_static_dS} \includegraphics[scale=0.2]{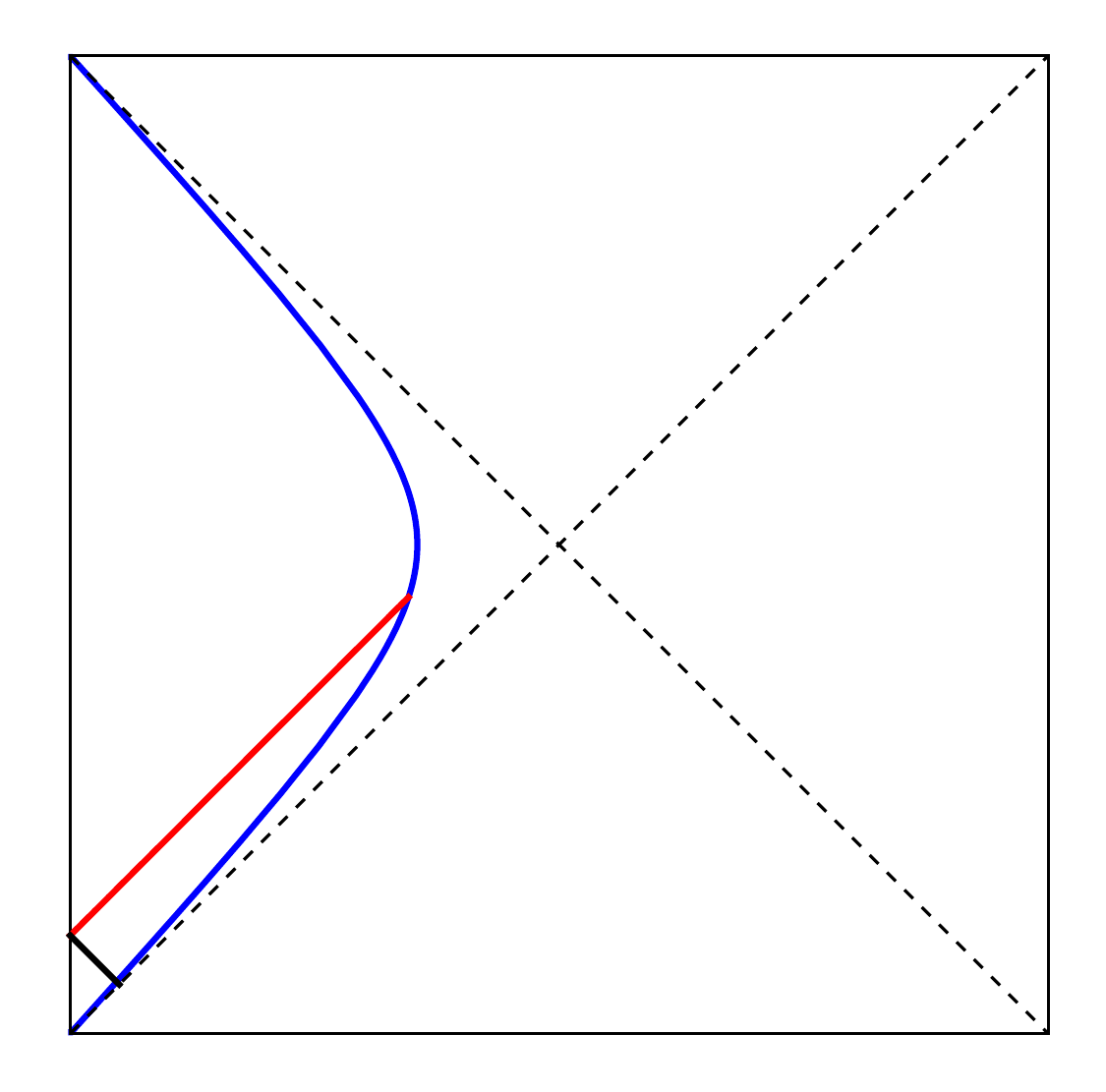}
\includegraphics[scale=0.2]{Figures/Penrose_static_o.pdf}}
    \caption{Trajectory of a radial almost-null geodesic in the Penrose diagram of the static bubble with either (a) AdS or (b) dS interior. The red line denotes an ingoing almost-null geodesic, while the black line is an outgoing geodesic.}
\end{figure}

The Penrose diagram and the trajectory of the radial almost-null geodesics in the case of an AdS (dS) interior are depicted in fig.~\ref{fig:nullgeo_static_AdS} (fig.~\ref{fig:nullgeo_static_dS}).
Here we only focus on the case where the domain wall is located inside the BH horizon, corresponding to region $C$ (or $A$) of the parameter space in fig.~\ref{Phase-Diagram}.
In case $E$ of the parameter space, the part behind the BH horizon is cut away when we glue the interior and exterior geometries. 
Since our goal is to employ bulk-cone singularities to study the interior structure of BHs, region $E$ in parameter space is less interesting, and we leave its investigation for the future.

We compute in appendix~\ref{app:bulk_cone_sing} the EF coordinates of radial almost-null geodesics leaving the right AdS boundary at time $t_{\rm in}$, and re-emerging at a time $t_{\rm fin}$.
The result reads 
\beq
\Delta t \equiv   t_{\rm fin} - t_{\rm in} =
2 \le -2 r_o^*(0) +R_o^* + \frac{1}{\bar{F}} R_i^* \ri  \, ,  \qquad
\bar{F} \equiv  \sqrt{\frac{f_o(R_{\rm static})}{f_i(R_{\rm static})}}  \, ,
\label{eq:diff_times_static}
\eeq
where $R_{o,i}^*=r^*_{i,o}(R_{\rm static})$.
In the above expression, we need to plug in the tortoise coordinates $r_o^*$ in eq.~\eqref{eq:ro_3d_exact} and $r_i^*$ in eq.~\eqref{eq:tortoise_interior}, together with the radius $R_{\rm static}$ of the static bubble taken from eq.~\eqref{eq:Rstatic_sol2_3d}.

\begin{figure}[ht]
    \centering
\includegraphics[scale=0.8]{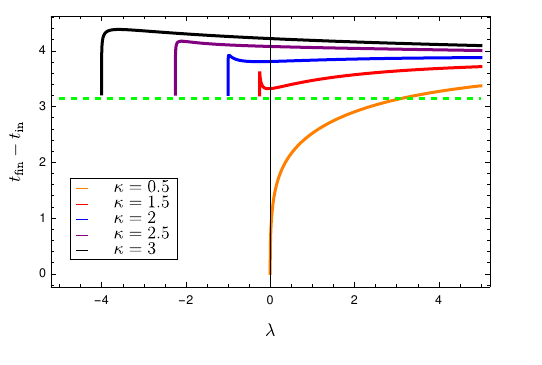}
    \caption{Time difference~\eqref{eq:diff_times_static} as a function of $\lambda$ for various choices of $\kappa$. 
  The part with $\lambda <0$ corresponds to region $C$ of the parameter space in fig.~\ref{Phase-Diagram}, while the part with $\lambda >0$ to region $A$. The dashed green curve corresponds to $\Delta t = \pi$. 
}
    \label{fig:plots_deltat_AdS3d_static}
\end{figure}

The plot with the dependence of $\Delta t$ on the parameters of the geometry, for both signs of the cosmological constant in the interior geometry, is depicted in fig.~\ref{fig:plots_deltat_AdS3d_static}.
When $0 < \kappa < 1$, we restrict to the case $\lambda >0$ because a negative cosmological constant in the interior geometry would correspond to region $B$ of the parameter space in fig.~\ref{Phase-Diagram}, where no static bubble exists.
For fixed $\kappa >1$, we need $\lambda>\l_2(\kappa)$ in eq.~\eqref{eq:critical_lambda} in order for the static bubble to exist (see region $C$ in fig.~\ref{Phase-Diagram}).
When the initial light ray corresponds to the red curve in figs.~\ref{fig:nullgeo_static_AdS}--\ref{fig:nullgeo_static_dS}, the  difference of times is always positive.

There are two interesting cases where analytic results can be achieved.
The first one is the limit $\lambda \to \lambda_2$ defined in eq.~\eqref{eq:critical_lambda},
which is the border between regions $C, D$ of the parameter space in fig.~\ref{Phase-Diagram}, approached from region $C$.
In this limit, we find that 
$r_h \propto m_{\rm static}^{1/3} \rightarrow \infty$
and the radius of the bubble is even larger, with $R_{\rm static} / r_h \rightarrow \infty$, as it can be inspected from eq.~\eqref{eq:Rstatic_sol2_3d}. 
The time difference in this limit reads
\beq
\lim_{\lambda \to \lambda_2} \Delta t = \pi \, .
\eeq
This is the same result as in bulk empty AdS spacetime \cite{Hubeny:2006yu}, and it is also the same time taken by a light ray to connect two antipodal points on the boundary of AdS.
Given an asymptotically AdS spacetimes satisfying the null energy and null generic conditions, Gao-Wald theorem~\cite{Gao:2000ga} states that there is a gravitational time delay of bulk null geodesics connecting antipodal boundary points.
In other words, the time difference between antipodal boundary points connected by a bulk null geodesic satisfies $\Delta t \geq \pi$, where the inequality is strict for any perturbation of empty AdS spacetime (satisfying the above conditions), and is saturated by empty AdS spacetime.
Note that our geodesics are not exactly null because they arise from the high energy of spacelike geodesics, therefore the Gao-Wald theorem does not strictly apply.
Looking at the plots in fig.~\ref{fig:plots_deltat_AdS3d_static},
we find numerical evidence that $\Delta t \geq \pi$ for $\l<0$, see the dashed green curve in the picture. Instead, for $\l>0$, it is possible to achieve the range $\Delta t < \pi$.

The second analytic treatable case is the limit $\l \rightarrow 0$ with $0 < \kappa <1$. At leading order, this implies that $r_h \rightarrow \infty$ but with the following fixed ratios
\beq
\frac{R_{\rm static}}{r_h}  \approx \frac{1}{(1-\kappa^2)^{1/3}} \, ,
\qquad
R_{\rm static}  \approx \frac{1}{\sqrt{\l} } \, ,
\eeq
see eq.~\eqref{eq:property-static-AB}. 
As a result, the time difference reads 
\beq
\lim_{\l \to 0^+} \Delta t = 0 \, ,
\eeq
where we used that $f_i(R_{\rm static}) \to 0$.
On physical grounds, this statement is the consequence of the domain wall sitting nearby the cosmological horizon of dS spacetime.
The previous limit is consistent with the orange curve in the numerical plot in fig.~\ref{fig:plots_deltat_AdS3d_static}.
As anticipated above, this curve admits a regime where $\Delta t<\pi$. 
This result is a signal of non-locality, because the bulk-cone singularity lies outside the CFT lightcone. 
Similar non-local features of bubble geometries with dS interior were discussed in Ref.~\cite{Freivogel:2005qh}.
Surprisingly, the analysis of section~\ref{sec:EE_bubbles} revealed that the HRT surface does not penetrate inside the static bubble, therefore the holographic EE is insensitive to the non-local aspects of this geometry.

In all cases, we observe that the time difference $|t_{\rm fin} - t_{\rm in}|$ is finite, in contrast with the expectations -- supported by ETH -- that bulk-cone singularities should disappear when the system thermalizes.
Rather, the static bubble shows properties typical of the quantum many-body \textit{scar states} (see, \eg Ref.~\cite{Serbyn:2020wys}). For studies of scar states in the context of the AdS/CFT correspondence,  see \eg Refs.~\cite{Dodelson:2022eiz,Milekhin:2023was}.
Even though most of the eigenstates of a non-integrable system look thermal, scar states are precisely an exception.
In particular, the static bubble admits bulk-cone singularities 
separated by a constant time $\Delta t$
during all the time evolution.


\subsubsection{Near-static bubble} 
\label{ssec:near_static_bubble}

It is important to emphasize that the static bubble require some amount of fine tuning of parameters. Let us illustrate this point with some numerical examples.

\begin{figure}[ht!]
\centering
\subfigure[collapsing bubble]{  \includegraphics[scale=0.28]{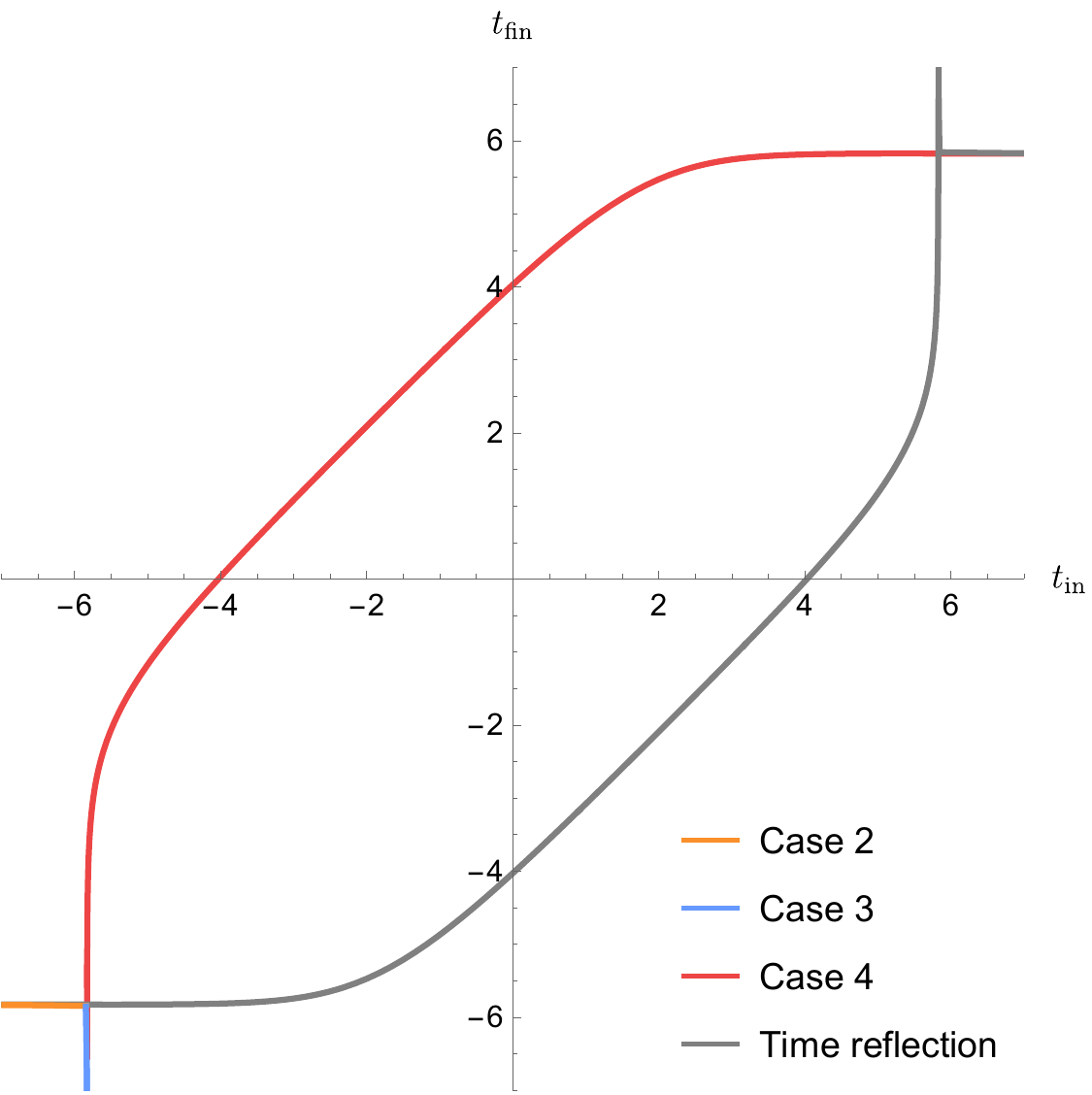}} \quad
\subfigure[expanding bubble]{\label{fig:expanding_bubble_nearstatic}  \includegraphics[scale=0.28]{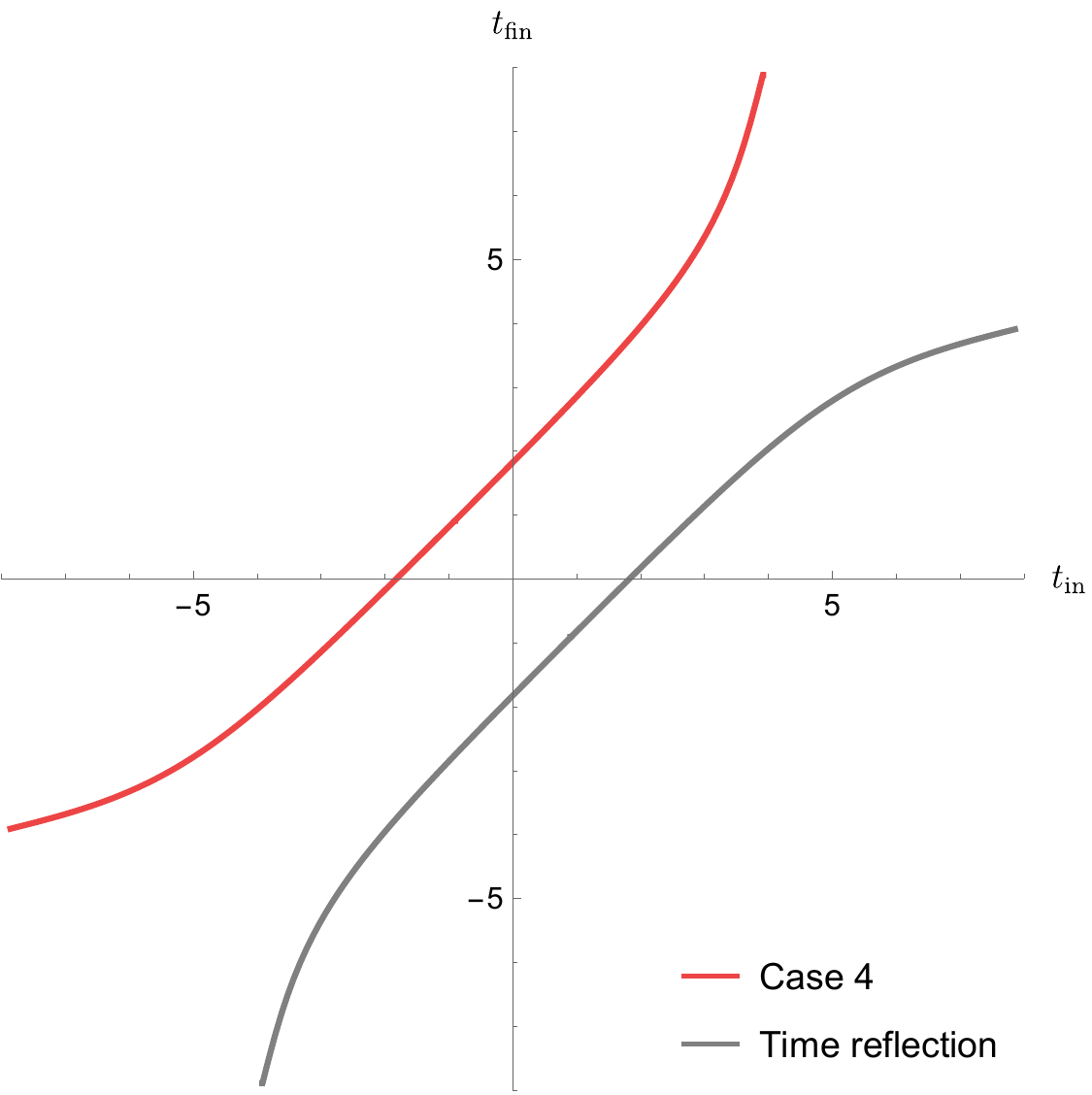}}
\caption{Plot of $t_{\rm fin}$ as a function of $t_{\rm in}$ for (a) a collapsing bubble and (b) an expanding bubble in the near-static limit. For both pictures, we set $\kappa=-2.5, \lambda=-1, r_{h} = (r_h)_{\rm st.} - \varepsilon, \varepsilon=10^{-5}$.
}
\label{fig:near_static_result}
\end{figure}

This limit can be realized in regions $A$ and $C$ of parameter space (see fig.~\ref{Phase-Diagram}),
both from the collapsing or expanding cases,by setting $r_h = (r_h)_{\rm st.} - \varepsilon$, where $(r_h)_{\rm st.}$ is the horizon radius of the static bubble, and we consider the limit $\varepsilon \rightarrow 0$. 
The dependence of $t_{\rm fin}$ on $t_{\rm in}$ in a numerical example
is shown in fig.~\ref{fig:near_static_result}.
For the collapsing case, we have that 
$t_{\rm fin}(t_{\rm in})$ is defined for all
values of $t_{\rm in}$ while for the expanding
case it is defined just for  $t_{\rm in1}<t_{\rm in}<t_{\rm in2})$.

We have shown in subsection~\ref{ssec:sing_static_bubble} that in the strict static limit, the bubble geometry is characterized by a finite value of the time difference. 
For the almost static case,
in some region centered nearby
$t_{\rm in}=0$,  $t_{\rm fin} (t_{\rm in})$ 
can be approximated by a 
constant value  of $|t_{\rm fin} - t_{\rm in}|$,
that we refer to as a \textit{plateau}.

\begin{figure}[ht!]
    \centering
\includegraphics[scale=0.35]{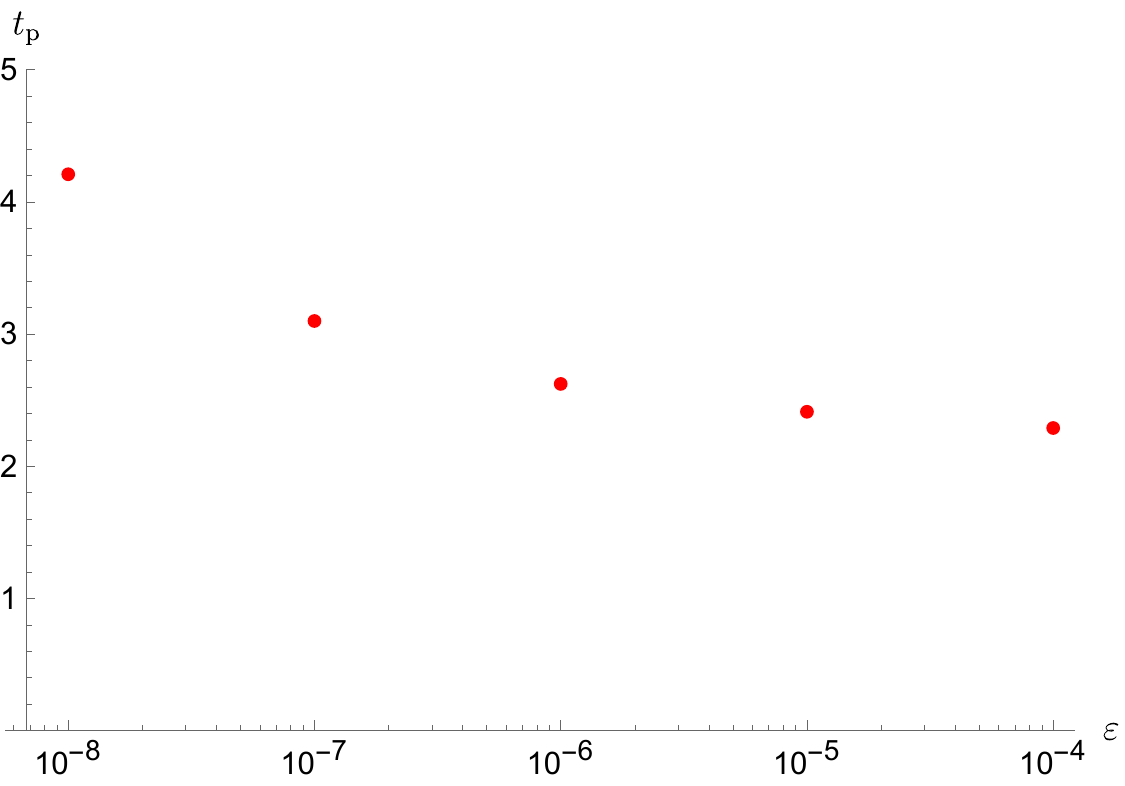}
    \caption{Plot of the plateau length $t_{\rm p}$ (where the value of $|t_{\rm fin} - t_{\rm in}|$ varies at most by $1\%$)
 as a function of $\varepsilon$ (parameterizing the difference between the horizon radius and its static value).
    }
\label{fig:near_static_length}
\end{figure}

In the near-static limit, the duration $t_{\rm p}$ of the above-mentioned plateau region is finite.
We depict in fig.~\ref{fig:near_static_length} the duration $t_{\rm p}$  as a function of the parameter $\varepsilon$ just introduced (which quantifies the difference between the horizon radius and its static value). By inspecting the red dots (obtained numerically), we notice that 
a very precise fine tuning in $r_h$ is needed
for a modest increase in $t_{\rm p}$.

%% file: Sections/Conclusions.tex
\section{Conclusions and discussion}
\label{sec:conclusions}

In this work, we considered and classified various kinds of Lorentzian asymptotically AdS geometries which include in their interior a spherically symmetric bubble with a different cosmological constant $\l$.  
In particular, we restricted to solutions of Einstein scalar gravity invariant under time reversal $t \to -t$.

Within the AdS$_{3}$/CFT$_{2}$ correspondence, minimal HRT geodesics are mapped to the entanglement entropy of a CFT state. 
Similarly, the investigation of almost-null spacelike geodesics in general spacetime dimension determines the bulk-cone singularities, which encode the divergences of two-point functions of boundary operators.
We used these two kinds of geodesics 
as a tool to probe the bulk structure from the perspective of a boundary observer.

\subsection{Phase diagram and parameter space}

In section~\ref{sec:preliminaries}, we determined the complete parameter space of Lorentzian bubble solutions, including both signs of the cosmological constant $\lambda$ in the interior geometry.
This study was initiated in Ref.~\cite{Freivogel:2005qh} for interior dS geometries ($\lambda >0$), and it was performed for arbitrary $\lambda$ in Euclidean signature in Ref.~\cite{Fu:2019oyc}.
We depicted in fig.~\ref{Phase-Diagram} the full parameter space as a function of the cosmological constant $\lambda$ and the tension of the domain wall $\kappa$. This phase diagram is divided in various regions. Region $A$ corresponds to a dS interior ($\l>0$), while regions $B,C,D,E$ to the AdS case ($\l<0$). 
The existing types of bubble solutions in each region of the parameter space are summarized in table~\ref{tab:parameter_space}.

\begin{table}[h!]   
\begin{center}   
\begin{tabular}  {|p{35mm}|c|c|c|c|} \hline  
 & \textbf{Static bubble} &\textbf{Expanding bubble} & \textbf{Collapsing bubble}   \\ \hline \rule{0pt}{4.9ex}
Region $A$ & \greencheck \, ($m=m_{\rm static}$)  & \greencheck \, ($m < m_{\rm static}$) & \greencheck \, ($m < m_{\rm static}$)   \\
\rule{0pt}{4.9ex} Region $B$ & \redmark & \greencheck & \greencheck  \\ 
\rule{0pt}{4.9ex} Region $C$ & \greencheck \, ($m=m_{\rm static}$)  & \greencheck \, ($m < m_{\rm static}$) & \greencheck \, ($m < m_{\rm static}$)   \\
\rule{0pt}{4.9ex} Region $D$ & \redmark & \redmark & \greencheck  \\
\rule{0pt}{4.9ex} Region $E$ & \greencheck \, ($m=m_{\rm static}$)  & \greencheck \, ($m < m_{\rm static}$) & \greencheck \, ($m < m_{\rm static}$)    \\[0.2cm]
\hline
\end{tabular}   
\caption{Allowed time-reversal symmetric solutions for the bubble geometry corresponding to the various regions of parameter space specified in fig.~\ref{Phase-Diagram}. The quantity $m_{\rm static}$ is a critical mass. For its explicit value in $d=2$, see eq.~\eqref{eq:mstatic_d2}; in $d=3$, see eq.~\eqref{eq:mstatic2_3d}.
 } 
\label{tab:parameter_space}
\end{center}
\end{table}

The different regions $A,B,C,D,E$ in fig.~\ref{Phase-Diagram} determine the causal structure of the expanding bubble, see the Penrose diagrams in fig.~\ref{Pen-Dia-Expanding-Bubble}.
We emphasize the following features:
\begin{itemize}
\item
In regions $A$ and $C$, the expanding bubbles are always located inside the BH bifurcation surface at bulk time $t=0$. 
In particular, region $A$ provides a model  to embed a cosmology inside an asymptotically AdS spacetime~\cite{Freivogel:2005qh}, because the interior of the bubble contains a portion of dS space with infinite spacetime volume, which resembles our expanding universe.
\item In region $B$, the geometry of the expanding bubble includes two disconnected AdS boundaries and resembles an eternal BH.
We can interpret this case as a false vacuum decay, where the false vacuum corresponds to the interior region with $\l<0$.
\item In region $D$, no expanding bubble solutions exist. 
\item In region $E$, the expanding bubble lies outside the BH horizon, and the interior of the BH is excluded from the bubble geometry. 
This case corresponds to a false vacuum decay, where the role of the false vacuum is played by the external AdS region.
\end{itemize}

\rep{It is important to emphasize that the dual interpretation
of the geometries  as genuine CFT states is not always clear.
As shown in~\cite{Fu:2019oyc}, in regions $A,B,C$ and $E$ the domain wall 
solution
does not admit a smooth Euclidean continuation with asymptotically AdS boundary,
providing an obstruction to a Euclidean  path integral preparation of the $t=0$ geometry.
Instead, the geometries in region $D$ 
admit a smooth Euclidean continuation,
which can be used as a saddle for 
a Euclidean CFT path integral.
}
Region $D$ is the only portion of the phase diagram that admits a precise holographic understanding of the dual field theory state, 
realized as a \textit{Holo-ween} quench (see Ref.~\cite{Simidzija:2020ukv}) between two different CFTs connected by an interface.
In this context, the gravitational description involves a gluing between AdS spacetimes with (possibly different) cosmological constant, separated by a thin wall.
Holographic tests in region $D$ can only be performed for collapsing bubbles, since this region does not admit any expanding solution.

\rep{Expanding bubble geometries in regions $A$ and $C$ have some pathological features where the domain wall approaches the asymptotic 
AdS boundaries, see fig.~\ref{Pen-Dia-Expanding-Bubble}.
In region $A$, a portion of AdS boundary
is attached to a portion of dS timelike infinite.
In region $C$, two distinct asymptotically 
AdS timelike boundaries bifurcate from the domain wall surface
 at $r \to \infty$.
It is unclear whether these curious features arise from the thin-wall approximation or are also realized in a full solution of the 
 gravity-scalar system.
It would be interesting to investigate the presence
of such pathologies with numerical methods.
These issues should not affect the geometry
at $t=0$, and consequently should not
 affect the calculation of the entanglement entropy 
at $t=0$. 
On the contrary, it is likely that a
resolution  of the pathological features 
might have an influence on the structure
 of the bulk-cone singularities
for the expanding bubbles,
since in some cases the relevant geodesics disappear 
(without bouncing back)
at the additional asymptotic boundaries.
}

The causal structure of the collapsing bubbles is characterized by three possible configurations of the Penrose diagram, denoted with I, II, and III in fig.~\ref{Pen-Dia-Collapsing-Bubble}.
In configurations I and II, the surface of the collapsing domain wall lies outside the BH bifurcation surface at $t=0$, while in case III it lies inside.
The difference between configurations I and II arises from the shape of the causal wedge, which can (or cannot) include the center of the bubble, respectively.
The phase diagram that distinguishes the various configurations of the collapsing bubble depends on the BH mass $m$ and on the boundary dimension $d$.
We studied the parameter space of these solutions in figs.~\ref{figure-beta-outside-d2} and \ref{figure-beta-outside-d3} for some representative examples in $d=2$ and $d=3$, respectively.

\subsection{Holographic entanglement entropy}

In section~\ref{sec:EE_bubbles}, we studied the holographic entanglement entropy $S(\Delta \theta)$ associated with a segment with opening angle $\Delta \theta$ on the time slice $t=0$ of a three-dimensional bubble geometry ($d=2$).
Using the HRT prescription, this computation required to determine the geodesic of minimal length among multiple  
extremal geodesics homologous to the given subregion.
Surprisingly, we found regimes in which the minimal geodesic can enter inside the domain wall even if
the latter is located  beyond the BH bifurcation surfaces.

With the exception of expanding bubbles in region $B$ (for which the causal structure is similar to the eternal BH), the homology constraint enforces the identity
\beq
S(\Delta \theta)=S(2 \pi- \Delta \theta) \, .
\label{eq:ee-conclu-segmento}
\eeq
We focused on the instructive case $\Delta \theta=\pi$, which corresponds to the largest opening angle whose entropy is not constrained by the relation~(\ref{eq:ee-conclu-segmento}).
The geodesic with $\Delta\theta=\pi$ is in general the deepest into the bulk. 
As such, it is crucial to detect phase transitions in the holographic entanglement entropy and the presence of Python's lunches.
The following distinct features emerge for collapsing and expanding bubbles:
\begin{itemize}
\item
For collapsing bubbles, we performed in fig.~\ref{figure-scan-EE} a numerical scan of the holographic entanglement entropy. In the case denoted with $e$, we found that
the minimal HRT surface enters the interior geometry, even though the bubble is located beyond the BH bifurcation surface (as seen from an observer at the boundary). 
For large BH mass, this configuration is realized in a large portion of the parameter space $(\l,\kappa)$
where the solution exists, see the bottom-right panel in fig.~\ref{figure-scan-EE}.\footnote{The HRT surface also detects the presence of the bubble in the remaining of the parameter space in the large-mass limit.
These cases, denoted with $b,c,d$, are physically more intuitive because the bubble is initially located outside the BH bifurcation surface.} 
\item
In the case of expanding bubbles,
we instead found  
numerical evidence that the minimal
HRT surface always lies outside the BH bifurcation surface, when it exist. In this case, the holographic entanglement entropy for $0 \leq \Delta \theta \leq \pi$ shows a thermal behavior, similar to an eternal BH.
\end{itemize}

\subsection{Bulk-cone singularities}

In section \ref{sec:sing_corr}, we studied the bulk-cone singularities of the two-point functions of a scalar primary operator with large conformal dimension, evaluated on the CFT state dual to the gravitational solution.
In the context of the AdS/CFT correspondence, these singularities
are detected by null and almost-null spacelike geodesics that leave the AdS boundary at time $t_{\rm in}$ and come back to it at time $t_{\rm fin}$.
We focused on $d=3$, which is the minimal dimension for which
almost-null spacelike geodesics are reflected by the BH singularity.
We computed the function $t_{\rm fin}(t_{\rm in})$ by considering a future-oriented almost-null radial geodesic inserted from the AdS boundary.
The case of a geodesic initially directed in the past direction was recovered using the time reflection symmetry in eq.~\eqref{eq:symm_time_rev}.
We found the following distinct behaviors:
\begin{itemize}
\item For any configuration (I, II, or III) of the collapsing bubble, the function $t_{\rm fin}(t_{\rm in})$ approaches a constant value $-t_h$ in the limit $t_{\rm in}\to -\infty$. 
The asymptotic behavior is 
\beq
t_{\rm fin} \approx
- t_{h} + W \, \exp \le 2 \pi T \, t_{\rm in} \ri \, ,
\eeq
where $T$ is the Hawking temperature, and $W$ is a constant
(which is positive for configuration I, and negative for configurations II and III).
For collapsing bubbles, we can then extract the BH temperature by
inspecting the large-time behavior of the singularities of two-point functions.
The qualitative structure of the function $t_{\rm fin}(t_{\rm in})$ is slightly different in configuration I compared to II and III, see fig.~\ref{fig:configurations-I-II}.
\begin{figure}[h!]
\centering
\subfigure[$\kappa = 0.25$]{\includegraphics[scale=0.24]{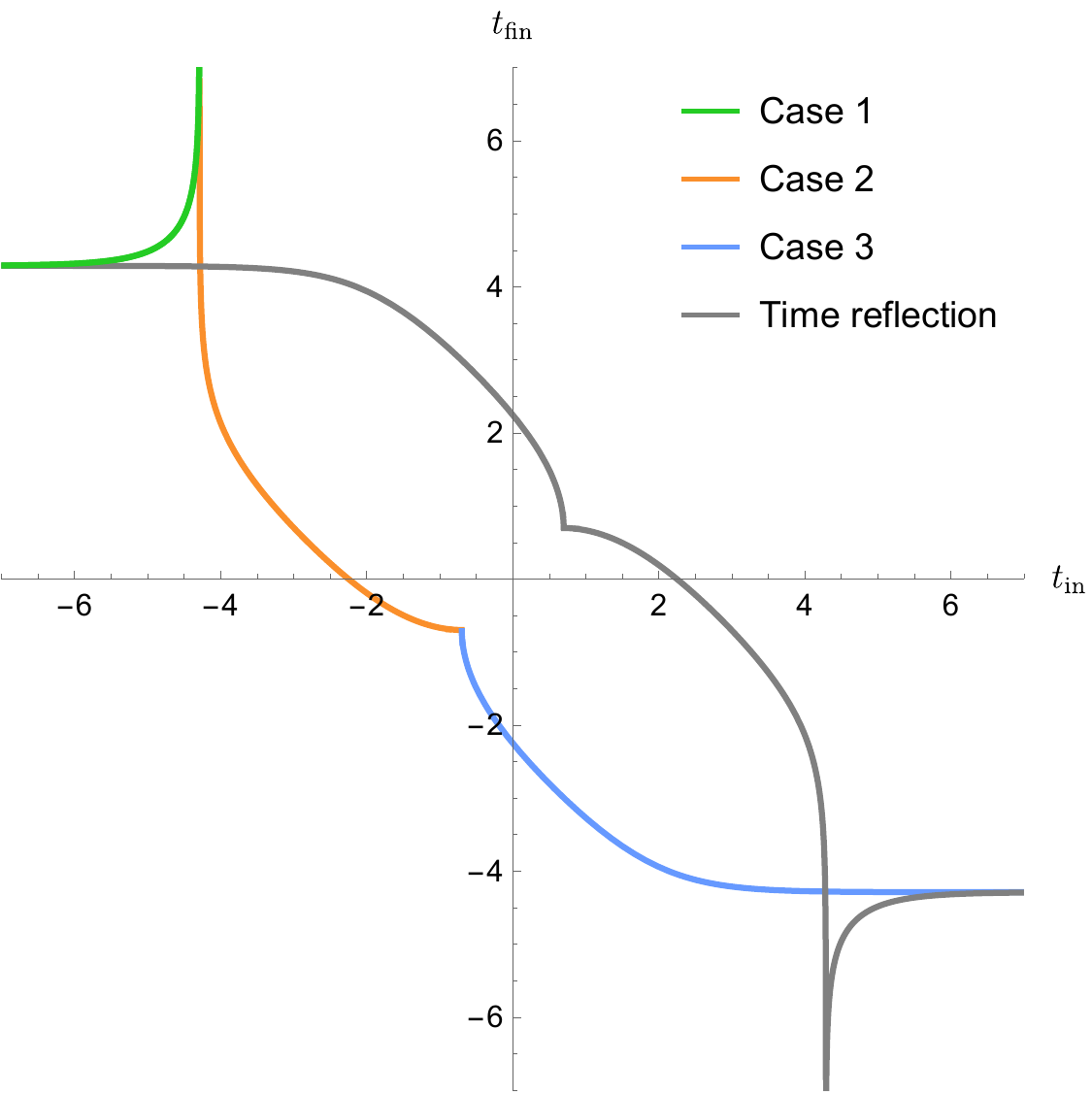}}
\qquad 
\subfigure[$\kappa \simeq 0.2529$]{\includegraphics[scale=0.24]{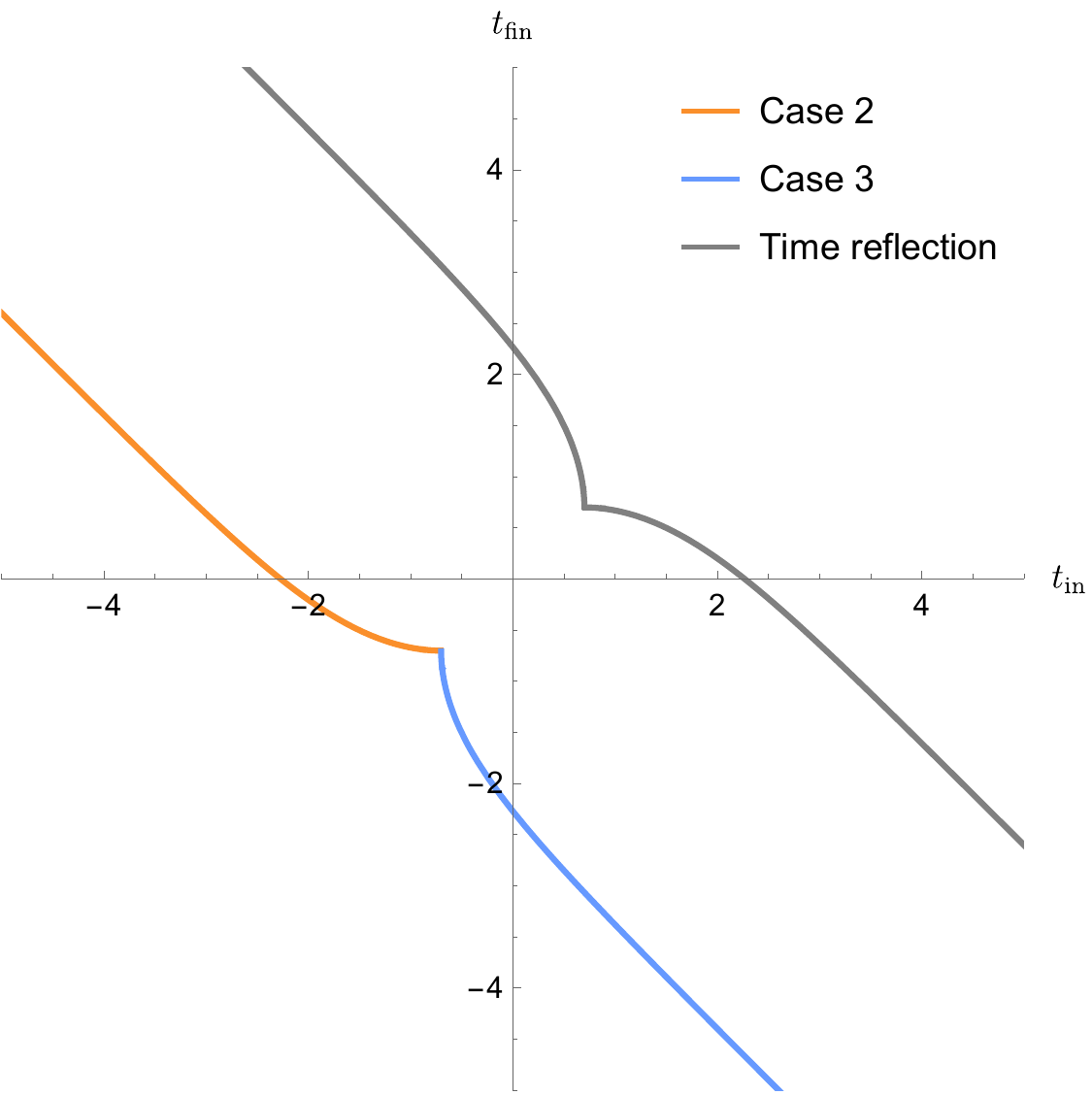}}
\qquad 
\subfigure[$\kappa=0.26$]{\includegraphics[scale=0.24]{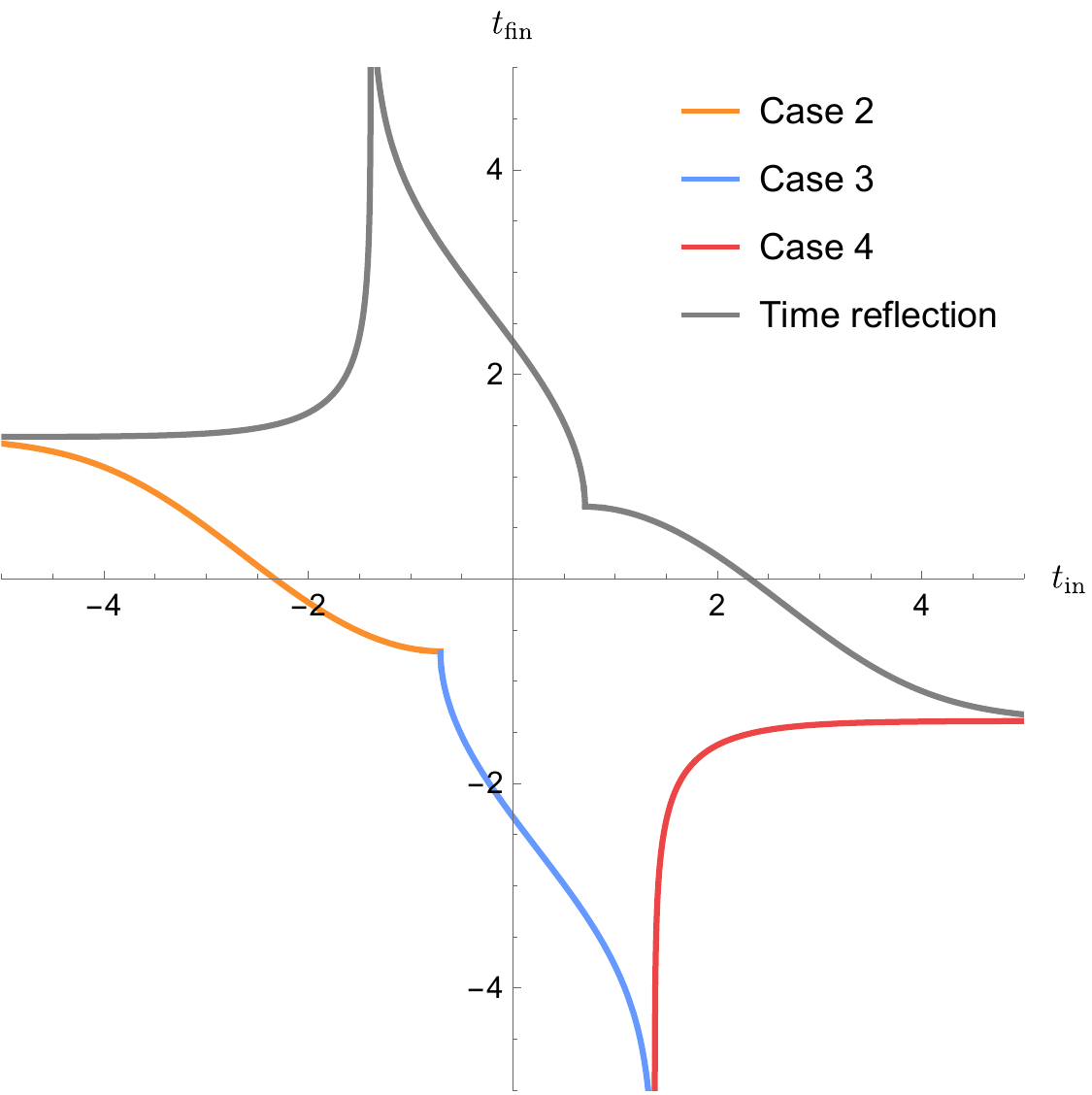}}
\caption{
The transition between bulk-cone singularities between configuration I (left panel), a fine-tuned transition case (central panel),
and configuration II (right panel). Here we fix $\l=-1$, $r_h=0.5$.
}
\label{fig:configurations-I-II}
\end{figure}

For configuration I, we interpret the value $t_h$ as the time of formation of the event horizon.
At the boundary in the parameter space between cases I and II, the quantity $t_h$ diverges to $-\infty$, see the central panel of
fig.~\ref{fig:configurations-I-II}.
The bulk-cone singularities have the same qualitative structure in configurations II and III.
In all the cases, the function $t_{\rm fin}(t_{\rm in})$ has a kink for $t_{\rm fin}=t_{\rm in}$, which is an artifact of the thin wall approximation.
The qualitative structure of $t_{\rm fin}(t_{\rm in})$ is independent of the sign of $\lambda$.
\item For the expanding bubbles, we found that the function $t_{\rm fin}(t_{\rm in})$ is not defined for large $t_{\rm in}$.
Geometrically, this fact happens because there are no almost-null radial geodesics which leave and come back to the AdS boundary.
In the $\l>0$ case, there exist situations in which the almost-null geodesic bounces at the timelike dS infinities of the interior geometry.
When this phenomenon happens, there is a discontinuity in the function $t_{\rm fin}(t_{\rm in})$, which might be artifact of the thin wall approximation.
We summarize these behavior in fig.~\ref{fig:expanding-summary}.
\begin{figure}[h!]
\centering
\includegraphics[scale=0.3]{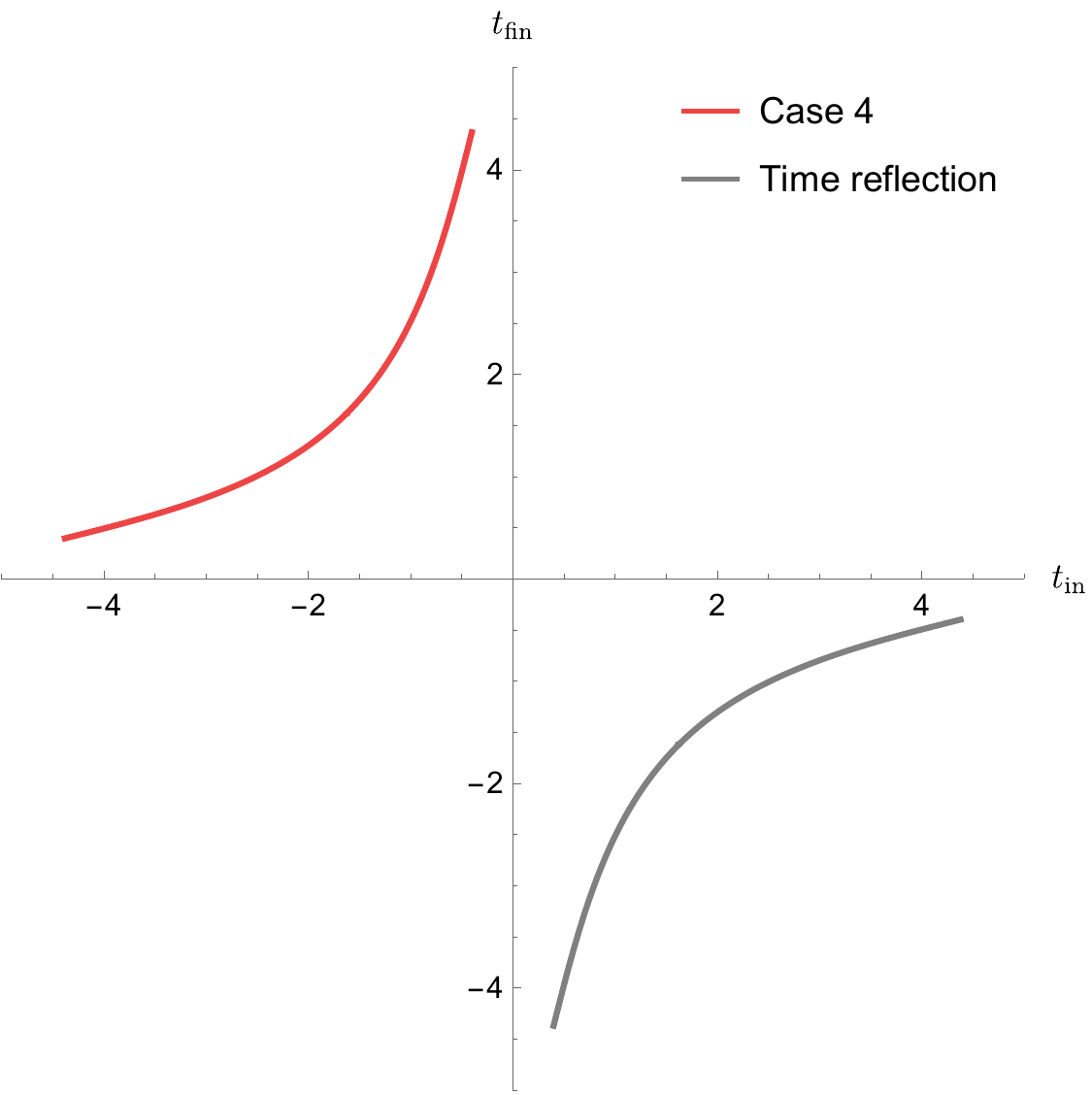}
\qquad 
\includegraphics[scale=0.3]{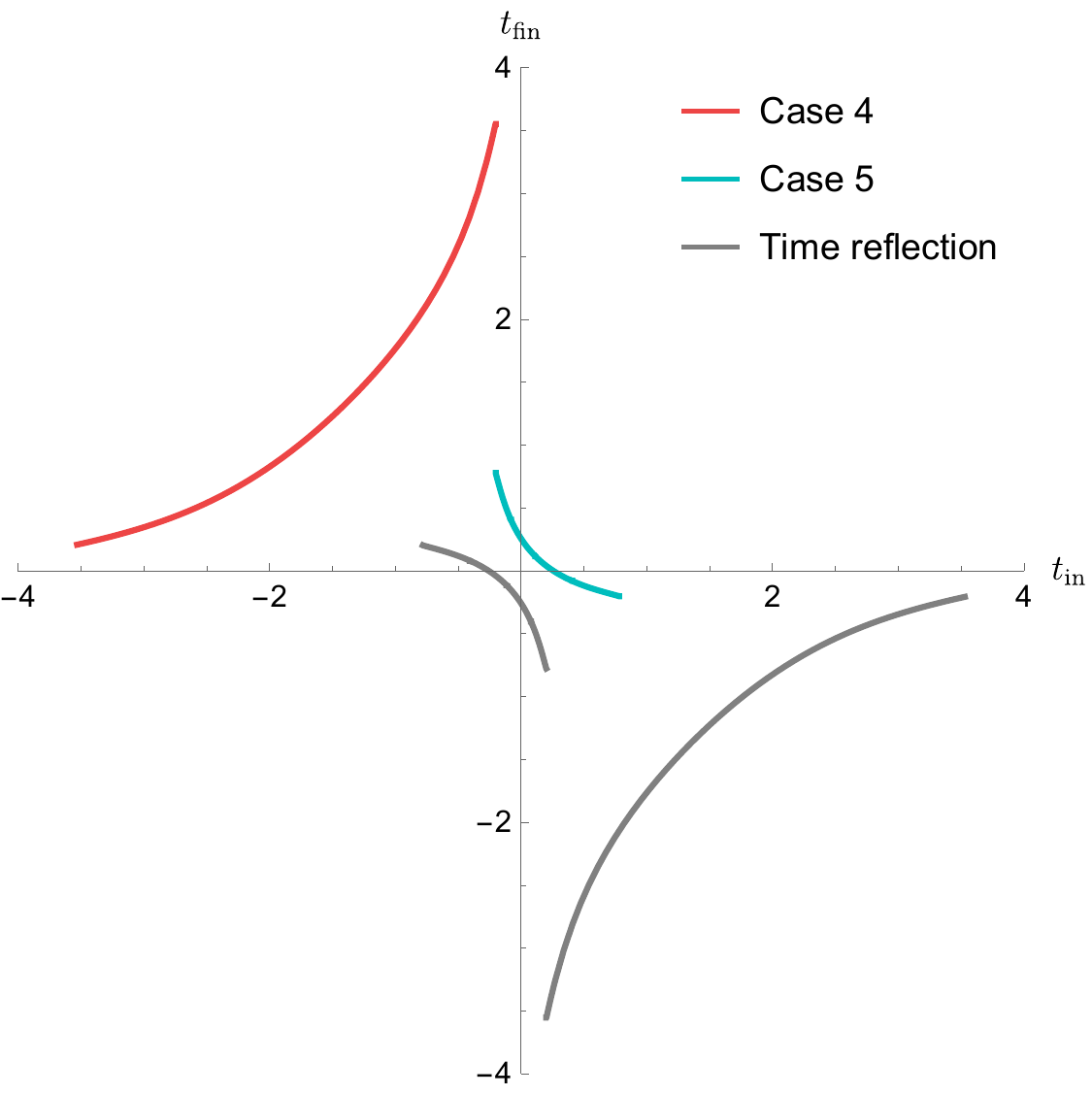}
\caption{
Bulk-cone singularities for two examples of expanding bubbles,
with either AdS (left panel) or dS (right panel) interiors. 
}
\label{fig:expanding-summary}
\end{figure}
\item For static bubbles, the function $t_{\rm fin}(t_{\rm in})$ is linear
\beq
t_{\rm fin}(t_{\rm in})=t_{\rm in}+\Delta t \, ,
\label{eq:static-conclu}
\eeq
where $\Delta t$ is a constant independent of the initial time.
In the full Kruskal extension of the eternal AdS BH, the bulk-cone singularities related to almost-null radial geodesics do not exist,
because such geodesics cannot come back to the original boundary. 
Due to ETH, we expect then that such bulk-cone singularities disappear 
also for the bubble geometries,
during the process of thermalization at large time. This means that, if thermalization holds, there are no bulk-cone singularities when $t_{{\rm in}}$ and $t_{{\rm fin}}$ both approach to infinity. Interestingly, the way in which the bulk-cone singularities disappear is different for the collapsing and the expanding bubble: in the former case they disappear because
  $|t_{{\rm fin}}-t_{{\rm in}}| \to \infty$ when $t_{{\rm in}} \to \infty$, while in the latter case because  no radial almost-null geodesics exist for large $t_{{\rm in}}$.
Equation~(\ref{eq:static-conclu}) shows that the bulk-cone singularities of static bubbles are inconsistent with the thermalization of the physical system, as diagnosed by two-point functions of scalar operators.
This property resembles quantum many-body scars, that are extensively studied in condensed matter physics (see Ref.~\cite{Serbyn:2020wys} for a review).

Gao-Wald theorem~\cite{Gao:2000ga} states that the time difference $\Delta t$ between antipodal boundary points connected by an exactly null bulk geodesic satisfies
 \beq
 \Delta t \geq \hat{L},
 \label{eq:GW}
 \eeq
where $\hat{L}=\pi$ is the length of the arc which connects the north and south poles of the sphere $S^{d-1}$ on the boundary. 
The inequality~(\ref{eq:GW}) strict holds for any perturbation of empty AdS spacetime (satisfying the null energy and null generic conditions), and is saturated by empty AdS spacetime.
In the case of an almost-null spacelike geodesic, the inequality~ (\ref{eq:GW}) can be violated~\cite{Hubeny:2006yu}.
Still, its violation points towards some source of non-locality.
We studied in fig.~\ref{fig:plots_deltat_AdS3d_static} the dependence of $\Delta t $ in eq.~(\ref{eq:static-conclu}) as a function of the parameters $(\lambda, \kappa)$.
We found numerical evidence that $\Delta t \geq \pi$ for $\l<0$, as shown by comparing with the dashed green curve in 
fig.~\ref{fig:plots_deltat_AdS3d_static}. 
Instead, it is possible to attain $\Delta t < \pi$ when $\l>0$.

The function $\Delta t$ admits two interesting limits:
\begin{itemize}
\item for  $\kappa<1$ 
\beq 
\lim_{\lambda \rightarrow 0^+} \Delta t = 0 \, .
\label{eq:disc_static_limits_1}
\eeq
Therefore, a static bubble with almost-flat interior present non-local features, as advocated in Ref.~\cite{Freivogel:2005qh}. 
Surprisingly, we observed in section \ref{sec:EE_bubbles}, this non-locality is not detected by the behavior of holographic entanglement entropy, since the HRT surface does not penetrate inside the static bubble.
\item for  $\kappa>1$ 
\beq
\lim_{ \lambda \rightarrow - \le 1 - \kappa \ri^2} \Delta t = \pi \, ,
\label{eq:disc_static_limits_2}
\eeq
where the limit in $\l$ is taken from above.
Therefore, eq.~\eqref{eq:disc_static_limits_2} saturates the time-delay predicted by Gao-Wald theorem~\cite{Gao:2000ga}. \end{itemize}

It is important to emphasize that the static bubble is a limit configuration which is realized with some
fine tuning  in parameter space, see fig.~\ref{fig:near_static_length}.

\end{itemize}

\subsection{Outlook}

In this work, we clarified several features on how we can detect the presence of a vacuum bubble in an asymptotically AdS spacetime from the boundary perspective.
We leave several interesting open questions for further investigations:
\begin{enumerate}
\item \textbf{Beyond the thin wall approximation.}
As discussed in section~\ref{ssec:causal_structure_expanding_bubble}, the expanding bubble solution in regions $A$ and $C$ of parameter space presents certain unconventional features.
For instance, the Penrose diagram admits a corner where the dS infinity is joined with an AdS boundary \cite{Freivogel:2005qh},
or where two AdS boundaries bifurcate from the intersection with the domain wall.
\rep{
Whether these features are artifacts of the thin-wall approximation remains unclear.
}
\rep{It would be interesting 
to perform a numerical study
in a full solution of the scalar-gravity system
to investigate the precise causal structure of the spacetime
in such a system. }
Possible modifications of the geometry might change some of the features of the bulk-cone singularities
for the expanding bubbles.
\item \textbf{Entanglement entropy at generic times.}
In this work, we focused on the computation of the holographic entanglement entropy
for $d=2$ and at boundary time $t_{b}=0$.
In the cases where the HRT surface has been found to explore the bubble interior (with the bubble either inside or outside the black hole bifurcation surface), we expect a time evolution of the holographic entanglement entropy towards thermalization.
It could be interesting to extend our analysis to arbitrary time in such situations.
Another possibility is to perform this computation in bubble geometries in dimensions $d \geq 3$. 
    \item \textbf{Relation to defects.}
Region $D$ in fig.~\ref{Phase-Diagram} is delimited by the curves in eq.~\eqref{eq:range_tensions_AdS}. In three-dimensional bottom-up models of bulk geometries with a thin brane, it was shown that the same region of parameter space define the regime where the dual interface CFT is well-defined~\cite{Karch:2000ct,Bachas:2001hpy}.
    In particular, Ref.~\cite{Simidzija:2020ukv} discussed the analytic continuation of the bubble geometries, and found the above bounds by computing the boundary entropy of a defect in $d=2$.
    \rep{In higher dimensions, the same inequalities were found in a different context in Ref.~\cite{May:2021xhz}.}
    Furthermore, the same bounds appear in the computation of the energy transport across a two-dimensional holographic interface~\cite{Bachas:2020yxv,Bachas:2021tnp,Baig:2022cnb,Bachas:2022etu,Baig:2023ahz,Baig:2024hfc,Gutperle:2024yiz,Liu:2025khw}.
    The transmission coefficient is related to a universal central charge entering the two-point functions of the stress tensors in different sides of the holographic interface.
    Therefore, it would be interesting to compute the transmission coefficient and the bulk-cone singularities associated with
    the stress tensor to provide complete information about this
    class of two-point functions.
    \item \textbf{Holographic complexity.}
    Reference~\cite{Auzzi:2023qbm} initiated the computation of volume complexity in three-dimensional bubble geometries (for recent reviews on quantum complexity, see~\cite{Susskind:2018pmk,Chapman:2021eyy,Nandy:2025ktk,Baiguera:2025dkc,Rabinovici:2025otw}).
    Holographic complexity provides a tool to investigate the interior of black holes, and it has been recently applied to diagnose properties of inflationary geometries, \eg see~\cite{Geng:2019ruz,Susskind:2021esx,Chapman:2021eyy,Jorstad:2022mls,Anegawa:2023wrk,Baiguera:2023tpt,Anegawa:2023dad,Aguilar-Gutierrez:2023zqm,Baiguera:2024xju,Aguilar-Gutierrez:2024rka,Faruk:2025bed,Aguilar-Gutierrez:2025mxf}.
    It would be interesting to study the holographic proposals in other bubble configurations, including higher dimensions and in the presence of subregions.
    One could then compare the results with the entanglement entropy, and study whether complexity provides another tool that distinguishes the various possible bubble configurations. 
    Furthermore, complexity also quantifies the difficulty to reconstruct information from the boundary, as discussed in the context of the Python's lunch (\eg see~\cite{Brown:2019rox}).
    \rep{It could be intriguing to understand whether the appearance of multiple extremal surfaces, observed in certain regimes considered in this paper, could be used to claim that the holographic dictionary is exponentially complex, as advocated in Refs.~\cite{Bouland:2019pvu,Akers:2024wre}.}
    \item \textbf{Chaos and OTOC.}
    In this work, we used the geodesic approximation to relate the singularities of the boundary two-point functions to the existence of null geodesics intersecting the AdS boundary twice.
    What about other correlators? 
    A particular class of four-point functions is provided by the out-of-time-order-correlators (OTOCs), which are used to distinguish chaotic from integrable systems~\cite{Maldacena:2015waa}.
    In holography, OTOCs can be studied by computing the length of geodesics in a geometry perturbed by shockwaves~\cite{Shenker:2013pqa,Shenker:2013yza}.
    It would be interesting to build bubble geometries perturbed by shockwaves, and then compute the OTOC in this background.
\end{enumerate}

%% file: Sections/Additional_mat.tex
\section{Details about the theoretical setting and the parameter space}

In this appendix, we collect additional details on the geometric setup and the parameter space of bubble geometries.

\subsection{Tortoise coordinate}
\label{ssec:tortoise}

In order to examine the causal structure of the geometry \eqref{metric-zero}, it is convenient to introduce the null coordinates~\eqref{u-v-def}, defined in terms of the following \textit{tortoise coordinate}:
\beq
r^*_{i,o}=\int \frac{d \tilde{r}}{f_{i,o}(\tilde{r}) } \, .
\label{tortoise}
\eeq
In a coordinate system adapted to the null directions as $(v,r,\dots)$ or $(u,r,\dots)$, the metric reads
\beq
\begin{aligned}
ds^2_{i,o} &=-f_{i,o}(r) \, dv_{i,o}^2+2 dr \, dv_{i,o} + r^2 d \Omega_{d-1}^2 \\ 
&= -f_{i,o}(r) \, du_{i,o}^2-2 dr \, du_{i,o} + r^2 d \Omega_{d-1}^2  \, .
\label{EF-metrica-esplicita}
\end{aligned}
\eeq
The tortoise coordinate can be analytically determined in the interior part of a bubble geometry with blackening factor~\eqref{f-dS}. Depending on the sign of the cosmological constant, it reads
\begin{subequations}
\beq
r^*_i(r) =\frac{1}{4 \sqrt{\l}} \log \left[ \left(\frac{1+ r \sqrt{\l}}{1- r \sqrt{\l}} \right)^2 \right]   \qquad   \text{if $\l>0$} \, , 
\label{eq:tortoise_interior_dS}
\eeq
\beq
r^*_i(r)= \frac{\arctan \le r \sqrt{| \l |} \ri}{\sqrt{|\l|}}  \qquad \text{if $\l<0$} \, .
\label{eq:tortoise_interior_AdS}
\eeq
\label{eq:tortoise_interior}
\end{subequations}
Notice that the case of a flat Minkoski interior $\l=0$ corresponds to $r_i^*(r)=r$, as can be seen by performing the limit $\l \rightarrow 0$ in the previous expressions, equivalently approached from above or below. 

For arbitrary dimension $d$ and blackening factor~\eqref{f-BH}, there is no closed-form expression for the tortoise coordinate $r_o^*$ in the exterior geometry.
However, analytic results can be achieved when fixing $d$ to a specific value.
In particular, this work focuses on the following cases:
\begin{itemize}
\item When $d=2$, the tortoise coordinate in the BTZ background reads
\beq
r^*_o(r)\Big|_{d=2} = \frac{1}{4 \sqrt{\mu} } \log \left[ \le \frac{r-\sqrt{\mu}}{r+\sqrt{\mu}} \ri^2 \right] \, .
\eeq
Notice that this solution satisfies $r_o^*(0) = r_o^*(\infty)=0$.
\item When $d=3$, the tortoise coordinate is given by
\beq
\begin{aligned}
r^*_{o}(r) \Big|_{d=3} & = 
 \frac{r_h}{2(3 r_h^2+1)} \log \left| \frac{(r-r_h)^2}{r^2+r_h r + r_h^2 +1} 
\right|  + \\
& + \frac{3 r_h^2 +2}{(3r_h^2 +1)\sqrt{3 r_h^2 +4}}  \arctan \le \frac{2r +r_h}{\sqrt{3 r_h^2 +4}} \ri -  \frac{\pi}{2} \, \frac{3 r_h^2 +2}{(3r_h^2 +1)\sqrt{3 r_h^2 +4}}  \, ,
\end{aligned}
\label{eq:ro_3d_exact}
\eeq
where the integration constant is fixed by imposing $r_o^*(\infty)=0.$
Notice that in higher dimensions $d \geq 3$, the blackening factor satisfies $r_o^*(0) <0$. In other words, a light ray sent from the AdS boundary along a trajectory at constant $v=0$ intersects the singularity in the right side of a Penrose diagram.
This phenomenon is represented by \textit{bending} the singularity towards the bottom of the Penrose diagram, \eg see figs.~\ref{Pen-Dia-Expanding-Bubble} and \ref{Pen-Dia-Collapsing-Bubble}.
\end{itemize}

\subsection{Small and large mass limits}
\label{app:small_large_mass_limits}

In this subsection, we investigate the regimes of small and large mass parameter $m$ in the parameter space.
Beginning with the small--$m$ limit, the expressions~\eqref{eq:effective_potential_ABC} and \eqref{eq:functions_ABC} imply that $R_{0} \to 1/\sqrt{\mathcal{A}}$ for an expanding bubble, while $R_{0} \to 0$ for a collapsing one.\footnote{When $d=2$, we need $m \geq 1$ for a BH horizon to exist. However, even in this case, we are still formally allowed to consider the regime $ 0 \leq m \leq 1$.} 
By inspection of eqs.~\eqref{eq-massa-bubble-2} and \eqref{eq:beta:discrimine}, we find that,
in the same limit, a collapsing bubble satisfies $\b_i(R_0)>0$, while an expanding bubble satisfies $\b_i(R_0)<0$.

Next, let us consider the large-mass limit. For fixed and finite values of $\l$ and $\kappa$, this regime can only be achieved when $R_0 \to \infty$ and $\lambda \leq 0$, as can be determined from eq.~\eqref{eq-massa-bubble}.
Let us focus on the two possible cases of a Minkowski or an empty AdS interior geometries.\footnote{In the case of an interior dS geometry, the mass of the external BH is bounded from above, therefore it cannot be parametrically large compared to the other scales.}
In the flat case, by plugging $\lambda =0$ inside eq.~\eqref{eq-massa-bubble}, we find that a large positive mass can only be achieved when $0<\kappa \leq 1$.
Referring to the parameter space in fig.~\ref{Phase-Diagram}, this setting corresponds to the boundary between regions $A$ and $B$, where both signs of $\b_i(R_0)$ (either positive or negative) are allowed.

In the case of an AdS interior, for large enough $R_0$ we can approximate the mass parameter~\eqref{eq-massa-bubble-2} as follows: 
\beq
\begin{aligned}
m_{\pm} \approx R_0^d \left( 1-(\kappa \mp \sqrt{-\l})^2 \right)
\pm \frac{\kappa}{\sqrt{-\l}} \, R_0^{d-2} \, .
\end{aligned}
\label{eq:leading_large_mass}
\eeq 
By requiring that the leading order term proportional to $R_0^d$ is positive, we end up in the following regions of parameter space in fig.~\ref{Phase-Diagram}:
\begin{itemize}
\item For $\b_i>0$, see eq.~\eqref{beta-i-beta}, we end up in regions $B$ and $D$.
\item For $\b_i <0$, we end up in region $B$.
\end{itemize}
The previous statements are consistent with the fact that region $D$ only admits a collapsing bubble solution for any value of $m$ (see table~\ref{tab:parameter_space}). 
Furthermore, the previous conclusion shows that in region
$B$, at large enough $m$, there exist two different solutions
(one collapsing, with $\b_i>0$, and one expanding, with $\b_i<0$).
The latter statement is consistent with the property that in region $B$
the static bubble does not exist, instead there are two solutions
(a collapsing and an expanding bubble) for every positive value of $m$.
In particular, the expanding bubble 
corresponds to a geometry with two disconnected boundaries where a Coleman-De Luccia bubble nucleates nearby the left boundary.

It is interesting to consider the fine-tuned values for which
the leading  $R_0^d$ term in the right-hand side of eq.~\eqref{eq:leading_large_mass} vanishes,  while the next-to-leading-order term is positive.
This can only happen when $\b_i>0$, and selects the border between regions $C, D$ and the border between regions $D, E$ in fig.~\ref{Phase-Diagram}.
Note that in the special case $d=2$, the next-to-leading term is finite.

\subsection{The sign of the curvature parameters $\b_i$ and $\b_o$}
\label{ssec:sign_curvature_parameters}

\begin{figure}[ht]
\center
\includegraphics[scale=0.55]{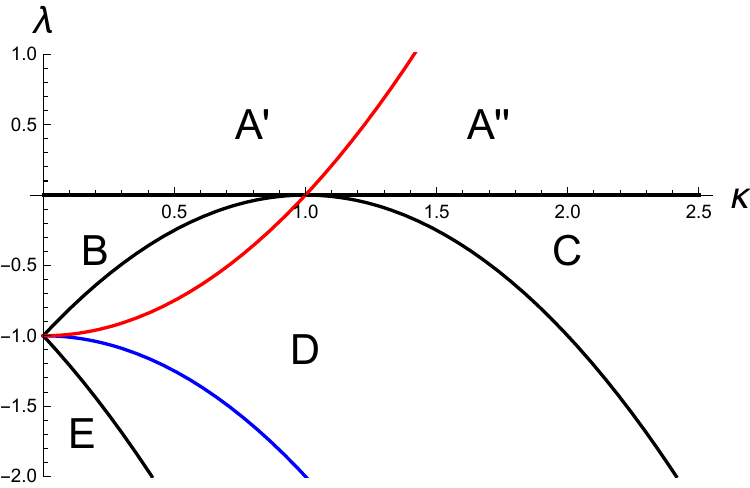}
\caption{Picture of the parameter space $(\kappa, \lambda)$ of a bubble geometry in the thin wall approximation.
In red (blue), we show the curves $\l=\kappa^2-1$ ($\l=-\kappa^2-1$) in the phase diagram, below which we find $\b_i>0 \, (\b_o>0)$ for every $m,d$ and $R(\tau)$.
}
\label{region-parameters-betai-betao}
\end{figure}

In this subsection, our goal is to study the signs of the parameters $\b_{i,o}(R_0)$, which allow to justify the causal structure discussed in
subsection~\ref{ssec:causal_structure_expanding_bubble} and~\ref{ssec:causal_structure_collapsing_bubble}.
To this aim, we depict two additional curves in the parameter space, see the red and blue lines in fig.~\ref{region-parameters-betai-betao}.

The red curve is defined by $\lambda = \kappa^2 -1$.
Let us discuss the case of a dS interior first. The red curve splits region $A$ in two subregions $A'$ and $A''$. 
When $\l<\kappa^2-1$ (region $A''$), for any values of the parameters $(m, d)$ and of the shell's radius $R(\tau)$, the condition $\beta_i >0$ holds, in particular at $R_0$, see eq.~\eqref{beta-i-beta}.
This fact shows that in region $A''$ the internal dS region 
can never contain a complete dS static patch. 
In the case of an interior AdS geometry, the above constraint 
$\beta_i >0$ applies to regions $C, E$ and to the part of region $D$ on the right of the red curve in fig.~\ref{region-parameters-betai-betao}.
Moreover, collapsing bubbles always satisfy $\beta_i(R_0)>0$ (see the discussion below eq.~\eqref{eq:beta:discrimine}), and so
this property is always satisfied in region $D$,
where no expanding bubble with positive mass $m$ exist.

In the regions $C, D, E$ and $A''$ of parameter space, for a fixed value of the initial radius $R_0$ and of the parameters $(\l,\kappa)$, there is only one allowed bubble solution, whose positive mass is given by $m_+(R_0)$ in eq.~\eqref{eq-massa-bubble-2}. In particular, we find that:
\begin{itemize}
\item In regions $A'', C$ and $E$, the function $m_+(R_0)$ has a maximum
in correspondence of the radius of the static bubble $R_{\rm static}$,
and it is defined only in the range $0 \leq R_0 \leq 1/\sqrt{\mathcal{A}}$.
For $0 < R < R_{\rm static}$, the solution corresponds to a collapsing bubble.
For $R_{\rm static} < R <  1/\sqrt{\mathcal{A}}$, the solution is an expanding bubble.
\item In region $D$, the function $m_+(R_0)$ is monotonic and defined for $0 \leq R_0 < \infty$. This solution always corresponds to a collapsing bubble.
\end{itemize}

In regions $A'$ and $B$, for a fixed value of $R_0$, it is possible to have two different bubbles whose mass is given by the two choices in eq.~\eqref{eq-massa-bubble-2}. In particular, we find that:
\begin{itemize}
\item In region $A'$, the function $m_{+}(R_0)$ is defined for $0 \leq R_0 \leq 1/\sqrt{\l}$, while $m_{-}(R_0)$ is defined for $1/\sqrt{\mathcal{A}} \leq R_0 \leq 1/\sqrt{\l}$.
The two functions coincide when $R_0=1/\sqrt{\l}$.
The function $m_+$ has a maximum at a value of $R_0$
which corresponds to the radius of the static bubble $R_{\rm static}$. For $0< R_0 < R_{\rm static}$,  $m_+(R_0)$ is the mass of a collapsing bubble. For $ R_{\rm static} < R_0 < 1/\sqrt{\l}$, $m_+(R_0)$ is the mass of an expanding bubble. 
The function $m_-(R_0)$ always corresponds to the mass of an expanding bubble.
\item In region $B$, both $m_{\pm}(R_0)$ are monotonic functions of $R_0$, and they diverge when $R_0 \to \infty$. 
The function $m_{+}(R_0)$ is defined for $0 \leq R_0 < \infty$, and 
identifies the mass of a collapsing bubble.
The function $m_{-}(R_0)$ is defined for $1/\sqrt{\mathcal{A}} \leq R_0 < \infty$ and corresponds to the mass of an expanding bubble.
\end{itemize}
Note that a collapsing bubble always satisfies the property $\b_i(R_0)>0$, while for an expanding bubble both signs of $\b_i(R_0)$ are possible, see the discussion below eq.~\eqref{eq:beta:discrimine}.
For convenience, the above classification of the existing bubble geometries in parameter space is collected in table~\ref{tab:parameter_space}.

The blue curve in fig.~\ref{region-parameters-betai-betao} is defined by $\lambda = -\kappa^2-1$.
When $\l<-\kappa^2-1$, we have $\b_o>0$, implying that the bubble is initially outside the BH horizon.   
In particular, in region $E$ we find $\b_o(R_0)>0$ for both collapsing and expanding bubbles. 
Therefore, in this case the expanding bubble is initially outside the horizon, and then reaches the AdS boundary in a finite amount of time (see configuration $E$ in fig.~\ref{Pen-Dia-Expanding-Bubble}).
In other words, almost all the spacetime is composed by the interior AdS geometry, the BH spacetime region becomes finite, and the dual CFT at $t=0$ becomes ill-defined.
This setting corresponds to the decay of a false vacuum as in Coleman-De Luccia \cite{Coleman:1980aw}, where the false vacuum corresponds to the exterior region.


\subsection{The domain wall trajectory}
\label{ssec:domain_wall_traj}

In this subsection, we determine the trajectory of the domain wall, that is employed to numerically investigate the bulk-cone singularities in section~\ref{sec:sing_corr}.
In general dimension $d$, we consider the velocity vector $\omega^{\mu}$ of the bubble at fixed angular coordinates, and appropriately normalized~\cite{Freivogel:2005qh}:
\beq
\omega^{\mu} = (\dot{T}, \dot{R}, 0) \, , \qquad
\omega_{\mu} \omega^{\mu} = -1 \, .
\eeq
In the previous expression, $(T, R)$ denote the time and radial Schwarzschild coordinates evaluated at the position of the domain wall. 
The above conditions can be equivalently recast into the identity
\beq
\dot{T}^2_{i, o} = \frac{1}{f_{i, o}(R)}  \le  1+ \frac{\dot{R}^2}{f_{i, o}(R)} \ri  \, .
\label{eq:matching2}
\eeq
Plugging \eqref{eq:matching2} inside the equations of motion \eqref{general-V-bubble} and using the parameter $\tau=R$, we obtain
\beq
\frac{dT_{i, o}}{dR} = \pm \frac{\sqrt{f_{i, o}(R) - V_{\rm eff}(R)}}{f_{i, o}(R) \sqrt{-V_{\rm eff}(R)}} \, .\label{eq:dTdR}
\eeq
In $d=2$, the previous formula reduces to eq.~(II.24) of Ref.~\cite{Auzzi:2023qbm}.
In general dimensions, after plugging the effective potential 
\eqref{general-V-bubble-potenziale} in eq.~\eqref{eq:dTdR}, we obtain
\begin{subequations}
\beq
\frac{dT_{i}}{dR}=\pm \frac{(\kappa^2-\l-1)R+ m \, R^{1-d} }{2\kappa(1-\l R^2)
\sqrt{\mathcal{A} \,R^2-1+ \mathcal{B} \,R^{2-d}+\mathcal{C} \, R^{2-2 d}}} \, ,
\label{eq:dynamical bubble differential equation 1}
\eeq
\beq
\frac{dT_{o}}{dR}=\pm \frac{(\kappa^2+\l+1)R-m\,R^{1-d}}{2\kappa(R^2+1-m \,R^{2-d})
\sqrt{\mathcal{A} \, R^2-1+ \mathcal{B} \,R^{2-d}+\mathcal{C} \, R^{2-2 d}}} \, .
\label{eq:dynamical bubble differential equation 2}
\eeq
\label{eq:dynamical_bubbles_diffeq}
\end{subequations}
Time reversal-invariant solutions to these equations are obtained by imposing the boundary condition $T_{i,o}(R_{\text{max}})=0$ ($T_{i,o}(R_{\text{min}})=0$) in the collapsing (expanding) case.
The above equations specify the detail of the spacetime geometry.

Next, we study how the time coordinate varies across a static shell.
Using the constraint $V_{\rm eff}(R_{\rm static})=0$ inside eq.~\eqref{eq:dTdR}, we obtain the following identity valid in general spacetime dimension:
\beq
\frac{dT_i}{dT_o} = \pm \sqrt{\frac{f_o(R_{\rm static})}{f_i(R_{\rm static})}} \, .
\label{eq:matching_genD}
\eeq


\section{Details on the computation of entanglement entropy}
\label{app:entropy}

In this appendix, we determine the length of the minimal HRT geodesic in a three-dimensional bubble geometry as a function of the opening angle $\Delta \theta$ associated with a boundary arc. Since the opening angle depends non-trivially on the conserved quantity $j$ introduced in eq.~\eqref{eq:j-def}, this step requires an analysis of the various configurations that the extremal curves can attain. We perform such analysis below.

For any value of $j$, it can be explicitly checked that
\begin{equation}
    \frac{dl(j)}{dj} = j \, \frac{d\Delta\theta(j)}{dj} \, .
    \label{eq:entropy-special-relation}
\end{equation}
This expression is valid both when the bubble is inside or outside the BH bifurcation surface.
It is useful to recast eq.~\eqref{eq:entropy-special-relation} as follows:
 \begin{equation}
     \frac{d \, l(\Delta\theta)}{d \, \Delta\theta} = j \, .
     \label{eq:entropy-special-relation-2}
 \end{equation}
This result can also be derived by the differential entropy formula in Ref.~\cite{Balasubramanian:2013lsa} applied to a circumference of radius $j$. The differential entropy formula is indeed known to hold in spacetimes where the HRT prescription applies \cite{Headrick:2014eia,Czech:2015qta}. 
From the positivity of $j$, eq.~\eqref{eq:entropy-special-relation-2} implies that the length of a geodesic increases monotonically with $\Delta\theta$. Nonetheless, the quantity $\frac{d \, l(\Delta\theta)}{d \, \Delta\theta}$
experiences a jump at the critical value when the HRT surface enters the bubble, because $j$ is indeed not continuous (there is a "first order phase transition" in the location of the HRT surface with minimal length). Equation~\eqref{eq:entropy-special-relation-2} will be useful in determining the geodesics with minimal length.

To this aim, let us analyze the behavior of the function $\Delta \theta (j)$. 
Let us denote by $j_w$ the critical value of $j$ for which the geodesic enters the bubble. The quantity $j_w$ can only take two possible values: $j_w = R_0$ or $j_w = r_h$, depending on whether the bubble is outside or inside the BH bifurcation surface, respectively.
Let us discuss the behavior of  $\Delta \theta (j)$ nearby $j=j_w$:
\begin{itemize}
 \item
For $j>j_w$, we find from eq.~\eqref{eq:Delta-theta-j-out} that
the function $\Delta \theta (j)$ is monotonically decreasing.
\item When the bubble is outside the BH bifurcation surface, let us consider the limit $j \to R_{0}^-$ 
in eq.~\eqref{eq:Delta-theta-j-full}:
\beq
\Delta \theta (j)  = \frac{2}{r_h} \mathrm{arctanh} \, \le \frac{r_h}{R_{0}} \ri 
+ \frac{2 \sqrt{2}}{\sqrt{R_{0}}} 
\le \frac{1}{\sqrt{f_i(R_0)}} - \frac{1}{\sqrt{f_o(R_0)}} \ri 
\sqrt{R_{0} -j} + \mathcal{O}(R_{0} -j) \, .
\eeq
Comparing with eq.~\eqref{eq:Delta-theta-j-out}, this expression shows that $\Delta \theta (j)$ is continuous at $j = R_{0}$.
Note that $\beta_{i,o}(R_{0}) = \pm \sqrt{f_{i,o}(R_{0})}$, see eq.~\eqref{general-V-bubble-1}. If the bubble is located outside the BH bifurcation surface, we have $\b_o(R_0)>0$. From eq. (\ref{beta-i-beta-o-gen}), we find  that $\b_o(R_0) \leq \b_i(R_0)$. This means that in this case we also have $\b_i(R_0)>0$.
Therefore, from eq.~\eqref{general-V-bubble-0} we get 
\beq
\sqrt{f_{i}(R_{0})}-\sqrt{f_{o}(R_{0})}= \kappa R_{0} \, .
\label{eq:kR_very_small}
\eeq
As a direct consequence of eq.~\eqref{eq:kR_very_small}, for $j \lesssim j_w=R_{0}$, the function $\Delta \theta (j)$ is monotonically increasing.
\item  When the bubble is located inside the BH bifurcation surface, the story is slightly different.
For $j>r_h$, the geodesic is external, and so $\Delta \theta(j)$ diverges logarithmically for $j \to r_h^+$,
see eq.~\eqref{eq:Delta-theta-j-out}.
On the other hand, for $j<r_h$ the geodesic extends inside the bubble.
Similarly, from eqs.~\eqref{eq:Delta-theta-j-other-side} and \eqref{eq:Delta-theta-j-other-side-full-static-patch} it can be checked that $\Delta \theta(j)$ also diverges logarithmically for $j \to r_h^-$. 
Note that also in this case, for $j \lesssim j_w = r_h$, the function $\Delta\theta(j)$ is monotonically increasing.
\end{itemize}
In summary, $\Delta \theta (j)$ has a non-smooth local maximum 
at the critical value $j=j_w$ for which the geodesic enters the bubble.
This occurs at a finite or infinite value of $\Delta \theta$
depending on whether the bubble is outside or inside the BH bifurcation surface, respectively.

\begin{figure}[ht]
\center
\includegraphics[scale=0.7]{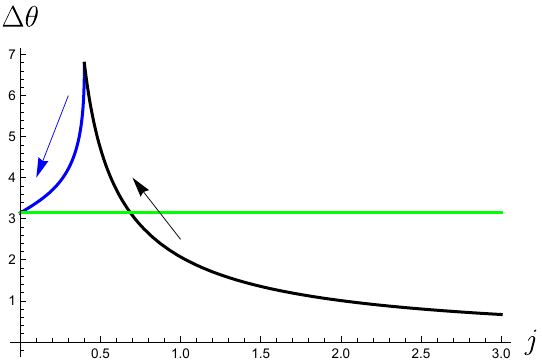}
\qquad
\includegraphics[scale=0.7]{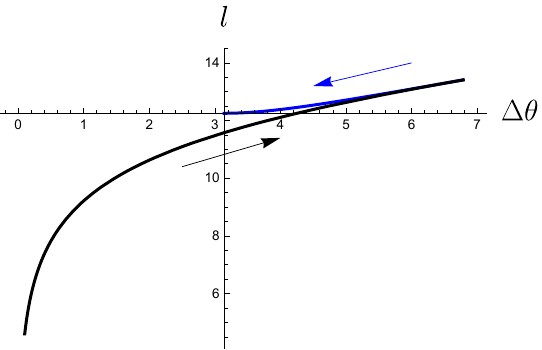}
\caption{
Example of case in which $\Delta \theta(j)$ is monotonic for $0 \leq j \leq j_w$, with $\l=-1$, $m=1.1$, $r_h=0.317$, $R_0=0.4$, $\kappa=2.08$, $R_c=100$.
Left: Opening angle $\Delta \theta$ of the boundary subregion as a function of the conserved momentum $j$ of the bulk geodesic. Right: length of the geodesic as a function of $\Delta \theta$. 
}
\label{fig:theta_j-nomaxmin}
\end{figure}

\begin{figure}[ht]
\center
\includegraphics[scale=0.7]{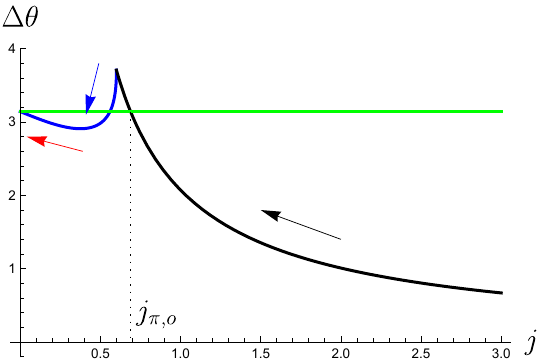}
\qquad
\includegraphics[scale=0.7]{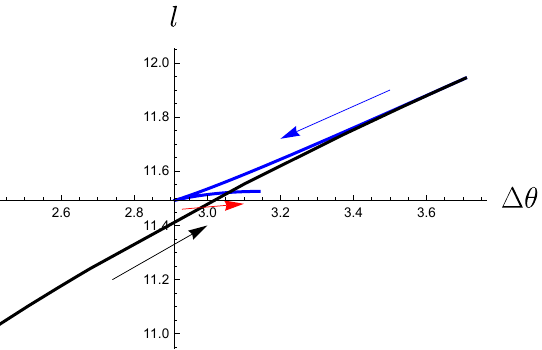}
\qquad
\includegraphics[scale=0.7]{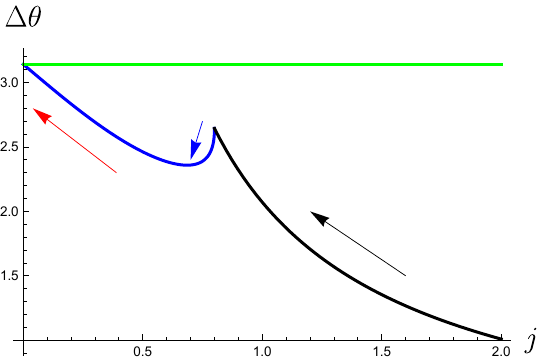}
\qquad
\includegraphics[scale=0.7]{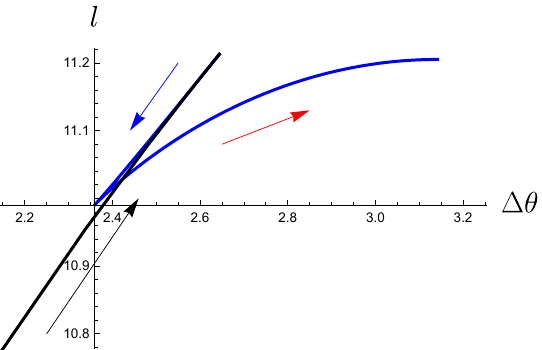}
\caption{
Examples where $\Delta \theta(j)$ has a local minimum 
and no local maximum for $0 \leq j \leq j_w$.
Upper panels: $\l=-1$, $m=1.1$,
$r_h=0.317$, $R_0=0.6$,
$\kappa=1.09$, $R_c=100$.
Lower panels:  $\l=-1$, $m=1.1$,
$r_h=0.317$, $R_0=0.8$,
$\kappa=0.68$, $R_c=100$.
}
\label{fig:theta_j-min-nomax}
\end{figure}

\begin{figure}[ht]
\center
\qquad
\includegraphics[scale=0.7]{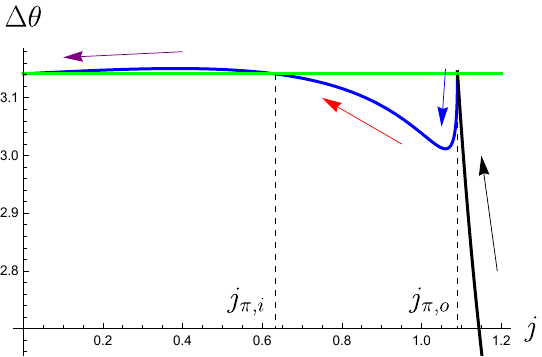}
\qquad
\includegraphics[scale=0.7]{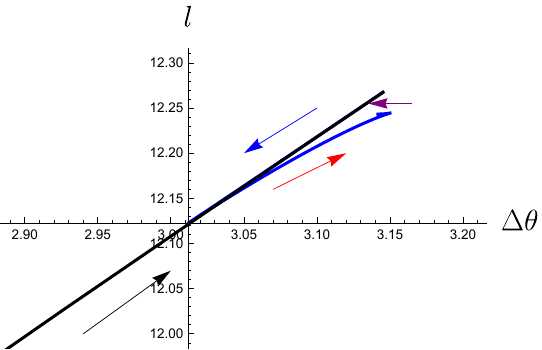}
\caption{
Example where $\Delta \theta(j)$ has a local minimum 
and a local maximum for $0 < j < j_w$,
with numerical values
$\l=0.5$, $m=2$, $r_h=1$, $R_0=1.09$, $\kappa=0.19$, $R_c=100$.
}
\label{fig:theta_maxmin}
\end{figure}

In order to determine the behavior of $\Delta \theta(j)$, it is useful to look for minima and maxima of the function  $\Delta \theta (j)$ in the window $0 \leq j \leq j_w$.
The condition $\Delta\theta'(j)=0$ 
for $0 \leq j \leq j_w$ takes a different form
depending on the position of the bubble with respect 
to the Bifurcation Surface (BS) and on the sign of $\b_i(R_0)$, i.e.,
\bea
\sqrt{R_0^2-j^2} &=& \frac{(1-\l j^2)}{\sqrt{1-\l R_0^2}}- \sqrt{R_0^2-r_h^2}
\qquad \text {Bubble outside the BS}
\nl
\sqrt{R_0^2-j^2}
&=& \frac{(1-\l j^2)}{\sqrt{1-\l R_0^2}}+ \sqrt{R_0^2-r_h^2}
\qquad \text {Bubble inside the BS, $\b_i(R_0)>0$}
\label{eq:minimo-lunghezza} \\
-\sqrt{R_0^2-j^2}
&=& \frac{(1-\l j^2)}{\sqrt{1-\l R_0^2}}- \sqrt{R_0^2-r_h^2}
\qquad \text {Bubble inside the BS, $\b_i(R_0)<0$}
\nonumber
\eea
By squaring eq.~\eqref{eq:minimo-lunghezza}, we get a second-order equation for $j^2$.  Depending on the location in the parameter space, some of these solutions can be spurious (\ie need to be discarded) when they correspond to squaring the left-hand side of the equation with a wrong sign.
Depending on the region in parameter space where the bubble solution belongs, eq.~(\ref{eq:minimo-lunghezza}) can admit zero, one or two solutions for $j^2$ in the window $0 \leq j^2 \leq j_w^2$. 
Accordingly, the plot of the function $\Delta\theta (j)$ has a different qualitative structure,
see figs.~\ref{fig:theta_j-nomaxmin},
\ref{fig:theta_j-min-nomax} and \ref{fig:theta_maxmin}.


In the following, we refer to the cases defined in subsection~\ref{subsec:geod-minimal}.
Using the property $\Delta\theta'(j)>0$ for $j \to j_w^-$, we find that:\footnote{We stress that in all the cases we discuss below, when $\Delta\theta > \pi$, there is always a shorter geodesic with $2\pi - \Delta\theta$ satisfying the same homology condition. Therefore, the minimal HRT surface always satisfies $l(\Delta\theta) = l(2\pi-\Delta\theta)$, see eq.~\eqref{eq:S_mirrored}. In other words, the actual plots for the minimal length are mirrored with respect to $\Delta\theta=\pi$, although this is not explicitly shown in the figures that follow.}
\begin{itemize}
\item If eq.~\eqref{eq:minimo-lunghezza} has no
solution in the interval $0 \leq j^2 \leq j_w^2$, then 
$\Delta \theta(j)$ is a monotonically increasing function
in the interval $0\leq j \leq j_w$ (see fig.~\ref{fig:theta_j-nomaxmin} for an example).
Note that as we decrease the conserved charge $j$ of the geodesic outside the bubble, $\Delta\theta$ increases.
As a consequence of eq.~\eqref{eq:entropy-special-relation-2}, the corresponding length $l(\Delta\theta)$ grows monotonically until $\Delta\theta(R_0)$ is reached (black arrow in fig.~\ref{fig:theta_j-nomaxmin}). On the other hand, as we decrease the charge $j$ of the geodesic exploring the bubble, $\Delta\theta$ also decreases. Then, eq.~\eqref{eq:entropy-special-relation-2} tells us that the corresponding length $l(\Delta\theta)$ decreases until $\Delta\theta =\pi$ is reached (blue arrow in fig.~\ref{fig:theta_j-nomaxmin}). 
It is important to stress that the decreasing rate of $l(\Delta\theta)$ inside the bubble is smaller than the increasing rate of $l(\Delta\theta)$ for a solution fully located outside the bubble, since $j$ is smaller in the former case. 
Therefore, we find that for a given $\Delta \theta$, the HRT surface with $j>R_0$ is always shorter compared to the HRT surface with $j<R_0$. In this case, for a given $0 \leq \Delta \theta \leq \pi $, the minimal geodesics are located outside the bubble (\textbf{case $a$}).
\item If eq.~\eqref{eq:minimo-lunghezza} has one solution
in the interval $0 \leq j^2 \leq j_w^2$, then it necessarily corresponds to a minimum of $\Delta \theta(j)$ (see fig.~\ref{fig:theta_j-min-nomax} for some examples).
As we decrease $j$ down to the minimum of $\Delta\theta(j)$, the length $l(\Delta\theta)$ behaves as in the previous case (which did not have a local minimum).
However, as we further decrease $j$, $\Delta\theta$ of the geodesic exploring the bubble grows, and so does the corresponding $l(\Delta\theta)$ (red arrow in fig.~\ref{fig:theta_j-min-nomax}).
The growth rate in this region is smaller, as can be observed from eq.~\eqref{eq:entropy-special-relation-2}.  
Let us denote by $j_{\pi,o}$ the solution to $\Delta \theta(j_{\pi,o})=\pi$,
with $j_{\pi,o}>j_w$.
Depending on the choice of parameters, two situations can arise:
either $l(j=0)>l(j=j_{\pi,o})$, such that the external geodesic with $j>j_w$ is always the one with minimal length for a given $\Delta \theta$ (\textbf{case $a$}),
or $l(j=0)<l(j=j_{\pi,o})$, so that there is a phase transition and the minimal geodesic jumps from outside to inside the bubble (\textbf{case $b$}).
When $\Delta\theta(j)$ has a local minimum in $0<j<R_0$ but $\Delta\theta(R_0)<\pi$, there is no geodesic with $\Delta\theta =\pi$ and located completely outside the bubble (\textbf{case $c$}). Consequently, the phase transition necessarily happens and the minimal geodesics eventually explore the bubble interior. This case is shown in the bottom panels of fig.~\ref{fig:theta_j-min-nomax}.
\item If eq.~\eqref{eq:minimo-lunghezza} admits two solutions
$j_{1,2}$ in the interval $0 \leq j^2 \leq j_w^2$,
assuming $j_1<j_2$, we have that $j_1$ is a local maximum
and $j_2$ is a local minimum of $\Delta \theta(j)$.
An example is shown in fig.~\ref{fig:theta_maxmin}.
Compared to the case in the first bullet point (which did not have a local maximum), when we decrease $j < j_1$ of the geodesic exploring the bubble, we find that $\Delta\theta(j)$ also decreases, and so does $l(\Delta\theta)$ (purple arrow in fig.~\ref{fig:theta_maxmin}).
Let us denote by $j_{\pi,i}$ the solution to $\Delta\theta(j_{\pi,i})=\pi$, with $j_1 < j_{\pi,i} < j_2$.
Two situations can be realized:
either $l(j=j_{\pi,i}) > l(j=j_{\pi,o})$, such that external geodesics with $j>j_w$ are always globally minimal for any $\Delta\theta$ (\textbf{case $a$}),
or $l(j=j_{\pi,i}) < l(j=j_{\pi,o})$, so that at large enough $\Delta\theta$ there is a phase transition and the globally minimal HRT has $j_{\pi,i} \leq j \leq j_2$ (\textbf{case $b$}).
When $\Delta\theta(R_0) < \pi$ and $j_{\pi,o}$ is not defined, the phase transition necessarily takes place (\textbf{case $c$}).
Fig.~\ref{fig:theta_maxmin} displays an example of \textbf{case $b$}.
\end{itemize}

We implemented a numerical scan in the parameter space $(\kappa,\lambda)$ with fixed $m$ to find out the location of the minimal HRT geodesic. The results are shown in fig.~\ref{figure-scan-EE} for collapsing bubbles.

\paragraph{Comments on the Python's lunch.}
Whenever at least two locally minimal surfaces are anchored at the same subregion $\Delta\theta$ and the global minimal surface is not the closest to the boundary subregion itself, a Python's lunch occurs.
Even though, from the right AdS boundary, it is possible to reconstruct the bulk entanglement wedge~\cite{Dong:2016eik} delimited by the global minimal surface, this task is conjectured to be exponentially complex for the region enclosed by the global minimal and the local minimal surfaces.
In the following discussion, we assume that the local minimal surfaces identifying a Python's lunch lie on the bulk slice $t=0$. 
We distinguish among various configurations:
\begin{itemize}
 \item In the case of fig.~\ref{fig:theta_j-nomaxmin}, where $\Delta\theta(j)$ is monotonically increasing in the interval $0 \leq j \leq j_w$, the global minimal geodesic is always characterized by a larger $j$. Here, there are no Python's lunches.
 \item In the cases of fig.~\ref{fig:theta_j-min-nomax}, where $\Delta\theta(j)$ has one minimum in $0 \leq j \leq j_w$, there exists a range of $\Delta\theta$ for which at least two minimal geodesics are homologous to the same boundary subregion. A Python's lunch takes place only when the deepest geodesic in the bulk (red arrow in the figure) is the shortest. This configuration can be realized for a range of $\Delta\theta$ including $\pi$ (\textbf{case $b$}, upper panel of fig.~\ref{fig:theta_j-min-nomax}) or not including $\pi$ (\textbf{case $c$}, lower panel of the fig.~\ref{fig:theta_j-min-nomax}).
Note that the geodesic on the increasing branch of $\Delta\theta(j)$ (blue arrow in the figure), which lies between the global minimum and the local minimum, always has a larger length than the other two. This is a candidate for the bulge surface of the Python's lunch. 
\item The case of fig.~\ref{fig:theta_maxmin}, where $\Delta\theta(j)$ has both a maximum ($j=j_1$) and a minimum ($j=j_2$) in $0 \leq j \leq j_w$, is analogous to the previous one. Notice that the geodesic with $0 \leq j < j_1$ (purple arrow in the figure), when it exists, is never the global minimal. So, it does not take part in the Python's lunch, whenever it occurs.
\end{itemize}

\section{Details on the computation of bulk-cone singularities}
\label{app:bulk_cone_sing}

In this appendix, we calculate the trajectories of radial almost-null geodesics that leave and come back to the right AdS boundary of a static bubble geometry.

\paragraph{Step 1: starting from the AdS boundary.}
The radial almost-null geodesic at constant $v=0$ is pushed away from the singularity $r=0$ (see, \eg Ref.~\cite{Fidkowski:2003nf}) before reaching the shell. In particular, the geodesic is reflected as a curve at constant $u = u_1$ coordinate, determined by
\beq
u_{1} = -2 r^*_o (0) \, ,
\label{eq:u1_static}
\eeq
where $u_1$ stands for step 1 of the present calculation.
Next, we follow this almost-null geodesic from the singularity towards larger values of the radial coordinate, until it meets the domain wall at $r= R_{\rm static}$ and time $T_o^{(1)}$ from the outside geometry.
By exploiting the fact that $u$ is constant along the almost-null geodesic (located in the exterior geometry), we obtain
\beq
T_o^{(1)} = u_1 + R^*_{o} \, , \qquad  \text{where} \quad
R^*_{o} \equiv r^*_o(R_{\rm static}) \, .
\label{eq:To1_staticbubble}
\eeq    
Next, we use the identity \eqref{eq:matching_genD} -- only valid in the case of a static bubble -- to obtain the time coordinate $T_i^{(1)}$ at the intersection with the domain wall on the interior side 
\beq
T_i^{(1)} = \bar{F} \, T_o^{(1)} \, , \qquad
\text{where} \quad  
\bar{F} \equiv  \sqrt{\frac{f_o(R_{\rm static})}{f_i(R_{\rm static})}} \, .
\label{eq:Ti1_staticbubble}
\eeq
The choice of the $+$ sign in \eqref{eq:matching_genD} follows from the fact that the orientation of the Killing vector needs to be the same on both sides of the domain wall, as we impose below.

\begin{figure}[ht]
    \centering
\subfigure[]{\includegraphics[scale=0.45]{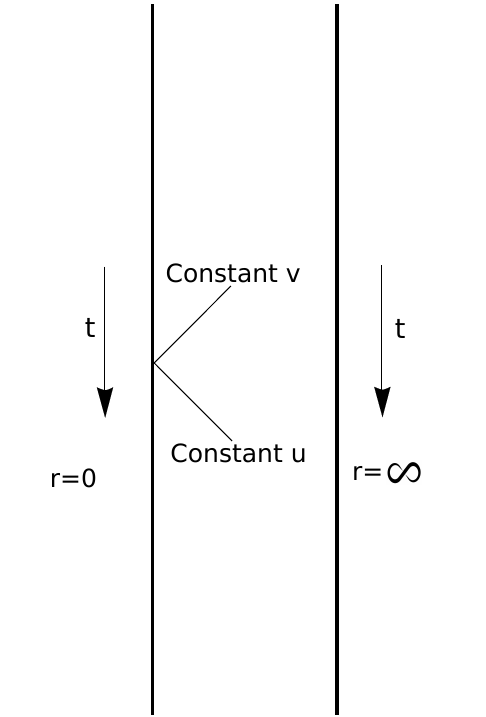}}\qquad \qquad
\subfigure[]{\includegraphics[scale=0.45]{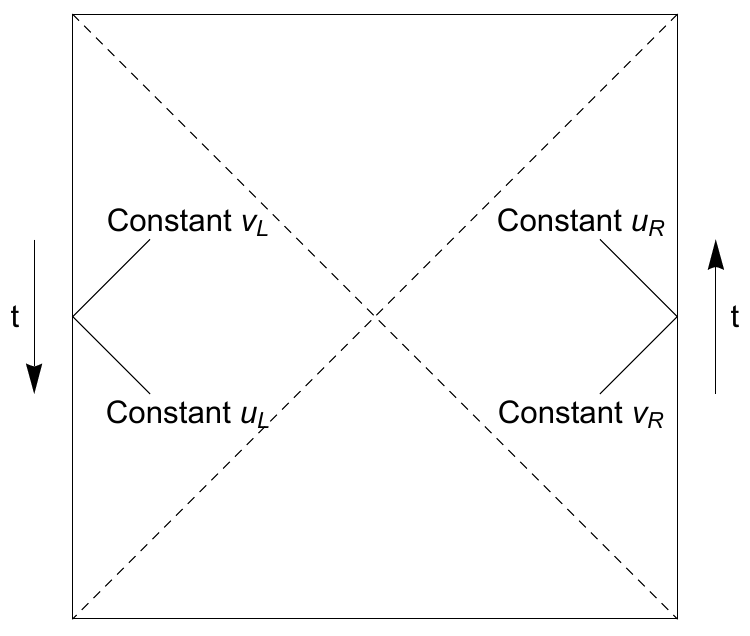}}
\\
\subfigure[]{\includegraphics[scale=0.45]{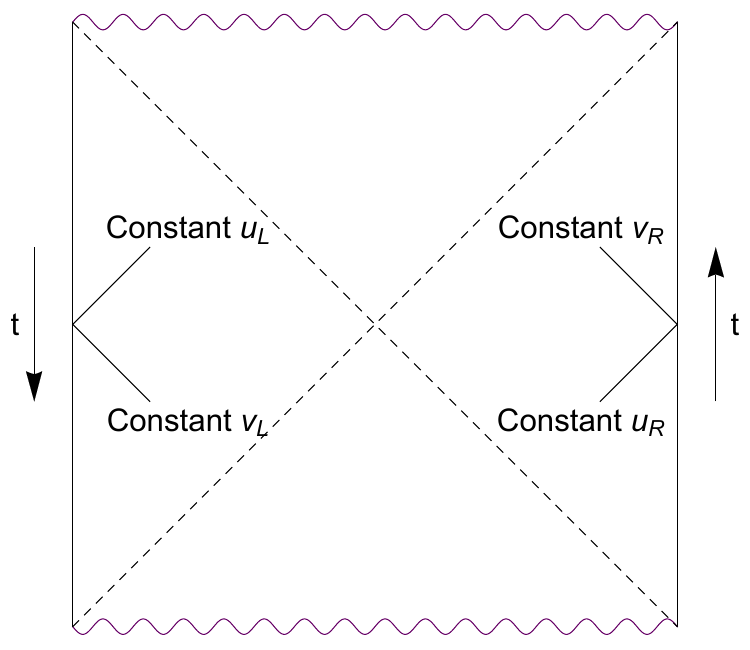}}
    \caption{Orientation of the null coordinates $u,v$ for an interior bubble geometry in the case of empty (a) AdS or (b) dS spacetime, and for the exterior AdS black hole solution (c).}
    \label{fig:orientation_null_coord}
\end{figure}

\paragraph{Step 2: entering the bubble.}
\rep{To determine the trajectory of the almost-null geodesic after it intersects the domain wall, we require that the bulk Killing vector $\partial_t$ is continuous across the domain wall.
By imposing this requirement, we find that the definition of the null coordinates $u,v$ in the interior geometry needs to be the same for 
both dS and AdS vacuum solutions.
From now on, we will work assuming that the null coordinates are oriented according to fig.~\ref{fig:orientation_null_coord}.}

As a consequence of this reasoning, the almost-null geodesic, that was at constant $u$ in the exterior geometry, is instead at constant $v$ in the interior. 
The value $v= v_2$ is obtained by evaluating the null coordinate
at $r=R_{\rm static}$ in the interior geometry. We find
\beq
v_2 = T_i^{(1)} + R_{i}^* = \bar{F} \le u_1 + R_{o}^*  \ri + R^*_{i} \, , \qquad  \text{where} \quad
R^*_{i} \equiv r^*_i(R_{\rm static}) \, ,
\eeq
where we used eqs.~\eqref{eq:To1_staticbubble}--\eqref{eq:Ti1_staticbubble} in the last step.
In general, this almost-null trajectory reaches a region of the interior geometry where it is reflected as a curve at constant $u=u_2$ after reaching $r=0$, corresponding to the center in the AdS case (or to the south pole in the dS case).
In both cases, the tortoise coordinate satisfies $r^*_i(0)=0$, therefore we can express the value of $u_2$ in a unified way as
\beq
u_2 = v_2 
\, .
\eeq
This almost-null geodesic meets the domain wall at the radial coordinate $r=R_{\rm static}$ with time coordinate $T_i^{(2)}$ such that
\beq
T_i^{(2)} = u_2 + R^*_i \, .
\eeq 
Using the identity~\eqref{eq:matching_genD}, we obtain the value of the time coordinate on the exterior side of the domain wall: 
\beq
T_o^{(2)}  = \frac{1}{\bar{F}} \, T_i^{(2)} \, .
\label{eq:To2_static}
\eeq

\paragraph{Step 3: coming back to the AdS boundary.}
The time coordinate $T_o^{(2)}$ in the external geometry allows us to identify the constant value $v=v_3$ of the null coordinate of the geodesic in the exterior geometry, after it crossed the domain wall: 
\beq
v_3 = T_o^{(2)} + R^*_o = u_1 + \frac{2}{\bar{F}} R_i^* + 2 R_o^* \, ,
\label{eq:v3_static}
\eeq
where we used eq.~\eqref{eq:To2_static} in the last step.
There is one final reflection at the BH singularity. This turns the trajectory to a curve at constant $u=u_3$, where
\beq
u_3 = v_3 - 2 r^*_o (0) \, .
\eeq
Using the definition of EF coordinate~\eqref{u-v-def}, together with the identities~\eqref{eq:u1_static} and \eqref{eq:v3_static}, we finally obtain the boundary time $t_{\rm fin}$ at which the geodesic intersects the AdS boundary again:
\beq
t_{\rm fin} = u_3 = 2 \le - 2 r^*_o (0) +R_o^* + \frac{1}{\bar{F}} R_i^* \ri \, .
\eeq
Computing $\Delta t = t_{\rm fin} - t_{\rm in}$, we get eq.~\eqref{eq:diff_times_static}.

\section{Vaidya limit}
\label{ssec:Vaidya}

In this appendix, we consider the limit where the bubble geometry becomes a Vaidya background (see, \eg Refs.~\cite{Vaidya:1999zz,Vaidya:1953zza,Poisson:2009pwt}). 
To achieve this setting, we need to consider $\l=-1$, such that the cosmological constant is the same (negative) value on both sides of the domain wall.
In this case, the mass of the collapsing bubble reads
\beq
m=m_+=-\kappa^2 R_0^d + 
2 \kappa R_0^{d-1} \sqrt{1+ R_0^2} \, .
\eeq
In the limit $\kappa \to 0$ and $R_0 \to \infty$, it is possible to tune the parameters such that the solution approaches a finite mass $m\approx  2 \kappa R_0^d$.
This special limit corresponds to the Vaidya geometry, which is dual to a global quench in the boundary CFT description. 
The dual CFT state is pure, since it can be obtained from the ground state of a CFT, which is later deformed by a quench in the Hamiltonian. 

We can a have a near-global-quench limit in the regimes of both configurations I and II of fig.~\ref{Pen-Dia-Collapsing-Bubble} (see also appendix~E of Ref.~\cite{Freivogel:2005qh}).
In terms of the mass $m_{\rm HP}=2$, corresponding to the Hawking-Page \cite{Hawking:1982dh} phase transition, we end up in configuration I when $m<m_{\rm HP}$, and in configuration II when $m>m_{\rm HP}$.

\begin{figure}[H]
    \centering
\subfigure{\label{fig:global_quench_AdS} \includegraphics[scale=0.18]{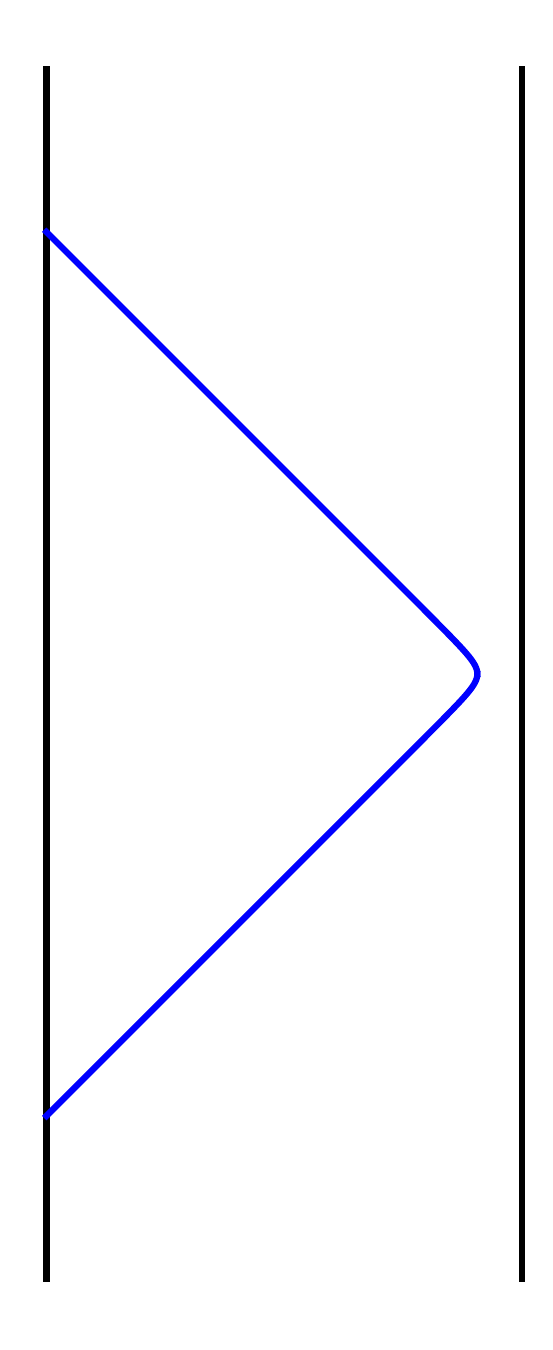}} \quad
\subfigure{ \label{fig:global_quench_AdS2} \includegraphics[scale=0.22]{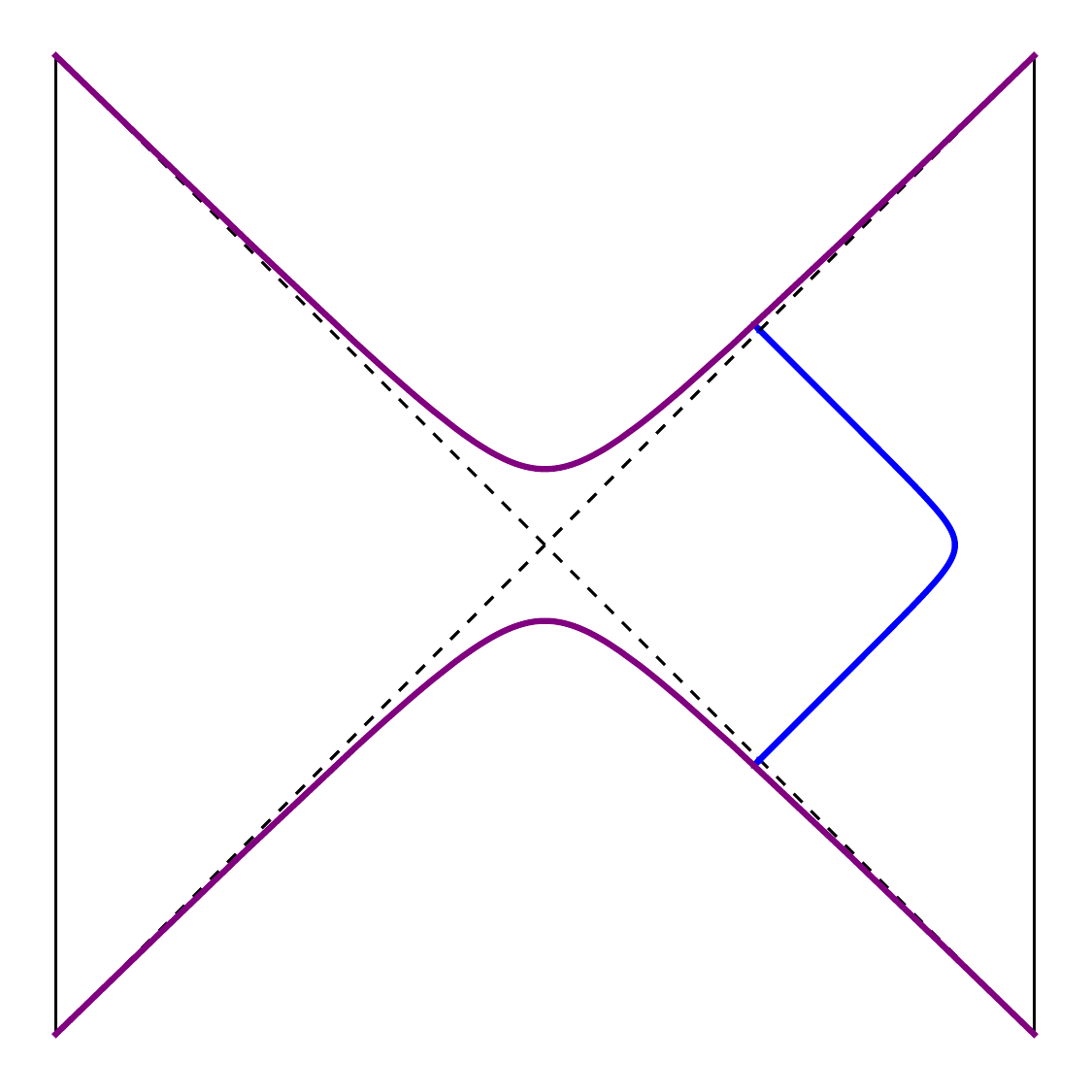}} \quad
\subfigure[]{ \label{Collapsing-Bubble-AdS-B-quench}   \includegraphics[scale=0.28]{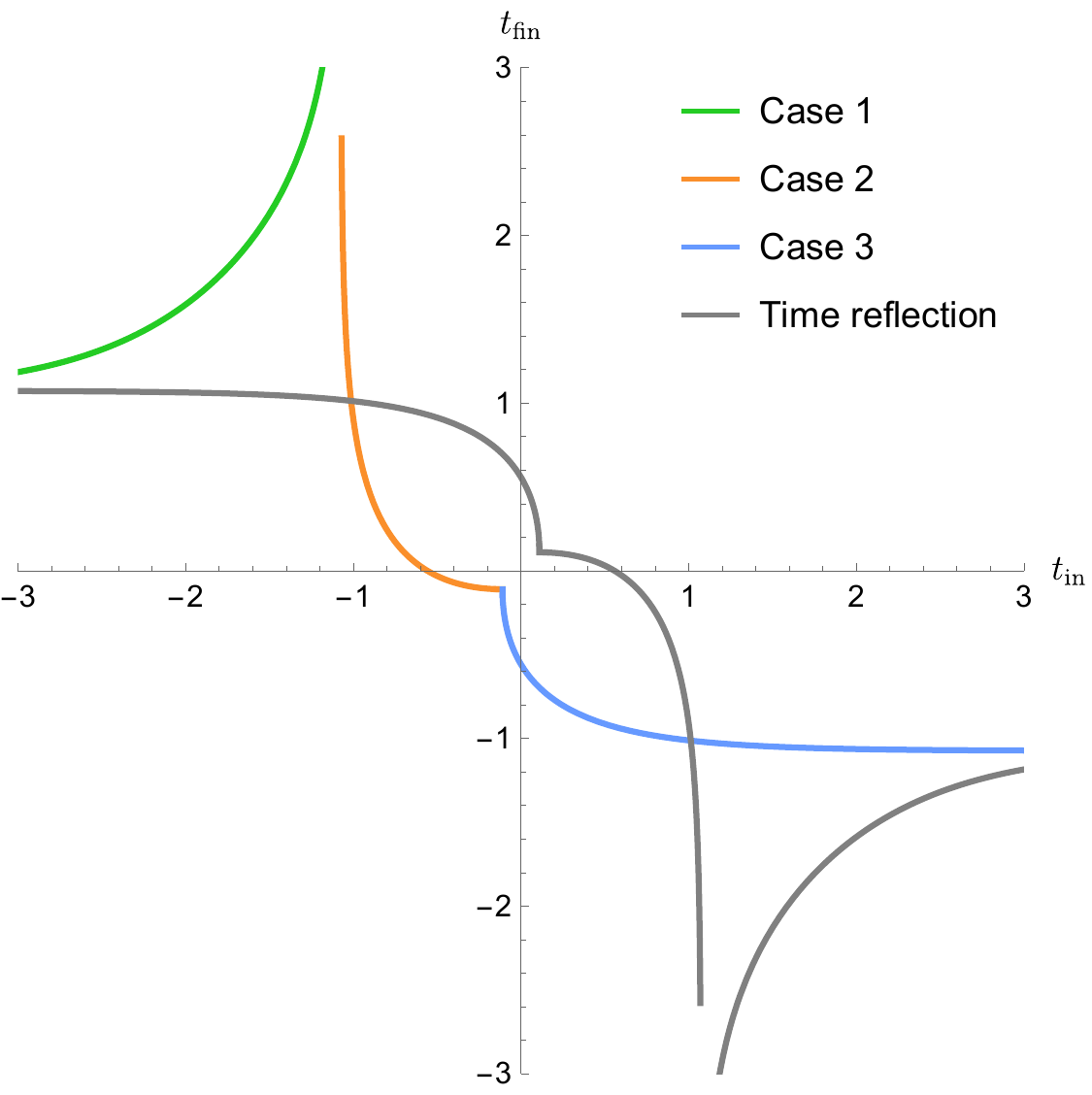}}
    \caption{Penrose diagram of an AdS collapsing bubble in the limit of a near-global-quench: 
    (a) interior AdS geometry, (b) exterior BH background.
    (c) Plot of $t_{\rm fin}$ as a function of $t_{\rm in}$ in the global quench limit.
The gray curve is obtained by applying the transformation~\eqref{eq:symm_time_rev} to the green, orange, and blue ones.
We fix $\kappa = 0.001, \l=-1, r_h=0.5$.}
\end{figure}

The Penrose diagram for an example involving configuration I is depicted in figs.~\ref{fig:global_quench_AdS}--\ref{fig:global_quench_AdS2}, while we collect in fig.~\ref{Collapsing-Bubble-AdS-B-quench} the dependence of $t_{\rm fin}$ on the initial boundary time.
The functional dependence of the blue curve is similar to the case -- referring to configuration I of fig.~\ref{Pen-Dia-Collapsing-Bubble} -- investigated in fig.~\ref{fig:very_small_AdS_a_cased}.
However, we notice that the kink (separating cases 2 and 3 in the evolution of radial almost-null geodesics) moved closer to the origin.
For an exact global quench (\eg see sections 5.1 and 5.3 of Ref.~\cite{Hubeny:2006yu}), we expect that the kink moves precisely at the origin.